\documentclass[preprint,nocopyrightspace]{sigplanconf}
\usepackage{fancyhdr}
\usepackage{hyperref}
\usepackage[utf8]{inputenc}
\usepackage{enumitem,algorithm}
\usepackage{color,graphicx}
\usepackage{amsmath,amssymb,eurosym,amsthm}

\setenumerate[0]{label=\textup{(\alph*)}}

\pdfglyphtounicode{A}{0041}
\pdfglyphtounicode{AE}{00C6}
\pdfglyphtounicode{AEacute}{01FC}
\pdfglyphtounicode{AEmacron}{01E2}
\pdfglyphtounicode{AEsmall}{00E6}
\pdfglyphtounicode{Aacute}{00C1}
\pdfglyphtounicode{Aacutesmall}{00E1}
\pdfglyphtounicode{Abreve}{0102}
\pdfglyphtounicode{Abreveacute}{1EAE}
\pdfglyphtounicode{Abrevecyrillic}{04D0}
\pdfglyphtounicode{Abrevedotbelow}{1EB6}
\pdfglyphtounicode{Abrevegrave}{1EB0}
\pdfglyphtounicode{Abrevehookabove}{1EB2}
\pdfglyphtounicode{Abrevetilde}{1EB4}
\pdfglyphtounicode{Acaron}{01CD}
\pdfglyphtounicode{Acircle}{24B6}
\pdfglyphtounicode{Acircumflex}{00C2}
\pdfglyphtounicode{Acircumflexacute}{1EA4}
\pdfglyphtounicode{Acircumflexdotbelow}{1EAC}
\pdfglyphtounicode{Acircumflexgrave}{1EA6}
\pdfglyphtounicode{Acircumflexhookabove}{1EA8}
\pdfglyphtounicode{Acircumflexsmall}{00E2}
\pdfglyphtounicode{Acircumflextilde}{1EAA}
\pdfglyphtounicode{Acute}{00B4}
\pdfglyphtounicode{Acutesmall}{00B4}
\pdfglyphtounicode{Acyrillic}{0410}
\pdfglyphtounicode{Adblgrave}{0200}
\pdfglyphtounicode{Adieresis}{00C4}
\pdfglyphtounicode{Adieresiscyrillic}{04D2}
\pdfglyphtounicode{Adieresismacron}{01DE}
\pdfglyphtounicode{Adieresissmall}{00E4}
\pdfglyphtounicode{Adotbelow}{1EA0}
\pdfglyphtounicode{Adotmacron}{01E0}
\pdfglyphtounicode{Agrave}{00C0}
\pdfglyphtounicode{Agravesmall}{00E0}
\pdfglyphtounicode{Ahookabove}{1EA2}
\pdfglyphtounicode{Aiecyrillic}{04D4}
\pdfglyphtounicode{Ainvertedbreve}{0202}
\pdfglyphtounicode{Alpha}{0391}
\pdfglyphtounicode{Alphatonos}{0386}
\pdfglyphtounicode{Amacron}{0100}
\pdfglyphtounicode{Amonospace}{FF21}
\pdfglyphtounicode{Aogonek}{0104}
\pdfglyphtounicode{Aring}{00C5}
\pdfglyphtounicode{Aringacute}{01FA}
\pdfglyphtounicode{Aringbelow}{1E00}
\pdfglyphtounicode{Aringsmall}{00E5}
\pdfglyphtounicode{Asmall}{0061}
\pdfglyphtounicode{Atilde}{00C3}
\pdfglyphtounicode{Atildesmall}{00E3}
\pdfglyphtounicode{Aybarmenian}{0531}
\pdfglyphtounicode{B}{0042}
\pdfglyphtounicode{Bcircle}{24B7}
\pdfglyphtounicode{Bdotaccent}{1E02}
\pdfglyphtounicode{Bdotbelow}{1E04}
\pdfglyphtounicode{Becyrillic}{0411}
\pdfglyphtounicode{Benarmenian}{0532}
\pdfglyphtounicode{Beta}{0392}
\pdfglyphtounicode{Bhook}{0181}
\pdfglyphtounicode{Blinebelow}{1E06}
\pdfglyphtounicode{Bmonospace}{FF22}
\pdfglyphtounicode{Brevesmall}{02D8}
\pdfglyphtounicode{Bsmall}{0062}
\pdfglyphtounicode{Btopbar}{0182}
\pdfglyphtounicode{C}{0043}
\pdfglyphtounicode{Caarmenian}{053E}
\pdfglyphtounicode{Cacute}{0106}
\pdfglyphtounicode{Caron}{02C7}
\pdfglyphtounicode{Caronsmall}{02C7}
\pdfglyphtounicode{Ccaron}{010C}
\pdfglyphtounicode{Ccedilla}{00C7}
\pdfglyphtounicode{Ccedillaacute}{1E08}
\pdfglyphtounicode{Ccedillasmall}{00E7}
\pdfglyphtounicode{Ccircle}{24B8}
\pdfglyphtounicode{Ccircumflex}{0108}
\pdfglyphtounicode{Cdot}{010A}
\pdfglyphtounicode{Cdotaccent}{010A}
\pdfglyphtounicode{Cedillasmall}{00B8}
\pdfglyphtounicode{Chaarmenian}{0549}
\pdfglyphtounicode{Cheabkhasiancyrillic}{04BC}
\pdfglyphtounicode{Checyrillic}{0427}
\pdfglyphtounicode{Chedescenderabkhasiancyrillic}{04BE}
\pdfglyphtounicode{Chedescendercyrillic}{04B6}
\pdfglyphtounicode{Chedieresiscyrillic}{04F4}
\pdfglyphtounicode{Cheharmenian}{0543}
\pdfglyphtounicode{Chekhakassiancyrillic}{04CB}
\pdfglyphtounicode{Cheverticalstrokecyrillic}{04B8}
\pdfglyphtounicode{Chi}{03A7}
\pdfglyphtounicode{Chook}{0187}
\pdfglyphtounicode{Circumflexsmall}{02C6}
\pdfglyphtounicode{Cmonospace}{FF23}
\pdfglyphtounicode{Coarmenian}{0551}
\pdfglyphtounicode{Csmall}{0063}
\pdfglyphtounicode{D}{0044}
\pdfglyphtounicode{DZ}{01F1}
\pdfglyphtounicode{DZcaron}{01C4}
\pdfglyphtounicode{Daarmenian}{0534}
\pdfglyphtounicode{Dafrican}{0189}
\pdfglyphtounicode{Dbar}{0110}
\pdfglyphtounicode{Dcaron}{010E}
\pdfglyphtounicode{Dcedilla}{1E10}
\pdfglyphtounicode{Dcircle}{24B9}
\pdfglyphtounicode{Dcircumflexbelow}{1E12}
\pdfglyphtounicode{Dcroat}{0110}
\pdfglyphtounicode{Ddotaccent}{1E0A}
\pdfglyphtounicode{Ddotbelow}{1E0C}
\pdfglyphtounicode{Decyrillic}{0414}
\pdfglyphtounicode{Deicoptic}{03EE}
\pdfglyphtounicode{Delta}{2206}
\pdfglyphtounicode{Deltagreek}{0394}
\pdfglyphtounicode{Dhook}{018A}
\pdfglyphtounicode{Dieresis}{00A8}
\pdfglyphtounicode{DieresisAcute}{F6CC}
\pdfglyphtounicode{DieresisGrave}{F6CD}
\pdfglyphtounicode{Dieresissmall}{00A8}
\pdfglyphtounicode{Digamma}{D875 DFCB}
\pdfglyphtounicode{Digammagreek}{03DC}
\pdfglyphtounicode{Djecyrillic}{0402}
\pdfglyphtounicode{Dlinebelow}{1E0E}
\pdfglyphtounicode{Dmonospace}{FF24}
\pdfglyphtounicode{Dotaccentsmall}{02D9}
\pdfglyphtounicode{Dslash}{0110}
\pdfglyphtounicode{Dsmall}{0064}
\pdfglyphtounicode{Dtopbar}{018B}
\pdfglyphtounicode{Dz}{01F2}
\pdfglyphtounicode{Dzcaron}{01C5}
\pdfglyphtounicode{Dzeabkhasiancyrillic}{04E0}
\pdfglyphtounicode{Dzecyrillic}{0405}
\pdfglyphtounicode{Dzhecyrillic}{040F}
\pdfglyphtounicode{E}{0045}
\pdfglyphtounicode{Eacute}{00C9}
\pdfglyphtounicode{Eacutesmall}{00E9}
\pdfglyphtounicode{Ebreve}{0114}
\pdfglyphtounicode{Ecaron}{011A}
\pdfglyphtounicode{Ecedillabreve}{1E1C}
\pdfglyphtounicode{Echarmenian}{0535}
\pdfglyphtounicode{Ecircle}{24BA}
\pdfglyphtounicode{Ecircumflex}{00CA}
\pdfglyphtounicode{Ecircumflexacute}{1EBE}
\pdfglyphtounicode{Ecircumflexbelow}{1E18}
\pdfglyphtounicode{Ecircumflexdotbelow}{1EC6}
\pdfglyphtounicode{Ecircumflexgrave}{1EC0}
\pdfglyphtounicode{Ecircumflexhookabove}{1EC2}
\pdfglyphtounicode{Ecircumflexsmall}{00EA}
\pdfglyphtounicode{Ecircumflextilde}{1EC4}
\pdfglyphtounicode{Ecyrillic}{0404}
\pdfglyphtounicode{Edblgrave}{0204}
\pdfglyphtounicode{Edieresis}{00CB}
\pdfglyphtounicode{Edieresissmall}{00EB}
\pdfglyphtounicode{Edot}{0116}
\pdfglyphtounicode{Edotaccent}{0116}
\pdfglyphtounicode{Edotbelow}{1EB8}
\pdfglyphtounicode{Efcyrillic}{0424}
\pdfglyphtounicode{Egrave}{00C8}
\pdfglyphtounicode{Egravesmall}{00E8}
\pdfglyphtounicode{Eharmenian}{0537}
\pdfglyphtounicode{Ehookabove}{1EBA}
\pdfglyphtounicode{Eightroman}{2167}
\pdfglyphtounicode{Einvertedbreve}{0206}
\pdfglyphtounicode{Eiotifiedcyrillic}{0464}
\pdfglyphtounicode{Elcyrillic}{041B}
\pdfglyphtounicode{Elevenroman}{216A}
\pdfglyphtounicode{Emacron}{0112}
\pdfglyphtounicode{Emacronacute}{1E16}
\pdfglyphtounicode{Emacrongrave}{1E14}
\pdfglyphtounicode{Emcyrillic}{041C}
\pdfglyphtounicode{Emonospace}{FF25}
\pdfglyphtounicode{Encyrillic}{041D}
\pdfglyphtounicode{Endescendercyrillic}{04A2}
\pdfglyphtounicode{Eng}{014A}
\pdfglyphtounicode{Enghecyrillic}{04A4}
\pdfglyphtounicode{Enhookcyrillic}{04C7}
\pdfglyphtounicode{Eogonek}{0118}
\pdfglyphtounicode{Eopen}{0190}
\pdfglyphtounicode{Epsilon}{0395}
\pdfglyphtounicode{Epsilontonos}{0388}
\pdfglyphtounicode{Ercyrillic}{0420}
\pdfglyphtounicode{Ereversed}{018E}
\pdfglyphtounicode{Ereversedcyrillic}{042D}
\pdfglyphtounicode{Escyrillic}{0421}
\pdfglyphtounicode{Esdescendercyrillic}{04AA}
\pdfglyphtounicode{Esh}{01A9}
\pdfglyphtounicode{Esmall}{0065}
\pdfglyphtounicode{Eta}{0397}
\pdfglyphtounicode{Etarmenian}{0538}
\pdfglyphtounicode{Etatonos}{0389}
\pdfglyphtounicode{Eth}{00D0}
\pdfglyphtounicode{Ethsmall}{00F0}
\pdfglyphtounicode{Etilde}{1EBC}
\pdfglyphtounicode{Etildebelow}{1E1A}
\pdfglyphtounicode{Euro}{20AC}
\pdfglyphtounicode{Ezh}{01B7}
\pdfglyphtounicode{Ezhcaron}{01EE}
\pdfglyphtounicode{Ezhreversed}{01B8}
\pdfglyphtounicode{F}{0046}
\pdfglyphtounicode{FFIsmall}{0066 0066 0069}
\pdfglyphtounicode{FFLsmall}{0066 0066 006C}
\pdfglyphtounicode{FFsmall}{0066 0066}
\pdfglyphtounicode{FIsmall}{0066 0069}
\pdfglyphtounicode{FLsmall}{0066 006C}
\pdfglyphtounicode{Fcircle}{24BB}
\pdfglyphtounicode{Fdotaccent}{1E1E}
\pdfglyphtounicode{Feharmenian}{0556}
\pdfglyphtounicode{Feicoptic}{03E4}
\pdfglyphtounicode{Fhook}{0191}
\pdfglyphtounicode{Finv}{2132}
\pdfglyphtounicode{Fitacyrillic}{0472}
\pdfglyphtounicode{Fiveroman}{2164}
\pdfglyphtounicode{Fmonospace}{FF26}
\pdfglyphtounicode{Fourroman}{2163}
\pdfglyphtounicode{Fsmall}{0066}
\pdfglyphtounicode{G}{0047}
\pdfglyphtounicode{GBsquare}{3387}
\pdfglyphtounicode{Gacute}{01F4}
\pdfglyphtounicode{Gamma}{0393}
\pdfglyphtounicode{Gammaafrican}{0194}
\pdfglyphtounicode{Gangiacoptic}{03EA}
\pdfglyphtounicode{Gbreve}{011E}
\pdfglyphtounicode{Gcaron}{01E6}
\pdfglyphtounicode{Gcedilla}{0122}
\pdfglyphtounicode{Gcircle}{24BC}
\pdfglyphtounicode{Gcircumflex}{011C}
\pdfglyphtounicode{Gcommaaccent}{0122}
\pdfglyphtounicode{Gdot}{0120}
\pdfglyphtounicode{Gdotaccent}{0120}
\pdfglyphtounicode{Gecyrillic}{0413}
\pdfglyphtounicode{Germandbls}{0053 0053}
\pdfglyphtounicode{Germandblssmall}{0073 0073}
\pdfglyphtounicode{Ghadarmenian}{0542}
\pdfglyphtounicode{Ghemiddlehookcyrillic}{0494}
\pdfglyphtounicode{Ghestrokecyrillic}{0492}
\pdfglyphtounicode{Gheupturncyrillic}{0490}
\pdfglyphtounicode{Ghook}{0193}
\pdfglyphtounicode{Gimarmenian}{0533}
\pdfglyphtounicode{Gjecyrillic}{0403}
\pdfglyphtounicode{Gmacron}{1E20}
\pdfglyphtounicode{Gmir}{2141}
\pdfglyphtounicode{Gmonospace}{FF27}
\pdfglyphtounicode{Grave}{0060}
\pdfglyphtounicode{Gravesmall}{0060}
\pdfglyphtounicode{Gsmall}{0067}
\pdfglyphtounicode{Gsmallhook}{029B}
\pdfglyphtounicode{Gstroke}{01E4}
\pdfglyphtounicode{H}{0048}
\pdfglyphtounicode{H18533}{25CF}
\pdfglyphtounicode{H18543}{25AA}
\pdfglyphtounicode{H18551}{25AB}
\pdfglyphtounicode{H22073}{25A1}
\pdfglyphtounicode{HPsquare}{33CB}
\pdfglyphtounicode{Haabkhasiancyrillic}{04A8}
\pdfglyphtounicode{Hadescendercyrillic}{04B2}
\pdfglyphtounicode{Hardsigncyrillic}{042A}
\pdfglyphtounicode{Hbar}{0126}
\pdfglyphtounicode{Hbrevebelow}{1E2A}
\pdfglyphtounicode{Hcedilla}{1E28}
\pdfglyphtounicode{Hcircle}{24BD}
\pdfglyphtounicode{Hcircumflex}{0124}
\pdfglyphtounicode{Hdieresis}{1E26}
\pdfglyphtounicode{Hdotaccent}{1E22}
\pdfglyphtounicode{Hdotbelow}{1E24}
\pdfglyphtounicode{Hmonospace}{FF28}
\pdfglyphtounicode{Hoarmenian}{0540}
\pdfglyphtounicode{Horicoptic}{03E8}
\pdfglyphtounicode{Hsmall}{0068}
\pdfglyphtounicode{Hungarumlaut}{02DD}
\pdfglyphtounicode{Hungarumlautsmall}{02DD}
\pdfglyphtounicode{Hzsquare}{3390}
\pdfglyphtounicode{I}{0049}
\pdfglyphtounicode{IAcyrillic}{042F}
\pdfglyphtounicode{IJ}{0132}
\pdfglyphtounicode{IUcyrillic}{042E}
\pdfglyphtounicode{Iacute}{00CD}
\pdfglyphtounicode{Iacutesmall}{00ED}
\pdfglyphtounicode{Ibreve}{012C}
\pdfglyphtounicode{Icaron}{01CF}
\pdfglyphtounicode{Icircle}{24BE}
\pdfglyphtounicode{Icircumflex}{00CE}
\pdfglyphtounicode{Icircumflexsmall}{00EE}
\pdfglyphtounicode{Icyrillic}{0406}
\pdfglyphtounicode{Idblgrave}{0208}
\pdfglyphtounicode{Idieresis}{00CF}
\pdfglyphtounicode{Idieresisacute}{1E2E}
\pdfglyphtounicode{Idieresiscyrillic}{04E4}
\pdfglyphtounicode{Idieresissmall}{00EF}
\pdfglyphtounicode{Idot}{0130}
\pdfglyphtounicode{Idotaccent}{0130}
\pdfglyphtounicode{Idotbelow}{1ECA}
\pdfglyphtounicode{Iebrevecyrillic}{04D6}
\pdfglyphtounicode{Iecyrillic}{0415}
\pdfglyphtounicode{Ifractur}{2111}
\pdfglyphtounicode{Ifraktur}{2111}
\pdfglyphtounicode{Igrave}{00CC}
\pdfglyphtounicode{Igravesmall}{00EC}
\pdfglyphtounicode{Ihookabove}{1EC8}
\pdfglyphtounicode{Iicyrillic}{0418}
\pdfglyphtounicode{Iinvertedbreve}{020A}
\pdfglyphtounicode{Iishortcyrillic}{0419}
\pdfglyphtounicode{Imacron}{012A}
\pdfglyphtounicode{Imacroncyrillic}{04E2}
\pdfglyphtounicode{Imonospace}{FF29}
\pdfglyphtounicode{Iniarmenian}{053B}
\pdfglyphtounicode{Iocyrillic}{0401}
\pdfglyphtounicode{Iogonek}{012E}
\pdfglyphtounicode{Iota}{0399}
\pdfglyphtounicode{Iotaafrican}{0196}
\pdfglyphtounicode{Iotadieresis}{03AA}
\pdfglyphtounicode{Iotatonos}{038A}
\pdfglyphtounicode{Ismall}{0069}
\pdfglyphtounicode{Istroke}{0197}
\pdfglyphtounicode{Itilde}{0128}
\pdfglyphtounicode{Itildebelow}{1E2C}
\pdfglyphtounicode{Izhitsacyrillic}{0474}
\pdfglyphtounicode{Izhitsadblgravecyrillic}{0476}
\pdfglyphtounicode{J}{004A}
\pdfglyphtounicode{Jaarmenian}{0541}
\pdfglyphtounicode{Jcircle}{24BF}
\pdfglyphtounicode{Jcircumflex}{0134}
\pdfglyphtounicode{Jecyrillic}{0408}
\pdfglyphtounicode{Jheharmenian}{054B}
\pdfglyphtounicode{Jmonospace}{FF2A}
\pdfglyphtounicode{Jsmall}{006A}
\pdfglyphtounicode{K}{004B}
\pdfglyphtounicode{KBsquare}{3385}
\pdfglyphtounicode{KKsquare}{33CD}
\pdfglyphtounicode{Kabashkircyrillic}{04A0}
\pdfglyphtounicode{Kacute}{1E30}
\pdfglyphtounicode{Kacyrillic}{041A}
\pdfglyphtounicode{Kadescendercyrillic}{049A}
\pdfglyphtounicode{Kahookcyrillic}{04C3}
\pdfglyphtounicode{Kappa}{039A}
\pdfglyphtounicode{Kastrokecyrillic}{049E}
\pdfglyphtounicode{Kaverticalstrokecyrillic}{049C}
\pdfglyphtounicode{Kcaron}{01E8}
\pdfglyphtounicode{Kcedilla}{0136}
\pdfglyphtounicode{Kcircle}{24C0}
\pdfglyphtounicode{Kcommaaccent}{0136}
\pdfglyphtounicode{Kdotbelow}{1E32}
\pdfglyphtounicode{Keharmenian}{0554}
\pdfglyphtounicode{Kenarmenian}{053F}
\pdfglyphtounicode{Khacyrillic}{0425}
\pdfglyphtounicode{Kheicoptic}{03E6}
\pdfglyphtounicode{Khook}{0198}
\pdfglyphtounicode{Kjecyrillic}{040C}
\pdfglyphtounicode{Klinebelow}{1E34}
\pdfglyphtounicode{Kmonospace}{FF2B}
\pdfglyphtounicode{Koppacyrillic}{0480}
\pdfglyphtounicode{Koppagreek}{03DE}
\pdfglyphtounicode{Ksicyrillic}{046E}
\pdfglyphtounicode{Ksmall}{006B}
\pdfglyphtounicode{L}{004C}
\pdfglyphtounicode{LJ}{01C7}
\pdfglyphtounicode{LL}{004C 004C}
\pdfglyphtounicode{Lacute}{0139}
\pdfglyphtounicode{Lambda}{039B}
\pdfglyphtounicode{Lcaron}{013D}
\pdfglyphtounicode{Lcedilla}{013B}
\pdfglyphtounicode{Lcircle}{24C1}
\pdfglyphtounicode{Lcircumflexbelow}{1E3C}
\pdfglyphtounicode{Lcommaaccent}{013B}
\pdfglyphtounicode{Ldot}{013F}
\pdfglyphtounicode{Ldotaccent}{013F}
\pdfglyphtounicode{Ldotbelow}{1E36}
\pdfglyphtounicode{Ldotbelowmacron}{1E38}
\pdfglyphtounicode{Liwnarmenian}{053C}
\pdfglyphtounicode{Lj}{01C8}
\pdfglyphtounicode{Ljecyrillic}{0409}
\pdfglyphtounicode{Llinebelow}{1E3A}
\pdfglyphtounicode{Lmonospace}{FF2C}
\pdfglyphtounicode{Lslash}{0141}
\pdfglyphtounicode{Lslashsmall}{0142}
\pdfglyphtounicode{Lsmall}{006C}
\pdfglyphtounicode{M}{004D}
\pdfglyphtounicode{MBsquare}{3386}
\pdfglyphtounicode{Macron}{00AF}
\pdfglyphtounicode{Macronsmall}{00AF}
\pdfglyphtounicode{Macute}{1E3E}
\pdfglyphtounicode{Mcircle}{24C2}
\pdfglyphtounicode{Mdotaccent}{1E40}
\pdfglyphtounicode{Mdotbelow}{1E42}
\pdfglyphtounicode{Menarmenian}{0544}
\pdfglyphtounicode{Mmonospace}{FF2D}
\pdfglyphtounicode{Msmall}{006D}
\pdfglyphtounicode{Mturned}{019C}
\pdfglyphtounicode{Mu}{039C}
\pdfglyphtounicode{N}{004E}
\pdfglyphtounicode{NJ}{01CA}
\pdfglyphtounicode{Nacute}{0143}
\pdfglyphtounicode{Ncaron}{0147}
\pdfglyphtounicode{Ncedilla}{0145}
\pdfglyphtounicode{Ncircle}{24C3}
\pdfglyphtounicode{Ncircumflexbelow}{1E4A}
\pdfglyphtounicode{Ncommaaccent}{0145}
\pdfglyphtounicode{Ndotaccent}{1E44}
\pdfglyphtounicode{Ndotbelow}{1E46}
\pdfglyphtounicode{Ng}{014A}
\pdfglyphtounicode{Nhookleft}{019D}
\pdfglyphtounicode{Nineroman}{2168}
\pdfglyphtounicode{Nj}{01CB}
\pdfglyphtounicode{Njecyrillic}{040A}
\pdfglyphtounicode{Nlinebelow}{1E48}
\pdfglyphtounicode{Nmonospace}{FF2E}
\pdfglyphtounicode{Nowarmenian}{0546}
\pdfglyphtounicode{Nsmall}{006E}
\pdfglyphtounicode{Ntilde}{00D1}
\pdfglyphtounicode{Ntildesmall}{00F1}
\pdfglyphtounicode{Nu}{039D}
\pdfglyphtounicode{O}{004F}
\pdfglyphtounicode{OE}{0152}
\pdfglyphtounicode{OEsmall}{0153}
\pdfglyphtounicode{Oacute}{00D3}
\pdfglyphtounicode{Oacutesmall}{00F3}
\pdfglyphtounicode{Obarredcyrillic}{04E8}
\pdfglyphtounicode{Obarreddieresiscyrillic}{04EA}
\pdfglyphtounicode{Obreve}{014E}
\pdfglyphtounicode{Ocaron}{01D1}
\pdfglyphtounicode{Ocenteredtilde}{019F}
\pdfglyphtounicode{Ocircle}{24C4}
\pdfglyphtounicode{Ocircumflex}{00D4}
\pdfglyphtounicode{Ocircumflexacute}{1ED0}
\pdfglyphtounicode{Ocircumflexdotbelow}{1ED8}
\pdfglyphtounicode{Ocircumflexgrave}{1ED2}
\pdfglyphtounicode{Ocircumflexhookabove}{1ED4}
\pdfglyphtounicode{Ocircumflexsmall}{00F4}
\pdfglyphtounicode{Ocircumflextilde}{1ED6}
\pdfglyphtounicode{Ocyrillic}{041E}
\pdfglyphtounicode{Odblacute}{0150}
\pdfglyphtounicode{Odblgrave}{020C}
\pdfglyphtounicode{Odieresis}{00D6}
\pdfglyphtounicode{Odieresiscyrillic}{04E6}
\pdfglyphtounicode{Odieresissmall}{00F6}
\pdfglyphtounicode{Odotbelow}{1ECC}
\pdfglyphtounicode{Ogoneksmall}{02DB}
\pdfglyphtounicode{Ograve}{00D2}
\pdfglyphtounicode{Ogravesmall}{00F2}
\pdfglyphtounicode{Oharmenian}{0555}
\pdfglyphtounicode{Ohm}{2126}
\pdfglyphtounicode{Ohookabove}{1ECE}
\pdfglyphtounicode{Ohorn}{01A0}
\pdfglyphtounicode{Ohornacute}{1EDA}
\pdfglyphtounicode{Ohorndotbelow}{1EE2}
\pdfglyphtounicode{Ohorngrave}{1EDC}
\pdfglyphtounicode{Ohornhookabove}{1EDE}
\pdfglyphtounicode{Ohorntilde}{1EE0}
\pdfglyphtounicode{Ohungarumlaut}{0150}
\pdfglyphtounicode{Oi}{01A2}
\pdfglyphtounicode{Oinvertedbreve}{020E}
\pdfglyphtounicode{Omacron}{014C}
\pdfglyphtounicode{Omacronacute}{1E52}
\pdfglyphtounicode{Omacrongrave}{1E50}
\pdfglyphtounicode{Omega}{2126}
\pdfglyphtounicode{Omegacyrillic}{0460}
\pdfglyphtounicode{Omegagreek}{03A9}
\pdfglyphtounicode{Omegainv}{2127}
\pdfglyphtounicode{Omegaroundcyrillic}{047A}
\pdfglyphtounicode{Omegatitlocyrillic}{047C}
\pdfglyphtounicode{Omegatonos}{038F}
\pdfglyphtounicode{Omicron}{039F}
\pdfglyphtounicode{Omicrontonos}{038C}
\pdfglyphtounicode{Omonospace}{FF2F}
\pdfglyphtounicode{Oneroman}{2160}
\pdfglyphtounicode{Oogonek}{01EA}
\pdfglyphtounicode{Oogonekmacron}{01EC}
\pdfglyphtounicode{Oopen}{0186}
\pdfglyphtounicode{Oslash}{00D8}
\pdfglyphtounicode{Oslashacute}{01FE}
\pdfglyphtounicode{Oslashsmall}{00F8}
\pdfglyphtounicode{Osmall}{006F}
\pdfglyphtounicode{Ostrokeacute}{01FE}
\pdfglyphtounicode{Otcyrillic}{047E}
\pdfglyphtounicode{Otilde}{00D5}
\pdfglyphtounicode{Otildeacute}{1E4C}
\pdfglyphtounicode{Otildedieresis}{1E4E}
\pdfglyphtounicode{Otildesmall}{00F5}
\pdfglyphtounicode{P}{0050}
\pdfglyphtounicode{Pacute}{1E54}
\pdfglyphtounicode{Pcircle}{24C5}
\pdfglyphtounicode{Pdotaccent}{1E56}
\pdfglyphtounicode{Pecyrillic}{041F}
\pdfglyphtounicode{Peharmenian}{054A}
\pdfglyphtounicode{Pemiddlehookcyrillic}{04A6}
\pdfglyphtounicode{Phi}{03A6}
\pdfglyphtounicode{Phook}{01A4}
\pdfglyphtounicode{Pi}{03A0}
\pdfglyphtounicode{Piwrarmenian}{0553}
\pdfglyphtounicode{Pmonospace}{FF30}
\pdfglyphtounicode{Psi}{03A8}
\pdfglyphtounicode{Psicyrillic}{0470}
\pdfglyphtounicode{Psmall}{0070}
\pdfglyphtounicode{Q}{0051}
\pdfglyphtounicode{Qcircle}{24C6}
\pdfglyphtounicode{Qmonospace}{FF31}
\pdfglyphtounicode{Qsmall}{0071}
\pdfglyphtounicode{R}{0052}
\pdfglyphtounicode{Raarmenian}{054C}
\pdfglyphtounicode{Racute}{0154}
\pdfglyphtounicode{Rcaron}{0158}
\pdfglyphtounicode{Rcedilla}{0156}
\pdfglyphtounicode{Rcircle}{24C7}
\pdfglyphtounicode{Rcommaaccent}{0156}
\pdfglyphtounicode{Rdblgrave}{0210}
\pdfglyphtounicode{Rdotaccent}{1E58}
\pdfglyphtounicode{Rdotbelow}{1E5A}
\pdfglyphtounicode{Rdotbelowmacron}{1E5C}
\pdfglyphtounicode{Reharmenian}{0550}
\pdfglyphtounicode{Rfractur}{211C}
\pdfglyphtounicode{Rfraktur}{211C}
\pdfglyphtounicode{Rho}{03A1}
\pdfglyphtounicode{Ringsmall}{02DA}
\pdfglyphtounicode{Rinvertedbreve}{0212}
\pdfglyphtounicode{Rlinebelow}{1E5E}
\pdfglyphtounicode{Rmonospace}{FF32}
\pdfglyphtounicode{Rsmall}{0072}
\pdfglyphtounicode{Rsmallinverted}{0281}
\pdfglyphtounicode{Rsmallinvertedsuperior}{02B6}
\pdfglyphtounicode{S}{0053}
\pdfglyphtounicode{SF010000}{250C}
\pdfglyphtounicode{SF020000}{2514}
\pdfglyphtounicode{SF030000}{2510}
\pdfglyphtounicode{SF040000}{2518}
\pdfglyphtounicode{SF050000}{253C}
\pdfglyphtounicode{SF060000}{252C}
\pdfglyphtounicode{SF070000}{2534}
\pdfglyphtounicode{SF080000}{251C}
\pdfglyphtounicode{SF090000}{2524}
\pdfglyphtounicode{SF100000}{2500}
\pdfglyphtounicode{SF110000}{2502}
\pdfglyphtounicode{SF190000}{2561}
\pdfglyphtounicode{SF200000}{2562}
\pdfglyphtounicode{SF210000}{2556}
\pdfglyphtounicode{SF220000}{2555}
\pdfglyphtounicode{SF230000}{2563}
\pdfglyphtounicode{SF240000}{2551}
\pdfglyphtounicode{SF250000}{2557}
\pdfglyphtounicode{SF260000}{255D}
\pdfglyphtounicode{SF270000}{255C}
\pdfglyphtounicode{SF280000}{255B}
\pdfglyphtounicode{SF360000}{255E}
\pdfglyphtounicode{SF370000}{255F}
\pdfglyphtounicode{SF380000}{255A}
\pdfglyphtounicode{SF390000}{2554}
\pdfglyphtounicode{SF400000}{2569}
\pdfglyphtounicode{SF410000}{2566}
\pdfglyphtounicode{SF420000}{2560}
\pdfglyphtounicode{SF430000}{2550}
\pdfglyphtounicode{SF440000}{256C}
\pdfglyphtounicode{SF450000}{2567}
\pdfglyphtounicode{SF460000}{2568}
\pdfglyphtounicode{SF470000}{2564}
\pdfglyphtounicode{SF480000}{2565}
\pdfglyphtounicode{SF490000}{2559}
\pdfglyphtounicode{SF500000}{2558}
\pdfglyphtounicode{SF510000}{2552}
\pdfglyphtounicode{SF520000}{2553}
\pdfglyphtounicode{SF530000}{256B}
\pdfglyphtounicode{SF540000}{256A}
\pdfglyphtounicode{SS}{0053 0053}
\pdfglyphtounicode{SSsmall}{0073 0073}
\pdfglyphtounicode{Sacute}{015A}
\pdfglyphtounicode{Sacutedotaccent}{1E64}
\pdfglyphtounicode{Sampigreek}{03E0}
\pdfglyphtounicode{Scaron}{0160}
\pdfglyphtounicode{Scarondotaccent}{1E66}
\pdfglyphtounicode{Scaronsmall}{0161}
\pdfglyphtounicode{Scedilla}{015E}
\pdfglyphtounicode{Schwa}{018F}
\pdfglyphtounicode{Schwacyrillic}{04D8}
\pdfglyphtounicode{Schwadieresiscyrillic}{04DA}
\pdfglyphtounicode{Scircle}{24C8}
\pdfglyphtounicode{Scircumflex}{015C}
\pdfglyphtounicode{Scommaaccent}{0218}
\pdfglyphtounicode{Sdotaccent}{1E60}
\pdfglyphtounicode{Sdotbelow}{1E62}
\pdfglyphtounicode{Sdotbelowdotaccent}{1E68}
\pdfglyphtounicode{Seharmenian}{054D}
\pdfglyphtounicode{Sevenroman}{2166}
\pdfglyphtounicode{Shaarmenian}{0547}
\pdfglyphtounicode{Shacyrillic}{0428}
\pdfglyphtounicode{Shchacyrillic}{0429}
\pdfglyphtounicode{Sheicoptic}{03E2}
\pdfglyphtounicode{Shhacyrillic}{04BA}
\pdfglyphtounicode{Shimacoptic}{03EC}
\pdfglyphtounicode{Sigma}{03A3}
\pdfglyphtounicode{Sixroman}{2165}
\pdfglyphtounicode{Smonospace}{FF33}
\pdfglyphtounicode{Softsigncyrillic}{042C}
\pdfglyphtounicode{Ssmall}{0073}
\pdfglyphtounicode{Stigmagreek}{03DA}
\pdfglyphtounicode{T}{0054}
\pdfglyphtounicode{Tau}{03A4}
\pdfglyphtounicode{Tbar}{0166}
\pdfglyphtounicode{Tcaron}{0164}
\pdfglyphtounicode{Tcedilla}{0162}
\pdfglyphtounicode{Tcircle}{24C9}
\pdfglyphtounicode{Tcircumflexbelow}{1E70}
\pdfglyphtounicode{Tcommaaccent}{0162}
\pdfglyphtounicode{Tdotaccent}{1E6A}
\pdfglyphtounicode{Tdotbelow}{1E6C}
\pdfglyphtounicode{Tecyrillic}{0422}
\pdfglyphtounicode{Tedescendercyrillic}{04AC}
\pdfglyphtounicode{Tenroman}{2169}
\pdfglyphtounicode{Tetsecyrillic}{04B4}
\pdfglyphtounicode{Theta}{0398}
\pdfglyphtounicode{Thook}{01AC}
\pdfglyphtounicode{Thorn}{00DE}
\pdfglyphtounicode{Thornsmall}{00FE}
\pdfglyphtounicode{Threeroman}{2162}
\pdfglyphtounicode{Tildesmall}{02DC}
\pdfglyphtounicode{Tiwnarmenian}{054F}
\pdfglyphtounicode{Tlinebelow}{1E6E}
\pdfglyphtounicode{Tmonospace}{FF34}
\pdfglyphtounicode{Toarmenian}{0539}
\pdfglyphtounicode{Tonefive}{01BC}
\pdfglyphtounicode{Tonesix}{0184}
\pdfglyphtounicode{Tonetwo}{01A7}
\pdfglyphtounicode{Tretroflexhook}{01AE}
\pdfglyphtounicode{Tsecyrillic}{0426}
\pdfglyphtounicode{Tshecyrillic}{040B}
\pdfglyphtounicode{Tsmall}{0074}
\pdfglyphtounicode{Twelveroman}{216B}
\pdfglyphtounicode{Tworoman}{2161}
\pdfglyphtounicode{U}{0055}
\pdfglyphtounicode{Uacute}{00DA}
\pdfglyphtounicode{Uacutesmall}{00FA}
\pdfglyphtounicode{Ubreve}{016C}
\pdfglyphtounicode{Ucaron}{01D3}
\pdfglyphtounicode{Ucircle}{24CA}
\pdfglyphtounicode{Ucircumflex}{00DB}
\pdfglyphtounicode{Ucircumflexbelow}{1E76}
\pdfglyphtounicode{Ucircumflexsmall}{00FB}
\pdfglyphtounicode{Ucyrillic}{0423}
\pdfglyphtounicode{Udblacute}{0170}
\pdfglyphtounicode{Udblgrave}{0214}
\pdfglyphtounicode{Udieresis}{00DC}
\pdfglyphtounicode{Udieresisacute}{01D7}
\pdfglyphtounicode{Udieresisbelow}{1E72}
\pdfglyphtounicode{Udieresiscaron}{01D9}
\pdfglyphtounicode{Udieresiscyrillic}{04F0}
\pdfglyphtounicode{Udieresisgrave}{01DB}
\pdfglyphtounicode{Udieresismacron}{01D5}
\pdfglyphtounicode{Udieresissmall}{00FC}
\pdfglyphtounicode{Udotbelow}{1EE4}
\pdfglyphtounicode{Ugrave}{00D9}
\pdfglyphtounicode{Ugravesmall}{00F9}
\pdfglyphtounicode{Uhookabove}{1EE6}
\pdfglyphtounicode{Uhorn}{01AF}
\pdfglyphtounicode{Uhornacute}{1EE8}
\pdfglyphtounicode{Uhorndotbelow}{1EF0}
\pdfglyphtounicode{Uhorngrave}{1EEA}
\pdfglyphtounicode{Uhornhookabove}{1EEC}
\pdfglyphtounicode{Uhorntilde}{1EEE}
\pdfglyphtounicode{Uhungarumlaut}{0170}
\pdfglyphtounicode{Uhungarumlautcyrillic}{04F2}
\pdfglyphtounicode{Uinvertedbreve}{0216}
\pdfglyphtounicode{Ukcyrillic}{0478}
\pdfglyphtounicode{Umacron}{016A}
\pdfglyphtounicode{Umacroncyrillic}{04EE}
\pdfglyphtounicode{Umacrondieresis}{1E7A}
\pdfglyphtounicode{Umonospace}{FF35}
\pdfglyphtounicode{Uogonek}{0172}
\pdfglyphtounicode{Upsilon}{03A5}
\pdfglyphtounicode{Upsilon1}{03D2}
\pdfglyphtounicode{Upsilonacutehooksymbolgreek}{03D3}
\pdfglyphtounicode{Upsilonafrican}{01B1}
\pdfglyphtounicode{Upsilondieresis}{03AB}
\pdfglyphtounicode{Upsilondieresishooksymbolgreek}{03D4}
\pdfglyphtounicode{Upsilonhooksymbol}{03D2}
\pdfglyphtounicode{Upsilontonos}{038E}
\pdfglyphtounicode{Uring}{016E}
\pdfglyphtounicode{Ushortcyrillic}{040E}
\pdfglyphtounicode{Usmall}{0075}
\pdfglyphtounicode{Ustraightcyrillic}{04AE}
\pdfglyphtounicode{Ustraightstrokecyrillic}{04B0}
\pdfglyphtounicode{Utilde}{0168}
\pdfglyphtounicode{Utildeacute}{1E78}
\pdfglyphtounicode{Utildebelow}{1E74}
\pdfglyphtounicode{V}{0056}
\pdfglyphtounicode{Vcircle}{24CB}
\pdfglyphtounicode{Vdotbelow}{1E7E}
\pdfglyphtounicode{Vecyrillic}{0412}
\pdfglyphtounicode{Vewarmenian}{054E}
\pdfglyphtounicode{Vhook}{01B2}
\pdfglyphtounicode{Vmonospace}{FF36}
\pdfglyphtounicode{Voarmenian}{0548}
\pdfglyphtounicode{Vsmall}{0076}
\pdfglyphtounicode{Vtilde}{1E7C}
\pdfglyphtounicode{W}{0057}
\pdfglyphtounicode{Wacute}{1E82}
\pdfglyphtounicode{Wcircle}{24CC}
\pdfglyphtounicode{Wcircumflex}{0174}
\pdfglyphtounicode{Wdieresis}{1E84}
\pdfglyphtounicode{Wdotaccent}{1E86}
\pdfglyphtounicode{Wdotbelow}{1E88}
\pdfglyphtounicode{Wgrave}{1E80}
\pdfglyphtounicode{Wmonospace}{FF37}
\pdfglyphtounicode{Wsmall}{0077}
\pdfglyphtounicode{X}{0058}
\pdfglyphtounicode{Xcircle}{24CD}
\pdfglyphtounicode{Xdieresis}{1E8C}
\pdfglyphtounicode{Xdotaccent}{1E8A}
\pdfglyphtounicode{Xeharmenian}{053D}
\pdfglyphtounicode{Xi}{039E}
\pdfglyphtounicode{Xmonospace}{FF38}
\pdfglyphtounicode{Xsmall}{0078}
\pdfglyphtounicode{Y}{0059}
\pdfglyphtounicode{Yacute}{00DD}
\pdfglyphtounicode{Yacutesmall}{00FD}
\pdfglyphtounicode{Yatcyrillic}{0462}
\pdfglyphtounicode{Ycircle}{24CE}
\pdfglyphtounicode{Ycircumflex}{0176}
\pdfglyphtounicode{Ydieresis}{0178}
\pdfglyphtounicode{Ydieresissmall}{00FF}
\pdfglyphtounicode{Ydotaccent}{1E8E}
\pdfglyphtounicode{Ydotbelow}{1EF4}
\pdfglyphtounicode{Yen}{00A5}
\pdfglyphtounicode{Yericyrillic}{042B}
\pdfglyphtounicode{Yerudieresiscyrillic}{04F8}
\pdfglyphtounicode{Ygrave}{1EF2}
\pdfglyphtounicode{Yhook}{01B3}
\pdfglyphtounicode{Yhookabove}{1EF6}
\pdfglyphtounicode{Yiarmenian}{0545}
\pdfglyphtounicode{Yicyrillic}{0407}
\pdfglyphtounicode{Yiwnarmenian}{0552}
\pdfglyphtounicode{Ymonospace}{FF39}
\pdfglyphtounicode{Ysmall}{0079}
\pdfglyphtounicode{Ytilde}{1EF8}
\pdfglyphtounicode{Yusbigcyrillic}{046A}
\pdfglyphtounicode{Yusbigiotifiedcyrillic}{046C}
\pdfglyphtounicode{Yuslittlecyrillic}{0466}
\pdfglyphtounicode{Yuslittleiotifiedcyrillic}{0468}
\pdfglyphtounicode{Z}{005A}
\pdfglyphtounicode{Zaarmenian}{0536}
\pdfglyphtounicode{Zacute}{0179}
\pdfglyphtounicode{Zcaron}{017D}
\pdfglyphtounicode{Zcaronsmall}{017E}
\pdfglyphtounicode{Zcircle}{24CF}
\pdfglyphtounicode{Zcircumflex}{1E90}
\pdfglyphtounicode{Zdot}{017B}
\pdfglyphtounicode{Zdotaccent}{017B}
\pdfglyphtounicode{Zdotbelow}{1E92}
\pdfglyphtounicode{Zecyrillic}{0417}
\pdfglyphtounicode{Zedescendercyrillic}{0498}
\pdfglyphtounicode{Zedieresiscyrillic}{04DE}
\pdfglyphtounicode{Zeta}{0396}
\pdfglyphtounicode{Zhearmenian}{053A}
\pdfglyphtounicode{Zhebrevecyrillic}{04C1}
\pdfglyphtounicode{Zhecyrillic}{0416}
\pdfglyphtounicode{Zhedescendercyrillic}{0496}
\pdfglyphtounicode{Zhedieresiscyrillic}{04DC}
\pdfglyphtounicode{Zlinebelow}{1E94}
\pdfglyphtounicode{Zmonospace}{FF3A}
\pdfglyphtounicode{Zsmall}{007A}
\pdfglyphtounicode{Zstroke}{01B5}
\pdfglyphtounicode{a}{0061}
\pdfglyphtounicode{aabengali}{0986}
\pdfglyphtounicode{aacute}{00E1}
\pdfglyphtounicode{aadeva}{0906}
\pdfglyphtounicode{aagujarati}{0A86}
\pdfglyphtounicode{aagurmukhi}{0A06}
\pdfglyphtounicode{aamatragurmukhi}{0A3E}
\pdfglyphtounicode{aarusquare}{3303}
\pdfglyphtounicode{aavowelsignbengali}{09BE}
\pdfglyphtounicode{aavowelsigndeva}{093E}
\pdfglyphtounicode{aavowelsigngujarati}{0ABE}
\pdfglyphtounicode{abbreviationmarkarmenian}{055F}
\pdfglyphtounicode{abbreviationsigndeva}{0970}
\pdfglyphtounicode{abengali}{0985}
\pdfglyphtounicode{abopomofo}{311A}
\pdfglyphtounicode{abreve}{0103}
\pdfglyphtounicode{abreveacute}{1EAF}
\pdfglyphtounicode{abrevecyrillic}{04D1}
\pdfglyphtounicode{abrevedotbelow}{1EB7}
\pdfglyphtounicode{abrevegrave}{1EB1}
\pdfglyphtounicode{abrevehookabove}{1EB3}
\pdfglyphtounicode{abrevetilde}{1EB5}
\pdfglyphtounicode{acaron}{01CE}
\pdfglyphtounicode{acircle}{24D0}
\pdfglyphtounicode{acircumflex}{00E2}
\pdfglyphtounicode{acircumflexacute}{1EA5}
\pdfglyphtounicode{acircumflexdotbelow}{1EAD}
\pdfglyphtounicode{acircumflexgrave}{1EA7}
\pdfglyphtounicode{acircumflexhookabove}{1EA9}
\pdfglyphtounicode{acircumflextilde}{1EAB}
\pdfglyphtounicode{acute}{00B4}
\pdfglyphtounicode{acutebelowcmb}{0317}
\pdfglyphtounicode{acutecmb}{0301}
\pdfglyphtounicode{acutecomb}{0301}
\pdfglyphtounicode{acutedeva}{0954}
\pdfglyphtounicode{acutelowmod}{02CF}
\pdfglyphtounicode{acutetonecmb}{0341}
\pdfglyphtounicode{acyrillic}{0430}
\pdfglyphtounicode{adblgrave}{0201}
\pdfglyphtounicode{addakgurmukhi}{0A71}
\pdfglyphtounicode{adeva}{0905}
\pdfglyphtounicode{adieresis}{00E4}
\pdfglyphtounicode{adieresiscyrillic}{04D3}
\pdfglyphtounicode{adieresismacron}{01DF}
\pdfglyphtounicode{adotbelow}{1EA1}
\pdfglyphtounicode{adotmacron}{01E1}
\pdfglyphtounicode{ae}{00E6}
\pdfglyphtounicode{aeacute}{01FD}
\pdfglyphtounicode{aekorean}{3150}
\pdfglyphtounicode{aemacron}{01E3}
\pdfglyphtounicode{afii00208}{2015}
\pdfglyphtounicode{afii08941}{20A4}
\pdfglyphtounicode{afii10017}{0410}
\pdfglyphtounicode{afii10018}{0411}
\pdfglyphtounicode{afii10019}{0412}
\pdfglyphtounicode{afii10020}{0413}
\pdfglyphtounicode{afii10021}{0414}
\pdfglyphtounicode{afii10022}{0415}
\pdfglyphtounicode{afii10023}{0401}
\pdfglyphtounicode{afii10024}{0416}
\pdfglyphtounicode{afii10025}{0417}
\pdfglyphtounicode{afii10026}{0418}
\pdfglyphtounicode{afii10027}{0419}
\pdfglyphtounicode{afii10028}{041A}
\pdfglyphtounicode{afii10029}{041B}
\pdfglyphtounicode{afii10030}{041C}
\pdfglyphtounicode{afii10031}{041D}
\pdfglyphtounicode{afii10032}{041E}
\pdfglyphtounicode{afii10033}{041F}
\pdfglyphtounicode{afii10034}{0420}
\pdfglyphtounicode{afii10035}{0421}
\pdfglyphtounicode{afii10036}{0422}
\pdfglyphtounicode{afii10037}{0423}
\pdfglyphtounicode{afii10038}{0424}
\pdfglyphtounicode{afii10039}{0425}
\pdfglyphtounicode{afii10040}{0426}
\pdfglyphtounicode{afii10041}{0427}
\pdfglyphtounicode{afii10042}{0428}
\pdfglyphtounicode{afii10043}{0429}
\pdfglyphtounicode{afii10044}{042A}
\pdfglyphtounicode{afii10045}{042B}
\pdfglyphtounicode{afii10046}{042C}
\pdfglyphtounicode{afii10047}{042D}
\pdfglyphtounicode{afii10048}{042E}
\pdfglyphtounicode{afii10049}{042F}
\pdfglyphtounicode{afii10050}{0490}
\pdfglyphtounicode{afii10051}{0402}
\pdfglyphtounicode{afii10052}{0403}
\pdfglyphtounicode{afii10053}{0404}
\pdfglyphtounicode{afii10054}{0405}
\pdfglyphtounicode{afii10055}{0406}
\pdfglyphtounicode{afii10056}{0407}
\pdfglyphtounicode{afii10057}{0408}
\pdfglyphtounicode{afii10058}{0409}
\pdfglyphtounicode{afii10059}{040A}
\pdfglyphtounicode{afii10060}{040B}
\pdfglyphtounicode{afii10061}{040C}
\pdfglyphtounicode{afii10062}{040E}
\pdfglyphtounicode{afii10063}{F6C4}
\pdfglyphtounicode{afii10064}{F6C5}
\pdfglyphtounicode{afii10065}{0430}
\pdfglyphtounicode{afii10066}{0431}
\pdfglyphtounicode{afii10067}{0432}
\pdfglyphtounicode{afii10068}{0433}
\pdfglyphtounicode{afii10069}{0434}
\pdfglyphtounicode{afii10070}{0435}
\pdfglyphtounicode{afii10071}{0451}
\pdfglyphtounicode{afii10072}{0436}
\pdfglyphtounicode{afii10073}{0437}
\pdfglyphtounicode{afii10074}{0438}
\pdfglyphtounicode{afii10075}{0439}
\pdfglyphtounicode{afii10076}{043A}
\pdfglyphtounicode{afii10077}{043B}
\pdfglyphtounicode{afii10078}{043C}
\pdfglyphtounicode{afii10079}{043D}
\pdfglyphtounicode{afii10080}{043E}
\pdfglyphtounicode{afii10081}{043F}
\pdfglyphtounicode{afii10082}{0440}
\pdfglyphtounicode{afii10083}{0441}
\pdfglyphtounicode{afii10084}{0442}
\pdfglyphtounicode{afii10085}{0443}
\pdfglyphtounicode{afii10086}{0444}
\pdfglyphtounicode{afii10087}{0445}
\pdfglyphtounicode{afii10088}{0446}
\pdfglyphtounicode{afii10089}{0447}
\pdfglyphtounicode{afii10090}{0448}
\pdfglyphtounicode{afii10091}{0449}
\pdfglyphtounicode{afii10092}{044A}
\pdfglyphtounicode{afii10093}{044B}
\pdfglyphtounicode{afii10094}{044C}
\pdfglyphtounicode{afii10095}{044D}
\pdfglyphtounicode{afii10096}{044E}
\pdfglyphtounicode{afii10097}{044F}
\pdfglyphtounicode{afii10098}{0491}
\pdfglyphtounicode{afii10099}{0452}
\pdfglyphtounicode{afii10100}{0453}
\pdfglyphtounicode{afii10101}{0454}
\pdfglyphtounicode{afii10102}{0455}
\pdfglyphtounicode{afii10103}{0456}
\pdfglyphtounicode{afii10104}{0457}
\pdfglyphtounicode{afii10105}{0458}
\pdfglyphtounicode{afii10106}{0459}
\pdfglyphtounicode{afii10107}{045A}
\pdfglyphtounicode{afii10108}{045B}
\pdfglyphtounicode{afii10109}{045C}
\pdfglyphtounicode{afii10110}{045E}
\pdfglyphtounicode{afii10145}{040F}
\pdfglyphtounicode{afii10146}{0462}
\pdfglyphtounicode{afii10147}{0472}
\pdfglyphtounicode{afii10148}{0474}
\pdfglyphtounicode{afii10192}{F6C6}
\pdfglyphtounicode{afii10193}{045F}
\pdfglyphtounicode{afii10194}{0463}
\pdfglyphtounicode{afii10195}{0473}
\pdfglyphtounicode{afii10196}{0475}
\pdfglyphtounicode{afii10831}{F6C7}
\pdfglyphtounicode{afii10832}{F6C8}
\pdfglyphtounicode{afii10846}{04D9}
\pdfglyphtounicode{afii299}{200E}
\pdfglyphtounicode{afii300}{200F}
\pdfglyphtounicode{afii301}{200D}
\pdfglyphtounicode{afii57381}{066A}
\pdfglyphtounicode{afii57388}{060C}
\pdfglyphtounicode{afii57392}{0660}
\pdfglyphtounicode{afii57393}{0661}
\pdfglyphtounicode{afii57394}{0662}
\pdfglyphtounicode{afii57395}{0663}
\pdfglyphtounicode{afii57396}{0664}
\pdfglyphtounicode{afii57397}{0665}
\pdfglyphtounicode{afii57398}{0666}
\pdfglyphtounicode{afii57399}{0667}
\pdfglyphtounicode{afii57400}{0668}
\pdfglyphtounicode{afii57401}{0669}
\pdfglyphtounicode{afii57403}{061B}
\pdfglyphtounicode{afii57407}{061F}
\pdfglyphtounicode{afii57409}{0621}
\pdfglyphtounicode{afii57410}{0622}
\pdfglyphtounicode{afii57411}{0623}
\pdfglyphtounicode{afii57412}{0624}
\pdfglyphtounicode{afii57413}{0625}
\pdfglyphtounicode{afii57414}{0626}
\pdfglyphtounicode{afii57415}{0627}
\pdfglyphtounicode{afii57416}{0628}
\pdfglyphtounicode{afii57417}{0629}
\pdfglyphtounicode{afii57418}{062A}
\pdfglyphtounicode{afii57419}{062B}
\pdfglyphtounicode{afii57420}{062C}
\pdfglyphtounicode{afii57421}{062D}
\pdfglyphtounicode{afii57422}{062E}
\pdfglyphtounicode{afii57423}{062F}
\pdfglyphtounicode{afii57424}{0630}
\pdfglyphtounicode{afii57425}{0631}
\pdfglyphtounicode{afii57426}{0632}
\pdfglyphtounicode{afii57427}{0633}
\pdfglyphtounicode{afii57428}{0634}
\pdfglyphtounicode{afii57429}{0635}
\pdfglyphtounicode{afii57430}{0636}
\pdfglyphtounicode{afii57431}{0637}
\pdfglyphtounicode{afii57432}{0638}
\pdfglyphtounicode{afii57433}{0639}
\pdfglyphtounicode{afii57434}{063A}
\pdfglyphtounicode{afii57440}{0640}
\pdfglyphtounicode{afii57441}{0641}
\pdfglyphtounicode{afii57442}{0642}
\pdfglyphtounicode{afii57443}{0643}
\pdfglyphtounicode{afii57444}{0644}
\pdfglyphtounicode{afii57445}{0645}
\pdfglyphtounicode{afii57446}{0646}
\pdfglyphtounicode{afii57448}{0648}
\pdfglyphtounicode{afii57449}{0649}
\pdfglyphtounicode{afii57450}{064A}
\pdfglyphtounicode{afii57451}{064B}
\pdfglyphtounicode{afii57452}{064C}
\pdfglyphtounicode{afii57453}{064D}
\pdfglyphtounicode{afii57454}{064E}
\pdfglyphtounicode{afii57455}{064F}
\pdfglyphtounicode{afii57456}{0650}
\pdfglyphtounicode{afii57457}{0651}
\pdfglyphtounicode{afii57458}{0652}
\pdfglyphtounicode{afii57470}{0647}
\pdfglyphtounicode{afii57505}{06A4}
\pdfglyphtounicode{afii57506}{067E}
\pdfglyphtounicode{afii57507}{0686}
\pdfglyphtounicode{afii57508}{0698}
\pdfglyphtounicode{afii57509}{06AF}
\pdfglyphtounicode{afii57511}{0679}
\pdfglyphtounicode{afii57512}{0688}
\pdfglyphtounicode{afii57513}{0691}
\pdfglyphtounicode{afii57514}{06BA}
\pdfglyphtounicode{afii57519}{06D2}
\pdfglyphtounicode{afii57534}{06D5}
\pdfglyphtounicode{afii57636}{20AA}
\pdfglyphtounicode{afii57645}{05BE}
\pdfglyphtounicode{afii57658}{05C3}
\pdfglyphtounicode{afii57664}{05D0}
\pdfglyphtounicode{afii57665}{05D1}
\pdfglyphtounicode{afii57666}{05D2}
\pdfglyphtounicode{afii57667}{05D3}
\pdfglyphtounicode{afii57668}{05D4}
\pdfglyphtounicode{afii57669}{05D5}
\pdfglyphtounicode{afii57670}{05D6}
\pdfglyphtounicode{afii57671}{05D7}
\pdfglyphtounicode{afii57672}{05D8}
\pdfglyphtounicode{afii57673}{05D9}
\pdfglyphtounicode{afii57674}{05DA}
\pdfglyphtounicode{afii57675}{05DB}
\pdfglyphtounicode{afii57676}{05DC}
\pdfglyphtounicode{afii57677}{05DD}
\pdfglyphtounicode{afii57678}{05DE}
\pdfglyphtounicode{afii57679}{05DF}
\pdfglyphtounicode{afii57680}{05E0}
\pdfglyphtounicode{afii57681}{05E1}
\pdfglyphtounicode{afii57682}{05E2}
\pdfglyphtounicode{afii57683}{05E3}
\pdfglyphtounicode{afii57684}{05E4}
\pdfglyphtounicode{afii57685}{05E5}
\pdfglyphtounicode{afii57686}{05E6}
\pdfglyphtounicode{afii57687}{05E7}
\pdfglyphtounicode{afii57688}{05E8}
\pdfglyphtounicode{afii57689}{05E9}
\pdfglyphtounicode{afii57690}{05EA}
\pdfglyphtounicode{afii57694}{FB2A}
\pdfglyphtounicode{afii57695}{FB2B}
\pdfglyphtounicode{afii57700}{FB4B}
\pdfglyphtounicode{afii57705}{FB1F}
\pdfglyphtounicode{afii57716}{05F0}
\pdfglyphtounicode{afii57717}{05F1}
\pdfglyphtounicode{afii57718}{05F2}
\pdfglyphtounicode{afii57723}{FB35}
\pdfglyphtounicode{afii57793}{05B4}
\pdfglyphtounicode{afii57794}{05B5}
\pdfglyphtounicode{afii57795}{05B6}
\pdfglyphtounicode{afii57796}{05BB}
\pdfglyphtounicode{afii57797}{05B8}
\pdfglyphtounicode{afii57798}{05B7}
\pdfglyphtounicode{afii57799}{05B0}
\pdfglyphtounicode{afii57800}{05B2}
\pdfglyphtounicode{afii57801}{05B1}
\pdfglyphtounicode{afii57802}{05B3}
\pdfglyphtounicode{afii57803}{05C2}
\pdfglyphtounicode{afii57804}{05C1}
\pdfglyphtounicode{afii57806}{05B9}
\pdfglyphtounicode{afii57807}{05BC}
\pdfglyphtounicode{afii57839}{05BD}
\pdfglyphtounicode{afii57841}{05BF}
\pdfglyphtounicode{afii57842}{05C0}
\pdfglyphtounicode{afii57929}{02BC}
\pdfglyphtounicode{afii61248}{2105}
\pdfglyphtounicode{afii61289}{2113}
\pdfglyphtounicode{afii61352}{2116}
\pdfglyphtounicode{afii61573}{202C}
\pdfglyphtounicode{afii61574}{202D}
\pdfglyphtounicode{afii61575}{202E}
\pdfglyphtounicode{afii61664}{200C}
\pdfglyphtounicode{afii63167}{066D}
\pdfglyphtounicode{afii64937}{02BD}
\pdfglyphtounicode{agrave}{00E0}
\pdfglyphtounicode{agujarati}{0A85}
\pdfglyphtounicode{agurmukhi}{0A05}
\pdfglyphtounicode{ahiragana}{3042}
\pdfglyphtounicode{ahookabove}{1EA3}
\pdfglyphtounicode{aibengali}{0990}
\pdfglyphtounicode{aibopomofo}{311E}
\pdfglyphtounicode{aideva}{0910}
\pdfglyphtounicode{aiecyrillic}{04D5}
\pdfglyphtounicode{aigujarati}{0A90}
\pdfglyphtounicode{aigurmukhi}{0A10}
\pdfglyphtounicode{aimatragurmukhi}{0A48}
\pdfglyphtounicode{ainarabic}{0639}
\pdfglyphtounicode{ainfinalarabic}{FECA}
\pdfglyphtounicode{aininitialarabic}{FECB}
\pdfglyphtounicode{ainmedialarabic}{FECC}
\pdfglyphtounicode{ainvertedbreve}{0203}
\pdfglyphtounicode{aivowelsignbengali}{09C8}
\pdfglyphtounicode{aivowelsigndeva}{0948}
\pdfglyphtounicode{aivowelsigngujarati}{0AC8}
\pdfglyphtounicode{akatakana}{30A2}
\pdfglyphtounicode{akatakanahalfwidth}{FF71}
\pdfglyphtounicode{akorean}{314F}
\pdfglyphtounicode{alef}{05D0}
\pdfglyphtounicode{alefarabic}{0627}
\pdfglyphtounicode{alefdageshhebrew}{FB30}
\pdfglyphtounicode{aleffinalarabic}{FE8E}
\pdfglyphtounicode{alefhamzaabovearabic}{0623}
\pdfglyphtounicode{alefhamzaabovefinalarabic}{FE84}
\pdfglyphtounicode{alefhamzabelowarabic}{0625}
\pdfglyphtounicode{alefhamzabelowfinalarabic}{FE88}
\pdfglyphtounicode{alefhebrew}{05D0}
\pdfglyphtounicode{aleflamedhebrew}{FB4F}
\pdfglyphtounicode{alefmaddaabovearabic}{0622}
\pdfglyphtounicode{alefmaddaabovefinalarabic}{FE82}
\pdfglyphtounicode{alefmaksuraarabic}{0649}
\pdfglyphtounicode{alefmaksurafinalarabic}{FEF0}
\pdfglyphtounicode{alefmaksurainitialarabic}{FEF3}
\pdfglyphtounicode{alefmaksuramedialarabic}{FEF4}
\pdfglyphtounicode{alefpatahhebrew}{FB2E}
\pdfglyphtounicode{alefqamatshebrew}{FB2F}
\pdfglyphtounicode{aleph}{2135}
\pdfglyphtounicode{allequal}{224C}
\pdfglyphtounicode{alpha}{03B1}
\pdfglyphtounicode{alphatonos}{03AC}
\pdfglyphtounicode{amacron}{0101}
\pdfglyphtounicode{amonospace}{FF41}
\pdfglyphtounicode{ampersand}{0026}
\pdfglyphtounicode{ampersandmonospace}{FF06}
\pdfglyphtounicode{ampersandsmall}{0026}
\pdfglyphtounicode{amsquare}{33C2}
\pdfglyphtounicode{anbopomofo}{3122}
\pdfglyphtounicode{angbopomofo}{3124}
\pdfglyphtounicode{angbracketleft}{27E8}
\pdfglyphtounicode{angbracketright}{27E9}
\pdfglyphtounicode{angkhankhuthai}{0E5A}
\pdfglyphtounicode{angle}{2220}
\pdfglyphtounicode{anglebracketleft}{3008}
\pdfglyphtounicode{anglebracketleftvertical}{FE3F}
\pdfglyphtounicode{anglebracketright}{3009}
\pdfglyphtounicode{anglebracketrightvertical}{FE40}
\pdfglyphtounicode{angleleft}{2329}
\pdfglyphtounicode{angleright}{232A}
\pdfglyphtounicode{angstrom}{212B}
\pdfglyphtounicode{anoteleia}{0387}
\pdfglyphtounicode{anticlockwise}{27F2}
\pdfglyphtounicode{anudattadeva}{0952}
\pdfglyphtounicode{anusvarabengali}{0982}
\pdfglyphtounicode{anusvaradeva}{0902}
\pdfglyphtounicode{anusvaragujarati}{0A82}
\pdfglyphtounicode{aogonek}{0105}
\pdfglyphtounicode{apaatosquare}{3300}
\pdfglyphtounicode{aparen}{249C}
\pdfglyphtounicode{apostrophearmenian}{055A}
\pdfglyphtounicode{apostrophemod}{02BC}
\pdfglyphtounicode{apple}{F8FF}
\pdfglyphtounicode{approaches}{2250}
\pdfglyphtounicode{approxequal}{2248}
\pdfglyphtounicode{approxequalorimage}{2252}
\pdfglyphtounicode{approximatelyequal}{2245}
\pdfglyphtounicode{approxorequal}{224A}
\pdfglyphtounicode{araeaekorean}{318E}
\pdfglyphtounicode{araeakorean}{318D}
\pdfglyphtounicode{arc}{2312}
\pdfglyphtounicode{archleftdown}{21B6}
\pdfglyphtounicode{archrightdown}{21B7}
\pdfglyphtounicode{arighthalfring}{1E9A}
\pdfglyphtounicode{aring}{00E5}
\pdfglyphtounicode{aringacute}{01FB}
\pdfglyphtounicode{aringbelow}{1E01}
\pdfglyphtounicode{arrowboth}{2194}
\pdfglyphtounicode{arrowbothv}{2195}
\pdfglyphtounicode{arrowdashdown}{21E3}
\pdfglyphtounicode{arrowdashleft}{21E0}
\pdfglyphtounicode{arrowdashright}{21E2}
\pdfglyphtounicode{arrowdashup}{21E1}
\pdfglyphtounicode{arrowdblboth}{21D4}
\pdfglyphtounicode{arrowdblbothv}{21D5}
\pdfglyphtounicode{arrowdbldown}{21D3}
\pdfglyphtounicode{arrowdblleft}{21D0}
\pdfglyphtounicode{arrowdblright}{21D2}
\pdfglyphtounicode{arrowdblup}{21D1}
\pdfglyphtounicode{arrowdown}{2193}
\pdfglyphtounicode{arrowdownleft}{2199}
\pdfglyphtounicode{arrowdownright}{2198}
\pdfglyphtounicode{arrowdownwhite}{21E9}
\pdfglyphtounicode{arrowheaddownmod}{02C5}
\pdfglyphtounicode{arrowheadleftmod}{02C2}
\pdfglyphtounicode{arrowheadrightmod}{02C3}
\pdfglyphtounicode{arrowheadupmod}{02C4}
\pdfglyphtounicode{arrowhorizex}{F8E7}
\pdfglyphtounicode{arrowleft}{2190}
\pdfglyphtounicode{arrowleftbothalf}{21BD}
\pdfglyphtounicode{arrowleftdbl}{21D0}
\pdfglyphtounicode{arrowleftdblstroke}{21CD}
\pdfglyphtounicode{arrowleftoverright}{21C6}
\pdfglyphtounicode{arrowlefttophalf}{21BC}
\pdfglyphtounicode{arrowleftwhite}{21E6}
\pdfglyphtounicode{arrownortheast}{2197}
\pdfglyphtounicode{arrownorthwest}{2196}
\pdfglyphtounicode{arrowparrleftright}{21C6}
\pdfglyphtounicode{arrowparrrightleft}{21C4}
\pdfglyphtounicode{arrowright}{2192}
\pdfglyphtounicode{arrowrightbothalf}{21C1}
\pdfglyphtounicode{arrowrightdblstroke}{21CF}
\pdfglyphtounicode{arrowrightheavy}{279E}
\pdfglyphtounicode{arrowrightoverleft}{21C4}
\pdfglyphtounicode{arrowrighttophalf}{21C0}
\pdfglyphtounicode{arrowrightwhite}{21E8}
\pdfglyphtounicode{arrowsoutheast}{2198}
\pdfglyphtounicode{arrowsouthwest}{2199}
\pdfglyphtounicode{arrowtableft}{21E4}
\pdfglyphtounicode{arrowtabright}{21E5}
\pdfglyphtounicode{arrowtailleft}{21A2}
\pdfglyphtounicode{arrowtailright}{21A3}
\pdfglyphtounicode{arrowtripleleft}{21DA}
\pdfglyphtounicode{arrowtripleright}{21DB}
\pdfglyphtounicode{arrowup}{2191}
\pdfglyphtounicode{arrowupdn}{2195}
\pdfglyphtounicode{arrowupdnbse}{21A8}
\pdfglyphtounicode{arrowupdownbase}{21A8}
\pdfglyphtounicode{arrowupleft}{2196}
\pdfglyphtounicode{arrowupleftofdown}{21C5}
\pdfglyphtounicode{arrowupright}{2197}
\pdfglyphtounicode{arrowupwhite}{21E7}
\pdfglyphtounicode{arrowvertex}{F8E6}
\pdfglyphtounicode{asciicircum}{005E}
\pdfglyphtounicode{asciicircummonospace}{FF3E}
\pdfglyphtounicode{asciitilde}{007E}
\pdfglyphtounicode{asciitildemonospace}{FF5E}
\pdfglyphtounicode{ascript}{0251}
\pdfglyphtounicode{ascriptturned}{0252}
\pdfglyphtounicode{asmallhiragana}{3041}
\pdfglyphtounicode{asmallkatakana}{30A1}
\pdfglyphtounicode{asmallkatakanahalfwidth}{FF67}
\pdfglyphtounicode{asterisk}{002A}
\pdfglyphtounicode{asteriskaltonearabic}{066D}
\pdfglyphtounicode{asteriskarabic}{066D}
\pdfglyphtounicode{asteriskcentered}{2217}
\pdfglyphtounicode{asteriskmath}{2217}
\pdfglyphtounicode{asteriskmonospace}{FF0A}
\pdfglyphtounicode{asterisksmall}{FE61}
\pdfglyphtounicode{asterism}{2042}
\pdfglyphtounicode{asuperior}{0061}
\pdfglyphtounicode{asymptoticallyequal}{2243}
\pdfglyphtounicode{at}{0040}
\pdfglyphtounicode{atilde}{00E3}
\pdfglyphtounicode{atmonospace}{FF20}
\pdfglyphtounicode{atsmall}{FE6B}
\pdfglyphtounicode{aturned}{0250}
\pdfglyphtounicode{aubengali}{0994}
\pdfglyphtounicode{aubopomofo}{3120}
\pdfglyphtounicode{audeva}{0914}
\pdfglyphtounicode{augujarati}{0A94}
\pdfglyphtounicode{augurmukhi}{0A14}
\pdfglyphtounicode{aulengthmarkbengali}{09D7}
\pdfglyphtounicode{aumatragurmukhi}{0A4C}
\pdfglyphtounicode{auvowelsignbengali}{09CC}
\pdfglyphtounicode{auvowelsigndeva}{094C}
\pdfglyphtounicode{auvowelsigngujarati}{0ACC}
\pdfglyphtounicode{avagrahadeva}{093D}
\pdfglyphtounicode{aybarmenian}{0561}
\pdfglyphtounicode{ayin}{05E2}
\pdfglyphtounicode{ayinaltonehebrew}{FB20}
\pdfglyphtounicode{ayinhebrew}{05E2}
\pdfglyphtounicode{b}{0062}
\pdfglyphtounicode{babengali}{09AC}
\pdfglyphtounicode{backslash}{005C}
\pdfglyphtounicode{backslashmonospace}{FF3C}
\pdfglyphtounicode{badeva}{092C}
\pdfglyphtounicode{bagujarati}{0AAC}
\pdfglyphtounicode{bagurmukhi}{0A2C}
\pdfglyphtounicode{bahiragana}{3070}
\pdfglyphtounicode{bahtthai}{0E3F}
\pdfglyphtounicode{bakatakana}{30D0}
\pdfglyphtounicode{bar}{007C}
\pdfglyphtounicode{bardbl}{2225}
\pdfglyphtounicode{barmonospace}{FF5C}
\pdfglyphtounicode{bbopomofo}{3105}
\pdfglyphtounicode{bcircle}{24D1}
\pdfglyphtounicode{bdotaccent}{1E03}
\pdfglyphtounicode{bdotbelow}{1E05}
\pdfglyphtounicode{beamedsixteenthnotes}{266C}
\pdfglyphtounicode{because}{2235}
\pdfglyphtounicode{becyrillic}{0431}
\pdfglyphtounicode{beharabic}{0628}
\pdfglyphtounicode{behfinalarabic}{FE90}
\pdfglyphtounicode{behinitialarabic}{FE91}
\pdfglyphtounicode{behiragana}{3079}
\pdfglyphtounicode{behmedialarabic}{FE92}
\pdfglyphtounicode{behmeeminitialarabic}{FC9F}
\pdfglyphtounicode{behmeemisolatedarabic}{FC08}
\pdfglyphtounicode{behnoonfinalarabic}{FC6D}
\pdfglyphtounicode{bekatakana}{30D9}
\pdfglyphtounicode{benarmenian}{0562}
\pdfglyphtounicode{bet}{05D1}
\pdfglyphtounicode{beta}{03B2}
\pdfglyphtounicode{betasymbolgreek}{03D0}
\pdfglyphtounicode{betdagesh}{FB31}
\pdfglyphtounicode{betdageshhebrew}{FB31}
\pdfglyphtounicode{beth}{2136}
\pdfglyphtounicode{bethebrew}{05D1}
\pdfglyphtounicode{betrafehebrew}{FB4C}
\pdfglyphtounicode{between}{226C}
\pdfglyphtounicode{bhabengali}{09AD}
\pdfglyphtounicode{bhadeva}{092D}
\pdfglyphtounicode{bhagujarati}{0AAD}
\pdfglyphtounicode{bhagurmukhi}{0A2D}
\pdfglyphtounicode{bhook}{0253}
\pdfglyphtounicode{bihiragana}{3073}
\pdfglyphtounicode{bikatakana}{30D3}
\pdfglyphtounicode{bilabialclick}{0298}
\pdfglyphtounicode{bindigurmukhi}{0A02}
\pdfglyphtounicode{birusquare}{3331}
\pdfglyphtounicode{blackcircle}{25CF}
\pdfglyphtounicode{blackdiamond}{25C6}
\pdfglyphtounicode{blackdownpointingtriangle}{25BC}
\pdfglyphtounicode{blackleftpointingpointer}{25C4}
\pdfglyphtounicode{blackleftpointingtriangle}{25C0}
\pdfglyphtounicode{blacklenticularbracketleft}{3010}
\pdfglyphtounicode{blacklenticularbracketleftvertical}{FE3B}
\pdfglyphtounicode{blacklenticularbracketright}{3011}
\pdfglyphtounicode{blacklenticularbracketrightvertical}{FE3C}
\pdfglyphtounicode{blacklowerlefttriangle}{25E3}
\pdfglyphtounicode{blacklowerrighttriangle}{25E2}
\pdfglyphtounicode{blackrectangle}{25AC}
\pdfglyphtounicode{blackrightpointingpointer}{25BA}
\pdfglyphtounicode{blackrightpointingtriangle}{25B6}
\pdfglyphtounicode{blacksmallsquare}{25AA}
\pdfglyphtounicode{blacksmilingface}{263B}
\pdfglyphtounicode{blacksquare}{25A0}
\pdfglyphtounicode{blackstar}{2605}
\pdfglyphtounicode{blackupperlefttriangle}{25E4}
\pdfglyphtounicode{blackupperrighttriangle}{25E5}
\pdfglyphtounicode{blackuppointingsmalltriangle}{25B4}
\pdfglyphtounicode{blackuppointingtriangle}{25B2}
\pdfglyphtounicode{blank}{2423}
\pdfglyphtounicode{blinebelow}{1E07}
\pdfglyphtounicode{block}{2588}
\pdfglyphtounicode{bmonospace}{FF42}
\pdfglyphtounicode{bobaimaithai}{0E1A}
\pdfglyphtounicode{bohiragana}{307C}
\pdfglyphtounicode{bokatakana}{30DC}
\pdfglyphtounicode{bparen}{249D}
\pdfglyphtounicode{bqsquare}{33C3}
\pdfglyphtounicode{braceex}{F8F4}
\pdfglyphtounicode{braceleft}{007B}
\pdfglyphtounicode{braceleftbt}{F8F3}
\pdfglyphtounicode{braceleftmid}{F8F2}
\pdfglyphtounicode{braceleftmonospace}{FF5B}
\pdfglyphtounicode{braceleftsmall}{FE5B}
\pdfglyphtounicode{bracelefttp}{F8F1}
\pdfglyphtounicode{braceleftvertical}{FE37}
\pdfglyphtounicode{braceright}{007D}
\pdfglyphtounicode{bracerightbt}{F8FE}
\pdfglyphtounicode{bracerightmid}{F8FD}
\pdfglyphtounicode{bracerightmonospace}{FF5D}
\pdfglyphtounicode{bracerightsmall}{FE5C}
\pdfglyphtounicode{bracerighttp}{F8FC}
\pdfglyphtounicode{bracerightvertical}{FE38}
\pdfglyphtounicode{bracketleft}{005B}
\pdfglyphtounicode{bracketleftbt}{F8F0}
\pdfglyphtounicode{bracketleftex}{F8EF}
\pdfglyphtounicode{bracketleftmonospace}{FF3B}
\pdfglyphtounicode{bracketlefttp}{F8EE}
\pdfglyphtounicode{bracketright}{005D}
\pdfglyphtounicode{bracketrightbt}{F8FB}
\pdfglyphtounicode{bracketrightex}{F8FA}
\pdfglyphtounicode{bracketrightmonospace}{FF3D}
\pdfglyphtounicode{bracketrighttp}{F8F9}
\pdfglyphtounicode{breve}{02D8}
\pdfglyphtounicode{brevebelowcmb}{032E}
\pdfglyphtounicode{brevecmb}{0306}
\pdfglyphtounicode{breveinvertedbelowcmb}{032F}
\pdfglyphtounicode{breveinvertedcmb}{0311}
\pdfglyphtounicode{breveinverteddoublecmb}{0361}
\pdfglyphtounicode{bridgebelowcmb}{032A}
\pdfglyphtounicode{bridgeinvertedbelowcmb}{033A}
\pdfglyphtounicode{brokenbar}{00A6}
\pdfglyphtounicode{bstroke}{0180}
\pdfglyphtounicode{bsuperior}{0062}
\pdfglyphtounicode{btopbar}{0183}
\pdfglyphtounicode{buhiragana}{3076}
\pdfglyphtounicode{bukatakana}{30D6}
\pdfglyphtounicode{bullet}{2022}
\pdfglyphtounicode{bulletinverse}{25D8}
\pdfglyphtounicode{bulletoperator}{2219}
\pdfglyphtounicode{bullseye}{25CE}
\pdfglyphtounicode{c}{0063}
\pdfglyphtounicode{caarmenian}{056E}
\pdfglyphtounicode{cabengali}{099A}
\pdfglyphtounicode{cacute}{0107}
\pdfglyphtounicode{cadeva}{091A}
\pdfglyphtounicode{cagujarati}{0A9A}
\pdfglyphtounicode{cagurmukhi}{0A1A}
\pdfglyphtounicode{calsquare}{3388}
\pdfglyphtounicode{candrabindubengali}{0981}
\pdfglyphtounicode{candrabinducmb}{0310}
\pdfglyphtounicode{candrabindudeva}{0901}
\pdfglyphtounicode{candrabindugujarati}{0A81}
\pdfglyphtounicode{capslock}{21EA}
\pdfglyphtounicode{careof}{2105}
\pdfglyphtounicode{caron}{02C7}
\pdfglyphtounicode{caronbelowcmb}{032C}
\pdfglyphtounicode{caroncmb}{030C}
\pdfglyphtounicode{carriagereturn}{21B5}
\pdfglyphtounicode{cbopomofo}{3118}
\pdfglyphtounicode{ccaron}{010D}
\pdfglyphtounicode{ccedilla}{00E7}
\pdfglyphtounicode{ccedillaacute}{1E09}
\pdfglyphtounicode{ccircle}{24D2}
\pdfglyphtounicode{ccircumflex}{0109}
\pdfglyphtounicode{ccurl}{0255}
\pdfglyphtounicode{cdot}{010B}
\pdfglyphtounicode{cdotaccent}{010B}
\pdfglyphtounicode{cdsquare}{33C5}
\pdfglyphtounicode{cedilla}{00B8}
\pdfglyphtounicode{cedillacmb}{0327}
\pdfglyphtounicode{ceilingleft}{2308}
\pdfglyphtounicode{ceilingright}{2309}
\pdfglyphtounicode{cent}{00A2}
\pdfglyphtounicode{centigrade}{2103}
\pdfglyphtounicode{centinferior}{00A2}
\pdfglyphtounicode{centmonospace}{FFE0}
\pdfglyphtounicode{centoldstyle}{00A2}
\pdfglyphtounicode{centsuperior}{00A2}
\pdfglyphtounicode{chaarmenian}{0579}
\pdfglyphtounicode{chabengali}{099B}
\pdfglyphtounicode{chadeva}{091B}
\pdfglyphtounicode{chagujarati}{0A9B}
\pdfglyphtounicode{chagurmukhi}{0A1B}
\pdfglyphtounicode{chbopomofo}{3114}
\pdfglyphtounicode{cheabkhasiancyrillic}{04BD}
\pdfglyphtounicode{check}{2713}
\pdfglyphtounicode{checkmark}{2713}
\pdfglyphtounicode{checyrillic}{0447}
\pdfglyphtounicode{chedescenderabkhasiancyrillic}{04BF}
\pdfglyphtounicode{chedescendercyrillic}{04B7}
\pdfglyphtounicode{chedieresiscyrillic}{04F5}
\pdfglyphtounicode{cheharmenian}{0573}
\pdfglyphtounicode{chekhakassiancyrillic}{04CC}
\pdfglyphtounicode{cheverticalstrokecyrillic}{04B9}
\pdfglyphtounicode{chi}{03C7}
\pdfglyphtounicode{chieuchacirclekorean}{3277}
\pdfglyphtounicode{chieuchaparenkorean}{3217}
\pdfglyphtounicode{chieuchcirclekorean}{3269}
\pdfglyphtounicode{chieuchkorean}{314A}
\pdfglyphtounicode{chieuchparenkorean}{3209}
\pdfglyphtounicode{chochangthai}{0E0A}
\pdfglyphtounicode{chochanthai}{0E08}
\pdfglyphtounicode{chochingthai}{0E09}
\pdfglyphtounicode{chochoethai}{0E0C}
\pdfglyphtounicode{chook}{0188}
\pdfglyphtounicode{cieucacirclekorean}{3276}
\pdfglyphtounicode{cieucaparenkorean}{3216}
\pdfglyphtounicode{cieuccirclekorean}{3268}
\pdfglyphtounicode{cieuckorean}{3148}
\pdfglyphtounicode{cieucparenkorean}{3208}
\pdfglyphtounicode{cieucuparenkorean}{321C}
\pdfglyphtounicode{circle}{25CB}
\pdfglyphtounicode{circleR}{00AE}
\pdfglyphtounicode{circleS}{24C8}
\pdfglyphtounicode{circleasterisk}{229B}
\pdfglyphtounicode{circlecopyrt}{20DD}
\pdfglyphtounicode{circledivide}{2298}
\pdfglyphtounicode{circledot}{2299}
\pdfglyphtounicode{circleequal}{229C}
\pdfglyphtounicode{circleminus}{2296}
\pdfglyphtounicode{circlemultiply}{2297}
\pdfglyphtounicode{circleot}{2299}
\pdfglyphtounicode{circleplus}{2295}
\pdfglyphtounicode{circlepostalmark}{3036}
\pdfglyphtounicode{circlering}{229A}
\pdfglyphtounicode{circlewithlefthalfblack}{25D0}
\pdfglyphtounicode{circlewithrighthalfblack}{25D1}
\pdfglyphtounicode{circumflex}{02C6}
\pdfglyphtounicode{circumflexbelowcmb}{032D}
\pdfglyphtounicode{circumflexcmb}{0302}
\pdfglyphtounicode{clear}{2327}
\pdfglyphtounicode{clickalveolar}{01C2}
\pdfglyphtounicode{clickdental}{01C0}
\pdfglyphtounicode{clicklateral}{01C1}
\pdfglyphtounicode{clickretroflex}{01C3}
\pdfglyphtounicode{clockwise}{27F3}
\pdfglyphtounicode{club}{2663}
\pdfglyphtounicode{clubsuitblack}{2663}
\pdfglyphtounicode{clubsuitwhite}{2667}
\pdfglyphtounicode{cmcubedsquare}{33A4}
\pdfglyphtounicode{cmonospace}{FF43}
\pdfglyphtounicode{cmsquaredsquare}{33A0}
\pdfglyphtounicode{coarmenian}{0581}
\pdfglyphtounicode{colon}{003A}
\pdfglyphtounicode{colonmonetary}{20A1}
\pdfglyphtounicode{colonmonospace}{FF1A}
\pdfglyphtounicode{colonsign}{20A1}
\pdfglyphtounicode{colonsmall}{FE55}
\pdfglyphtounicode{colontriangularhalfmod}{02D1}
\pdfglyphtounicode{colontriangularmod}{02D0}
\pdfglyphtounicode{comma}{002C}
\pdfglyphtounicode{commaabovecmb}{0313}
\pdfglyphtounicode{commaaboverightcmb}{0315}
\pdfglyphtounicode{commaaccent}{F6C3}
\pdfglyphtounicode{commaarabic}{060C}
\pdfglyphtounicode{commaarmenian}{055D}
\pdfglyphtounicode{commainferior}{002C}
\pdfglyphtounicode{commamonospace}{FF0C}
\pdfglyphtounicode{commareversedabovecmb}{0314}
\pdfglyphtounicode{commareversedmod}{02BD}
\pdfglyphtounicode{commasmall}{FE50}
\pdfglyphtounicode{commasuperior}{002C}
\pdfglyphtounicode{commaturnedabovecmb}{0312}
\pdfglyphtounicode{commaturnedmod}{02BB}
\pdfglyphtounicode{compass}{263C}
\pdfglyphtounicode{complement}{2201}
\pdfglyphtounicode{compwordmark}{200C}
\pdfglyphtounicode{congruent}{2245}
\pdfglyphtounicode{contourintegral}{222E}
\pdfglyphtounicode{control}{2303}
\pdfglyphtounicode{controlACK}{0006}
\pdfglyphtounicode{controlBEL}{0007}
\pdfglyphtounicode{controlBS}{0008}
\pdfglyphtounicode{controlCAN}{0018}
\pdfglyphtounicode{controlCR}{000D}
\pdfglyphtounicode{controlDC1}{0011}
\pdfglyphtounicode{controlDC2}{0012}
\pdfglyphtounicode{controlDC3}{0013}
\pdfglyphtounicode{controlDC4}{0014}
\pdfglyphtounicode{controlDEL}{007F}
\pdfglyphtounicode{controlDLE}{0010}
\pdfglyphtounicode{controlEM}{0019}
\pdfglyphtounicode{controlENQ}{0005}
\pdfglyphtounicode{controlEOT}{0004}
\pdfglyphtounicode{controlESC}{001B}
\pdfglyphtounicode{controlETB}{0017}
\pdfglyphtounicode{controlETX}{0003}
\pdfglyphtounicode{controlFF}{000C}
\pdfglyphtounicode{controlFS}{001C}
\pdfglyphtounicode{controlGS}{001D}
\pdfglyphtounicode{controlHT}{0009}
\pdfglyphtounicode{controlLF}{000A}
\pdfglyphtounicode{controlNAK}{0015}
\pdfglyphtounicode{controlRS}{001E}
\pdfglyphtounicode{controlSI}{000F}
\pdfglyphtounicode{controlSO}{000E}
\pdfglyphtounicode{controlSOT}{0002}
\pdfglyphtounicode{controlSTX}{0001}
\pdfglyphtounicode{controlSUB}{001A}
\pdfglyphtounicode{controlSYN}{0016}
\pdfglyphtounicode{controlUS}{001F}
\pdfglyphtounicode{controlVT}{000B}
\pdfglyphtounicode{coproduct}{2A3F}
\pdfglyphtounicode{copyright}{00A9}
\pdfglyphtounicode{copyrightsans}{00A9}
\pdfglyphtounicode{copyrightserif}{00A9}
\pdfglyphtounicode{cornerbracketleft}{300C}
\pdfglyphtounicode{cornerbracketlefthalfwidth}{FF62}
\pdfglyphtounicode{cornerbracketleftvertical}{FE41}
\pdfglyphtounicode{cornerbracketright}{300D}
\pdfglyphtounicode{cornerbracketrighthalfwidth}{FF63}
\pdfglyphtounicode{cornerbracketrightvertical}{FE42}
\pdfglyphtounicode{corporationsquare}{337F}
\pdfglyphtounicode{cosquare}{33C7}
\pdfglyphtounicode{coverkgsquare}{33C6}
\pdfglyphtounicode{cparen}{249E}
\pdfglyphtounicode{cruzeiro}{20A2}
\pdfglyphtounicode{cstretched}{0297}
\pdfglyphtounicode{ct}{0063 0074}
\pdfglyphtounicode{curlyand}{22CF}
\pdfglyphtounicode{curlyleft}{21AB}
\pdfglyphtounicode{curlyor}{22CE}
\pdfglyphtounicode{curlyright}{21AC}
\pdfglyphtounicode{currency}{00A4}
\pdfglyphtounicode{cwm}{200C}
\pdfglyphtounicode{cyrBreve}{02D8}
\pdfglyphtounicode{cyrFlex}{00A0 0311}
\pdfglyphtounicode{cyrbreve}{02D8}
\pdfglyphtounicode{cyrflex}{00A0 0311}
\pdfglyphtounicode{d}{0064}
\pdfglyphtounicode{daarmenian}{0564}
\pdfglyphtounicode{dabengali}{09A6}
\pdfglyphtounicode{dadarabic}{0636}
\pdfglyphtounicode{dadeva}{0926}
\pdfglyphtounicode{dadfinalarabic}{FEBE}
\pdfglyphtounicode{dadinitialarabic}{FEBF}
\pdfglyphtounicode{dadmedialarabic}{FEC0}
\pdfglyphtounicode{dagesh}{05BC}
\pdfglyphtounicode{dageshhebrew}{05BC}
\pdfglyphtounicode{dagger}{2020}
\pdfglyphtounicode{daggerdbl}{2021}
\pdfglyphtounicode{dagujarati}{0AA6}
\pdfglyphtounicode{dagurmukhi}{0A26}
\pdfglyphtounicode{dahiragana}{3060}
\pdfglyphtounicode{dakatakana}{30C0}
\pdfglyphtounicode{dalarabic}{062F}
\pdfglyphtounicode{dalet}{05D3}
\pdfglyphtounicode{daletdagesh}{FB33}
\pdfglyphtounicode{daletdageshhebrew}{FB33}
\pdfglyphtounicode{daleth}{2138}
\pdfglyphtounicode{dalethatafpatah}{05D3 05B2}
\pdfglyphtounicode{dalethatafpatahhebrew}{05D3 05B2}
\pdfglyphtounicode{dalethatafsegol}{05D3 05B1}
\pdfglyphtounicode{dalethatafsegolhebrew}{05D3 05B1}
\pdfglyphtounicode{dalethebrew}{05D3}
\pdfglyphtounicode{dalethiriq}{05D3 05B4}
\pdfglyphtounicode{dalethiriqhebrew}{05D3 05B4}
\pdfglyphtounicode{daletholam}{05D3 05B9}
\pdfglyphtounicode{daletholamhebrew}{05D3 05B9}
\pdfglyphtounicode{daletpatah}{05D3 05B7}
\pdfglyphtounicode{daletpatahhebrew}{05D3 05B7}
\pdfglyphtounicode{daletqamats}{05D3 05B8}
\pdfglyphtounicode{daletqamatshebrew}{05D3 05B8}
\pdfglyphtounicode{daletqubuts}{05D3 05BB}
\pdfglyphtounicode{daletqubutshebrew}{05D3 05BB}
\pdfglyphtounicode{daletsegol}{05D3 05B6}
\pdfglyphtounicode{daletsegolhebrew}{05D3 05B6}
\pdfglyphtounicode{daletsheva}{05D3 05B0}
\pdfglyphtounicode{daletshevahebrew}{05D3 05B0}
\pdfglyphtounicode{dalettsere}{05D3 05B5}
\pdfglyphtounicode{dalettserehebrew}{05D3 05B5}
\pdfglyphtounicode{dalfinalarabic}{FEAA}
\pdfglyphtounicode{dammaarabic}{064F}
\pdfglyphtounicode{dammalowarabic}{064F}
\pdfglyphtounicode{dammatanaltonearabic}{064C}
\pdfglyphtounicode{dammatanarabic}{064C}
\pdfglyphtounicode{danda}{0964}
\pdfglyphtounicode{dargahebrew}{05A7}
\pdfglyphtounicode{dargalefthebrew}{05A7}
\pdfglyphtounicode{dasiapneumatacyrilliccmb}{0485}
\pdfglyphtounicode{dbar}{0111}
\pdfglyphtounicode{dblGrave}{00A0 030F}
\pdfglyphtounicode{dblanglebracketleft}{300A}
\pdfglyphtounicode{dblanglebracketleftvertical}{FE3D}
\pdfglyphtounicode{dblanglebracketright}{300B}
\pdfglyphtounicode{dblanglebracketrightvertical}{FE3E}
\pdfglyphtounicode{dblarchinvertedbelowcmb}{032B}
\pdfglyphtounicode{dblarrowdwn}{21CA}
\pdfglyphtounicode{dblarrowheadleft}{219E}
\pdfglyphtounicode{dblarrowheadright}{21A0}
\pdfglyphtounicode{dblarrowleft}{21D4}
\pdfglyphtounicode{dblarrowright}{21D2}
\pdfglyphtounicode{dblarrowup}{21C8}
\pdfglyphtounicode{dblbracketleft}{27E6}
\pdfglyphtounicode{dblbracketright}{27E7}
\pdfglyphtounicode{dbldanda}{0965}
\pdfglyphtounicode{dblgrave}{00A0 030F}
\pdfglyphtounicode{dblgravecmb}{030F}
\pdfglyphtounicode{dblintegral}{222C}
\pdfglyphtounicode{dbllowline}{2017}
\pdfglyphtounicode{dbllowlinecmb}{0333}
\pdfglyphtounicode{dbloverlinecmb}{033F}
\pdfglyphtounicode{dblprimemod}{02BA}
\pdfglyphtounicode{dblverticalbar}{2016}
\pdfglyphtounicode{dblverticallineabovecmb}{030E}
\pdfglyphtounicode{dbopomofo}{3109}
\pdfglyphtounicode{dbsquare}{33C8}
\pdfglyphtounicode{dcaron}{010F}
\pdfglyphtounicode{dcedilla}{1E11}
\pdfglyphtounicode{dcircle}{24D3}
\pdfglyphtounicode{dcircumflexbelow}{1E13}
\pdfglyphtounicode{dcroat}{0111}
\pdfglyphtounicode{ddabengali}{09A1}
\pdfglyphtounicode{ddadeva}{0921}
\pdfglyphtounicode{ddagujarati}{0AA1}
\pdfglyphtounicode{ddagurmukhi}{0A21}
\pdfglyphtounicode{ddalarabic}{0688}
\pdfglyphtounicode{ddalfinalarabic}{FB89}
\pdfglyphtounicode{dddhadeva}{095C}
\pdfglyphtounicode{ddhabengali}{09A2}
\pdfglyphtounicode{ddhadeva}{0922}
\pdfglyphtounicode{ddhagujarati}{0AA2}
\pdfglyphtounicode{ddhagurmukhi}{0A22}
\pdfglyphtounicode{ddotaccent}{1E0B}
\pdfglyphtounicode{ddotbelow}{1E0D}
\pdfglyphtounicode{decimalseparatorarabic}{066B}
\pdfglyphtounicode{decimalseparatorpersian}{066B}
\pdfglyphtounicode{decyrillic}{0434}
\pdfglyphtounicode{defines}{225C}
\pdfglyphtounicode{degree}{00B0}
\pdfglyphtounicode{dehihebrew}{05AD}
\pdfglyphtounicode{dehiragana}{3067}
\pdfglyphtounicode{deicoptic}{03EF}
\pdfglyphtounicode{dekatakana}{30C7}
\pdfglyphtounicode{deleteleft}{232B}
\pdfglyphtounicode{deleteright}{2326}
\pdfglyphtounicode{delta}{03B4}
\pdfglyphtounicode{deltaturned}{018D}
\pdfglyphtounicode{denominatorminusonenumeratorbengali}{09F8}
\pdfglyphtounicode{dezh}{02A4}
\pdfglyphtounicode{dhabengali}{09A7}
\pdfglyphtounicode{dhadeva}{0927}
\pdfglyphtounicode{dhagujarati}{0AA7}
\pdfglyphtounicode{dhagurmukhi}{0A27}
\pdfglyphtounicode{dhook}{0257}
\pdfglyphtounicode{dialytikatonos}{0385}
\pdfglyphtounicode{dialytikatonoscmb}{0344}
\pdfglyphtounicode{diamond}{2666}
\pdfglyphtounicode{diamondmath}{22C4}
\pdfglyphtounicode{diamondsolid}{2666}
\pdfglyphtounicode{diamondsuitwhite}{2662}
\pdfglyphtounicode{dieresis}{00A8}
\pdfglyphtounicode{dieresisacute}{00A0 0308 0301}
\pdfglyphtounicode{dieresisbelowcmb}{0324}
\pdfglyphtounicode{dieresiscmb}{0308}
\pdfglyphtounicode{dieresisgrave}{00A0 0308 0300}
\pdfglyphtounicode{dieresistonos}{0385}
\pdfglyphtounicode{difference}{224F}
\pdfglyphtounicode{dihiragana}{3062}
\pdfglyphtounicode{dikatakana}{30C2}
\pdfglyphtounicode{dittomark}{3003}
\pdfglyphtounicode{divide}{00F7}
\pdfglyphtounicode{dividemultiply}{22C7}
\pdfglyphtounicode{divides}{2223}
\pdfglyphtounicode{divisionslash}{2215}
\pdfglyphtounicode{djecyrillic}{0452}
\pdfglyphtounicode{dkshade}{2593}
\pdfglyphtounicode{dlinebelow}{1E0F}
\pdfglyphtounicode{dlsquare}{3397}
\pdfglyphtounicode{dmacron}{0111}
\pdfglyphtounicode{dmonospace}{FF44}
\pdfglyphtounicode{dnblock}{2584}
\pdfglyphtounicode{dochadathai}{0E0E}
\pdfglyphtounicode{dodekthai}{0E14}
\pdfglyphtounicode{dohiragana}{3069}
\pdfglyphtounicode{dokatakana}{30C9}
\pdfglyphtounicode{dollar}{0024}
\pdfglyphtounicode{dollarinferior}{0024}
\pdfglyphtounicode{dollarmonospace}{FF04}
\pdfglyphtounicode{dollaroldstyle}{0024}
\pdfglyphtounicode{dollarsmall}{FE69}
\pdfglyphtounicode{dollarsuperior}{0024}
\pdfglyphtounicode{dong}{20AB}
\pdfglyphtounicode{dorusquare}{3326}
\pdfglyphtounicode{dotaccent}{02D9}
\pdfglyphtounicode{dotaccentcmb}{0307}
\pdfglyphtounicode{dotbelowcmb}{0323}
\pdfglyphtounicode{dotbelowcomb}{0323}
\pdfglyphtounicode{dotkatakana}{30FB}
\pdfglyphtounicode{dotlessi}{0131}
\pdfglyphtounicode{dotlessj}{0237}
\pdfglyphtounicode{dotlessjstrokehook}{0284}
\pdfglyphtounicode{dotmath}{22C5}
\pdfglyphtounicode{dotplus}{2214}
\pdfglyphtounicode{dottedcircle}{25CC}
\pdfglyphtounicode{doubleyodpatah}{FB1F}
\pdfglyphtounicode{doubleyodpatahhebrew}{FB1F}
\pdfglyphtounicode{downfall}{22CE}
\pdfglyphtounicode{downslope}{29F9}
\pdfglyphtounicode{downtackbelowcmb}{031E}
\pdfglyphtounicode{downtackmod}{02D5}
\pdfglyphtounicode{dparen}{249F}
\pdfglyphtounicode{dsuperior}{0064}
\pdfglyphtounicode{dtail}{0256}
\pdfglyphtounicode{dtopbar}{018C}
\pdfglyphtounicode{duhiragana}{3065}
\pdfglyphtounicode{dukatakana}{30C5}
\pdfglyphtounicode{dz}{01F3}
\pdfglyphtounicode{dzaltone}{02A3}
\pdfglyphtounicode{dzcaron}{01C6}
\pdfglyphtounicode{dzcurl}{02A5}
\pdfglyphtounicode{dzeabkhasiancyrillic}{04E1}
\pdfglyphtounicode{dzecyrillic}{0455}
\pdfglyphtounicode{dzhecyrillic}{045F}
\pdfglyphtounicode{e}{0065}
\pdfglyphtounicode{eacute}{00E9}
\pdfglyphtounicode{earth}{2641}
\pdfglyphtounicode{ebengali}{098F}
\pdfglyphtounicode{ebopomofo}{311C}
\pdfglyphtounicode{ebreve}{0115}
\pdfglyphtounicode{ecandradeva}{090D}
\pdfglyphtounicode{ecandragujarati}{0A8D}
\pdfglyphtounicode{ecandravowelsigndeva}{0945}
\pdfglyphtounicode{ecandravowelsigngujarati}{0AC5}
\pdfglyphtounicode{ecaron}{011B}
\pdfglyphtounicode{ecedillabreve}{1E1D}
\pdfglyphtounicode{echarmenian}{0565}
\pdfglyphtounicode{echyiwnarmenian}{0587}
\pdfglyphtounicode{ecircle}{24D4}
\pdfglyphtounicode{ecircumflex}{00EA}
\pdfglyphtounicode{ecircumflexacute}{1EBF}
\pdfglyphtounicode{ecircumflexbelow}{1E19}
\pdfglyphtounicode{ecircumflexdotbelow}{1EC7}
\pdfglyphtounicode{ecircumflexgrave}{1EC1}
\pdfglyphtounicode{ecircumflexhookabove}{1EC3}
\pdfglyphtounicode{ecircumflextilde}{1EC5}
\pdfglyphtounicode{ecyrillic}{0454}
\pdfglyphtounicode{edblgrave}{0205}
\pdfglyphtounicode{edeva}{090F}
\pdfglyphtounicode{edieresis}{00EB}
\pdfglyphtounicode{edot}{0117}
\pdfglyphtounicode{edotaccent}{0117}
\pdfglyphtounicode{edotbelow}{1EB9}
\pdfglyphtounicode{eegurmukhi}{0A0F}
\pdfglyphtounicode{eematragurmukhi}{0A47}
\pdfglyphtounicode{efcyrillic}{0444}
\pdfglyphtounicode{egrave}{00E8}
\pdfglyphtounicode{egujarati}{0A8F}
\pdfglyphtounicode{eharmenian}{0567}
\pdfglyphtounicode{ehbopomofo}{311D}
\pdfglyphtounicode{ehiragana}{3048}
\pdfglyphtounicode{ehookabove}{1EBB}
\pdfglyphtounicode{eibopomofo}{311F}
\pdfglyphtounicode{eight}{0038}
\pdfglyphtounicode{eightarabic}{0668}
\pdfglyphtounicode{eightbengali}{09EE}
\pdfglyphtounicode{eightcircle}{2467}
\pdfglyphtounicode{eightcircleinversesansserif}{2791}
\pdfglyphtounicode{eightdeva}{096E}
\pdfglyphtounicode{eighteencircle}{2471}
\pdfglyphtounicode{eighteenparen}{2485}
\pdfglyphtounicode{eighteenperiod}{2499}
\pdfglyphtounicode{eightgujarati}{0AEE}
\pdfglyphtounicode{eightgurmukhi}{0A6E}
\pdfglyphtounicode{eighthackarabic}{0668}
\pdfglyphtounicode{eighthangzhou}{3028}
\pdfglyphtounicode{eighthnotebeamed}{266B}
\pdfglyphtounicode{eightideographicparen}{3227}
\pdfglyphtounicode{eightinferior}{2088}
\pdfglyphtounicode{eightmonospace}{FF18}
\pdfglyphtounicode{eightoldstyle}{0038}
\pdfglyphtounicode{eightparen}{247B}
\pdfglyphtounicode{eightperiod}{248F}
\pdfglyphtounicode{eightpersian}{06F8}
\pdfglyphtounicode{eightroman}{2177}
\pdfglyphtounicode{eightsuperior}{2078}
\pdfglyphtounicode{eightthai}{0E58}
\pdfglyphtounicode{einvertedbreve}{0207}
\pdfglyphtounicode{eiotifiedcyrillic}{0465}
\pdfglyphtounicode{ekatakana}{30A8}
\pdfglyphtounicode{ekatakanahalfwidth}{FF74}
\pdfglyphtounicode{ekonkargurmukhi}{0A74}
\pdfglyphtounicode{ekorean}{3154}
\pdfglyphtounicode{elcyrillic}{043B}
\pdfglyphtounicode{element}{2208}
\pdfglyphtounicode{elevencircle}{246A}
\pdfglyphtounicode{elevenparen}{247E}
\pdfglyphtounicode{elevenperiod}{2492}
\pdfglyphtounicode{elevenroman}{217A}
\pdfglyphtounicode{ellipsis}{2026}
\pdfglyphtounicode{ellipsisvertical}{22EE}
\pdfglyphtounicode{emacron}{0113}
\pdfglyphtounicode{emacronacute}{1E17}
\pdfglyphtounicode{emacrongrave}{1E15}
\pdfglyphtounicode{emcyrillic}{043C}
\pdfglyphtounicode{emdash}{2014}
\pdfglyphtounicode{emdashvertical}{FE31}
\pdfglyphtounicode{emonospace}{FF45}
\pdfglyphtounicode{emphasismarkarmenian}{055B}
\pdfglyphtounicode{emptyset}{2205}
\pdfglyphtounicode{enbopomofo}{3123}
\pdfglyphtounicode{encyrillic}{043D}
\pdfglyphtounicode{endash}{2013}
\pdfglyphtounicode{endashvertical}{FE32}
\pdfglyphtounicode{endescendercyrillic}{04A3}
\pdfglyphtounicode{eng}{014B}
\pdfglyphtounicode{engbopomofo}{3125}
\pdfglyphtounicode{enghecyrillic}{04A5}
\pdfglyphtounicode{enhookcyrillic}{04C8}
\pdfglyphtounicode{enspace}{2002}
\pdfglyphtounicode{eogonek}{0119}
\pdfglyphtounicode{eokorean}{3153}
\pdfglyphtounicode{eopen}{025B}
\pdfglyphtounicode{eopenclosed}{029A}
\pdfglyphtounicode{eopenreversed}{025C}
\pdfglyphtounicode{eopenreversedclosed}{025E}
\pdfglyphtounicode{eopenreversedhook}{025D}
\pdfglyphtounicode{eparen}{24A0}
\pdfglyphtounicode{epsilon}{03B5}
\pdfglyphtounicode{epsilon1}{03F5}
\pdfglyphtounicode{epsiloninv}{03F6}
\pdfglyphtounicode{epsilontonos}{03AD}
\pdfglyphtounicode{equal}{003D}
\pdfglyphtounicode{equaldotleftright}{2252}
\pdfglyphtounicode{equaldotrightleft}{2253}
\pdfglyphtounicode{equalmonospace}{FF1D}
\pdfglyphtounicode{equalorfollows}{22DF}
\pdfglyphtounicode{equalorgreater}{2A96}
\pdfglyphtounicode{equalorless}{2A95}
\pdfglyphtounicode{equalorprecedes}{22DE}
\pdfglyphtounicode{equalorsimilar}{2242}
\pdfglyphtounicode{equalsdots}{2251}
\pdfglyphtounicode{equalsmall}{FE66}
\pdfglyphtounicode{equalsuperior}{207C}
\pdfglyphtounicode{equivalence}{2261}
\pdfglyphtounicode{equivasymptotic}{224D}
\pdfglyphtounicode{erbopomofo}{3126}
\pdfglyphtounicode{ercyrillic}{0440}
\pdfglyphtounicode{ereversed}{0258}
\pdfglyphtounicode{ereversedcyrillic}{044D}
\pdfglyphtounicode{escyrillic}{0441}
\pdfglyphtounicode{esdescendercyrillic}{04AB}
\pdfglyphtounicode{esh}{0283}
\pdfglyphtounicode{eshcurl}{0286}
\pdfglyphtounicode{eshortdeva}{090E}
\pdfglyphtounicode{eshortvowelsigndeva}{0946}
\pdfglyphtounicode{eshreversedloop}{01AA}
\pdfglyphtounicode{eshsquatreversed}{0285}
\pdfglyphtounicode{esmallhiragana}{3047}
\pdfglyphtounicode{esmallkatakana}{30A7}
\pdfglyphtounicode{esmallkatakanahalfwidth}{FF6A}
\pdfglyphtounicode{estimated}{212E}
\pdfglyphtounicode{esuperior}{0065}
\pdfglyphtounicode{eta}{03B7}
\pdfglyphtounicode{etarmenian}{0568}
\pdfglyphtounicode{etatonos}{03AE}
\pdfglyphtounicode{eth}{00F0}
\pdfglyphtounicode{etilde}{1EBD}
\pdfglyphtounicode{etildebelow}{1E1B}
\pdfglyphtounicode{etnahtafoukhhebrew}{0591}
\pdfglyphtounicode{etnahtafoukhlefthebrew}{0591}
\pdfglyphtounicode{etnahtahebrew}{0591}
\pdfglyphtounicode{etnahtalefthebrew}{0591}
\pdfglyphtounicode{eturned}{01DD}
\pdfglyphtounicode{eukorean}{3161}
\pdfglyphtounicode{euro}{20AC}
\pdfglyphtounicode{evowelsignbengali}{09C7}
\pdfglyphtounicode{evowelsigndeva}{0947}
\pdfglyphtounicode{evowelsigngujarati}{0AC7}
\pdfglyphtounicode{exclam}{0021}
\pdfglyphtounicode{exclamarmenian}{055C}
\pdfglyphtounicode{exclamdbl}{203C}
\pdfglyphtounicode{exclamdown}{00A1}
\pdfglyphtounicode{exclamdownsmall}{00A1}
\pdfglyphtounicode{exclammonospace}{FF01}
\pdfglyphtounicode{exclamsmall}{0021}
\pdfglyphtounicode{existential}{2203}
\pdfglyphtounicode{ezh}{0292}
\pdfglyphtounicode{ezhcaron}{01EF}
\pdfglyphtounicode{ezhcurl}{0293}
\pdfglyphtounicode{ezhreversed}{01B9}
\pdfglyphtounicode{ezhtail}{01BA}
\pdfglyphtounicode{f}{0066}
\pdfglyphtounicode{fadeva}{095E}
\pdfglyphtounicode{fagurmukhi}{0A5E}
\pdfglyphtounicode{fahrenheit}{2109}
\pdfglyphtounicode{fathaarabic}{064E}
\pdfglyphtounicode{fathalowarabic}{064E}
\pdfglyphtounicode{fathatanarabic}{064B}
\pdfglyphtounicode{fbopomofo}{3108}
\pdfglyphtounicode{fcircle}{24D5}
\pdfglyphtounicode{fdotaccent}{1E1F}
\pdfglyphtounicode{feharabic}{0641}
\pdfglyphtounicode{feharmenian}{0586}
\pdfglyphtounicode{fehfinalarabic}{FED2}
\pdfglyphtounicode{fehinitialarabic}{FED3}
\pdfglyphtounicode{fehmedialarabic}{FED4}
\pdfglyphtounicode{feicoptic}{03E5}
\pdfglyphtounicode{female}{2640}
\pdfglyphtounicode{ff}{0066 0066}
\pdfglyphtounicode{ffi}{0066 0066 0069}
\pdfglyphtounicode{ffl}{0066 0066 006C}
\pdfglyphtounicode{fi}{0066 0069}
\pdfglyphtounicode{fifteencircle}{246E}
\pdfglyphtounicode{fifteenparen}{2482}
\pdfglyphtounicode{fifteenperiod}{2496}
\pdfglyphtounicode{figuredash}{2012}
\pdfglyphtounicode{filledbox}{25A0}
\pdfglyphtounicode{filledrect}{25AC}
\pdfglyphtounicode{finalkaf}{05DA}
\pdfglyphtounicode{finalkafdagesh}{FB3A}
\pdfglyphtounicode{finalkafdageshhebrew}{FB3A}
\pdfglyphtounicode{finalkafhebrew}{05DA}
\pdfglyphtounicode{finalkafqamats}{05DA 05B8}
\pdfglyphtounicode{finalkafqamatshebrew}{05DA 05B8}
\pdfglyphtounicode{finalkafsheva}{05DA 05B0}
\pdfglyphtounicode{finalkafshevahebrew}{05DA 05B0}
\pdfglyphtounicode{finalmem}{05DD}
\pdfglyphtounicode{finalmemhebrew}{05DD}
\pdfglyphtounicode{finalnun}{05DF}
\pdfglyphtounicode{finalnunhebrew}{05DF}
\pdfglyphtounicode{finalpe}{05E3}
\pdfglyphtounicode{finalpehebrew}{05E3}
\pdfglyphtounicode{finaltsadi}{05E5}
\pdfglyphtounicode{finaltsadihebrew}{05E5}
\pdfglyphtounicode{firsttonechinese}{02C9}
\pdfglyphtounicode{fisheye}{25C9}
\pdfglyphtounicode{fitacyrillic}{0473}
\pdfglyphtounicode{five}{0035}
\pdfglyphtounicode{fivearabic}{0665}
\pdfglyphtounicode{fivebengali}{09EB}
\pdfglyphtounicode{fivecircle}{2464}
\pdfglyphtounicode{fivecircleinversesansserif}{278E}
\pdfglyphtounicode{fivedeva}{096B}
\pdfglyphtounicode{fiveeighths}{215D}
\pdfglyphtounicode{fivegujarati}{0AEB}
\pdfglyphtounicode{fivegurmukhi}{0A6B}
\pdfglyphtounicode{fivehackarabic}{0665}
\pdfglyphtounicode{fivehangzhou}{3025}
\pdfglyphtounicode{fiveideographicparen}{3224}
\pdfglyphtounicode{fiveinferior}{2085}
\pdfglyphtounicode{fivemonospace}{FF15}
\pdfglyphtounicode{fiveoldstyle}{0035}
\pdfglyphtounicode{fiveparen}{2478}
\pdfglyphtounicode{fiveperiod}{248C}
\pdfglyphtounicode{fivepersian}{06F5}
\pdfglyphtounicode{fiveroman}{2174}
\pdfglyphtounicode{fivesuperior}{2075}
\pdfglyphtounicode{fivethai}{0E55}
\pdfglyphtounicode{fl}{0066 006C}
\pdfglyphtounicode{flat}{266D}
\pdfglyphtounicode{floorleft}{230A}
\pdfglyphtounicode{floorright}{230B}
\pdfglyphtounicode{florin}{0192}
\pdfglyphtounicode{fmonospace}{FF46}
\pdfglyphtounicode{fmsquare}{3399}
\pdfglyphtounicode{fofanthai}{0E1F}
\pdfglyphtounicode{fofathai}{0E1D}
\pdfglyphtounicode{follownotdbleqv}{2ABA}
\pdfglyphtounicode{follownotslnteql}{2AB6}
\pdfglyphtounicode{followornoteqvlnt}{22E9}
\pdfglyphtounicode{follows}{227B}
\pdfglyphtounicode{followsequal}{2AB0}
\pdfglyphtounicode{followsorcurly}{227D}
\pdfglyphtounicode{followsorequal}{227F}
\pdfglyphtounicode{fongmanthai}{0E4F}
\pdfglyphtounicode{forall}{2200}
\pdfglyphtounicode{forces}{22A9}
\pdfglyphtounicode{forcesbar}{22AA}
\pdfglyphtounicode{fork}{22D4}
\pdfglyphtounicode{four}{0034}
\pdfglyphtounicode{fourarabic}{0664}
\pdfglyphtounicode{fourbengali}{09EA}
\pdfglyphtounicode{fourcircle}{2463}
\pdfglyphtounicode{fourcircleinversesansserif}{278D}
\pdfglyphtounicode{fourdeva}{096A}
\pdfglyphtounicode{fourgujarati}{0AEA}
\pdfglyphtounicode{fourgurmukhi}{0A6A}
\pdfglyphtounicode{fourhackarabic}{0664}
\pdfglyphtounicode{fourhangzhou}{3024}
\pdfglyphtounicode{fourideographicparen}{3223}
\pdfglyphtounicode{fourinferior}{2084}
\pdfglyphtounicode{fourmonospace}{FF14}
\pdfglyphtounicode{fournumeratorbengali}{09F7}
\pdfglyphtounicode{fouroldstyle}{0034}
\pdfglyphtounicode{fourparen}{2477}
\pdfglyphtounicode{fourperiod}{248B}
\pdfglyphtounicode{fourpersian}{06F4}
\pdfglyphtounicode{fourroman}{2173}
\pdfglyphtounicode{foursuperior}{2074}
\pdfglyphtounicode{fourteencircle}{246D}
\pdfglyphtounicode{fourteenparen}{2481}
\pdfglyphtounicode{fourteenperiod}{2495}
\pdfglyphtounicode{fourthai}{0E54}
\pdfglyphtounicode{fourthtonechinese}{02CB}
\pdfglyphtounicode{fparen}{24A1}
\pdfglyphtounicode{fraction}{2044}
\pdfglyphtounicode{franc}{20A3}
\pdfglyphtounicode{frown}{2322}
\pdfglyphtounicode{g}{0067}
\pdfglyphtounicode{gabengali}{0997}
\pdfglyphtounicode{gacute}{01F5}
\pdfglyphtounicode{gadeva}{0917}
\pdfglyphtounicode{gafarabic}{06AF}
\pdfglyphtounicode{gaffinalarabic}{FB93}
\pdfglyphtounicode{gafinitialarabic}{FB94}
\pdfglyphtounicode{gafmedialarabic}{FB95}
\pdfglyphtounicode{gagujarati}{0A97}
\pdfglyphtounicode{gagurmukhi}{0A17}
\pdfglyphtounicode{gahiragana}{304C}
\pdfglyphtounicode{gakatakana}{30AC}
\pdfglyphtounicode{gamma}{03B3}
\pdfglyphtounicode{gammalatinsmall}{0263}
\pdfglyphtounicode{gammasuperior}{02E0}
\pdfglyphtounicode{gangiacoptic}{03EB}
\pdfglyphtounicode{gbopomofo}{310D}
\pdfglyphtounicode{gbreve}{011F}
\pdfglyphtounicode{gcaron}{01E7}
\pdfglyphtounicode{gcedilla}{0123}
\pdfglyphtounicode{gcircle}{24D6}
\pdfglyphtounicode{gcircumflex}{011D}
\pdfglyphtounicode{gcommaaccent}{0123}
\pdfglyphtounicode{gdot}{0121}
\pdfglyphtounicode{gdotaccent}{0121}
\pdfglyphtounicode{gecyrillic}{0433}
\pdfglyphtounicode{gehiragana}{3052}
\pdfglyphtounicode{gekatakana}{30B2}
\pdfglyphtounicode{geomequivalent}{224E}
\pdfglyphtounicode{geometricallyequal}{2251}
\pdfglyphtounicode{gereshaccenthebrew}{059C}
\pdfglyphtounicode{gereshhebrew}{05F3}
\pdfglyphtounicode{gereshmuqdamhebrew}{059D}
\pdfglyphtounicode{germandbls}{00DF}
\pdfglyphtounicode{gershayimaccenthebrew}{059E}
\pdfglyphtounicode{gershayimhebrew}{05F4}
\pdfglyphtounicode{getamark}{3013}
\pdfglyphtounicode{ghabengali}{0998}
\pdfglyphtounicode{ghadarmenian}{0572}
\pdfglyphtounicode{ghadeva}{0918}
\pdfglyphtounicode{ghagujarati}{0A98}
\pdfglyphtounicode{ghagurmukhi}{0A18}
\pdfglyphtounicode{ghainarabic}{063A}
\pdfglyphtounicode{ghainfinalarabic}{FECE}
\pdfglyphtounicode{ghaininitialarabic}{FECF}
\pdfglyphtounicode{ghainmedialarabic}{FED0}
\pdfglyphtounicode{ghemiddlehookcyrillic}{0495}
\pdfglyphtounicode{ghestrokecyrillic}{0493}
\pdfglyphtounicode{gheupturncyrillic}{0491}
\pdfglyphtounicode{ghhadeva}{095A}
\pdfglyphtounicode{ghhagurmukhi}{0A5A}
\pdfglyphtounicode{ghook}{0260}
\pdfglyphtounicode{ghzsquare}{3393}
\pdfglyphtounicode{gihiragana}{304E}
\pdfglyphtounicode{gikatakana}{30AE}
\pdfglyphtounicode{gimarmenian}{0563}
\pdfglyphtounicode{gimel}{05D2}
\pdfglyphtounicode{gimeldagesh}{FB32}
\pdfglyphtounicode{gimeldageshhebrew}{FB32}
\pdfglyphtounicode{gimelhebrew}{05D2}
\pdfglyphtounicode{gjecyrillic}{0453}
\pdfglyphtounicode{glottalinvertedstroke}{01BE}
\pdfglyphtounicode{glottalstop}{0294}
\pdfglyphtounicode{glottalstopinverted}{0296}
\pdfglyphtounicode{glottalstopmod}{02C0}
\pdfglyphtounicode{glottalstopreversed}{0295}
\pdfglyphtounicode{glottalstopreversedmod}{02C1}
\pdfglyphtounicode{glottalstopreversedsuperior}{02E4}
\pdfglyphtounicode{glottalstopstroke}{02A1}
\pdfglyphtounicode{glottalstopstrokereversed}{02A2}
\pdfglyphtounicode{gmacron}{1E21}
\pdfglyphtounicode{gmonospace}{FF47}
\pdfglyphtounicode{gohiragana}{3054}
\pdfglyphtounicode{gokatakana}{30B4}
\pdfglyphtounicode{gparen}{24A2}
\pdfglyphtounicode{gpasquare}{33AC}
\pdfglyphtounicode{gradient}{2207}
\pdfglyphtounicode{grave}{0060}
\pdfglyphtounicode{gravebelowcmb}{0316}
\pdfglyphtounicode{gravecmb}{0300}
\pdfglyphtounicode{gravecomb}{0300}
\pdfglyphtounicode{gravedeva}{0953}
\pdfglyphtounicode{gravelowmod}{02CE}
\pdfglyphtounicode{gravemonospace}{FF40}
\pdfglyphtounicode{gravetonecmb}{0340}
\pdfglyphtounicode{greater}{003E}
\pdfglyphtounicode{greaterdbleqlless}{2A8C}
\pdfglyphtounicode{greaterdblequal}{2267}
\pdfglyphtounicode{greaterdot}{22D7}
\pdfglyphtounicode{greaterequal}{2265}
\pdfglyphtounicode{greaterequalorless}{22DB}
\pdfglyphtounicode{greaterlessequal}{22DB}
\pdfglyphtounicode{greatermonospace}{FF1E}
\pdfglyphtounicode{greatermuch}{226B}
\pdfglyphtounicode{greaternotdblequal}{2A8A}
\pdfglyphtounicode{greaternotequal}{2A88}
\pdfglyphtounicode{greaterorapproxeql}{2A86}
\pdfglyphtounicode{greaterorequalslant}{2A7E}
\pdfglyphtounicode{greaterorequivalent}{2273}
\pdfglyphtounicode{greaterorless}{2277}
\pdfglyphtounicode{greaterornotdbleql}{2269}
\pdfglyphtounicode{greaterornotequal}{2269}
\pdfglyphtounicode{greaterorsimilar}{2273}
\pdfglyphtounicode{greateroverequal}{2267}
\pdfglyphtounicode{greatersmall}{FE65}
\pdfglyphtounicode{gscript}{0261}
\pdfglyphtounicode{gstroke}{01E5}
\pdfglyphtounicode{guhiragana}{3050}
\pdfglyphtounicode{guillemotleft}{00AB}
\pdfglyphtounicode{guillemotright}{00BB}
\pdfglyphtounicode{guilsinglleft}{2039}
\pdfglyphtounicode{guilsinglright}{203A}
\pdfglyphtounicode{gukatakana}{30B0}
\pdfglyphtounicode{guramusquare}{3318}
\pdfglyphtounicode{gysquare}{33C9}
\pdfglyphtounicode{h}{0068}
\pdfglyphtounicode{haabkhasiancyrillic}{04A9}
\pdfglyphtounicode{haaltonearabic}{06C1}
\pdfglyphtounicode{habengali}{09B9}
\pdfglyphtounicode{hadescendercyrillic}{04B3}
\pdfglyphtounicode{hadeva}{0939}
\pdfglyphtounicode{hagujarati}{0AB9}
\pdfglyphtounicode{hagurmukhi}{0A39}
\pdfglyphtounicode{haharabic}{062D}
\pdfglyphtounicode{hahfinalarabic}{FEA2}
\pdfglyphtounicode{hahinitialarabic}{FEA3}
\pdfglyphtounicode{hahiragana}{306F}
\pdfglyphtounicode{hahmedialarabic}{FEA4}
\pdfglyphtounicode{haitusquare}{332A}
\pdfglyphtounicode{hakatakana}{30CF}
\pdfglyphtounicode{hakatakanahalfwidth}{FF8A}
\pdfglyphtounicode{halantgurmukhi}{0A4D}
\pdfglyphtounicode{hamzaarabic}{0621}
\pdfglyphtounicode{hamzadammaarabic}{0621 064F}
\pdfglyphtounicode{hamzadammatanarabic}{0621 064C}
\pdfglyphtounicode{hamzafathaarabic}{0621 064E}
\pdfglyphtounicode{hamzafathatanarabic}{0621 064B}
\pdfglyphtounicode{hamzalowarabic}{0621}
\pdfglyphtounicode{hamzalowkasraarabic}{0621 0650}
\pdfglyphtounicode{hamzalowkasratanarabic}{0621 064D}
\pdfglyphtounicode{hamzasukunarabic}{0621 0652}
\pdfglyphtounicode{hangulfiller}{3164}
\pdfglyphtounicode{hardsigncyrillic}{044A}
\pdfglyphtounicode{harpoondownleft}{21C3}
\pdfglyphtounicode{harpoondownright}{21C2}
\pdfglyphtounicode{harpoonleftbarbup}{21BC}
\pdfglyphtounicode{harpoonleftright}{21CC}
\pdfglyphtounicode{harpoonrightbarbup}{21C0}
\pdfglyphtounicode{harpoonrightleft}{21CB}
\pdfglyphtounicode{harpoonupleft}{21BF}
\pdfglyphtounicode{harpoonupright}{21BE}
\pdfglyphtounicode{hasquare}{33CA}
\pdfglyphtounicode{hatafpatah}{05B2}
\pdfglyphtounicode{hatafpatah16}{05B2}
\pdfglyphtounicode{hatafpatah23}{05B2}
\pdfglyphtounicode{hatafpatah2f}{05B2}
\pdfglyphtounicode{hatafpatahhebrew}{05B2}
\pdfglyphtounicode{hatafpatahnarrowhebrew}{05B2}
\pdfglyphtounicode{hatafpatahquarterhebrew}{05B2}
\pdfglyphtounicode{hatafpatahwidehebrew}{05B2}
\pdfglyphtounicode{hatafqamats}{05B3}
\pdfglyphtounicode{hatafqamats1b}{05B3}
\pdfglyphtounicode{hatafqamats28}{05B3}
\pdfglyphtounicode{hatafqamats34}{05B3}
\pdfglyphtounicode{hatafqamatshebrew}{05B3}
\pdfglyphtounicode{hatafqamatsnarrowhebrew}{05B3}
\pdfglyphtounicode{hatafqamatsquarterhebrew}{05B3}
\pdfglyphtounicode{hatafqamatswidehebrew}{05B3}
\pdfglyphtounicode{hatafsegol}{05B1}
\pdfglyphtounicode{hatafsegol17}{05B1}
\pdfglyphtounicode{hatafsegol24}{05B1}
\pdfglyphtounicode{hatafsegol30}{05B1}
\pdfglyphtounicode{hatafsegolhebrew}{05B1}
\pdfglyphtounicode{hatafsegolnarrowhebrew}{05B1}
\pdfglyphtounicode{hatafsegolquarterhebrew}{05B1}
\pdfglyphtounicode{hatafsegolwidehebrew}{05B1}
\pdfglyphtounicode{hbar}{0127}
\pdfglyphtounicode{hbopomofo}{310F}
\pdfglyphtounicode{hbrevebelow}{1E2B}
\pdfglyphtounicode{hcedilla}{1E29}
\pdfglyphtounicode{hcircle}{24D7}
\pdfglyphtounicode{hcircumflex}{0125}
\pdfglyphtounicode{hdieresis}{1E27}
\pdfglyphtounicode{hdotaccent}{1E23}
\pdfglyphtounicode{hdotbelow}{1E25}
\pdfglyphtounicode{he}{05D4}
\pdfglyphtounicode{heart}{2665}
\pdfglyphtounicode{heartsuitblack}{2665}
\pdfglyphtounicode{heartsuitwhite}{2661}
\pdfglyphtounicode{hedagesh}{FB34}
\pdfglyphtounicode{hedageshhebrew}{FB34}
\pdfglyphtounicode{hehaltonearabic}{06C1}
\pdfglyphtounicode{heharabic}{0647}
\pdfglyphtounicode{hehebrew}{05D4}
\pdfglyphtounicode{hehfinalaltonearabic}{FBA7}
\pdfglyphtounicode{hehfinalalttwoarabic}{FEEA}
\pdfglyphtounicode{hehfinalarabic}{FEEA}
\pdfglyphtounicode{hehhamzaabovefinalarabic}{FBA5}
\pdfglyphtounicode{hehhamzaaboveisolatedarabic}{FBA4}
\pdfglyphtounicode{hehinitialaltonearabic}{FBA8}
\pdfglyphtounicode{hehinitialarabic}{FEEB}
\pdfglyphtounicode{hehiragana}{3078}
\pdfglyphtounicode{hehmedialaltonearabic}{FBA9}
\pdfglyphtounicode{hehmedialarabic}{FEEC}
\pdfglyphtounicode{heiseierasquare}{337B}
\pdfglyphtounicode{hekatakana}{30D8}
\pdfglyphtounicode{hekatakanahalfwidth}{FF8D}
\pdfglyphtounicode{hekutaarusquare}{3336}
\pdfglyphtounicode{henghook}{0267}
\pdfglyphtounicode{herutusquare}{3339}
\pdfglyphtounicode{het}{05D7}
\pdfglyphtounicode{hethebrew}{05D7}
\pdfglyphtounicode{hhook}{0266}
\pdfglyphtounicode{hhooksuperior}{02B1}
\pdfglyphtounicode{hieuhacirclekorean}{327B}
\pdfglyphtounicode{hieuhaparenkorean}{321B}
\pdfglyphtounicode{hieuhcirclekorean}{326D}
\pdfglyphtounicode{hieuhkorean}{314E}
\pdfglyphtounicode{hieuhparenkorean}{320D}
\pdfglyphtounicode{hihiragana}{3072}
\pdfglyphtounicode{hikatakana}{30D2}
\pdfglyphtounicode{hikatakanahalfwidth}{FF8B}
\pdfglyphtounicode{hiriq}{05B4}
\pdfglyphtounicode{hiriq14}{05B4}
\pdfglyphtounicode{hiriq21}{05B4}
\pdfglyphtounicode{hiriq2d}{05B4}
\pdfglyphtounicode{hiriqhebrew}{05B4}
\pdfglyphtounicode{hiriqnarrowhebrew}{05B4}
\pdfglyphtounicode{hiriqquarterhebrew}{05B4}
\pdfglyphtounicode{hiriqwidehebrew}{05B4}
\pdfglyphtounicode{hlinebelow}{1E96}
\pdfglyphtounicode{hmonospace}{FF48}
\pdfglyphtounicode{hoarmenian}{0570}
\pdfglyphtounicode{hohipthai}{0E2B}
\pdfglyphtounicode{hohiragana}{307B}
\pdfglyphtounicode{hokatakana}{30DB}
\pdfglyphtounicode{hokatakanahalfwidth}{FF8E}
\pdfglyphtounicode{holam}{05B9}
\pdfglyphtounicode{holam19}{05B9}
\pdfglyphtounicode{holam26}{05B9}
\pdfglyphtounicode{holam32}{05B9}
\pdfglyphtounicode{holamhebrew}{05B9}
\pdfglyphtounicode{holamnarrowhebrew}{05B9}
\pdfglyphtounicode{holamquarterhebrew}{05B9}
\pdfglyphtounicode{holamwidehebrew}{05B9}
\pdfglyphtounicode{honokhukthai}{0E2E}
\pdfglyphtounicode{hookabovecomb}{0309}
\pdfglyphtounicode{hookcmb}{0309}
\pdfglyphtounicode{hookpalatalizedbelowcmb}{0321}
\pdfglyphtounicode{hookretroflexbelowcmb}{0322}
\pdfglyphtounicode{hoonsquare}{3342}
\pdfglyphtounicode{horicoptic}{03E9}
\pdfglyphtounicode{horizontalbar}{2015}
\pdfglyphtounicode{horncmb}{031B}
\pdfglyphtounicode{hotsprings}{2668}
\pdfglyphtounicode{house}{2302}
\pdfglyphtounicode{hparen}{24A3}
\pdfglyphtounicode{hsuperior}{02B0}
\pdfglyphtounicode{hturned}{0265}
\pdfglyphtounicode{huhiragana}{3075}
\pdfglyphtounicode{huiitosquare}{3333}
\pdfglyphtounicode{hukatakana}{30D5}
\pdfglyphtounicode{hukatakanahalfwidth}{FF8C}
\pdfglyphtounicode{hungarumlaut}{02DD}
\pdfglyphtounicode{hungarumlautcmb}{030B}
\pdfglyphtounicode{hv}{0195}
\pdfglyphtounicode{hyphen}{002D}
\pdfglyphtounicode{hyphenchar}{002D}
\pdfglyphtounicode{hypheninferior}{002D}
\pdfglyphtounicode{hyphenmonospace}{FF0D}
\pdfglyphtounicode{hyphensmall}{FE63}
\pdfglyphtounicode{hyphensuperior}{002D}
\pdfglyphtounicode{hyphentwo}{2010}
\pdfglyphtounicode{i}{0069}
\pdfglyphtounicode{iacute}{00ED}
\pdfglyphtounicode{iacyrillic}{044F}
\pdfglyphtounicode{ibengali}{0987}
\pdfglyphtounicode{ibopomofo}{3127}
\pdfglyphtounicode{ibreve}{012D}
\pdfglyphtounicode{icaron}{01D0}
\pdfglyphtounicode{icircle}{24D8}
\pdfglyphtounicode{icircumflex}{00EE}
\pdfglyphtounicode{icyrillic}{0456}
\pdfglyphtounicode{idblgrave}{0209}
\pdfglyphtounicode{ideographearthcircle}{328F}
\pdfglyphtounicode{ideographfirecircle}{328B}
\pdfglyphtounicode{ideographicallianceparen}{323F}
\pdfglyphtounicode{ideographiccallparen}{323A}
\pdfglyphtounicode{ideographiccentrecircle}{32A5}
\pdfglyphtounicode{ideographicclose}{3006}
\pdfglyphtounicode{ideographiccomma}{3001}
\pdfglyphtounicode{ideographiccommaleft}{FF64}
\pdfglyphtounicode{ideographiccongratulationparen}{3237}
\pdfglyphtounicode{ideographiccorrectcircle}{32A3}
\pdfglyphtounicode{ideographicearthparen}{322F}
\pdfglyphtounicode{ideographicenterpriseparen}{323D}
\pdfglyphtounicode{ideographicexcellentcircle}{329D}
\pdfglyphtounicode{ideographicfestivalparen}{3240}
\pdfglyphtounicode{ideographicfinancialcircle}{3296}
\pdfglyphtounicode{ideographicfinancialparen}{3236}
\pdfglyphtounicode{ideographicfireparen}{322B}
\pdfglyphtounicode{ideographichaveparen}{3232}
\pdfglyphtounicode{ideographichighcircle}{32A4}
\pdfglyphtounicode{ideographiciterationmark}{3005}
\pdfglyphtounicode{ideographiclaborcircle}{3298}
\pdfglyphtounicode{ideographiclaborparen}{3238}
\pdfglyphtounicode{ideographicleftcircle}{32A7}
\pdfglyphtounicode{ideographiclowcircle}{32A6}
\pdfglyphtounicode{ideographicmedicinecircle}{32A9}
\pdfglyphtounicode{ideographicmetalparen}{322E}
\pdfglyphtounicode{ideographicmoonparen}{322A}
\pdfglyphtounicode{ideographicnameparen}{3234}
\pdfglyphtounicode{ideographicperiod}{3002}
\pdfglyphtounicode{ideographicprintcircle}{329E}
\pdfglyphtounicode{ideographicreachparen}{3243}
\pdfglyphtounicode{ideographicrepresentparen}{3239}
\pdfglyphtounicode{ideographicresourceparen}{323E}
\pdfglyphtounicode{ideographicrightcircle}{32A8}
\pdfglyphtounicode{ideographicsecretcircle}{3299}
\pdfglyphtounicode{ideographicselfparen}{3242}
\pdfglyphtounicode{ideographicsocietyparen}{3233}
\pdfglyphtounicode{ideographicspace}{3000}
\pdfglyphtounicode{ideographicspecialparen}{3235}
\pdfglyphtounicode{ideographicstockparen}{3231}
\pdfglyphtounicode{ideographicstudyparen}{323B}
\pdfglyphtounicode{ideographicsunparen}{3230}
\pdfglyphtounicode{ideographicsuperviseparen}{323C}
\pdfglyphtounicode{ideographicwaterparen}{322C}
\pdfglyphtounicode{ideographicwoodparen}{322D}
\pdfglyphtounicode{ideographiczero}{3007}
\pdfglyphtounicode{ideographmetalcircle}{328E}
\pdfglyphtounicode{ideographmooncircle}{328A}
\pdfglyphtounicode{ideographnamecircle}{3294}
\pdfglyphtounicode{ideographsuncircle}{3290}
\pdfglyphtounicode{ideographwatercircle}{328C}
\pdfglyphtounicode{ideographwoodcircle}{328D}
\pdfglyphtounicode{ideva}{0907}
\pdfglyphtounicode{idieresis}{00EF}
\pdfglyphtounicode{idieresisacute}{1E2F}
\pdfglyphtounicode{idieresiscyrillic}{04E5}
\pdfglyphtounicode{idotbelow}{1ECB}
\pdfglyphtounicode{iebrevecyrillic}{04D7}
\pdfglyphtounicode{iecyrillic}{0435}
\pdfglyphtounicode{ieungacirclekorean}{3275}
\pdfglyphtounicode{ieungaparenkorean}{3215}
\pdfglyphtounicode{ieungcirclekorean}{3267}
\pdfglyphtounicode{ieungkorean}{3147}
\pdfglyphtounicode{ieungparenkorean}{3207}
\pdfglyphtounicode{igrave}{00EC}
\pdfglyphtounicode{igujarati}{0A87}
\pdfglyphtounicode{igurmukhi}{0A07}
\pdfglyphtounicode{ihiragana}{3044}
\pdfglyphtounicode{ihookabove}{1EC9}
\pdfglyphtounicode{iibengali}{0988}
\pdfglyphtounicode{iicyrillic}{0438}
\pdfglyphtounicode{iideva}{0908}
\pdfglyphtounicode{iigujarati}{0A88}
\pdfglyphtounicode{iigurmukhi}{0A08}
\pdfglyphtounicode{iimatragurmukhi}{0A40}
\pdfglyphtounicode{iinvertedbreve}{020B}
\pdfglyphtounicode{iishortcyrillic}{0439}
\pdfglyphtounicode{iivowelsignbengali}{09C0}
\pdfglyphtounicode{iivowelsigndeva}{0940}
\pdfglyphtounicode{iivowelsigngujarati}{0AC0}
\pdfglyphtounicode{ij}{0133}
\pdfglyphtounicode{ikatakana}{30A4}
\pdfglyphtounicode{ikatakanahalfwidth}{FF72}
\pdfglyphtounicode{ikorean}{3163}
\pdfglyphtounicode{ilde}{02DC}
\pdfglyphtounicode{iluyhebrew}{05AC}
\pdfglyphtounicode{imacron}{012B}
\pdfglyphtounicode{imacroncyrillic}{04E3}
\pdfglyphtounicode{imageorapproximatelyequal}{2253}
\pdfglyphtounicode{imatragurmukhi}{0A3F}
\pdfglyphtounicode{imonospace}{FF49}
\pdfglyphtounicode{increment}{2206}
\pdfglyphtounicode{infinity}{221E}
\pdfglyphtounicode{iniarmenian}{056B}
\pdfglyphtounicode{integerdivide}{2216}
\pdfglyphtounicode{integral}{222B}
\pdfglyphtounicode{integralbottom}{2321}
\pdfglyphtounicode{integralbt}{2321}
\pdfglyphtounicode{integralex}{F8F5}
\pdfglyphtounicode{integraltop}{2320}
\pdfglyphtounicode{integraltp}{2320}
\pdfglyphtounicode{intercal}{22BA}
\pdfglyphtounicode{interrobang}{203D}
\pdfglyphtounicode{interrobangdown}{2E18}
\pdfglyphtounicode{intersection}{2229}
\pdfglyphtounicode{intersectiondbl}{22D2}
\pdfglyphtounicode{intersectionsq}{2293}
\pdfglyphtounicode{intisquare}{3305}
\pdfglyphtounicode{invbullet}{25D8}
\pdfglyphtounicode{invcircle}{25D9}
\pdfglyphtounicode{invsmileface}{263B}
\pdfglyphtounicode{iocyrillic}{0451}
\pdfglyphtounicode{iogonek}{012F}
\pdfglyphtounicode{iota}{03B9}
\pdfglyphtounicode{iotadieresis}{03CA}
\pdfglyphtounicode{iotadieresistonos}{0390}
\pdfglyphtounicode{iotalatin}{0269}
\pdfglyphtounicode{iotatonos}{03AF}
\pdfglyphtounicode{iparen}{24A4}
\pdfglyphtounicode{irigurmukhi}{0A72}
\pdfglyphtounicode{ismallhiragana}{3043}
\pdfglyphtounicode{ismallkatakana}{30A3}
\pdfglyphtounicode{ismallkatakanahalfwidth}{FF68}
\pdfglyphtounicode{issharbengali}{09FA}
\pdfglyphtounicode{istroke}{0268}
\pdfglyphtounicode{isuperior}{0069}
\pdfglyphtounicode{iterationhiragana}{309D}
\pdfglyphtounicode{iterationkatakana}{30FD}
\pdfglyphtounicode{itilde}{0129}
\pdfglyphtounicode{itildebelow}{1E2D}
\pdfglyphtounicode{iubopomofo}{3129}
\pdfglyphtounicode{iucyrillic}{044E}
\pdfglyphtounicode{ivowelsignbengali}{09BF}
\pdfglyphtounicode{ivowelsigndeva}{093F}
\pdfglyphtounicode{ivowelsigngujarati}{0ABF}
\pdfglyphtounicode{izhitsacyrillic}{0475}
\pdfglyphtounicode{izhitsadblgravecyrillic}{0477}
\pdfglyphtounicode{j}{006A}
\pdfglyphtounicode{jaarmenian}{0571}
\pdfglyphtounicode{jabengali}{099C}
\pdfglyphtounicode{jadeva}{091C}
\pdfglyphtounicode{jagujarati}{0A9C}
\pdfglyphtounicode{jagurmukhi}{0A1C}
\pdfglyphtounicode{jbopomofo}{3110}
\pdfglyphtounicode{jcaron}{01F0}
\pdfglyphtounicode{jcircle}{24D9}
\pdfglyphtounicode{jcircumflex}{0135}
\pdfglyphtounicode{jcrossedtail}{029D}
\pdfglyphtounicode{jdotlessstroke}{025F}
\pdfglyphtounicode{jecyrillic}{0458}
\pdfglyphtounicode{jeemarabic}{062C}
\pdfglyphtounicode{jeemfinalarabic}{FE9E}
\pdfglyphtounicode{jeeminitialarabic}{FE9F}
\pdfglyphtounicode{jeemmedialarabic}{FEA0}
\pdfglyphtounicode{jeharabic}{0698}
\pdfglyphtounicode{jehfinalarabic}{FB8B}
\pdfglyphtounicode{jhabengali}{099D}
\pdfglyphtounicode{jhadeva}{091D}
\pdfglyphtounicode{jhagujarati}{0A9D}
\pdfglyphtounicode{jhagurmukhi}{0A1D}
\pdfglyphtounicode{jheharmenian}{057B}
\pdfglyphtounicode{jis}{3004}
\pdfglyphtounicode{jmonospace}{FF4A}
\pdfglyphtounicode{jparen}{24A5}
\pdfglyphtounicode{jsuperior}{02B2}
\pdfglyphtounicode{k}{006B}
\pdfglyphtounicode{kabashkircyrillic}{04A1}
\pdfglyphtounicode{kabengali}{0995}
\pdfglyphtounicode{kacute}{1E31}
\pdfglyphtounicode{kacyrillic}{043A}
\pdfglyphtounicode{kadescendercyrillic}{049B}
\pdfglyphtounicode{kadeva}{0915}
\pdfglyphtounicode{kaf}{05DB}
\pdfglyphtounicode{kafarabic}{0643}
\pdfglyphtounicode{kafdagesh}{FB3B}
\pdfglyphtounicode{kafdageshhebrew}{FB3B}
\pdfglyphtounicode{kaffinalarabic}{FEDA}
\pdfglyphtounicode{kafhebrew}{05DB}
\pdfglyphtounicode{kafinitialarabic}{FEDB}
\pdfglyphtounicode{kafmedialarabic}{FEDC}
\pdfglyphtounicode{kafrafehebrew}{FB4D}
\pdfglyphtounicode{kagujarati}{0A95}
\pdfglyphtounicode{kagurmukhi}{0A15}
\pdfglyphtounicode{kahiragana}{304B}
\pdfglyphtounicode{kahookcyrillic}{04C4}
\pdfglyphtounicode{kakatakana}{30AB}
\pdfglyphtounicode{kakatakanahalfwidth}{FF76}
\pdfglyphtounicode{kappa}{03BA}
\pdfglyphtounicode{kappasymbolgreek}{03F0}
\pdfglyphtounicode{kapyeounmieumkorean}{3171}
\pdfglyphtounicode{kapyeounphieuphkorean}{3184}
\pdfglyphtounicode{kapyeounpieupkorean}{3178}
\pdfglyphtounicode{kapyeounssangpieupkorean}{3179}
\pdfglyphtounicode{karoriisquare}{330D}
\pdfglyphtounicode{kashidaautoarabic}{0640}
\pdfglyphtounicode{kashidaautonosidebearingarabic}{0640}
\pdfglyphtounicode{kasmallkatakana}{30F5}
\pdfglyphtounicode{kasquare}{3384}
\pdfglyphtounicode{kasraarabic}{0650}
\pdfglyphtounicode{kasratanarabic}{064D}
\pdfglyphtounicode{kastrokecyrillic}{049F}
\pdfglyphtounicode{katahiraprolongmarkhalfwidth}{FF70}
\pdfglyphtounicode{kaverticalstrokecyrillic}{049D}
\pdfglyphtounicode{kbopomofo}{310E}
\pdfglyphtounicode{kcalsquare}{3389}
\pdfglyphtounicode{kcaron}{01E9}
\pdfglyphtounicode{kcedilla}{0137}
\pdfglyphtounicode{kcircle}{24DA}
\pdfglyphtounicode{kcommaaccent}{0137}
\pdfglyphtounicode{kdotbelow}{1E33}
\pdfglyphtounicode{keharmenian}{0584}
\pdfglyphtounicode{kehiragana}{3051}
\pdfglyphtounicode{kekatakana}{30B1}
\pdfglyphtounicode{kekatakanahalfwidth}{FF79}
\pdfglyphtounicode{kenarmenian}{056F}
\pdfglyphtounicode{kesmallkatakana}{30F6}
\pdfglyphtounicode{kgreenlandic}{0138}
\pdfglyphtounicode{khabengali}{0996}
\pdfglyphtounicode{khacyrillic}{0445}
\pdfglyphtounicode{khadeva}{0916}
\pdfglyphtounicode{khagujarati}{0A96}
\pdfglyphtounicode{khagurmukhi}{0A16}
\pdfglyphtounicode{khaharabic}{062E}
\pdfglyphtounicode{khahfinalarabic}{FEA6}
\pdfglyphtounicode{khahinitialarabic}{FEA7}
\pdfglyphtounicode{khahmedialarabic}{FEA8}
\pdfglyphtounicode{kheicoptic}{03E7}
\pdfglyphtounicode{khhadeva}{0959}
\pdfglyphtounicode{khhagurmukhi}{0A59}
\pdfglyphtounicode{khieukhacirclekorean}{3278}
\pdfglyphtounicode{khieukhaparenkorean}{3218}
\pdfglyphtounicode{khieukhcirclekorean}{326A}
\pdfglyphtounicode{khieukhkorean}{314B}
\pdfglyphtounicode{khieukhparenkorean}{320A}
\pdfglyphtounicode{khokhaithai}{0E02}
\pdfglyphtounicode{khokhonthai}{0E05}
\pdfglyphtounicode{khokhuatthai}{0E03}
\pdfglyphtounicode{khokhwaithai}{0E04}
\pdfglyphtounicode{khomutthai}{0E5B}
\pdfglyphtounicode{khook}{0199}
\pdfglyphtounicode{khorakhangthai}{0E06}
\pdfglyphtounicode{khzsquare}{3391}
\pdfglyphtounicode{kihiragana}{304D}
\pdfglyphtounicode{kikatakana}{30AD}
\pdfglyphtounicode{kikatakanahalfwidth}{FF77}
\pdfglyphtounicode{kiroguramusquare}{3315}
\pdfglyphtounicode{kiromeetorusquare}{3316}
\pdfglyphtounicode{kirosquare}{3314}
\pdfglyphtounicode{kiyeokacirclekorean}{326E}
\pdfglyphtounicode{kiyeokaparenkorean}{320E}
\pdfglyphtounicode{kiyeokcirclekorean}{3260}
\pdfglyphtounicode{kiyeokkorean}{3131}
\pdfglyphtounicode{kiyeokparenkorean}{3200}
\pdfglyphtounicode{kiyeoksioskorean}{3133}
\pdfglyphtounicode{kjecyrillic}{045C}
\pdfglyphtounicode{klinebelow}{1E35}
\pdfglyphtounicode{klsquare}{3398}
\pdfglyphtounicode{kmcubedsquare}{33A6}
\pdfglyphtounicode{kmonospace}{FF4B}
\pdfglyphtounicode{kmsquaredsquare}{33A2}
\pdfglyphtounicode{kohiragana}{3053}
\pdfglyphtounicode{kohmsquare}{33C0}
\pdfglyphtounicode{kokaithai}{0E01}
\pdfglyphtounicode{kokatakana}{30B3}
\pdfglyphtounicode{kokatakanahalfwidth}{FF7A}
\pdfglyphtounicode{kooposquare}{331E}
\pdfglyphtounicode{koppacyrillic}{0481}
\pdfglyphtounicode{koreanstandardsymbol}{327F}
\pdfglyphtounicode{koroniscmb}{0343}
\pdfglyphtounicode{kparen}{24A6}
\pdfglyphtounicode{kpasquare}{33AA}
\pdfglyphtounicode{ksicyrillic}{046F}
\pdfglyphtounicode{ktsquare}{33CF}
\pdfglyphtounicode{kturned}{029E}
\pdfglyphtounicode{kuhiragana}{304F}
\pdfglyphtounicode{kukatakana}{30AF}
\pdfglyphtounicode{kukatakanahalfwidth}{FF78}
\pdfglyphtounicode{kvsquare}{33B8}
\pdfglyphtounicode{kwsquare}{33BE}
\pdfglyphtounicode{l}{006C}
\pdfglyphtounicode{labengali}{09B2}
\pdfglyphtounicode{lacute}{013A}
\pdfglyphtounicode{ladeva}{0932}
\pdfglyphtounicode{lagujarati}{0AB2}
\pdfglyphtounicode{lagurmukhi}{0A32}
\pdfglyphtounicode{lakkhangyaothai}{0E45}
\pdfglyphtounicode{lamaleffinalarabic}{FEFC}
\pdfglyphtounicode{lamalefhamzaabovefinalarabic}{FEF8}
\pdfglyphtounicode{lamalefhamzaaboveisolatedarabic}{FEF7}
\pdfglyphtounicode{lamalefhamzabelowfinalarabic}{FEFA}
\pdfglyphtounicode{lamalefhamzabelowisolatedarabic}{FEF9}
\pdfglyphtounicode{lamalefisolatedarabic}{FEFB}
\pdfglyphtounicode{lamalefmaddaabovefinalarabic}{FEF6}
\pdfglyphtounicode{lamalefmaddaaboveisolatedarabic}{FEF5}
\pdfglyphtounicode{lamarabic}{0644}
\pdfglyphtounicode{lambda}{03BB}
\pdfglyphtounicode{lambdastroke}{019B}
\pdfglyphtounicode{lamed}{05DC}
\pdfglyphtounicode{lameddagesh}{FB3C}
\pdfglyphtounicode{lameddageshhebrew}{FB3C}
\pdfglyphtounicode{lamedhebrew}{05DC}
\pdfglyphtounicode{lamedholam}{05DC 05B9}
\pdfglyphtounicode{lamedholamdagesh}{05DC 05B9 05BC}
\pdfglyphtounicode{lamedholamdageshhebrew}{05DC 05B9 05BC}
\pdfglyphtounicode{lamedholamhebrew}{05DC 05B9}
\pdfglyphtounicode{lamfinalarabic}{FEDE}
\pdfglyphtounicode{lamhahinitialarabic}{FCCA}
\pdfglyphtounicode{laminitialarabic}{FEDF}
\pdfglyphtounicode{lamjeeminitialarabic}{FCC9}
\pdfglyphtounicode{lamkhahinitialarabic}{FCCB}
\pdfglyphtounicode{lamlamhehisolatedarabic}{FDF2}
\pdfglyphtounicode{lammedialarabic}{FEE0}
\pdfglyphtounicode{lammeemhahinitialarabic}{FD88}
\pdfglyphtounicode{lammeeminitialarabic}{FCCC}
\pdfglyphtounicode{lammeemjeeminitialarabic}{FEDF FEE4 FEA0}
\pdfglyphtounicode{lammeemkhahinitialarabic}{FEDF FEE4 FEA8}
\pdfglyphtounicode{largecircle}{25EF}
\pdfglyphtounicode{latticetop}{22A4}
\pdfglyphtounicode{lbar}{019A}
\pdfglyphtounicode{lbelt}{026C}
\pdfglyphtounicode{lbopomofo}{310C}
\pdfglyphtounicode{lcaron}{013E}
\pdfglyphtounicode{lcedilla}{013C}
\pdfglyphtounicode{lcircle}{24DB}
\pdfglyphtounicode{lcircumflexbelow}{1E3D}
\pdfglyphtounicode{lcommaaccent}{013C}
\pdfglyphtounicode{ldot}{0140}
\pdfglyphtounicode{ldotaccent}{0140}
\pdfglyphtounicode{ldotbelow}{1E37}
\pdfglyphtounicode{ldotbelowmacron}{1E39}
\pdfglyphtounicode{leftangleabovecmb}{031A}
\pdfglyphtounicode{lefttackbelowcmb}{0318}
\pdfglyphtounicode{less}{003C}
\pdfglyphtounicode{lessdbleqlgreater}{2A8B}
\pdfglyphtounicode{lessdblequal}{2266}
\pdfglyphtounicode{lessdot}{22D6}
\pdfglyphtounicode{lessequal}{2264}
\pdfglyphtounicode{lessequalgreater}{22DA}
\pdfglyphtounicode{lessequalorgreater}{22DA}
\pdfglyphtounicode{lessmonospace}{FF1C}
\pdfglyphtounicode{lessmuch}{226A}
\pdfglyphtounicode{lessnotdblequal}{2A89}
\pdfglyphtounicode{lessnotequal}{2A87}
\pdfglyphtounicode{lessorapproxeql}{2A85}
\pdfglyphtounicode{lessorequalslant}{2A7D}
\pdfglyphtounicode{lessorequivalent}{2272}
\pdfglyphtounicode{lessorgreater}{2276}
\pdfglyphtounicode{lessornotdbleql}{2268}
\pdfglyphtounicode{lessornotequal}{2268}
\pdfglyphtounicode{lessorsimilar}{2272}
\pdfglyphtounicode{lessoverequal}{2266}
\pdfglyphtounicode{lesssmall}{FE64}
\pdfglyphtounicode{lezh}{026E}
\pdfglyphtounicode{lfblock}{258C}
\pdfglyphtounicode{lhookretroflex}{026D}
\pdfglyphtounicode{lira}{20A4}
\pdfglyphtounicode{liwnarmenian}{056C}
\pdfglyphtounicode{lj}{01C9}
\pdfglyphtounicode{ljecyrillic}{0459}
\pdfglyphtounicode{ll}{006C 006C}
\pdfglyphtounicode{lladeva}{0933}
\pdfglyphtounicode{llagujarati}{0AB3}
\pdfglyphtounicode{llinebelow}{1E3B}
\pdfglyphtounicode{llladeva}{0934}
\pdfglyphtounicode{llvocalicbengali}{09E1}
\pdfglyphtounicode{llvocalicdeva}{0961}
\pdfglyphtounicode{llvocalicvowelsignbengali}{09E3}
\pdfglyphtounicode{llvocalicvowelsigndeva}{0963}
\pdfglyphtounicode{lmiddletilde}{026B}
\pdfglyphtounicode{lmonospace}{FF4C}
\pdfglyphtounicode{lmsquare}{33D0}
\pdfglyphtounicode{lochulathai}{0E2C}
\pdfglyphtounicode{logicaland}{2227}
\pdfglyphtounicode{logicalnot}{00AC}
\pdfglyphtounicode{logicalnotreversed}{2310}
\pdfglyphtounicode{logicalor}{2228}
\pdfglyphtounicode{lolingthai}{0E25}
\pdfglyphtounicode{longdbls}{017F 017F}
\pdfglyphtounicode{longs}{017F}
\pdfglyphtounicode{longsh}{017F 0068}
\pdfglyphtounicode{longsi}{017F 0069}
\pdfglyphtounicode{longsl}{017F 006C}
\pdfglyphtounicode{longst}{017F 0074}
\pdfglyphtounicode{lowlinecenterline}{FE4E}
\pdfglyphtounicode{lowlinecmb}{0332}
\pdfglyphtounicode{lowlinedashed}{FE4D}
\pdfglyphtounicode{lozenge}{25CA}
\pdfglyphtounicode{lparen}{24A7}
\pdfglyphtounicode{lscript}{2113}
\pdfglyphtounicode{lslash}{0142}
\pdfglyphtounicode{lsquare}{2113}
\pdfglyphtounicode{lsuperior}{006C}
\pdfglyphtounicode{ltshade}{2591}
\pdfglyphtounicode{luthai}{0E26}
\pdfglyphtounicode{lvocalicbengali}{098C}
\pdfglyphtounicode{lvocalicdeva}{090C}
\pdfglyphtounicode{lvocalicvowelsignbengali}{09E2}
\pdfglyphtounicode{lvocalicvowelsigndeva}{0962}
\pdfglyphtounicode{lxsquare}{33D3}
\pdfglyphtounicode{m}{006D}
\pdfglyphtounicode{mabengali}{09AE}
\pdfglyphtounicode{macron}{00AF}
\pdfglyphtounicode{macronbelowcmb}{0331}
\pdfglyphtounicode{macroncmb}{0304}
\pdfglyphtounicode{macronlowmod}{02CD}
\pdfglyphtounicode{macronmonospace}{FFE3}
\pdfglyphtounicode{macute}{1E3F}
\pdfglyphtounicode{madeva}{092E}
\pdfglyphtounicode{magujarati}{0AAE}
\pdfglyphtounicode{magurmukhi}{0A2E}
\pdfglyphtounicode{mahapakhhebrew}{05A4}
\pdfglyphtounicode{mahapakhlefthebrew}{05A4}
\pdfglyphtounicode{mahiragana}{307E}
\pdfglyphtounicode{maichattawalowleftthai}{F895}
\pdfglyphtounicode{maichattawalowrightthai}{F894}
\pdfglyphtounicode{maichattawathai}{0E4B}
\pdfglyphtounicode{maichattawaupperleftthai}{F893}
\pdfglyphtounicode{maieklowleftthai}{F88C}
\pdfglyphtounicode{maieklowrightthai}{F88B}
\pdfglyphtounicode{maiekthai}{0E48}
\pdfglyphtounicode{maiekupperleftthai}{F88A}
\pdfglyphtounicode{maihanakatleftthai}{F884}
\pdfglyphtounicode{maihanakatthai}{0E31}
\pdfglyphtounicode{maitaikhuleftthai}{F889}
\pdfglyphtounicode{maitaikhuthai}{0E47}
\pdfglyphtounicode{maitholowleftthai}{F88F}
\pdfglyphtounicode{maitholowrightthai}{F88E}
\pdfglyphtounicode{maithothai}{0E49}
\pdfglyphtounicode{maithoupperleftthai}{F88D}
\pdfglyphtounicode{maitrilowleftthai}{F892}
\pdfglyphtounicode{maitrilowrightthai}{F891}
\pdfglyphtounicode{maitrithai}{0E4A}
\pdfglyphtounicode{maitriupperleftthai}{F890}
\pdfglyphtounicode{maiyamokthai}{0E46}
\pdfglyphtounicode{makatakana}{30DE}
\pdfglyphtounicode{makatakanahalfwidth}{FF8F}
\pdfglyphtounicode{male}{2642}
\pdfglyphtounicode{maltesecross}{2720}
\pdfglyphtounicode{mansyonsquare}{3347}
\pdfglyphtounicode{maqafhebrew}{05BE}
\pdfglyphtounicode{mars}{2642}
\pdfglyphtounicode{masoracirclehebrew}{05AF}
\pdfglyphtounicode{masquare}{3383}
\pdfglyphtounicode{mbopomofo}{3107}
\pdfglyphtounicode{mbsquare}{33D4}
\pdfglyphtounicode{mcircle}{24DC}
\pdfglyphtounicode{mcubedsquare}{33A5}
\pdfglyphtounicode{mdotaccent}{1E41}
\pdfglyphtounicode{mdotbelow}{1E43}
\pdfglyphtounicode{measuredangle}{2221}
\pdfglyphtounicode{meemarabic}{0645}
\pdfglyphtounicode{meemfinalarabic}{FEE2}
\pdfglyphtounicode{meeminitialarabic}{FEE3}
\pdfglyphtounicode{meemmedialarabic}{FEE4}
\pdfglyphtounicode{meemmeeminitialarabic}{FCD1}
\pdfglyphtounicode{meemmeemisolatedarabic}{FC48}
\pdfglyphtounicode{meetorusquare}{334D}
\pdfglyphtounicode{mehiragana}{3081}
\pdfglyphtounicode{meizierasquare}{337E}
\pdfglyphtounicode{mekatakana}{30E1}
\pdfglyphtounicode{mekatakanahalfwidth}{FF92}
\pdfglyphtounicode{mem}{05DE}
\pdfglyphtounicode{memdagesh}{FB3E}
\pdfglyphtounicode{memdageshhebrew}{FB3E}
\pdfglyphtounicode{memhebrew}{05DE}
\pdfglyphtounicode{menarmenian}{0574}
\pdfglyphtounicode{merkhahebrew}{05A5}
\pdfglyphtounicode{merkhakefulahebrew}{05A6}
\pdfglyphtounicode{merkhakefulalefthebrew}{05A6}
\pdfglyphtounicode{merkhalefthebrew}{05A5}
\pdfglyphtounicode{mhook}{0271}
\pdfglyphtounicode{mhzsquare}{3392}
\pdfglyphtounicode{middledotkatakanahalfwidth}{FF65}
\pdfglyphtounicode{middot}{00B7}
\pdfglyphtounicode{mieumacirclekorean}{3272}
\pdfglyphtounicode{mieumaparenkorean}{3212}
\pdfglyphtounicode{mieumcirclekorean}{3264}
\pdfglyphtounicode{mieumkorean}{3141}
\pdfglyphtounicode{mieumpansioskorean}{3170}
\pdfglyphtounicode{mieumparenkorean}{3204}
\pdfglyphtounicode{mieumpieupkorean}{316E}
\pdfglyphtounicode{mieumsioskorean}{316F}
\pdfglyphtounicode{mihiragana}{307F}
\pdfglyphtounicode{mikatakana}{30DF}
\pdfglyphtounicode{mikatakanahalfwidth}{FF90}
\pdfglyphtounicode{minus}{2212}
\pdfglyphtounicode{minusbelowcmb}{0320}
\pdfglyphtounicode{minuscircle}{2296}
\pdfglyphtounicode{minusmod}{02D7}
\pdfglyphtounicode{minusplus}{2213}
\pdfglyphtounicode{minute}{2032}
\pdfglyphtounicode{miribaarusquare}{334A}
\pdfglyphtounicode{mirisquare}{3349}
\pdfglyphtounicode{mlonglegturned}{0270}
\pdfglyphtounicode{mlsquare}{3396}
\pdfglyphtounicode{mmcubedsquare}{33A3}
\pdfglyphtounicode{mmonospace}{FF4D}
\pdfglyphtounicode{mmsquaredsquare}{339F}
\pdfglyphtounicode{mohiragana}{3082}
\pdfglyphtounicode{mohmsquare}{33C1}
\pdfglyphtounicode{mokatakana}{30E2}
\pdfglyphtounicode{mokatakanahalfwidth}{FF93}
\pdfglyphtounicode{molsquare}{33D6}
\pdfglyphtounicode{momathai}{0E21}
\pdfglyphtounicode{moverssquare}{33A7}
\pdfglyphtounicode{moverssquaredsquare}{33A8}
\pdfglyphtounicode{mparen}{24A8}
\pdfglyphtounicode{mpasquare}{33AB}
\pdfglyphtounicode{mssquare}{33B3}
\pdfglyphtounicode{msuperior}{006D}
\pdfglyphtounicode{mturned}{026F}
\pdfglyphtounicode{mu}{00B5}
\pdfglyphtounicode{mu1}{00B5}
\pdfglyphtounicode{muasquare}{3382}
\pdfglyphtounicode{muchgreater}{226B}
\pdfglyphtounicode{muchless}{226A}
\pdfglyphtounicode{mufsquare}{338C}
\pdfglyphtounicode{mugreek}{03BC}
\pdfglyphtounicode{mugsquare}{338D}
\pdfglyphtounicode{muhiragana}{3080}
\pdfglyphtounicode{mukatakana}{30E0}
\pdfglyphtounicode{mukatakanahalfwidth}{FF91}
\pdfglyphtounicode{mulsquare}{3395}
\pdfglyphtounicode{multicloseleft}{22C9}
\pdfglyphtounicode{multicloseright}{22CA}
\pdfglyphtounicode{multimap}{22B8}
\pdfglyphtounicode{multiopenleft}{22CB}
\pdfglyphtounicode{multiopenright}{22CC}
\pdfglyphtounicode{multiply}{00D7}
\pdfglyphtounicode{mumsquare}{339B}
\pdfglyphtounicode{munahhebrew}{05A3}
\pdfglyphtounicode{munahlefthebrew}{05A3}
\pdfglyphtounicode{musicalnote}{266A}
\pdfglyphtounicode{musicalnotedbl}{266B}
\pdfglyphtounicode{musicflatsign}{266D}
\pdfglyphtounicode{musicsharpsign}{266F}
\pdfglyphtounicode{mussquare}{33B2}
\pdfglyphtounicode{muvsquare}{33B6}
\pdfglyphtounicode{muwsquare}{33BC}
\pdfglyphtounicode{mvmegasquare}{33B9}
\pdfglyphtounicode{mvsquare}{33B7}
\pdfglyphtounicode{mwmegasquare}{33BF}
\pdfglyphtounicode{mwsquare}{33BD}
\pdfglyphtounicode{n}{006E}
\pdfglyphtounicode{nabengali}{09A8}
\pdfglyphtounicode{nabla}{2207}
\pdfglyphtounicode{nacute}{0144}
\pdfglyphtounicode{nadeva}{0928}
\pdfglyphtounicode{nagujarati}{0AA8}
\pdfglyphtounicode{nagurmukhi}{0A28}
\pdfglyphtounicode{nahiragana}{306A}
\pdfglyphtounicode{nakatakana}{30CA}
\pdfglyphtounicode{nakatakanahalfwidth}{FF85}
\pdfglyphtounicode{nand}{22BC}
\pdfglyphtounicode{napostrophe}{0149}
\pdfglyphtounicode{nasquare}{3381}
\pdfglyphtounicode{natural}{266E}
\pdfglyphtounicode{nbopomofo}{310B}
\pdfglyphtounicode{nbspace}{00A0}
\pdfglyphtounicode{ncaron}{0148}
\pdfglyphtounicode{ncedilla}{0146}
\pdfglyphtounicode{ncircle}{24DD}
\pdfglyphtounicode{ncircumflexbelow}{1E4B}
\pdfglyphtounicode{ncommaaccent}{0146}
\pdfglyphtounicode{ndotaccent}{1E45}
\pdfglyphtounicode{ndotbelow}{1E47}
\pdfglyphtounicode{negationslash}{0338}
\pdfglyphtounicode{nehiragana}{306D}
\pdfglyphtounicode{nekatakana}{30CD}
\pdfglyphtounicode{nekatakanahalfwidth}{FF88}
\pdfglyphtounicode{newsheqelsign}{20AA}
\pdfglyphtounicode{nfsquare}{338B}
\pdfglyphtounicode{ng}{014B}
\pdfglyphtounicode{ngabengali}{0999}
\pdfglyphtounicode{ngadeva}{0919}
\pdfglyphtounicode{ngagujarati}{0A99}
\pdfglyphtounicode{ngagurmukhi}{0A19}
\pdfglyphtounicode{ngonguthai}{0E07}
\pdfglyphtounicode{nhiragana}{3093}
\pdfglyphtounicode{nhookleft}{0272}
\pdfglyphtounicode{nhookretroflex}{0273}
\pdfglyphtounicode{nieunacirclekorean}{326F}
\pdfglyphtounicode{nieunaparenkorean}{320F}
\pdfglyphtounicode{nieuncieuckorean}{3135}
\pdfglyphtounicode{nieuncirclekorean}{3261}
\pdfglyphtounicode{nieunhieuhkorean}{3136}
\pdfglyphtounicode{nieunkorean}{3134}
\pdfglyphtounicode{nieunpansioskorean}{3168}
\pdfglyphtounicode{nieunparenkorean}{3201}
\pdfglyphtounicode{nieunsioskorean}{3167}
\pdfglyphtounicode{nieuntikeutkorean}{3166}
\pdfglyphtounicode{nihiragana}{306B}
\pdfglyphtounicode{nikatakana}{30CB}
\pdfglyphtounicode{nikatakanahalfwidth}{FF86}
\pdfglyphtounicode{nikhahitleftthai}{F899}
\pdfglyphtounicode{nikhahitthai}{0E4D}
\pdfglyphtounicode{nine}{0039}
\pdfglyphtounicode{ninearabic}{0669}
\pdfglyphtounicode{ninebengali}{09EF}
\pdfglyphtounicode{ninecircle}{2468}
\pdfglyphtounicode{ninecircleinversesansserif}{2792}
\pdfglyphtounicode{ninedeva}{096F}
\pdfglyphtounicode{ninegujarati}{0AEF}
\pdfglyphtounicode{ninegurmukhi}{0A6F}
\pdfglyphtounicode{ninehackarabic}{0669}
\pdfglyphtounicode{ninehangzhou}{3029}
\pdfglyphtounicode{nineideographicparen}{3228}
\pdfglyphtounicode{nineinferior}{2089}
\pdfglyphtounicode{ninemonospace}{FF19}
\pdfglyphtounicode{nineoldstyle}{0039}
\pdfglyphtounicode{nineparen}{247C}
\pdfglyphtounicode{nineperiod}{2490}
\pdfglyphtounicode{ninepersian}{06F9}
\pdfglyphtounicode{nineroman}{2178}
\pdfglyphtounicode{ninesuperior}{2079}
\pdfglyphtounicode{nineteencircle}{2472}
\pdfglyphtounicode{nineteenparen}{2486}
\pdfglyphtounicode{nineteenperiod}{249A}
\pdfglyphtounicode{ninethai}{0E59}
\pdfglyphtounicode{nj}{01CC}
\pdfglyphtounicode{njecyrillic}{045A}
\pdfglyphtounicode{nkatakana}{30F3}
\pdfglyphtounicode{nkatakanahalfwidth}{FF9D}
\pdfglyphtounicode{nlegrightlong}{019E}
\pdfglyphtounicode{nlinebelow}{1E49}
\pdfglyphtounicode{nmonospace}{FF4E}
\pdfglyphtounicode{nmsquare}{339A}
\pdfglyphtounicode{nnabengali}{09A3}
\pdfglyphtounicode{nnadeva}{0923}
\pdfglyphtounicode{nnagujarati}{0AA3}
\pdfglyphtounicode{nnagurmukhi}{0A23}
\pdfglyphtounicode{nnnadeva}{0929}
\pdfglyphtounicode{nohiragana}{306E}
\pdfglyphtounicode{nokatakana}{30CE}
\pdfglyphtounicode{nokatakanahalfwidth}{FF89}
\pdfglyphtounicode{nonbreakingspace}{00A0}
\pdfglyphtounicode{nonenthai}{0E13}
\pdfglyphtounicode{nonuthai}{0E19}
\pdfglyphtounicode{noonarabic}{0646}
\pdfglyphtounicode{noonfinalarabic}{FEE6}
\pdfglyphtounicode{noonghunnaarabic}{06BA}
\pdfglyphtounicode{noonghunnafinalarabic}{FB9F}
\pdfglyphtounicode{noonhehinitialarabic}{FEE7 FEEC}
\pdfglyphtounicode{nooninitialarabic}{FEE7}
\pdfglyphtounicode{noonjeeminitialarabic}{FCD2}
\pdfglyphtounicode{noonjeemisolatedarabic}{FC4B}
\pdfglyphtounicode{noonmedialarabic}{FEE8}
\pdfglyphtounicode{noonmeeminitialarabic}{FCD5}
\pdfglyphtounicode{noonmeemisolatedarabic}{FC4E}
\pdfglyphtounicode{noonnoonfinalarabic}{FC8D}
\pdfglyphtounicode{notapproxequal}{2247}
\pdfglyphtounicode{notarrowboth}{21AE}
\pdfglyphtounicode{notarrowleft}{219A}
\pdfglyphtounicode{notarrowright}{219B}
\pdfglyphtounicode{notbar}{2224}
\pdfglyphtounicode{notcontains}{220C}
\pdfglyphtounicode{notdblarrowboth}{21CE}
\pdfglyphtounicode{notdblarrowleft}{21CD}
\pdfglyphtounicode{notdblarrowright}{21CF}
\pdfglyphtounicode{notelement}{2209}
\pdfglyphtounicode{notelementof}{2209}
\pdfglyphtounicode{notequal}{2260}
\pdfglyphtounicode{notexistential}{2204}
\pdfglyphtounicode{notfollows}{2281}
\pdfglyphtounicode{notfollowsoreql}{2AB0 0338}
\pdfglyphtounicode{notforces}{22AE}
\pdfglyphtounicode{notforcesextra}{22AF}
\pdfglyphtounicode{notgreater}{226F}
\pdfglyphtounicode{notgreaterdblequal}{2267 0338}
\pdfglyphtounicode{notgreaterequal}{2271}
\pdfglyphtounicode{notgreaternorequal}{2271}
\pdfglyphtounicode{notgreaternorless}{2279}
\pdfglyphtounicode{notgreaterorslnteql}{2A7E 0338}
\pdfglyphtounicode{notidentical}{2262}
\pdfglyphtounicode{notless}{226E}
\pdfglyphtounicode{notlessdblequal}{2266 0338}
\pdfglyphtounicode{notlessequal}{2270}
\pdfglyphtounicode{notlessnorequal}{2270}
\pdfglyphtounicode{notlessorslnteql}{2A7D 0338}
\pdfglyphtounicode{notparallel}{2226}
\pdfglyphtounicode{notprecedes}{2280}
\pdfglyphtounicode{notprecedesoreql}{2AAF 0338}
\pdfglyphtounicode{notsatisfies}{22AD}
\pdfglyphtounicode{notsimilar}{2241}
\pdfglyphtounicode{notsubset}{2284}
\pdfglyphtounicode{notsubseteql}{2288}
\pdfglyphtounicode{notsubsetordbleql}{2AC5 0338}
\pdfglyphtounicode{notsubsetoreql}{228A}
\pdfglyphtounicode{notsucceeds}{2281}
\pdfglyphtounicode{notsuperset}{2285}
\pdfglyphtounicode{notsuperseteql}{2289}
\pdfglyphtounicode{notsupersetordbleql}{2AC6 0338}
\pdfglyphtounicode{notsupersetoreql}{228B}
\pdfglyphtounicode{nottriangeqlleft}{22EC}
\pdfglyphtounicode{nottriangeqlright}{22ED}
\pdfglyphtounicode{nottriangleleft}{22EA}
\pdfglyphtounicode{nottriangleright}{22EB}
\pdfglyphtounicode{notturnstile}{22AC}
\pdfglyphtounicode{nowarmenian}{0576}
\pdfglyphtounicode{nparen}{24A9}
\pdfglyphtounicode{nssquare}{33B1}
\pdfglyphtounicode{nsuperior}{207F}
\pdfglyphtounicode{ntilde}{00F1}
\pdfglyphtounicode{nu}{03BD}
\pdfglyphtounicode{nuhiragana}{306C}
\pdfglyphtounicode{nukatakana}{30CC}
\pdfglyphtounicode{nukatakanahalfwidth}{FF87}
\pdfglyphtounicode{nuktabengali}{09BC}
\pdfglyphtounicode{nuktadeva}{093C}
\pdfglyphtounicode{nuktagujarati}{0ABC}
\pdfglyphtounicode{nuktagurmukhi}{0A3C}
\pdfglyphtounicode{numbersign}{0023}
\pdfglyphtounicode{numbersignmonospace}{FF03}
\pdfglyphtounicode{numbersignsmall}{FE5F}
\pdfglyphtounicode{numeralsigngreek}{0374}
\pdfglyphtounicode{numeralsignlowergreek}{0375}
\pdfglyphtounicode{numero}{2116}
\pdfglyphtounicode{nun}{05E0}
\pdfglyphtounicode{nundagesh}{FB40}
\pdfglyphtounicode{nundageshhebrew}{FB40}
\pdfglyphtounicode{nunhebrew}{05E0}
\pdfglyphtounicode{nvsquare}{33B5}
\pdfglyphtounicode{nwsquare}{33BB}
\pdfglyphtounicode{nyabengali}{099E}
\pdfglyphtounicode{nyadeva}{091E}
\pdfglyphtounicode{nyagujarati}{0A9E}
\pdfglyphtounicode{nyagurmukhi}{0A1E}
\pdfglyphtounicode{o}{006F}
\pdfglyphtounicode{oacute}{00F3}
\pdfglyphtounicode{oangthai}{0E2D}
\pdfglyphtounicode{obarred}{0275}
\pdfglyphtounicode{obarredcyrillic}{04E9}
\pdfglyphtounicode{obarreddieresiscyrillic}{04EB}
\pdfglyphtounicode{obengali}{0993}
\pdfglyphtounicode{obopomofo}{311B}
\pdfglyphtounicode{obreve}{014F}
\pdfglyphtounicode{ocandradeva}{0911}
\pdfglyphtounicode{ocandragujarati}{0A91}
\pdfglyphtounicode{ocandravowelsigndeva}{0949}
\pdfglyphtounicode{ocandravowelsigngujarati}{0AC9}
\pdfglyphtounicode{ocaron}{01D2}
\pdfglyphtounicode{ocircle}{24DE}
\pdfglyphtounicode{ocircumflex}{00F4}
\pdfglyphtounicode{ocircumflexacute}{1ED1}
\pdfglyphtounicode{ocircumflexdotbelow}{1ED9}
\pdfglyphtounicode{ocircumflexgrave}{1ED3}
\pdfglyphtounicode{ocircumflexhookabove}{1ED5}
\pdfglyphtounicode{ocircumflextilde}{1ED7}
\pdfglyphtounicode{ocyrillic}{043E}
\pdfglyphtounicode{odblacute}{0151}
\pdfglyphtounicode{odblgrave}{020D}
\pdfglyphtounicode{odeva}{0913}
\pdfglyphtounicode{odieresis}{00F6}
\pdfglyphtounicode{odieresiscyrillic}{04E7}
\pdfglyphtounicode{odotbelow}{1ECD}
\pdfglyphtounicode{oe}{0153}
\pdfglyphtounicode{oekorean}{315A}
\pdfglyphtounicode{ogonek}{02DB}
\pdfglyphtounicode{ogonekcmb}{0328}
\pdfglyphtounicode{ograve}{00F2}
\pdfglyphtounicode{ogujarati}{0A93}
\pdfglyphtounicode{oharmenian}{0585}
\pdfglyphtounicode{ohiragana}{304A}
\pdfglyphtounicode{ohookabove}{1ECF}
\pdfglyphtounicode{ohorn}{01A1}
\pdfglyphtounicode{ohornacute}{1EDB}
\pdfglyphtounicode{ohorndotbelow}{1EE3}
\pdfglyphtounicode{ohorngrave}{1EDD}
\pdfglyphtounicode{ohornhookabove}{1EDF}
\pdfglyphtounicode{ohorntilde}{1EE1}
\pdfglyphtounicode{ohungarumlaut}{0151}
\pdfglyphtounicode{oi}{01A3}
\pdfglyphtounicode{oinvertedbreve}{020F}
\pdfglyphtounicode{okatakana}{30AA}
\pdfglyphtounicode{okatakanahalfwidth}{FF75}
\pdfglyphtounicode{okorean}{3157}
\pdfglyphtounicode{olehebrew}{05AB}
\pdfglyphtounicode{omacron}{014D}
\pdfglyphtounicode{omacronacute}{1E53}
\pdfglyphtounicode{omacrongrave}{1E51}
\pdfglyphtounicode{omdeva}{0950}
\pdfglyphtounicode{omega}{03C9}
\pdfglyphtounicode{omega1}{03D6}
\pdfglyphtounicode{omegacyrillic}{0461}
\pdfglyphtounicode{omegalatinclosed}{0277}
\pdfglyphtounicode{omegaroundcyrillic}{047B}
\pdfglyphtounicode{omegatitlocyrillic}{047D}
\pdfglyphtounicode{omegatonos}{03CE}
\pdfglyphtounicode{omgujarati}{0AD0}
\pdfglyphtounicode{omicron}{03BF}
\pdfglyphtounicode{omicrontonos}{03CC}
\pdfglyphtounicode{omonospace}{FF4F}
\pdfglyphtounicode{one}{0031}
\pdfglyphtounicode{onearabic}{0661}
\pdfglyphtounicode{onebengali}{09E7}
\pdfglyphtounicode{onecircle}{2460}
\pdfglyphtounicode{onecircleinversesansserif}{278A}
\pdfglyphtounicode{onedeva}{0967}
\pdfglyphtounicode{onedotenleader}{2024}
\pdfglyphtounicode{oneeighth}{215B}
\pdfglyphtounicode{onefitted}{0031}
\pdfglyphtounicode{onegujarati}{0AE7}
\pdfglyphtounicode{onegurmukhi}{0A67}
\pdfglyphtounicode{onehackarabic}{0661}
\pdfglyphtounicode{onehalf}{00BD}
\pdfglyphtounicode{onehangzhou}{3021}
\pdfglyphtounicode{oneideographicparen}{3220}
\pdfglyphtounicode{oneinferior}{2081}
\pdfglyphtounicode{onemonospace}{FF11}
\pdfglyphtounicode{onenumeratorbengali}{09F4}
\pdfglyphtounicode{oneoldstyle}{0031}
\pdfglyphtounicode{oneparen}{2474}
\pdfglyphtounicode{oneperiod}{2488}
\pdfglyphtounicode{onepersian}{06F1}
\pdfglyphtounicode{onequarter}{00BC}
\pdfglyphtounicode{oneroman}{2170}
\pdfglyphtounicode{onesuperior}{00B9}
\pdfglyphtounicode{onethai}{0E51}
\pdfglyphtounicode{onethird}{2153}
\pdfglyphtounicode{oogonek}{01EB}
\pdfglyphtounicode{oogonekmacron}{01ED}
\pdfglyphtounicode{oogurmukhi}{0A13}
\pdfglyphtounicode{oomatragurmukhi}{0A4B}
\pdfglyphtounicode{oopen}{0254}
\pdfglyphtounicode{oparen}{24AA}
\pdfglyphtounicode{openbullet}{25E6}
\pdfglyphtounicode{option}{2325}
\pdfglyphtounicode{ordfeminine}{00AA}
\pdfglyphtounicode{ordmasculine}{00BA}
\pdfglyphtounicode{orthogonal}{221F}
\pdfglyphtounicode{orunderscore}{22BB}
\pdfglyphtounicode{oshortdeva}{0912}
\pdfglyphtounicode{oshortvowelsigndeva}{094A}
\pdfglyphtounicode{oslash}{00F8}
\pdfglyphtounicode{oslashacute}{01FF}
\pdfglyphtounicode{osmallhiragana}{3049}
\pdfglyphtounicode{osmallkatakana}{30A9}
\pdfglyphtounicode{osmallkatakanahalfwidth}{FF6B}
\pdfglyphtounicode{ostrokeacute}{01FF}
\pdfglyphtounicode{osuperior}{006F}
\pdfglyphtounicode{otcyrillic}{047F}
\pdfglyphtounicode{otilde}{00F5}
\pdfglyphtounicode{otildeacute}{1E4D}
\pdfglyphtounicode{otildedieresis}{1E4F}
\pdfglyphtounicode{oubopomofo}{3121}
\pdfglyphtounicode{overline}{203E}
\pdfglyphtounicode{overlinecenterline}{FE4A}
\pdfglyphtounicode{overlinecmb}{0305}
\pdfglyphtounicode{overlinedashed}{FE49}
\pdfglyphtounicode{overlinedblwavy}{FE4C}
\pdfglyphtounicode{overlinewavy}{FE4B}
\pdfglyphtounicode{overscore}{00AF}
\pdfglyphtounicode{ovowelsignbengali}{09CB}
\pdfglyphtounicode{ovowelsigndeva}{094B}
\pdfglyphtounicode{ovowelsigngujarati}{0ACB}
\pdfglyphtounicode{owner}{220B}
\pdfglyphtounicode{p}{0070}
\pdfglyphtounicode{paampssquare}{3380}
\pdfglyphtounicode{paasentosquare}{332B}
\pdfglyphtounicode{pabengali}{09AA}
\pdfglyphtounicode{pacute}{1E55}
\pdfglyphtounicode{padeva}{092A}
\pdfglyphtounicode{pagedown}{21DF}
\pdfglyphtounicode{pageup}{21DE}
\pdfglyphtounicode{pagujarati}{0AAA}
\pdfglyphtounicode{pagurmukhi}{0A2A}
\pdfglyphtounicode{pahiragana}{3071}
\pdfglyphtounicode{paiyannoithai}{0E2F}
\pdfglyphtounicode{pakatakana}{30D1}
\pdfglyphtounicode{palatalizationcyrilliccmb}{0484}
\pdfglyphtounicode{palochkacyrillic}{04C0}
\pdfglyphtounicode{pansioskorean}{317F}
\pdfglyphtounicode{paragraph}{00B6}
\pdfglyphtounicode{parallel}{2225}
\pdfglyphtounicode{parenleft}{0028}
\pdfglyphtounicode{parenleftaltonearabic}{FD3E}
\pdfglyphtounicode{parenleftbt}{F8ED}
\pdfglyphtounicode{parenleftex}{F8EC}
\pdfglyphtounicode{parenleftinferior}{208D}
\pdfglyphtounicode{parenleftmonospace}{FF08}
\pdfglyphtounicode{parenleftsmall}{FE59}
\pdfglyphtounicode{parenleftsuperior}{207D}
\pdfglyphtounicode{parenlefttp}{F8EB}
\pdfglyphtounicode{parenleftvertical}{FE35}
\pdfglyphtounicode{parenright}{0029}
\pdfglyphtounicode{parenrightaltonearabic}{FD3F}
\pdfglyphtounicode{parenrightbt}{F8F8}
\pdfglyphtounicode{parenrightex}{F8F7}
\pdfglyphtounicode{parenrightinferior}{208E}
\pdfglyphtounicode{parenrightmonospace}{FF09}
\pdfglyphtounicode{parenrightsmall}{FE5A}
\pdfglyphtounicode{parenrightsuperior}{207E}
\pdfglyphtounicode{parenrighttp}{F8F6}
\pdfglyphtounicode{parenrightvertical}{FE36}
\pdfglyphtounicode{partialdiff}{2202}
\pdfglyphtounicode{paseqhebrew}{05C0}
\pdfglyphtounicode{pashtahebrew}{0599}
\pdfglyphtounicode{pasquare}{33A9}
\pdfglyphtounicode{patah}{05B7}
\pdfglyphtounicode{patah11}{05B7}
\pdfglyphtounicode{patah1d}{05B7}
\pdfglyphtounicode{patah2a}{05B7}
\pdfglyphtounicode{patahhebrew}{05B7}
\pdfglyphtounicode{patahnarrowhebrew}{05B7}
\pdfglyphtounicode{patahquarterhebrew}{05B7}
\pdfglyphtounicode{patahwidehebrew}{05B7}
\pdfglyphtounicode{pazerhebrew}{05A1}
\pdfglyphtounicode{pbopomofo}{3106}
\pdfglyphtounicode{pcircle}{24DF}
\pdfglyphtounicode{pdotaccent}{1E57}
\pdfglyphtounicode{pe}{05E4}
\pdfglyphtounicode{pecyrillic}{043F}
\pdfglyphtounicode{pedagesh}{FB44}
\pdfglyphtounicode{pedageshhebrew}{FB44}
\pdfglyphtounicode{peezisquare}{333B}
\pdfglyphtounicode{pefinaldageshhebrew}{FB43}
\pdfglyphtounicode{peharabic}{067E}
\pdfglyphtounicode{peharmenian}{057A}
\pdfglyphtounicode{pehebrew}{05E4}
\pdfglyphtounicode{pehfinalarabic}{FB57}
\pdfglyphtounicode{pehinitialarabic}{FB58}
\pdfglyphtounicode{pehiragana}{307A}
\pdfglyphtounicode{pehmedialarabic}{FB59}
\pdfglyphtounicode{pekatakana}{30DA}
\pdfglyphtounicode{pemiddlehookcyrillic}{04A7}
\pdfglyphtounicode{perafehebrew}{FB4E}
\pdfglyphtounicode{percent}{0025}
\pdfglyphtounicode{percentarabic}{066A}
\pdfglyphtounicode{percentmonospace}{FF05}
\pdfglyphtounicode{percentsmall}{FE6A}
\pdfglyphtounicode{period}{002E}
\pdfglyphtounicode{periodarmenian}{0589}
\pdfglyphtounicode{periodcentered}{00B7}
\pdfglyphtounicode{periodhalfwidth}{FF61}
\pdfglyphtounicode{periodinferior}{002E}
\pdfglyphtounicode{periodmonospace}{FF0E}
\pdfglyphtounicode{periodsmall}{FE52}
\pdfglyphtounicode{periodsuperior}{002E}
\pdfglyphtounicode{perispomenigreekcmb}{0342}
\pdfglyphtounicode{perpcorrespond}{2A5E}
\pdfglyphtounicode{perpendicular}{22A5}
\pdfglyphtounicode{pertenthousand}{2031}
\pdfglyphtounicode{perthousand}{2030}
\pdfglyphtounicode{peseta}{20A7}
\pdfglyphtounicode{pfsquare}{338A}
\pdfglyphtounicode{phabengali}{09AB}
\pdfglyphtounicode{phadeva}{092B}
\pdfglyphtounicode{phagujarati}{0AAB}
\pdfglyphtounicode{phagurmukhi}{0A2B}
\pdfglyphtounicode{phi}{03C6}
\pdfglyphtounicode{phi1}{03D5}
\pdfglyphtounicode{phieuphacirclekorean}{327A}
\pdfglyphtounicode{phieuphaparenkorean}{321A}
\pdfglyphtounicode{phieuphcirclekorean}{326C}
\pdfglyphtounicode{phieuphkorean}{314D}
\pdfglyphtounicode{phieuphparenkorean}{320C}
\pdfglyphtounicode{philatin}{0278}
\pdfglyphtounicode{phinthuthai}{0E3A}
\pdfglyphtounicode{phisymbolgreek}{03D5}
\pdfglyphtounicode{phook}{01A5}
\pdfglyphtounicode{phophanthai}{0E1E}
\pdfglyphtounicode{phophungthai}{0E1C}
\pdfglyphtounicode{phosamphaothai}{0E20}
\pdfglyphtounicode{pi}{03C0}
\pdfglyphtounicode{pi1}{03D6}
\pdfglyphtounicode{pieupacirclekorean}{3273}
\pdfglyphtounicode{pieupaparenkorean}{3213}
\pdfglyphtounicode{pieupcieuckorean}{3176}
\pdfglyphtounicode{pieupcirclekorean}{3265}
\pdfglyphtounicode{pieupkiyeokkorean}{3172}
\pdfglyphtounicode{pieupkorean}{3142}
\pdfglyphtounicode{pieupparenkorean}{3205}
\pdfglyphtounicode{pieupsioskiyeokkorean}{3174}
\pdfglyphtounicode{pieupsioskorean}{3144}
\pdfglyphtounicode{pieupsiostikeutkorean}{3175}
\pdfglyphtounicode{pieupthieuthkorean}{3177}
\pdfglyphtounicode{pieuptikeutkorean}{3173}
\pdfglyphtounicode{pihiragana}{3074}
\pdfglyphtounicode{pikatakana}{30D4}
\pdfglyphtounicode{pisymbolgreek}{03D6}
\pdfglyphtounicode{piwrarmenian}{0583}
\pdfglyphtounicode{planckover2pi}{210F}
\pdfglyphtounicode{planckover2pi1}{210F}
\pdfglyphtounicode{plus}{002B}
\pdfglyphtounicode{plusbelowcmb}{031F}
\pdfglyphtounicode{pluscircle}{2295}
\pdfglyphtounicode{plusminus}{00B1}
\pdfglyphtounicode{plusmod}{02D6}
\pdfglyphtounicode{plusmonospace}{FF0B}
\pdfglyphtounicode{plussmall}{FE62}
\pdfglyphtounicode{plussuperior}{207A}
\pdfglyphtounicode{pmonospace}{FF50}
\pdfglyphtounicode{pmsquare}{33D8}
\pdfglyphtounicode{pohiragana}{307D}
\pdfglyphtounicode{pointingindexdownwhite}{261F}
\pdfglyphtounicode{pointingindexleftwhite}{261C}
\pdfglyphtounicode{pointingindexrightwhite}{261E}
\pdfglyphtounicode{pointingindexupwhite}{261D}
\pdfglyphtounicode{pokatakana}{30DD}
\pdfglyphtounicode{poplathai}{0E1B}
\pdfglyphtounicode{postalmark}{3012}
\pdfglyphtounicode{postalmarkface}{3020}
\pdfglyphtounicode{pparen}{24AB}
\pdfglyphtounicode{precedenotdbleqv}{2AB9}
\pdfglyphtounicode{precedenotslnteql}{2AB5}
\pdfglyphtounicode{precedeornoteqvlnt}{22E8}
\pdfglyphtounicode{precedes}{227A}
\pdfglyphtounicode{precedesequal}{2AAF}
\pdfglyphtounicode{precedesorcurly}{227C}
\pdfglyphtounicode{precedesorequal}{227E}
\pdfglyphtounicode{prescription}{211E}
\pdfglyphtounicode{prime}{2032}
\pdfglyphtounicode{primemod}{02B9}
\pdfglyphtounicode{primereverse}{2035}
\pdfglyphtounicode{primereversed}{2035}
\pdfglyphtounicode{product}{220F}
\pdfglyphtounicode{projective}{2305}
\pdfglyphtounicode{prolongedkana}{30FC}
\pdfglyphtounicode{propellor}{2318}
\pdfglyphtounicode{propersubset}{2282}
\pdfglyphtounicode{propersuperset}{2283}
\pdfglyphtounicode{proportion}{2237}
\pdfglyphtounicode{proportional}{221D}
\pdfglyphtounicode{psi}{03C8}
\pdfglyphtounicode{psicyrillic}{0471}
\pdfglyphtounicode{psilipneumatacyrilliccmb}{0486}
\pdfglyphtounicode{pssquare}{33B0}
\pdfglyphtounicode{puhiragana}{3077}
\pdfglyphtounicode{pukatakana}{30D7}
\pdfglyphtounicode{punctdash}{2014}
\pdfglyphtounicode{pvsquare}{33B4}
\pdfglyphtounicode{pwsquare}{33BA}
\pdfglyphtounicode{q}{0071}
\pdfglyphtounicode{qadeva}{0958}
\pdfglyphtounicode{qadmahebrew}{05A8}
\pdfglyphtounicode{qafarabic}{0642}
\pdfglyphtounicode{qaffinalarabic}{FED6}
\pdfglyphtounicode{qafinitialarabic}{FED7}
\pdfglyphtounicode{qafmedialarabic}{FED8}
\pdfglyphtounicode{qamats}{05B8}
\pdfglyphtounicode{qamats10}{05B8}
\pdfglyphtounicode{qamats1a}{05B8}
\pdfglyphtounicode{qamats1c}{05B8}
\pdfglyphtounicode{qamats27}{05B8}
\pdfglyphtounicode{qamats29}{05B8}
\pdfglyphtounicode{qamats33}{05B8}
\pdfglyphtounicode{qamatsde}{05B8}
\pdfglyphtounicode{qamatshebrew}{05B8}
\pdfglyphtounicode{qamatsnarrowhebrew}{05B8}
\pdfglyphtounicode{qamatsqatanhebrew}{05B8}
\pdfglyphtounicode{qamatsqatannarrowhebrew}{05B8}
\pdfglyphtounicode{qamatsqatanquarterhebrew}{05B8}
\pdfglyphtounicode{qamatsqatanwidehebrew}{05B8}
\pdfglyphtounicode{qamatsquarterhebrew}{05B8}
\pdfglyphtounicode{qamatswidehebrew}{05B8}
\pdfglyphtounicode{qarneyparahebrew}{059F}
\pdfglyphtounicode{qbopomofo}{3111}
\pdfglyphtounicode{qcircle}{24E0}
\pdfglyphtounicode{qhook}{02A0}
\pdfglyphtounicode{qmonospace}{FF51}
\pdfglyphtounicode{qof}{05E7}
\pdfglyphtounicode{qofdagesh}{FB47}
\pdfglyphtounicode{qofdageshhebrew}{FB47}
\pdfglyphtounicode{qofhatafpatah}{05E7 05B2}
\pdfglyphtounicode{qofhatafpatahhebrew}{05E7 05B2}
\pdfglyphtounicode{qofhatafsegol}{05E7 05B1}
\pdfglyphtounicode{qofhatafsegolhebrew}{05E7 05B1}
\pdfglyphtounicode{qofhebrew}{05E7}
\pdfglyphtounicode{qofhiriq}{05E7 05B4}
\pdfglyphtounicode{qofhiriqhebrew}{05E7 05B4}
\pdfglyphtounicode{qofholam}{05E7 05B9}
\pdfglyphtounicode{qofholamhebrew}{05E7 05B9}
\pdfglyphtounicode{qofpatah}{05E7 05B7}
\pdfglyphtounicode{qofpatahhebrew}{05E7 05B7}
\pdfglyphtounicode{qofqamats}{05E7 05B8}
\pdfglyphtounicode{qofqamatshebrew}{05E7 05B8}
\pdfglyphtounicode{qofqubuts}{05E7 05BB}
\pdfglyphtounicode{qofqubutshebrew}{05E7 05BB}
\pdfglyphtounicode{qofsegol}{05E7 05B6}
\pdfglyphtounicode{qofsegolhebrew}{05E7 05B6}
\pdfglyphtounicode{qofsheva}{05E7 05B0}
\pdfglyphtounicode{qofshevahebrew}{05E7 05B0}
\pdfglyphtounicode{qoftsere}{05E7 05B5}
\pdfglyphtounicode{qoftserehebrew}{05E7 05B5}
\pdfglyphtounicode{qparen}{24AC}
\pdfglyphtounicode{quarternote}{2669}
\pdfglyphtounicode{qubuts}{05BB}
\pdfglyphtounicode{qubuts18}{05BB}
\pdfglyphtounicode{qubuts25}{05BB}
\pdfglyphtounicode{qubuts31}{05BB}
\pdfglyphtounicode{qubutshebrew}{05BB}
\pdfglyphtounicode{qubutsnarrowhebrew}{05BB}
\pdfglyphtounicode{qubutsquarterhebrew}{05BB}
\pdfglyphtounicode{qubutswidehebrew}{05BB}
\pdfglyphtounicode{question}{003F}
\pdfglyphtounicode{questionarabic}{061F}
\pdfglyphtounicode{questionarmenian}{055E}
\pdfglyphtounicode{questiondown}{00BF}
\pdfglyphtounicode{questiondownsmall}{00BF}
\pdfglyphtounicode{questiongreek}{037E}
\pdfglyphtounicode{questionmonospace}{FF1F}
\pdfglyphtounicode{questionsmall}{003F}
\pdfglyphtounicode{quotedbl}{0022}
\pdfglyphtounicode{quotedblbase}{201E}
\pdfglyphtounicode{quotedblleft}{201C}
\pdfglyphtounicode{quotedblmonospace}{FF02}
\pdfglyphtounicode{quotedblprime}{301E}
\pdfglyphtounicode{quotedblprimereversed}{301D}
\pdfglyphtounicode{quotedblright}{201D}
\pdfglyphtounicode{quoteleft}{2018}
\pdfglyphtounicode{quoteleftreversed}{201B}
\pdfglyphtounicode{quotereversed}{201B}
\pdfglyphtounicode{quoteright}{2019}
\pdfglyphtounicode{quoterightn}{0149}
\pdfglyphtounicode{quotesinglbase}{201A}
\pdfglyphtounicode{quotesingle}{0027}
\pdfglyphtounicode{quotesinglemonospace}{FF07}
\pdfglyphtounicode{r}{0072}
\pdfglyphtounicode{raarmenian}{057C}
\pdfglyphtounicode{rabengali}{09B0}
\pdfglyphtounicode{racute}{0155}
\pdfglyphtounicode{radeva}{0930}
\pdfglyphtounicode{radical}{221A}
\pdfglyphtounicode{radicalex}{F8E5}
\pdfglyphtounicode{radoverssquare}{33AE}
\pdfglyphtounicode{radoverssquaredsquare}{33AF}
\pdfglyphtounicode{radsquare}{33AD}
\pdfglyphtounicode{rafe}{05BF}
\pdfglyphtounicode{rafehebrew}{05BF}
\pdfglyphtounicode{ragujarati}{0AB0}
\pdfglyphtounicode{ragurmukhi}{0A30}
\pdfglyphtounicode{rahiragana}{3089}
\pdfglyphtounicode{rakatakana}{30E9}
\pdfglyphtounicode{rakatakanahalfwidth}{FF97}
\pdfglyphtounicode{ralowerdiagonalbengali}{09F1}
\pdfglyphtounicode{ramiddlediagonalbengali}{09F0}
\pdfglyphtounicode{ramshorn}{0264}
\pdfglyphtounicode{rangedash}{2013}
\pdfglyphtounicode{ratio}{2236}
\pdfglyphtounicode{rbopomofo}{3116}
\pdfglyphtounicode{rcaron}{0159}
\pdfglyphtounicode{rcedilla}{0157}
\pdfglyphtounicode{rcircle}{24E1}
\pdfglyphtounicode{rcommaaccent}{0157}
\pdfglyphtounicode{rdblgrave}{0211}
\pdfglyphtounicode{rdotaccent}{1E59}
\pdfglyphtounicode{rdotbelow}{1E5B}
\pdfglyphtounicode{rdotbelowmacron}{1E5D}
\pdfglyphtounicode{referencemark}{203B}
\pdfglyphtounicode{reflexsubset}{2286}
\pdfglyphtounicode{reflexsuperset}{2287}
\pdfglyphtounicode{registered}{00AE}
\pdfglyphtounicode{registersans}{00AE}
\pdfglyphtounicode{registerserif}{00AE}
\pdfglyphtounicode{reharabic}{0631}
\pdfglyphtounicode{reharmenian}{0580}
\pdfglyphtounicode{rehfinalarabic}{FEAE}
\pdfglyphtounicode{rehiragana}{308C}
\pdfglyphtounicode{rehyehaleflamarabic}{0631 FEF3 FE8E 0644}
\pdfglyphtounicode{rekatakana}{30EC}
\pdfglyphtounicode{rekatakanahalfwidth}{FF9A}
\pdfglyphtounicode{resh}{05E8}
\pdfglyphtounicode{reshdageshhebrew}{FB48}
\pdfglyphtounicode{reshhatafpatah}{05E8 05B2}
\pdfglyphtounicode{reshhatafpatahhebrew}{05E8 05B2}
\pdfglyphtounicode{reshhatafsegol}{05E8 05B1}
\pdfglyphtounicode{reshhatafsegolhebrew}{05E8 05B1}
\pdfglyphtounicode{reshhebrew}{05E8}
\pdfglyphtounicode{reshhiriq}{05E8 05B4}
\pdfglyphtounicode{reshhiriqhebrew}{05E8 05B4}
\pdfglyphtounicode{reshholam}{05E8 05B9}
\pdfglyphtounicode{reshholamhebrew}{05E8 05B9}
\pdfglyphtounicode{reshpatah}{05E8 05B7}
\pdfglyphtounicode{reshpatahhebrew}{05E8 05B7}
\pdfglyphtounicode{reshqamats}{05E8 05B8}
\pdfglyphtounicode{reshqamatshebrew}{05E8 05B8}
\pdfglyphtounicode{reshqubuts}{05E8 05BB}
\pdfglyphtounicode{reshqubutshebrew}{05E8 05BB}
\pdfglyphtounicode{reshsegol}{05E8 05B6}
\pdfglyphtounicode{reshsegolhebrew}{05E8 05B6}
\pdfglyphtounicode{reshsheva}{05E8 05B0}
\pdfglyphtounicode{reshshevahebrew}{05E8 05B0}
\pdfglyphtounicode{reshtsere}{05E8 05B5}
\pdfglyphtounicode{reshtserehebrew}{05E8 05B5}
\pdfglyphtounicode{revasymptequal}{22CD}
\pdfglyphtounicode{reversedtilde}{223D}
\pdfglyphtounicode{reviahebrew}{0597}
\pdfglyphtounicode{reviamugrashhebrew}{0597}
\pdfglyphtounicode{revlogicalnot}{2310}
\pdfglyphtounicode{revsimilar}{223D}
\pdfglyphtounicode{rfishhook}{027E}
\pdfglyphtounicode{rfishhookreversed}{027F}
\pdfglyphtounicode{rhabengali}{09DD}
\pdfglyphtounicode{rhadeva}{095D}
\pdfglyphtounicode{rho}{03C1}
\pdfglyphtounicode{rho1}{03F1}
\pdfglyphtounicode{rhook}{027D}
\pdfglyphtounicode{rhookturned}{027B}
\pdfglyphtounicode{rhookturnedsuperior}{02B5}
\pdfglyphtounicode{rhosymbolgreek}{03F1}
\pdfglyphtounicode{rhotichookmod}{02DE}
\pdfglyphtounicode{rieulacirclekorean}{3271}
\pdfglyphtounicode{rieulaparenkorean}{3211}
\pdfglyphtounicode{rieulcirclekorean}{3263}
\pdfglyphtounicode{rieulhieuhkorean}{3140}
\pdfglyphtounicode{rieulkiyeokkorean}{313A}
\pdfglyphtounicode{rieulkiyeoksioskorean}{3169}
\pdfglyphtounicode{rieulkorean}{3139}
\pdfglyphtounicode{rieulmieumkorean}{313B}
\pdfglyphtounicode{rieulpansioskorean}{316C}
\pdfglyphtounicode{rieulparenkorean}{3203}
\pdfglyphtounicode{rieulphieuphkorean}{313F}
\pdfglyphtounicode{rieulpieupkorean}{313C}
\pdfglyphtounicode{rieulpieupsioskorean}{316B}
\pdfglyphtounicode{rieulsioskorean}{313D}
\pdfglyphtounicode{rieulthieuthkorean}{313E}
\pdfglyphtounicode{rieultikeutkorean}{316A}
\pdfglyphtounicode{rieulyeorinhieuhkorean}{316D}
\pdfglyphtounicode{rightangle}{221F}
\pdfglyphtounicode{rightanglene}{231D}
\pdfglyphtounicode{rightanglenw}{231C}
\pdfglyphtounicode{rightanglese}{231F}
\pdfglyphtounicode{rightanglesw}{231E}
\pdfglyphtounicode{righttackbelowcmb}{0319}
\pdfglyphtounicode{righttriangle}{22BF}
\pdfglyphtounicode{rihiragana}{308A}
\pdfglyphtounicode{rikatakana}{30EA}
\pdfglyphtounicode{rikatakanahalfwidth}{FF98}
\pdfglyphtounicode{ring}{02DA}
\pdfglyphtounicode{ringbelowcmb}{0325}
\pdfglyphtounicode{ringcmb}{030A}
\pdfglyphtounicode{ringhalfleft}{02BF}
\pdfglyphtounicode{ringhalfleftarmenian}{0559}
\pdfglyphtounicode{ringhalfleftbelowcmb}{031C}
\pdfglyphtounicode{ringhalfleftcentered}{02D3}
\pdfglyphtounicode{ringhalfright}{02BE}
\pdfglyphtounicode{ringhalfrightbelowcmb}{0339}
\pdfglyphtounicode{ringhalfrightcentered}{02D2}
\pdfglyphtounicode{ringinequal}{2256}
\pdfglyphtounicode{rinvertedbreve}{0213}
\pdfglyphtounicode{rittorusquare}{3351}
\pdfglyphtounicode{rlinebelow}{1E5F}
\pdfglyphtounicode{rlongleg}{027C}
\pdfglyphtounicode{rlonglegturned}{027A}
\pdfglyphtounicode{rmonospace}{FF52}
\pdfglyphtounicode{rohiragana}{308D}
\pdfglyphtounicode{rokatakana}{30ED}
\pdfglyphtounicode{rokatakanahalfwidth}{FF9B}
\pdfglyphtounicode{roruathai}{0E23}
\pdfglyphtounicode{rparen}{24AD}
\pdfglyphtounicode{rrabengali}{09DC}
\pdfglyphtounicode{rradeva}{0931}
\pdfglyphtounicode{rragurmukhi}{0A5C}
\pdfglyphtounicode{rreharabic}{0691}
\pdfglyphtounicode{rrehfinalarabic}{FB8D}
\pdfglyphtounicode{rrvocalicbengali}{09E0}
\pdfglyphtounicode{rrvocalicdeva}{0960}
\pdfglyphtounicode{rrvocalicgujarati}{0AE0}
\pdfglyphtounicode{rrvocalicvowelsignbengali}{09C4}
\pdfglyphtounicode{rrvocalicvowelsigndeva}{0944}
\pdfglyphtounicode{rrvocalicvowelsigngujarati}{0AC4}
\pdfglyphtounicode{rsuperior}{0072}
\pdfglyphtounicode{rtblock}{2590}
\pdfglyphtounicode{rturned}{0279}
\pdfglyphtounicode{rturnedsuperior}{02B4}
\pdfglyphtounicode{ruhiragana}{308B}
\pdfglyphtounicode{rukatakana}{30EB}
\pdfglyphtounicode{rukatakanahalfwidth}{FF99}
\pdfglyphtounicode{rupeemarkbengali}{09F2}
\pdfglyphtounicode{rupeesignbengali}{09F3}
\pdfglyphtounicode{rupiah}{20A8}
\pdfglyphtounicode{ruthai}{0E24}
\pdfglyphtounicode{rvocalicbengali}{098B}
\pdfglyphtounicode{rvocalicdeva}{090B}
\pdfglyphtounicode{rvocalicgujarati}{0A8B}
\pdfglyphtounicode{rvocalicvowelsignbengali}{09C3}
\pdfglyphtounicode{rvocalicvowelsigndeva}{0943}
\pdfglyphtounicode{rvocalicvowelsigngujarati}{0AC3}
\pdfglyphtounicode{s}{0073}
\pdfglyphtounicode{sabengali}{09B8}
\pdfglyphtounicode{sacute}{015B}
\pdfglyphtounicode{sacutedotaccent}{1E65}
\pdfglyphtounicode{sadarabic}{0635}
\pdfglyphtounicode{sadeva}{0938}
\pdfglyphtounicode{sadfinalarabic}{FEBA}
\pdfglyphtounicode{sadinitialarabic}{FEBB}
\pdfglyphtounicode{sadmedialarabic}{FEBC}
\pdfglyphtounicode{sagujarati}{0AB8}
\pdfglyphtounicode{sagurmukhi}{0A38}
\pdfglyphtounicode{sahiragana}{3055}
\pdfglyphtounicode{sakatakana}{30B5}
\pdfglyphtounicode{sakatakanahalfwidth}{FF7B}
\pdfglyphtounicode{sallallahoualayhewasallamarabic}{FDFA}
\pdfglyphtounicode{samekh}{05E1}
\pdfglyphtounicode{samekhdagesh}{FB41}
\pdfglyphtounicode{samekhdageshhebrew}{FB41}
\pdfglyphtounicode{samekhhebrew}{05E1}
\pdfglyphtounicode{saraaathai}{0E32}
\pdfglyphtounicode{saraaethai}{0E41}
\pdfglyphtounicode{saraaimaimalaithai}{0E44}
\pdfglyphtounicode{saraaimaimuanthai}{0E43}
\pdfglyphtounicode{saraamthai}{0E33}
\pdfglyphtounicode{saraathai}{0E30}
\pdfglyphtounicode{saraethai}{0E40}
\pdfglyphtounicode{saraiileftthai}{F886}
\pdfglyphtounicode{saraiithai}{0E35}
\pdfglyphtounicode{saraileftthai}{F885}
\pdfglyphtounicode{saraithai}{0E34}
\pdfglyphtounicode{saraothai}{0E42}
\pdfglyphtounicode{saraueeleftthai}{F888}
\pdfglyphtounicode{saraueethai}{0E37}
\pdfglyphtounicode{saraueleftthai}{F887}
\pdfglyphtounicode{sarauethai}{0E36}
\pdfglyphtounicode{sarauthai}{0E38}
\pdfglyphtounicode{sarauuthai}{0E39}
\pdfglyphtounicode{satisfies}{22A8}
\pdfglyphtounicode{sbopomofo}{3119}
\pdfglyphtounicode{scaron}{0161}
\pdfglyphtounicode{scarondotaccent}{1E67}
\pdfglyphtounicode{scedilla}{015F}
\pdfglyphtounicode{schwa}{0259}
\pdfglyphtounicode{schwacyrillic}{04D9}
\pdfglyphtounicode{schwadieresiscyrillic}{04DB}
\pdfglyphtounicode{schwahook}{025A}
\pdfglyphtounicode{scircle}{24E2}
\pdfglyphtounicode{scircumflex}{015D}
\pdfglyphtounicode{scommaaccent}{0219}
\pdfglyphtounicode{sdotaccent}{1E61}
\pdfglyphtounicode{sdotbelow}{1E63}
\pdfglyphtounicode{sdotbelowdotaccent}{1E69}
\pdfglyphtounicode{seagullbelowcmb}{033C}
\pdfglyphtounicode{second}{2033}
\pdfglyphtounicode{secondtonechinese}{02CA}
\pdfglyphtounicode{section}{00A7}
\pdfglyphtounicode{seenarabic}{0633}
\pdfglyphtounicode{seenfinalarabic}{FEB2}
\pdfglyphtounicode{seeninitialarabic}{FEB3}
\pdfglyphtounicode{seenmedialarabic}{FEB4}
\pdfglyphtounicode{segol}{05B6}
\pdfglyphtounicode{segol13}{05B6}
\pdfglyphtounicode{segol1f}{05B6}
\pdfglyphtounicode{segol2c}{05B6}
\pdfglyphtounicode{segolhebrew}{05B6}
\pdfglyphtounicode{segolnarrowhebrew}{05B6}
\pdfglyphtounicode{segolquarterhebrew}{05B6}
\pdfglyphtounicode{segoltahebrew}{0592}
\pdfglyphtounicode{segolwidehebrew}{05B6}
\pdfglyphtounicode{seharmenian}{057D}
\pdfglyphtounicode{sehiragana}{305B}
\pdfglyphtounicode{sekatakana}{30BB}
\pdfglyphtounicode{sekatakanahalfwidth}{FF7E}
\pdfglyphtounicode{semicolon}{003B}
\pdfglyphtounicode{semicolonarabic}{061B}
\pdfglyphtounicode{semicolonmonospace}{FF1B}
\pdfglyphtounicode{semicolonsmall}{FE54}
\pdfglyphtounicode{semivoicedmarkkana}{309C}
\pdfglyphtounicode{semivoicedmarkkanahalfwidth}{FF9F}
\pdfglyphtounicode{sentisquare}{3322}
\pdfglyphtounicode{sentosquare}{3323}
\pdfglyphtounicode{seven}{0037}
\pdfglyphtounicode{sevenarabic}{0667}
\pdfglyphtounicode{sevenbengali}{09ED}
\pdfglyphtounicode{sevencircle}{2466}
\pdfglyphtounicode{sevencircleinversesansserif}{2790}
\pdfglyphtounicode{sevendeva}{096D}
\pdfglyphtounicode{seveneighths}{215E}
\pdfglyphtounicode{sevengujarati}{0AED}
\pdfglyphtounicode{sevengurmukhi}{0A6D}
\pdfglyphtounicode{sevenhackarabic}{0667}
\pdfglyphtounicode{sevenhangzhou}{3027}
\pdfglyphtounicode{sevenideographicparen}{3226}
\pdfglyphtounicode{seveninferior}{2087}
\pdfglyphtounicode{sevenmonospace}{FF17}
\pdfglyphtounicode{sevenoldstyle}{0037}
\pdfglyphtounicode{sevenparen}{247A}
\pdfglyphtounicode{sevenperiod}{248E}
\pdfglyphtounicode{sevenpersian}{06F7}
\pdfglyphtounicode{sevenroman}{2176}
\pdfglyphtounicode{sevensuperior}{2077}
\pdfglyphtounicode{seventeencircle}{2470}
\pdfglyphtounicode{seventeenparen}{2484}
\pdfglyphtounicode{seventeenperiod}{2498}
\pdfglyphtounicode{seventhai}{0E57}
\pdfglyphtounicode{sfthyphen}{00AD}
\pdfglyphtounicode{shaarmenian}{0577}
\pdfglyphtounicode{shabengali}{09B6}
\pdfglyphtounicode{shacyrillic}{0448}
\pdfglyphtounicode{shaddaarabic}{0651}
\pdfglyphtounicode{shaddadammaarabic}{FC61}
\pdfglyphtounicode{shaddadammatanarabic}{FC5E}
\pdfglyphtounicode{shaddafathaarabic}{FC60}
\pdfglyphtounicode{shaddafathatanarabic}{0651 064B}
\pdfglyphtounicode{shaddakasraarabic}{FC62}
\pdfglyphtounicode{shaddakasratanarabic}{FC5F}
\pdfglyphtounicode{shade}{2592}
\pdfglyphtounicode{shadedark}{2593}
\pdfglyphtounicode{shadelight}{2591}
\pdfglyphtounicode{shademedium}{2592}
\pdfglyphtounicode{shadeva}{0936}
\pdfglyphtounicode{shagujarati}{0AB6}
\pdfglyphtounicode{shagurmukhi}{0A36}
\pdfglyphtounicode{shalshelethebrew}{0593}
\pdfglyphtounicode{sharp}{266F}
\pdfglyphtounicode{shbopomofo}{3115}
\pdfglyphtounicode{shchacyrillic}{0449}
\pdfglyphtounicode{sheenarabic}{0634}
\pdfglyphtounicode{sheenfinalarabic}{FEB6}
\pdfglyphtounicode{sheeninitialarabic}{FEB7}
\pdfglyphtounicode{sheenmedialarabic}{FEB8}
\pdfglyphtounicode{sheicoptic}{03E3}
\pdfglyphtounicode{sheqel}{20AA}
\pdfglyphtounicode{sheqelhebrew}{20AA}
\pdfglyphtounicode{sheva}{05B0}
\pdfglyphtounicode{sheva115}{05B0}
\pdfglyphtounicode{sheva15}{05B0}
\pdfglyphtounicode{sheva22}{05B0}
\pdfglyphtounicode{sheva2e}{05B0}
\pdfglyphtounicode{shevahebrew}{05B0}
\pdfglyphtounicode{shevanarrowhebrew}{05B0}
\pdfglyphtounicode{shevaquarterhebrew}{05B0}
\pdfglyphtounicode{shevawidehebrew}{05B0}
\pdfglyphtounicode{shhacyrillic}{04BB}
\pdfglyphtounicode{shiftleft}{21B0}
\pdfglyphtounicode{shiftright}{21B1}
\pdfglyphtounicode{shimacoptic}{03ED}
\pdfglyphtounicode{shin}{05E9}
\pdfglyphtounicode{shindagesh}{FB49}
\pdfglyphtounicode{shindageshhebrew}{FB49}
\pdfglyphtounicode{shindageshshindot}{FB2C}
\pdfglyphtounicode{shindageshshindothebrew}{FB2C}
\pdfglyphtounicode{shindageshsindot}{FB2D}
\pdfglyphtounicode{shindageshsindothebrew}{FB2D}
\pdfglyphtounicode{shindothebrew}{05C1}
\pdfglyphtounicode{shinhebrew}{05E9}
\pdfglyphtounicode{shinshindot}{FB2A}
\pdfglyphtounicode{shinshindothebrew}{FB2A}
\pdfglyphtounicode{shinsindot}{FB2B}
\pdfglyphtounicode{shinsindothebrew}{FB2B}
\pdfglyphtounicode{shook}{0282}
\pdfglyphtounicode{sigma}{03C3}
\pdfglyphtounicode{sigma1}{03C2}
\pdfglyphtounicode{sigmafinal}{03C2}
\pdfglyphtounicode{sigmalunatesymbolgreek}{03F2}
\pdfglyphtounicode{sihiragana}{3057}
\pdfglyphtounicode{sikatakana}{30B7}
\pdfglyphtounicode{sikatakanahalfwidth}{FF7C}
\pdfglyphtounicode{siluqhebrew}{05BD}
\pdfglyphtounicode{siluqlefthebrew}{05BD}
\pdfglyphtounicode{similar}{223C}
\pdfglyphtounicode{similarequal}{2243}
\pdfglyphtounicode{sindothebrew}{05C2}
\pdfglyphtounicode{siosacirclekorean}{3274}
\pdfglyphtounicode{siosaparenkorean}{3214}
\pdfglyphtounicode{sioscieuckorean}{317E}
\pdfglyphtounicode{sioscirclekorean}{3266}
\pdfglyphtounicode{sioskiyeokkorean}{317A}
\pdfglyphtounicode{sioskorean}{3145}
\pdfglyphtounicode{siosnieunkorean}{317B}
\pdfglyphtounicode{siosparenkorean}{3206}
\pdfglyphtounicode{siospieupkorean}{317D}
\pdfglyphtounicode{siostikeutkorean}{317C}
\pdfglyphtounicode{six}{0036}
\pdfglyphtounicode{sixarabic}{0666}
\pdfglyphtounicode{sixbengali}{09EC}
\pdfglyphtounicode{sixcircle}{2465}
\pdfglyphtounicode{sixcircleinversesansserif}{278F}
\pdfglyphtounicode{sixdeva}{096C}
\pdfglyphtounicode{sixgujarati}{0AEC}
\pdfglyphtounicode{sixgurmukhi}{0A6C}
\pdfglyphtounicode{sixhackarabic}{0666}
\pdfglyphtounicode{sixhangzhou}{3026}
\pdfglyphtounicode{sixideographicparen}{3225}
\pdfglyphtounicode{sixinferior}{2086}
\pdfglyphtounicode{sixmonospace}{FF16}
\pdfglyphtounicode{sixoldstyle}{0036}
\pdfglyphtounicode{sixparen}{2479}
\pdfglyphtounicode{sixperiod}{248D}
\pdfglyphtounicode{sixpersian}{06F6}
\pdfglyphtounicode{sixroman}{2175}
\pdfglyphtounicode{sixsuperior}{2076}
\pdfglyphtounicode{sixteencircle}{246F}
\pdfglyphtounicode{sixteencurrencydenominatorbengali}{09F9}
\pdfglyphtounicode{sixteenparen}{2483}
\pdfglyphtounicode{sixteenperiod}{2497}
\pdfglyphtounicode{sixthai}{0E56}
\pdfglyphtounicode{slash}{002F}
\pdfglyphtounicode{slashmonospace}{FF0F}
\pdfglyphtounicode{slong}{017F}
\pdfglyphtounicode{slongdotaccent}{1E9B}
\pdfglyphtounicode{slurabove}{2322}
\pdfglyphtounicode{slurbelow}{2323}
\pdfglyphtounicode{smile}{2323}
\pdfglyphtounicode{smileface}{263A}
\pdfglyphtounicode{smonospace}{FF53}
\pdfglyphtounicode{sofpasuqhebrew}{05C3}
\pdfglyphtounicode{softhyphen}{00AD}
\pdfglyphtounicode{softsigncyrillic}{044C}
\pdfglyphtounicode{sohiragana}{305D}
\pdfglyphtounicode{sokatakana}{30BD}
\pdfglyphtounicode{sokatakanahalfwidth}{FF7F}
\pdfglyphtounicode{soliduslongoverlaycmb}{0338}
\pdfglyphtounicode{solidusshortoverlaycmb}{0337}
\pdfglyphtounicode{sorusithai}{0E29}
\pdfglyphtounicode{sosalathai}{0E28}
\pdfglyphtounicode{sosothai}{0E0B}
\pdfglyphtounicode{sosuathai}{0E2A}
\pdfglyphtounicode{space}{0020}
\pdfglyphtounicode{spacehackarabic}{0020}
\pdfglyphtounicode{spade}{2660}
\pdfglyphtounicode{spadesuitblack}{2660}
\pdfglyphtounicode{spadesuitwhite}{2664}
\pdfglyphtounicode{sparen}{24AE}
\pdfglyphtounicode{sphericalangle}{2222}
\pdfglyphtounicode{square}{25A1}
\pdfglyphtounicode{squarebelowcmb}{033B}
\pdfglyphtounicode{squarecc}{33C4}
\pdfglyphtounicode{squarecm}{339D}
\pdfglyphtounicode{squarediagonalcrosshatchfill}{25A9}
\pdfglyphtounicode{squaredot}{22A1}
\pdfglyphtounicode{squarehorizontalfill}{25A4}
\pdfglyphtounicode{squareimage}{228F}
\pdfglyphtounicode{squarekg}{338F}
\pdfglyphtounicode{squarekm}{339E}
\pdfglyphtounicode{squarekmcapital}{33CE}
\pdfglyphtounicode{squareln}{33D1}
\pdfglyphtounicode{squarelog}{33D2}
\pdfglyphtounicode{squaremg}{338E}
\pdfglyphtounicode{squaremil}{33D5}
\pdfglyphtounicode{squareminus}{229F}
\pdfglyphtounicode{squaremm}{339C}
\pdfglyphtounicode{squaremsquared}{33A1}
\pdfglyphtounicode{squaremultiply}{22A0}
\pdfglyphtounicode{squareoriginal}{2290}
\pdfglyphtounicode{squareorthogonalcrosshatchfill}{25A6}
\pdfglyphtounicode{squareplus}{229E}
\pdfglyphtounicode{squaresolid}{25A0}
\pdfglyphtounicode{squareupperlefttolowerrightfill}{25A7}
\pdfglyphtounicode{squareupperrighttolowerleftfill}{25A8}
\pdfglyphtounicode{squareverticalfill}{25A5}
\pdfglyphtounicode{squarewhitewithsmallblack}{25A3}
\pdfglyphtounicode{squiggleleftright}{21AD}
\pdfglyphtounicode{squiggleright}{21DD}
\pdfglyphtounicode{srsquare}{33DB}
\pdfglyphtounicode{ssabengali}{09B7}
\pdfglyphtounicode{ssadeva}{0937}
\pdfglyphtounicode{ssagujarati}{0AB7}
\pdfglyphtounicode{ssangcieuckorean}{3149}
\pdfglyphtounicode{ssanghieuhkorean}{3185}
\pdfglyphtounicode{ssangieungkorean}{3180}
\pdfglyphtounicode{ssangkiyeokkorean}{3132}
\pdfglyphtounicode{ssangnieunkorean}{3165}
\pdfglyphtounicode{ssangpieupkorean}{3143}
\pdfglyphtounicode{ssangsioskorean}{3146}
\pdfglyphtounicode{ssangtikeutkorean}{3138}
\pdfglyphtounicode{ssuperior}{0073}
\pdfglyphtounicode{st}{0073 0074}
\pdfglyphtounicode{star}{22C6}
\pdfglyphtounicode{sterling}{00A3}
\pdfglyphtounicode{sterlingmonospace}{FFE1}
\pdfglyphtounicode{strokelongoverlaycmb}{0336}
\pdfglyphtounicode{strokeshortoverlaycmb}{0335}
\pdfglyphtounicode{subset}{2282}
\pdfglyphtounicode{subsetdbl}{22D0}
\pdfglyphtounicode{subsetdblequal}{2AC5}
\pdfglyphtounicode{subsetnoteql}{228A}
\pdfglyphtounicode{subsetnotequal}{228A}
\pdfglyphtounicode{subsetorequal}{2286}
\pdfglyphtounicode{subsetornotdbleql}{2ACB}
\pdfglyphtounicode{subsetsqequal}{2291}
\pdfglyphtounicode{succeeds}{227B}
\pdfglyphtounicode{suchthat}{220B}
\pdfglyphtounicode{suhiragana}{3059}
\pdfglyphtounicode{sukatakana}{30B9}
\pdfglyphtounicode{sukatakanahalfwidth}{FF7D}
\pdfglyphtounicode{sukunarabic}{0652}
\pdfglyphtounicode{summation}{2211}
\pdfglyphtounicode{sun}{263C}
\pdfglyphtounicode{superset}{2283}
\pdfglyphtounicode{supersetdbl}{22D1}
\pdfglyphtounicode{supersetdblequal}{2AC6}
\pdfglyphtounicode{supersetnoteql}{228B}
\pdfglyphtounicode{supersetnotequal}{228B}
\pdfglyphtounicode{supersetorequal}{2287}
\pdfglyphtounicode{supersetornotdbleql}{2ACC}
\pdfglyphtounicode{supersetsqequal}{2292}
\pdfglyphtounicode{svsquare}{33DC}
\pdfglyphtounicode{syouwaerasquare}{337C}
\pdfglyphtounicode{t}{0074}
\pdfglyphtounicode{tabengali}{09A4}
\pdfglyphtounicode{tackdown}{22A4}
\pdfglyphtounicode{tackleft}{22A3}
\pdfglyphtounicode{tadeva}{0924}
\pdfglyphtounicode{tagujarati}{0AA4}
\pdfglyphtounicode{tagurmukhi}{0A24}
\pdfglyphtounicode{taharabic}{0637}
\pdfglyphtounicode{tahfinalarabic}{FEC2}
\pdfglyphtounicode{tahinitialarabic}{FEC3}
\pdfglyphtounicode{tahiragana}{305F}
\pdfglyphtounicode{tahmedialarabic}{FEC4}
\pdfglyphtounicode{taisyouerasquare}{337D}
\pdfglyphtounicode{takatakana}{30BF}
\pdfglyphtounicode{takatakanahalfwidth}{FF80}
\pdfglyphtounicode{tatweelarabic}{0640}
\pdfglyphtounicode{tau}{03C4}
\pdfglyphtounicode{tav}{05EA}
\pdfglyphtounicode{tavdages}{FB4A}
\pdfglyphtounicode{tavdagesh}{FB4A}
\pdfglyphtounicode{tavdageshhebrew}{FB4A}
\pdfglyphtounicode{tavhebrew}{05EA}
\pdfglyphtounicode{tbar}{0167}
\pdfglyphtounicode{tbopomofo}{310A}
\pdfglyphtounicode{tcaron}{0165}
\pdfglyphtounicode{tccurl}{02A8}
\pdfglyphtounicode{tcedilla}{0163}
\pdfglyphtounicode{tcheharabic}{0686}
\pdfglyphtounicode{tchehfinalarabic}{FB7B}
\pdfglyphtounicode{tchehinitialarabic}{FB7C}
\pdfglyphtounicode{tchehmedialarabic}{FB7D}
\pdfglyphtounicode{tchehmeeminitialarabic}{FB7C FEE4}
\pdfglyphtounicode{tcircle}{24E3}
\pdfglyphtounicode{tcircumflexbelow}{1E71}
\pdfglyphtounicode{tcommaaccent}{0163}
\pdfglyphtounicode{tdieresis}{1E97}
\pdfglyphtounicode{tdotaccent}{1E6B}
\pdfglyphtounicode{tdotbelow}{1E6D}
\pdfglyphtounicode{tecyrillic}{0442}
\pdfglyphtounicode{tedescendercyrillic}{04AD}
\pdfglyphtounicode{teharabic}{062A}
\pdfglyphtounicode{tehfinalarabic}{FE96}
\pdfglyphtounicode{tehhahinitialarabic}{FCA2}
\pdfglyphtounicode{tehhahisolatedarabic}{FC0C}
\pdfglyphtounicode{tehinitialarabic}{FE97}
\pdfglyphtounicode{tehiragana}{3066}
\pdfglyphtounicode{tehjeeminitialarabic}{FCA1}
\pdfglyphtounicode{tehjeemisolatedarabic}{FC0B}
\pdfglyphtounicode{tehmarbutaarabic}{0629}
\pdfglyphtounicode{tehmarbutafinalarabic}{FE94}
\pdfglyphtounicode{tehmedialarabic}{FE98}
\pdfglyphtounicode{tehmeeminitialarabic}{FCA4}
\pdfglyphtounicode{tehmeemisolatedarabic}{FC0E}
\pdfglyphtounicode{tehnoonfinalarabic}{FC73}
\pdfglyphtounicode{tekatakana}{30C6}
\pdfglyphtounicode{tekatakanahalfwidth}{FF83}
\pdfglyphtounicode{telephone}{2121}
\pdfglyphtounicode{telephoneblack}{260E}
\pdfglyphtounicode{telishagedolahebrew}{05A0}
\pdfglyphtounicode{telishaqetanahebrew}{05A9}
\pdfglyphtounicode{tencircle}{2469}
\pdfglyphtounicode{tenideographicparen}{3229}
\pdfglyphtounicode{tenparen}{247D}
\pdfglyphtounicode{tenperiod}{2491}
\pdfglyphtounicode{tenroman}{2179}
\pdfglyphtounicode{tesh}{02A7}
\pdfglyphtounicode{tet}{05D8}
\pdfglyphtounicode{tetdagesh}{FB38}
\pdfglyphtounicode{tetdageshhebrew}{FB38}
\pdfglyphtounicode{tethebrew}{05D8}
\pdfglyphtounicode{tetsecyrillic}{04B5}
\pdfglyphtounicode{tevirhebrew}{059B}
\pdfglyphtounicode{tevirlefthebrew}{059B}
\pdfglyphtounicode{tfm:cmbsy10/diamond}{2662}
\pdfglyphtounicode{tfm:cmbsy10/heart}{2661}
\pdfglyphtounicode{tfm:cmbsy5/diamond}{2662}
\pdfglyphtounicode{tfm:cmbsy5/heart}{2661}
\pdfglyphtounicode{tfm:cmbsy6/diamond}{2662}
\pdfglyphtounicode{tfm:cmbsy6/heart}{2661}
\pdfglyphtounicode{tfm:cmbsy7/diamond}{2662}
\pdfglyphtounicode{tfm:cmbsy7/heart}{2661}
\pdfglyphtounicode{tfm:cmbsy8/diamond}{2662}
\pdfglyphtounicode{tfm:cmbsy8/heart}{2661}
\pdfglyphtounicode{tfm:cmbsy9/diamond}{2662}
\pdfglyphtounicode{tfm:cmbsy9/heart}{2661}
\pdfglyphtounicode{tfm:cmmi10/phi}{03D5}
\pdfglyphtounicode{tfm:cmmi10/phi1}{03C6}
\pdfglyphtounicode{tfm:cmmi12/phi}{03D5}
\pdfglyphtounicode{tfm:cmmi12/phi1}{03C6}
\pdfglyphtounicode{tfm:cmmi5/phi}{03D5}
\pdfglyphtounicode{tfm:cmmi5/phi1}{03C6}
\pdfglyphtounicode{tfm:cmmi6/phi}{03D5}
\pdfglyphtounicode{tfm:cmmi6/phi1}{03C6}
\pdfglyphtounicode{tfm:cmmi7/phi}{03D5}
\pdfglyphtounicode{tfm:cmmi7/phi1}{03C6}
\pdfglyphtounicode{tfm:cmmi8/phi}{03D5}
\pdfglyphtounicode{tfm:cmmi8/phi1}{03C6}
\pdfglyphtounicode{tfm:cmmi9/phi}{03D5}
\pdfglyphtounicode{tfm:cmmi9/phi1}{03C6}
\pdfglyphtounicode{tfm:cmmib10/phi}{03D5}
\pdfglyphtounicode{tfm:cmmib10/phi1}{03C6}
\pdfglyphtounicode{tfm:cmmib5/phi}{03D5}
\pdfglyphtounicode{tfm:cmmib5/phi1}{03C6}
\pdfglyphtounicode{tfm:cmmib6/phi}{03D5}
\pdfglyphtounicode{tfm:cmmib6/phi1}{03C6}
\pdfglyphtounicode{tfm:cmmib7/phi}{03D5}
\pdfglyphtounicode{tfm:cmmib7/phi1}{03C6}
\pdfglyphtounicode{tfm:cmmib8/phi}{03D5}
\pdfglyphtounicode{tfm:cmmib8/phi1}{03C6}
\pdfglyphtounicode{tfm:cmmib9/phi}{03D5}
\pdfglyphtounicode{tfm:cmmib9/phi1}{03C6}
\pdfglyphtounicode{tfm:cmsy10/diamond}{2662}
\pdfglyphtounicode{tfm:cmsy10/heart}{2661}
\pdfglyphtounicode{tfm:cmsy5/heart}{2661}
\pdfglyphtounicode{tfm:cmsy6/diamond}{2662}
\pdfglyphtounicode{tfm:cmsy6/heart}{2661}
\pdfglyphtounicode{tfm:cmsy7/diamond}{2662}
\pdfglyphtounicode{tfm:cmsy7/heart}{2661}
\pdfglyphtounicode{tfm:cmsy8/diamond}{2662}
\pdfglyphtounicode{tfm:cmsy8/heart}{2661}
\pdfglyphtounicode{tfm:cmsy9/diamond}{2662}
\pdfglyphtounicode{tfm:cmsy9/heart}{2661}
\pdfglyphtounicode{tfm:eurb10/phi}{03D5}
\pdfglyphtounicode{tfm:eurb10/phi1}{03C6}
\pdfglyphtounicode{tfm:eurb5/phi}{03D5}
\pdfglyphtounicode{tfm:eurb5/phi1}{03C6}
\pdfglyphtounicode{tfm:eurb6/phi}{03D5}
\pdfglyphtounicode{tfm:eurb6/phi1}{03C6}
\pdfglyphtounicode{tfm:eurb7/phi}{03D5}
\pdfglyphtounicode{tfm:eurb7/phi1}{03C6}
\pdfglyphtounicode{tfm:eurb8/phi}{03D5}
\pdfglyphtounicode{tfm:eurb8/phi1}{03C6}
\pdfglyphtounicode{tfm:eurb9/phi}{03D5}
\pdfglyphtounicode{tfm:eurb9/phi1}{03C6}
\pdfglyphtounicode{tfm:eurm10/phi}{03D5}
\pdfglyphtounicode{tfm:eurm10/phi1}{03C6}
\pdfglyphtounicode{tfm:eurm5/phi}{03D5}
\pdfglyphtounicode{tfm:eurm5/phi1}{03C6}
\pdfglyphtounicode{tfm:eurm6/phi}{03D5}
\pdfglyphtounicode{tfm:eurm6/phi1}{03C6}
\pdfglyphtounicode{tfm:eurm7/phi}{03D5}
\pdfglyphtounicode{tfm:eurm7/phi1}{03C6}
\pdfglyphtounicode{tfm:eurm8/phi}{03D5}
\pdfglyphtounicode{tfm:eurm8/phi1}{03C6}
\pdfglyphtounicode{tfm:eurm9/phi}{03D5}
\pdfglyphtounicode{tfm:eurm9/phi1}{03C6}
\pdfglyphtounicode{tfm:fplmbi/phi}{03D5}
\pdfglyphtounicode{tfm:fplmbi/phi1}{03C6}
\pdfglyphtounicode{tfm:fplmri/phi}{03D5}
\pdfglyphtounicode{tfm:fplmri/phi1}{03C6}
\pdfglyphtounicode{tfm:lmbsy10/diamond}{2662}
\pdfglyphtounicode{tfm:lmbsy10/heart}{2661}
\pdfglyphtounicode{tfm:lmbsy5/diamond}{2662}
\pdfglyphtounicode{tfm:lmbsy5/heart}{2661}
\pdfglyphtounicode{tfm:lmbsy7/diamond}{2662}
\pdfglyphtounicode{tfm:lmbsy7/heart}{2661}
\pdfglyphtounicode{tfm:lmmi10/phi}{03D5}
\pdfglyphtounicode{tfm:lmmi10/phi1}{03C6}
\pdfglyphtounicode{tfm:lmmi12/phi}{03D5}
\pdfglyphtounicode{tfm:lmmi12/phi1}{03C6}
\pdfglyphtounicode{tfm:lmmi5/phi}{03D5}
\pdfglyphtounicode{tfm:lmmi5/phi1}{03C6}
\pdfglyphtounicode{tfm:lmmi6/phi}{03D5}
\pdfglyphtounicode{tfm:lmmi6/phi1}{03C6}
\pdfglyphtounicode{tfm:lmmi7/phi}{03D5}
\pdfglyphtounicode{tfm:lmmi7/phi1}{03C6}
\pdfglyphtounicode{tfm:lmmi8/phi}{03D5}
\pdfglyphtounicode{tfm:lmmi8/phi1}{03C6}
\pdfglyphtounicode{tfm:lmmi9/phi}{03D5}
\pdfglyphtounicode{tfm:lmmi9/phi1}{03C6}
\pdfglyphtounicode{tfm:lmmib10/phi}{03D5}
\pdfglyphtounicode{tfm:lmmib10/phi1}{03C6}
\pdfglyphtounicode{tfm:lmmib5/phi}{03D5}
\pdfglyphtounicode{tfm:lmmib5/phi1}{03C6}
\pdfglyphtounicode{tfm:lmmib7/phi}{03D5}
\pdfglyphtounicode{tfm:lmmib7/phi1}{03C6}
\pdfglyphtounicode{tfm:lmsy10/diamond}{2662}
\pdfglyphtounicode{tfm:lmsy10/heart}{2661}
\pdfglyphtounicode{tfm:lmsy5/diamond}{2662}
\pdfglyphtounicode{tfm:lmsy5/heart}{2661}
\pdfglyphtounicode{tfm:lmsy6/diamond}{2662}
\pdfglyphtounicode{tfm:lmsy6/heart}{2661}
\pdfglyphtounicode{tfm:lmsy7/diamond}{2662}
\pdfglyphtounicode{tfm:lmsy7/heart}{2661}
\pdfglyphtounicode{tfm:lmsy8/diamond}{2662}
\pdfglyphtounicode{tfm:lmsy8/heart}{2661}
\pdfglyphtounicode{tfm:lmsy9/diamond}{2662}
\pdfglyphtounicode{tfm:lmsy9/heart}{2661}
\pdfglyphtounicode{tfm:msam10/diamond}{2662}
\pdfglyphtounicode{tfm:msam5/diamond}{2662}
\pdfglyphtounicode{tfm:msam6/diamond}{2662}
\pdfglyphtounicode{tfm:msam7/diamond}{2662}
\pdfglyphtounicode{tfm:msam8/diamond}{2662}
\pdfglyphtounicode{tfm:msam9/diamond}{2662}
\pdfglyphtounicode{tfm:pxbmia/phi}{03D5}
\pdfglyphtounicode{tfm:pxbmia/phi1}{03C6}
\pdfglyphtounicode{tfm:pxbsy/diamond}{2662}
\pdfglyphtounicode{tfm:pxbsy/heart}{2661}
\pdfglyphtounicode{tfm:pxbsya/diamond}{2662}
\pdfglyphtounicode{tfm:pxmia/phi}{03D5}
\pdfglyphtounicode{tfm:pxmia/phi1}{03C6}
\pdfglyphtounicode{tfm:pxsy/diamond}{2662}
\pdfglyphtounicode{tfm:pxsy/heart}{2661}
\pdfglyphtounicode{tfm:pxsya/diamond}{2662}
\pdfglyphtounicode{tfm:pzdr/a1}{2701}
\pdfglyphtounicode{tfm:pzdr/a10}{2721}
\pdfglyphtounicode{tfm:pzdr/a100}{275E}
\pdfglyphtounicode{tfm:pzdr/a101}{2761}
\pdfglyphtounicode{tfm:pzdr/a102}{2762}
\pdfglyphtounicode{tfm:pzdr/a103}{2763}
\pdfglyphtounicode{tfm:pzdr/a104}{2764}
\pdfglyphtounicode{tfm:pzdr/a105}{2710}
\pdfglyphtounicode{tfm:pzdr/a106}{2765}
\pdfglyphtounicode{tfm:pzdr/a107}{2766}
\pdfglyphtounicode{tfm:pzdr/a108}{2767}
\pdfglyphtounicode{tfm:pzdr/a109}{2660}
\pdfglyphtounicode{tfm:pzdr/a11}{261B}
\pdfglyphtounicode{tfm:pzdr/a110}{2665}
\pdfglyphtounicode{tfm:pzdr/a111}{2666}
\pdfglyphtounicode{tfm:pzdr/a112}{2663}
\pdfglyphtounicode{tfm:pzdr/a117}{2709}
\pdfglyphtounicode{tfm:pzdr/a118}{2708}
\pdfglyphtounicode{tfm:pzdr/a119}{2707}
\pdfglyphtounicode{tfm:pzdr/a12}{261E}
\pdfglyphtounicode{tfm:pzdr/a120}{2460}
\pdfglyphtounicode{tfm:pzdr/a121}{2461}
\pdfglyphtounicode{tfm:pzdr/a122}{2462}
\pdfglyphtounicode{tfm:pzdr/a123}{2463}
\pdfglyphtounicode{tfm:pzdr/a124}{2464}
\pdfglyphtounicode{tfm:pzdr/a125}{2465}
\pdfglyphtounicode{tfm:pzdr/a126}{2466}
\pdfglyphtounicode{tfm:pzdr/a127}{2467}
\pdfglyphtounicode{tfm:pzdr/a128}{2468}
\pdfglyphtounicode{tfm:pzdr/a129}{2469}
\pdfglyphtounicode{tfm:pzdr/a13}{270C}
\pdfglyphtounicode{tfm:pzdr/a130}{2776}
\pdfglyphtounicode{tfm:pzdr/a131}{2777}
\pdfglyphtounicode{tfm:pzdr/a132}{2778}
\pdfglyphtounicode{tfm:pzdr/a133}{2779}
\pdfglyphtounicode{tfm:pzdr/a134}{277A}
\pdfglyphtounicode{tfm:pzdr/a135}{277B}
\pdfglyphtounicode{tfm:pzdr/a136}{277C}
\pdfglyphtounicode{tfm:pzdr/a137}{277D}
\pdfglyphtounicode{tfm:pzdr/a138}{277E}
\pdfglyphtounicode{tfm:pzdr/a139}{277F}
\pdfglyphtounicode{tfm:pzdr/a14}{270D}
\pdfglyphtounicode{tfm:pzdr/a140}{2780}
\pdfglyphtounicode{tfm:pzdr/a141}{2781}
\pdfglyphtounicode{tfm:pzdr/a142}{2782}
\pdfglyphtounicode{tfm:pzdr/a143}{2783}
\pdfglyphtounicode{tfm:pzdr/a144}{2784}
\pdfglyphtounicode{tfm:pzdr/a145}{2785}
\pdfglyphtounicode{tfm:pzdr/a146}{2786}
\pdfglyphtounicode{tfm:pzdr/a147}{2787}
\pdfglyphtounicode{tfm:pzdr/a148}{2788}
\pdfglyphtounicode{tfm:pzdr/a149}{2789}
\pdfglyphtounicode{tfm:pzdr/a15}{270E}
\pdfglyphtounicode{tfm:pzdr/a150}{278A}
\pdfglyphtounicode{tfm:pzdr/a151}{278B}
\pdfglyphtounicode{tfm:pzdr/a152}{278C}
\pdfglyphtounicode{tfm:pzdr/a153}{278D}
\pdfglyphtounicode{tfm:pzdr/a154}{278E}
\pdfglyphtounicode{tfm:pzdr/a155}{278F}
\pdfglyphtounicode{tfm:pzdr/a156}{2790}
\pdfglyphtounicode{tfm:pzdr/a157}{2791}
\pdfglyphtounicode{tfm:pzdr/a158}{2792}
\pdfglyphtounicode{tfm:pzdr/a159}{2793}
\pdfglyphtounicode{tfm:pzdr/a16}{270F}
\pdfglyphtounicode{tfm:pzdr/a160}{2794}
\pdfglyphtounicode{tfm:pzdr/a161}{2192}
\pdfglyphtounicode{tfm:pzdr/a162}{27A3}
\pdfglyphtounicode{tfm:pzdr/a163}{2194}
\pdfglyphtounicode{tfm:pzdr/a164}{2195}
\pdfglyphtounicode{tfm:pzdr/a165}{2799}
\pdfglyphtounicode{tfm:pzdr/a166}{279B}
\pdfglyphtounicode{tfm:pzdr/a167}{279C}
\pdfglyphtounicode{tfm:pzdr/a168}{279D}
\pdfglyphtounicode{tfm:pzdr/a169}{279E}
\pdfglyphtounicode{tfm:pzdr/a17}{2711}
\pdfglyphtounicode{tfm:pzdr/a170}{279F}
\pdfglyphtounicode{tfm:pzdr/a171}{27A0}
\pdfglyphtounicode{tfm:pzdr/a172}{27A1}
\pdfglyphtounicode{tfm:pzdr/a173}{27A2}
\pdfglyphtounicode{tfm:pzdr/a174}{27A4}
\pdfglyphtounicode{tfm:pzdr/a175}{27A5}
\pdfglyphtounicode{tfm:pzdr/a176}{27A6}
\pdfglyphtounicode{tfm:pzdr/a177}{27A7}
\pdfglyphtounicode{tfm:pzdr/a178}{27A8}
\pdfglyphtounicode{tfm:pzdr/a179}{27A9}
\pdfglyphtounicode{tfm:pzdr/a18}{2712}
\pdfglyphtounicode{tfm:pzdr/a180}{27AB}
\pdfglyphtounicode{tfm:pzdr/a181}{27AD}
\pdfglyphtounicode{tfm:pzdr/a182}{27AF}
\pdfglyphtounicode{tfm:pzdr/a183}{27B2}
\pdfglyphtounicode{tfm:pzdr/a184}{27B3}
\pdfglyphtounicode{tfm:pzdr/a185}{27B5}
\pdfglyphtounicode{tfm:pzdr/a186}{27B8}
\pdfglyphtounicode{tfm:pzdr/a187}{27BA}
\pdfglyphtounicode{tfm:pzdr/a188}{27BB}
\pdfglyphtounicode{tfm:pzdr/a189}{27BC}
\pdfglyphtounicode{tfm:pzdr/a19}{2713}
\pdfglyphtounicode{tfm:pzdr/a190}{27BD}
\pdfglyphtounicode{tfm:pzdr/a191}{27BE}
\pdfglyphtounicode{tfm:pzdr/a192}{279A}
\pdfglyphtounicode{tfm:pzdr/a193}{27AA}
\pdfglyphtounicode{tfm:pzdr/a194}{27B6}
\pdfglyphtounicode{tfm:pzdr/a195}{27B9}
\pdfglyphtounicode{tfm:pzdr/a196}{2798}
\pdfglyphtounicode{tfm:pzdr/a197}{27B4}
\pdfglyphtounicode{tfm:pzdr/a198}{27B7}
\pdfglyphtounicode{tfm:pzdr/a199}{27AC}
\pdfglyphtounicode{tfm:pzdr/a2}{2702}
\pdfglyphtounicode{tfm:pzdr/a20}{2714}
\pdfglyphtounicode{tfm:pzdr/a200}{27AE}
\pdfglyphtounicode{tfm:pzdr/a201}{27B1}
\pdfglyphtounicode{tfm:pzdr/a202}{2703}
\pdfglyphtounicode{tfm:pzdr/a203}{2750}
\pdfglyphtounicode{tfm:pzdr/a204}{2752}
\pdfglyphtounicode{tfm:pzdr/a205}{276E}
\pdfglyphtounicode{tfm:pzdr/a206}{2770}
\pdfglyphtounicode{tfm:pzdr/a21}{2715}
\pdfglyphtounicode{tfm:pzdr/a22}{2716}
\pdfglyphtounicode{tfm:pzdr/a23}{2717}
\pdfglyphtounicode{tfm:pzdr/a24}{2718}
\pdfglyphtounicode{tfm:pzdr/a25}{2719}
\pdfglyphtounicode{tfm:pzdr/a26}{271A}
\pdfglyphtounicode{tfm:pzdr/a27}{271B}
\pdfglyphtounicode{tfm:pzdr/a28}{271C}
\pdfglyphtounicode{tfm:pzdr/a29}{2722}
\pdfglyphtounicode{tfm:pzdr/a3}{2704}
\pdfglyphtounicode{tfm:pzdr/a30}{2723}
\pdfglyphtounicode{tfm:pzdr/a31}{2724}
\pdfglyphtounicode{tfm:pzdr/a32}{2725}
\pdfglyphtounicode{tfm:pzdr/a33}{2726}
\pdfglyphtounicode{tfm:pzdr/a34}{2727}
\pdfglyphtounicode{tfm:pzdr/a35}{2605}
\pdfglyphtounicode{tfm:pzdr/a36}{2729}
\pdfglyphtounicode{tfm:pzdr/a37}{272A}
\pdfglyphtounicode{tfm:pzdr/a38}{272B}
\pdfglyphtounicode{tfm:pzdr/a39}{272C}
\pdfglyphtounicode{tfm:pzdr/a4}{260E}
\pdfglyphtounicode{tfm:pzdr/a40}{272D}
\pdfglyphtounicode{tfm:pzdr/a41}{272E}
\pdfglyphtounicode{tfm:pzdr/a42}{272F}
\pdfglyphtounicode{tfm:pzdr/a43}{2730}
\pdfglyphtounicode{tfm:pzdr/a44}{2731}
\pdfglyphtounicode{tfm:pzdr/a45}{2732}
\pdfglyphtounicode{tfm:pzdr/a46}{2733}
\pdfglyphtounicode{tfm:pzdr/a47}{2734}
\pdfglyphtounicode{tfm:pzdr/a48}{2735}
\pdfglyphtounicode{tfm:pzdr/a49}{2736}
\pdfglyphtounicode{tfm:pzdr/a5}{2706}
\pdfglyphtounicode{tfm:pzdr/a50}{2737}
\pdfglyphtounicode{tfm:pzdr/a51}{2738}
\pdfglyphtounicode{tfm:pzdr/a52}{2739}
\pdfglyphtounicode{tfm:pzdr/a53}{273A}
\pdfglyphtounicode{tfm:pzdr/a54}{273B}
\pdfglyphtounicode{tfm:pzdr/a55}{273C}
\pdfglyphtounicode{tfm:pzdr/a56}{273D}
\pdfglyphtounicode{tfm:pzdr/a57}{273E}
\pdfglyphtounicode{tfm:pzdr/a58}{273F}
\pdfglyphtounicode{tfm:pzdr/a59}{2740}
\pdfglyphtounicode{tfm:pzdr/a6}{271D}
\pdfglyphtounicode{tfm:pzdr/a60}{2741}
\pdfglyphtounicode{tfm:pzdr/a61}{2742}
\pdfglyphtounicode{tfm:pzdr/a62}{2743}
\pdfglyphtounicode{tfm:pzdr/a63}{2744}
\pdfglyphtounicode{tfm:pzdr/a64}{2745}
\pdfglyphtounicode{tfm:pzdr/a65}{2746}
\pdfglyphtounicode{tfm:pzdr/a66}{2747}
\pdfglyphtounicode{tfm:pzdr/a67}{2748}
\pdfglyphtounicode{tfm:pzdr/a68}{2749}
\pdfglyphtounicode{tfm:pzdr/a69}{274A}
\pdfglyphtounicode{tfm:pzdr/a7}{271E}
\pdfglyphtounicode{tfm:pzdr/a70}{274B}
\pdfglyphtounicode{tfm:pzdr/a71}{25CF}
\pdfglyphtounicode{tfm:pzdr/a72}{274D}
\pdfglyphtounicode{tfm:pzdr/a73}{25A0}
\pdfglyphtounicode{tfm:pzdr/a74}{274F}
\pdfglyphtounicode{tfm:pzdr/a75}{2751}
\pdfglyphtounicode{tfm:pzdr/a76}{25B2}
\pdfglyphtounicode{tfm:pzdr/a77}{25BC}
\pdfglyphtounicode{tfm:pzdr/a78}{25C6}
\pdfglyphtounicode{tfm:pzdr/a79}{2756}
\pdfglyphtounicode{tfm:pzdr/a8}{271F}
\pdfglyphtounicode{tfm:pzdr/a81}{25D7}
\pdfglyphtounicode{tfm:pzdr/a82}{2758}
\pdfglyphtounicode{tfm:pzdr/a83}{2759}
\pdfglyphtounicode{tfm:pzdr/a84}{275A}
\pdfglyphtounicode{tfm:pzdr/a85}{276F}
\pdfglyphtounicode{tfm:pzdr/a86}{2771}
\pdfglyphtounicode{tfm:pzdr/a87}{2772}
\pdfglyphtounicode{tfm:pzdr/a88}{2773}
\pdfglyphtounicode{tfm:pzdr/a89}{2768}
\pdfglyphtounicode{tfm:pzdr/a9}{2720}
\pdfglyphtounicode{tfm:pzdr/a90}{2769}
\pdfglyphtounicode{tfm:pzdr/a91}{276C}
\pdfglyphtounicode{tfm:pzdr/a92}{276D}
\pdfglyphtounicode{tfm:pzdr/a93}{276A}
\pdfglyphtounicode{tfm:pzdr/a94}{276B}
\pdfglyphtounicode{tfm:pzdr/a95}{2774}
\pdfglyphtounicode{tfm:pzdr/a96}{2775}
\pdfglyphtounicode{tfm:pzdr/a97}{275B}
\pdfglyphtounicode{tfm:pzdr/a98}{275C}
\pdfglyphtounicode{tfm:pzdr/a99}{275D}
\pdfglyphtounicode{tfm:rpxbmi/phi}{03D5}
\pdfglyphtounicode{tfm:rpxbmi/phi1}{03C6}
\pdfglyphtounicode{tfm:rpxmi/phi}{03D5}
\pdfglyphtounicode{tfm:rpxmi/phi1}{03C6}
\pdfglyphtounicode{tfm:rpzdr/a1}{2701}
\pdfglyphtounicode{tfm:rpzdr/a10}{2721}
\pdfglyphtounicode{tfm:rpzdr/a100}{275E}
\pdfglyphtounicode{tfm:rpzdr/a101}{2761}
\pdfglyphtounicode{tfm:rpzdr/a102}{2762}
\pdfglyphtounicode{tfm:rpzdr/a103}{2763}
\pdfglyphtounicode{tfm:rpzdr/a104}{2764}
\pdfglyphtounicode{tfm:rpzdr/a105}{2710}
\pdfglyphtounicode{tfm:rpzdr/a106}{2765}
\pdfglyphtounicode{tfm:rpzdr/a107}{2766}
\pdfglyphtounicode{tfm:rpzdr/a108}{2767}
\pdfglyphtounicode{tfm:rpzdr/a109}{2660}
\pdfglyphtounicode{tfm:rpzdr/a11}{261B}
\pdfglyphtounicode{tfm:rpzdr/a110}{2665}
\pdfglyphtounicode{tfm:rpzdr/a111}{2666}
\pdfglyphtounicode{tfm:rpzdr/a112}{2663}
\pdfglyphtounicode{tfm:rpzdr/a117}{2709}
\pdfglyphtounicode{tfm:rpzdr/a118}{2708}
\pdfglyphtounicode{tfm:rpzdr/a119}{2707}
\pdfglyphtounicode{tfm:rpzdr/a12}{261E}
\pdfglyphtounicode{tfm:rpzdr/a120}{2460}
\pdfglyphtounicode{tfm:rpzdr/a121}{2461}
\pdfglyphtounicode{tfm:rpzdr/a122}{2462}
\pdfglyphtounicode{tfm:rpzdr/a123}{2463}
\pdfglyphtounicode{tfm:rpzdr/a124}{2464}
\pdfglyphtounicode{tfm:rpzdr/a125}{2465}
\pdfglyphtounicode{tfm:rpzdr/a126}{2466}
\pdfglyphtounicode{tfm:rpzdr/a127}{2467}
\pdfglyphtounicode{tfm:rpzdr/a128}{2468}
\pdfglyphtounicode{tfm:rpzdr/a129}{2469}
\pdfglyphtounicode{tfm:rpzdr/a13}{270C}
\pdfglyphtounicode{tfm:rpzdr/a130}{2776}
\pdfglyphtounicode{tfm:rpzdr/a131}{2777}
\pdfglyphtounicode{tfm:rpzdr/a132}{2778}
\pdfglyphtounicode{tfm:rpzdr/a133}{2779}
\pdfglyphtounicode{tfm:rpzdr/a134}{277A}
\pdfglyphtounicode{tfm:rpzdr/a135}{277B}
\pdfglyphtounicode{tfm:rpzdr/a136}{277C}
\pdfglyphtounicode{tfm:rpzdr/a137}{277D}
\pdfglyphtounicode{tfm:rpzdr/a138}{277E}
\pdfglyphtounicode{tfm:rpzdr/a139}{277F}
\pdfglyphtounicode{tfm:rpzdr/a14}{270D}
\pdfglyphtounicode{tfm:rpzdr/a140}{2780}
\pdfglyphtounicode{tfm:rpzdr/a141}{2781}
\pdfglyphtounicode{tfm:rpzdr/a142}{2782}
\pdfglyphtounicode{tfm:rpzdr/a143}{2783}
\pdfglyphtounicode{tfm:rpzdr/a144}{2784}
\pdfglyphtounicode{tfm:rpzdr/a145}{2785}
\pdfglyphtounicode{tfm:rpzdr/a146}{2786}
\pdfglyphtounicode{tfm:rpzdr/a147}{2787}
\pdfglyphtounicode{tfm:rpzdr/a148}{2788}
\pdfglyphtounicode{tfm:rpzdr/a149}{2789}
\pdfglyphtounicode{tfm:rpzdr/a15}{270E}
\pdfglyphtounicode{tfm:rpzdr/a150}{278A}
\pdfglyphtounicode{tfm:rpzdr/a151}{278B}
\pdfglyphtounicode{tfm:rpzdr/a152}{278C}
\pdfglyphtounicode{tfm:rpzdr/a153}{278D}
\pdfglyphtounicode{tfm:rpzdr/a154}{278E}
\pdfglyphtounicode{tfm:rpzdr/a155}{278F}
\pdfglyphtounicode{tfm:rpzdr/a156}{2790}
\pdfglyphtounicode{tfm:rpzdr/a157}{2791}
\pdfglyphtounicode{tfm:rpzdr/a158}{2792}
\pdfglyphtounicode{tfm:rpzdr/a159}{2793}
\pdfglyphtounicode{tfm:rpzdr/a16}{270F}
\pdfglyphtounicode{tfm:rpzdr/a160}{2794}
\pdfglyphtounicode{tfm:rpzdr/a161}{2192}
\pdfglyphtounicode{tfm:rpzdr/a162}{27A3}
\pdfglyphtounicode{tfm:rpzdr/a163}{2194}
\pdfglyphtounicode{tfm:rpzdr/a164}{2195}
\pdfglyphtounicode{tfm:rpzdr/a165}{2799}
\pdfglyphtounicode{tfm:rpzdr/a166}{279B}
\pdfglyphtounicode{tfm:rpzdr/a167}{279C}
\pdfglyphtounicode{tfm:rpzdr/a168}{279D}
\pdfglyphtounicode{tfm:rpzdr/a169}{279E}
\pdfglyphtounicode{tfm:rpzdr/a17}{2711}
\pdfglyphtounicode{tfm:rpzdr/a170}{279F}
\pdfglyphtounicode{tfm:rpzdr/a171}{27A0}
\pdfglyphtounicode{tfm:rpzdr/a172}{27A1}
\pdfglyphtounicode{tfm:rpzdr/a173}{27A2}
\pdfglyphtounicode{tfm:rpzdr/a174}{27A4}
\pdfglyphtounicode{tfm:rpzdr/a175}{27A5}
\pdfglyphtounicode{tfm:rpzdr/a176}{27A6}
\pdfglyphtounicode{tfm:rpzdr/a177}{27A7}
\pdfglyphtounicode{tfm:rpzdr/a178}{27A8}
\pdfglyphtounicode{tfm:rpzdr/a179}{27A9}
\pdfglyphtounicode{tfm:rpzdr/a18}{2712}
\pdfglyphtounicode{tfm:rpzdr/a180}{27AB}
\pdfglyphtounicode{tfm:rpzdr/a181}{27AD}
\pdfglyphtounicode{tfm:rpzdr/a182}{27AF}
\pdfglyphtounicode{tfm:rpzdr/a183}{27B2}
\pdfglyphtounicode{tfm:rpzdr/a184}{27B3}
\pdfglyphtounicode{tfm:rpzdr/a185}{27B5}
\pdfglyphtounicode{tfm:rpzdr/a186}{27B8}
\pdfglyphtounicode{tfm:rpzdr/a187}{27BA}
\pdfglyphtounicode{tfm:rpzdr/a188}{27BB}
\pdfglyphtounicode{tfm:rpzdr/a189}{27BC}
\pdfglyphtounicode{tfm:rpzdr/a19}{2713}
\pdfglyphtounicode{tfm:rpzdr/a190}{27BD}
\pdfglyphtounicode{tfm:rpzdr/a191}{27BE}
\pdfglyphtounicode{tfm:rpzdr/a192}{279A}
\pdfglyphtounicode{tfm:rpzdr/a193}{27AA}
\pdfglyphtounicode{tfm:rpzdr/a194}{27B6}
\pdfglyphtounicode{tfm:rpzdr/a195}{27B9}
\pdfglyphtounicode{tfm:rpzdr/a196}{2798}
\pdfglyphtounicode{tfm:rpzdr/a197}{27B4}
\pdfglyphtounicode{tfm:rpzdr/a198}{27B7}
\pdfglyphtounicode{tfm:rpzdr/a199}{27AC}
\pdfglyphtounicode{tfm:rpzdr/a2}{2702}
\pdfglyphtounicode{tfm:rpzdr/a20}{2714}
\pdfglyphtounicode{tfm:rpzdr/a200}{27AE}
\pdfglyphtounicode{tfm:rpzdr/a201}{27B1}
\pdfglyphtounicode{tfm:rpzdr/a202}{2703}
\pdfglyphtounicode{tfm:rpzdr/a203}{2750}
\pdfglyphtounicode{tfm:rpzdr/a204}{2752}
\pdfglyphtounicode{tfm:rpzdr/a205}{276E}
\pdfglyphtounicode{tfm:rpzdr/a206}{2770}
\pdfglyphtounicode{tfm:rpzdr/a21}{2715}
\pdfglyphtounicode{tfm:rpzdr/a22}{2716}
\pdfglyphtounicode{tfm:rpzdr/a23}{2717}
\pdfglyphtounicode{tfm:rpzdr/a24}{2718}
\pdfglyphtounicode{tfm:rpzdr/a25}{2719}
\pdfglyphtounicode{tfm:rpzdr/a26}{271A}
\pdfglyphtounicode{tfm:rpzdr/a27}{271B}
\pdfglyphtounicode{tfm:rpzdr/a28}{271C}
\pdfglyphtounicode{tfm:rpzdr/a29}{2722}
\pdfglyphtounicode{tfm:rpzdr/a3}{2704}
\pdfglyphtounicode{tfm:rpzdr/a30}{2723}
\pdfglyphtounicode{tfm:rpzdr/a31}{2724}
\pdfglyphtounicode{tfm:rpzdr/a32}{2725}
\pdfglyphtounicode{tfm:rpzdr/a33}{2726}
\pdfglyphtounicode{tfm:rpzdr/a34}{2727}
\pdfglyphtounicode{tfm:rpzdr/a35}{2605}
\pdfglyphtounicode{tfm:rpzdr/a36}{2729}
\pdfglyphtounicode{tfm:rpzdr/a37}{272A}
\pdfglyphtounicode{tfm:rpzdr/a38}{272B}
\pdfglyphtounicode{tfm:rpzdr/a39}{272C}
\pdfglyphtounicode{tfm:rpzdr/a4}{260E}
\pdfglyphtounicode{tfm:rpzdr/a40}{272D}
\pdfglyphtounicode{tfm:rpzdr/a41}{272E}
\pdfglyphtounicode{tfm:rpzdr/a42}{272F}
\pdfglyphtounicode{tfm:rpzdr/a43}{2730}
\pdfglyphtounicode{tfm:rpzdr/a44}{2731}
\pdfglyphtounicode{tfm:rpzdr/a45}{2732}
\pdfglyphtounicode{tfm:rpzdr/a46}{2733}
\pdfglyphtounicode{tfm:rpzdr/a47}{2734}
\pdfglyphtounicode{tfm:rpzdr/a48}{2735}
\pdfglyphtounicode{tfm:rpzdr/a49}{2736}
\pdfglyphtounicode{tfm:rpzdr/a5}{2706}
\pdfglyphtounicode{tfm:rpzdr/a50}{2737}
\pdfglyphtounicode{tfm:rpzdr/a51}{2738}
\pdfglyphtounicode{tfm:rpzdr/a52}{2739}
\pdfglyphtounicode{tfm:rpzdr/a53}{273A}
\pdfglyphtounicode{tfm:rpzdr/a54}{273B}
\pdfglyphtounicode{tfm:rpzdr/a55}{273C}
\pdfglyphtounicode{tfm:rpzdr/a56}{273D}
\pdfglyphtounicode{tfm:rpzdr/a57}{273E}
\pdfglyphtounicode{tfm:rpzdr/a58}{273F}
\pdfglyphtounicode{tfm:rpzdr/a59}{2740}
\pdfglyphtounicode{tfm:rpzdr/a6}{271D}
\pdfglyphtounicode{tfm:rpzdr/a60}{2741}
\pdfglyphtounicode{tfm:rpzdr/a61}{2742}
\pdfglyphtounicode{tfm:rpzdr/a62}{2743}
\pdfglyphtounicode{tfm:rpzdr/a63}{2744}
\pdfglyphtounicode{tfm:rpzdr/a64}{2745}
\pdfglyphtounicode{tfm:rpzdr/a65}{2746}
\pdfglyphtounicode{tfm:rpzdr/a66}{2747}
\pdfglyphtounicode{tfm:rpzdr/a67}{2748}
\pdfglyphtounicode{tfm:rpzdr/a68}{2749}
\pdfglyphtounicode{tfm:rpzdr/a69}{274A}
\pdfglyphtounicode{tfm:rpzdr/a7}{271E}
\pdfglyphtounicode{tfm:rpzdr/a70}{274B}
\pdfglyphtounicode{tfm:rpzdr/a71}{25CF}
\pdfglyphtounicode{tfm:rpzdr/a72}{274D}
\pdfglyphtounicode{tfm:rpzdr/a73}{25A0}
\pdfglyphtounicode{tfm:rpzdr/a74}{274F}
\pdfglyphtounicode{tfm:rpzdr/a75}{2751}
\pdfglyphtounicode{tfm:rpzdr/a76}{25B2}
\pdfglyphtounicode{tfm:rpzdr/a77}{25BC}
\pdfglyphtounicode{tfm:rpzdr/a78}{25C6}
\pdfglyphtounicode{tfm:rpzdr/a79}{2756}
\pdfglyphtounicode{tfm:rpzdr/a8}{271F}
\pdfglyphtounicode{tfm:rpzdr/a81}{25D7}
\pdfglyphtounicode{tfm:rpzdr/a82}{2758}
\pdfglyphtounicode{tfm:rpzdr/a83}{2759}
\pdfglyphtounicode{tfm:rpzdr/a84}{275A}
\pdfglyphtounicode{tfm:rpzdr/a85}{276F}
\pdfglyphtounicode{tfm:rpzdr/a86}{2771}
\pdfglyphtounicode{tfm:rpzdr/a87}{2772}
\pdfglyphtounicode{tfm:rpzdr/a88}{2773}
\pdfglyphtounicode{tfm:rpzdr/a89}{2768}
\pdfglyphtounicode{tfm:rpzdr/a9}{2720}
\pdfglyphtounicode{tfm:rpzdr/a90}{2769}
\pdfglyphtounicode{tfm:rpzdr/a91}{276C}
\pdfglyphtounicode{tfm:rpzdr/a92}{276D}
\pdfglyphtounicode{tfm:rpzdr/a93}{276A}
\pdfglyphtounicode{tfm:rpzdr/a94}{276B}
\pdfglyphtounicode{tfm:rpzdr/a95}{2774}
\pdfglyphtounicode{tfm:rpzdr/a96}{2775}
\pdfglyphtounicode{tfm:rpzdr/a97}{275B}
\pdfglyphtounicode{tfm:rpzdr/a98}{275C}
\pdfglyphtounicode{tfm:rpzdr/a99}{275D}
\pdfglyphtounicode{tfm:rtxbmi/phi}{03D5}
\pdfglyphtounicode{tfm:rtxbmi/phi1}{03C6}
\pdfglyphtounicode{tfm:rtxmi/phi}{03D5}
\pdfglyphtounicode{tfm:rtxmi/phi1}{03C6}
\pdfglyphtounicode{tfm:txbmia/phi}{03D5}
\pdfglyphtounicode{tfm:txbmia/phi1}{03C6}
\pdfglyphtounicode{tfm:txbsy/diamond}{2662}
\pdfglyphtounicode{tfm:txbsy/heart}{2661}
\pdfglyphtounicode{tfm:txbsya/diamond}{2662}
\pdfglyphtounicode{tfm:txmia/phi}{03D5}
\pdfglyphtounicode{tfm:txmia/phi1}{03C6}
\pdfglyphtounicode{tfm:txsy/diamond}{2662}
\pdfglyphtounicode{tfm:txsy/heart}{2661}
\pdfglyphtounicode{tfm:txsya/diamond}{2662}
\pdfglyphtounicode{tfm:zd/a1}{2701}
\pdfglyphtounicode{tfm:zd/a10}{2721}
\pdfglyphtounicode{tfm:zd/a100}{275E}
\pdfglyphtounicode{tfm:zd/a101}{2761}
\pdfglyphtounicode{tfm:zd/a102}{2762}
\pdfglyphtounicode{tfm:zd/a103}{2763}
\pdfglyphtounicode{tfm:zd/a104}{2764}
\pdfglyphtounicode{tfm:zd/a105}{2710}
\pdfglyphtounicode{tfm:zd/a106}{2765}
\pdfglyphtounicode{tfm:zd/a107}{2766}
\pdfglyphtounicode{tfm:zd/a108}{2767}
\pdfglyphtounicode{tfm:zd/a109}{2660}
\pdfglyphtounicode{tfm:zd/a11}{261B}
\pdfglyphtounicode{tfm:zd/a110}{2665}
\pdfglyphtounicode{tfm:zd/a111}{2666}
\pdfglyphtounicode{tfm:zd/a112}{2663}
\pdfglyphtounicode{tfm:zd/a117}{2709}
\pdfglyphtounicode{tfm:zd/a118}{2708}
\pdfglyphtounicode{tfm:zd/a119}{2707}
\pdfglyphtounicode{tfm:zd/a12}{261E}
\pdfglyphtounicode{tfm:zd/a120}{2460}
\pdfglyphtounicode{tfm:zd/a121}{2461}
\pdfglyphtounicode{tfm:zd/a122}{2462}
\pdfglyphtounicode{tfm:zd/a123}{2463}
\pdfglyphtounicode{tfm:zd/a124}{2464}
\pdfglyphtounicode{tfm:zd/a125}{2465}
\pdfglyphtounicode{tfm:zd/a126}{2466}
\pdfglyphtounicode{tfm:zd/a127}{2467}
\pdfglyphtounicode{tfm:zd/a128}{2468}
\pdfglyphtounicode{tfm:zd/a129}{2469}
\pdfglyphtounicode{tfm:zd/a13}{270C}
\pdfglyphtounicode{tfm:zd/a130}{2776}
\pdfglyphtounicode{tfm:zd/a131}{2777}
\pdfglyphtounicode{tfm:zd/a132}{2778}
\pdfglyphtounicode{tfm:zd/a133}{2779}
\pdfglyphtounicode{tfm:zd/a134}{277A}
\pdfglyphtounicode{tfm:zd/a135}{277B}
\pdfglyphtounicode{tfm:zd/a136}{277C}
\pdfglyphtounicode{tfm:zd/a137}{277D}
\pdfglyphtounicode{tfm:zd/a138}{277E}
\pdfglyphtounicode{tfm:zd/a139}{277F}
\pdfglyphtounicode{tfm:zd/a14}{270D}
\pdfglyphtounicode{tfm:zd/a140}{2780}
\pdfglyphtounicode{tfm:zd/a141}{2781}
\pdfglyphtounicode{tfm:zd/a142}{2782}
\pdfglyphtounicode{tfm:zd/a143}{2783}
\pdfglyphtounicode{tfm:zd/a144}{2784}
\pdfglyphtounicode{tfm:zd/a145}{2785}
\pdfglyphtounicode{tfm:zd/a146}{2786}
\pdfglyphtounicode{tfm:zd/a147}{2787}
\pdfglyphtounicode{tfm:zd/a148}{2788}
\pdfglyphtounicode{tfm:zd/a149}{2789}
\pdfglyphtounicode{tfm:zd/a15}{270E}
\pdfglyphtounicode{tfm:zd/a150}{278A}
\pdfglyphtounicode{tfm:zd/a151}{278B}
\pdfglyphtounicode{tfm:zd/a152}{278C}
\pdfglyphtounicode{tfm:zd/a153}{278D}
\pdfglyphtounicode{tfm:zd/a154}{278E}
\pdfglyphtounicode{tfm:zd/a155}{278F}
\pdfglyphtounicode{tfm:zd/a156}{2790}
\pdfglyphtounicode{tfm:zd/a157}{2791}
\pdfglyphtounicode{tfm:zd/a158}{2792}
\pdfglyphtounicode{tfm:zd/a159}{2793}
\pdfglyphtounicode{tfm:zd/a16}{270F}
\pdfglyphtounicode{tfm:zd/a160}{2794}
\pdfglyphtounicode{tfm:zd/a161}{2192}
\pdfglyphtounicode{tfm:zd/a162}{27A3}
\pdfglyphtounicode{tfm:zd/a163}{2194}
\pdfglyphtounicode{tfm:zd/a164}{2195}
\pdfglyphtounicode{tfm:zd/a165}{2799}
\pdfglyphtounicode{tfm:zd/a166}{279B}
\pdfglyphtounicode{tfm:zd/a167}{279C}
\pdfglyphtounicode{tfm:zd/a168}{279D}
\pdfglyphtounicode{tfm:zd/a169}{279E}
\pdfglyphtounicode{tfm:zd/a17}{2711}
\pdfglyphtounicode{tfm:zd/a170}{279F}
\pdfglyphtounicode{tfm:zd/a171}{27A0}
\pdfglyphtounicode{tfm:zd/a172}{27A1}
\pdfglyphtounicode{tfm:zd/a173}{27A2}
\pdfglyphtounicode{tfm:zd/a174}{27A4}
\pdfglyphtounicode{tfm:zd/a175}{27A5}
\pdfglyphtounicode{tfm:zd/a176}{27A6}
\pdfglyphtounicode{tfm:zd/a177}{27A7}
\pdfglyphtounicode{tfm:zd/a178}{27A8}
\pdfglyphtounicode{tfm:zd/a179}{27A9}
\pdfglyphtounicode{tfm:zd/a18}{2712}
\pdfglyphtounicode{tfm:zd/a180}{27AB}
\pdfglyphtounicode{tfm:zd/a181}{27AD}
\pdfglyphtounicode{tfm:zd/a182}{27AF}
\pdfglyphtounicode{tfm:zd/a183}{27B2}
\pdfglyphtounicode{tfm:zd/a184}{27B3}
\pdfglyphtounicode{tfm:zd/a185}{27B5}
\pdfglyphtounicode{tfm:zd/a186}{27B8}
\pdfglyphtounicode{tfm:zd/a187}{27BA}
\pdfglyphtounicode{tfm:zd/a188}{27BB}
\pdfglyphtounicode{tfm:zd/a189}{27BC}
\pdfglyphtounicode{tfm:zd/a19}{2713}
\pdfglyphtounicode{tfm:zd/a190}{27BD}
\pdfglyphtounicode{tfm:zd/a191}{27BE}
\pdfglyphtounicode{tfm:zd/a192}{279A}
\pdfglyphtounicode{tfm:zd/a193}{27AA}
\pdfglyphtounicode{tfm:zd/a194}{27B6}
\pdfglyphtounicode{tfm:zd/a195}{27B9}
\pdfglyphtounicode{tfm:zd/a196}{2798}
\pdfglyphtounicode{tfm:zd/a197}{27B4}
\pdfglyphtounicode{tfm:zd/a198}{27B7}
\pdfglyphtounicode{tfm:zd/a199}{27AC}
\pdfglyphtounicode{tfm:zd/a2}{2702}
\pdfglyphtounicode{tfm:zd/a20}{2714}
\pdfglyphtounicode{tfm:zd/a200}{27AE}
\pdfglyphtounicode{tfm:zd/a201}{27B1}
\pdfglyphtounicode{tfm:zd/a202}{2703}
\pdfglyphtounicode{tfm:zd/a203}{2750}
\pdfglyphtounicode{tfm:zd/a204}{2752}
\pdfglyphtounicode{tfm:zd/a205}{276E}
\pdfglyphtounicode{tfm:zd/a206}{2770}
\pdfglyphtounicode{tfm:zd/a21}{2715}
\pdfglyphtounicode{tfm:zd/a22}{2716}
\pdfglyphtounicode{tfm:zd/a23}{2717}
\pdfglyphtounicode{tfm:zd/a24}{2718}
\pdfglyphtounicode{tfm:zd/a25}{2719}
\pdfglyphtounicode{tfm:zd/a26}{271A}
\pdfglyphtounicode{tfm:zd/a27}{271B}
\pdfglyphtounicode{tfm:zd/a28}{271C}
\pdfglyphtounicode{tfm:zd/a29}{2722}
\pdfglyphtounicode{tfm:zd/a3}{2704}
\pdfglyphtounicode{tfm:zd/a30}{2723}
\pdfglyphtounicode{tfm:zd/a31}{2724}
\pdfglyphtounicode{tfm:zd/a32}{2725}
\pdfglyphtounicode{tfm:zd/a33}{2726}
\pdfglyphtounicode{tfm:zd/a34}{2727}
\pdfglyphtounicode{tfm:zd/a35}{2605}
\pdfglyphtounicode{tfm:zd/a36}{2729}
\pdfglyphtounicode{tfm:zd/a37}{272A}
\pdfglyphtounicode{tfm:zd/a38}{272B}
\pdfglyphtounicode{tfm:zd/a39}{272C}
\pdfglyphtounicode{tfm:zd/a4}{260E}
\pdfglyphtounicode{tfm:zd/a40}{272D}
\pdfglyphtounicode{tfm:zd/a41}{272E}
\pdfglyphtounicode{tfm:zd/a42}{272F}
\pdfglyphtounicode{tfm:zd/a43}{2730}
\pdfglyphtounicode{tfm:zd/a44}{2731}
\pdfglyphtounicode{tfm:zd/a45}{2732}
\pdfglyphtounicode{tfm:zd/a46}{2733}
\pdfglyphtounicode{tfm:zd/a47}{2734}
\pdfglyphtounicode{tfm:zd/a48}{2735}
\pdfglyphtounicode{tfm:zd/a49}{2736}
\pdfglyphtounicode{tfm:zd/a5}{2706}
\pdfglyphtounicode{tfm:zd/a50}{2737}
\pdfglyphtounicode{tfm:zd/a51}{2738}
\pdfglyphtounicode{tfm:zd/a52}{2739}
\pdfglyphtounicode{tfm:zd/a53}{273A}
\pdfglyphtounicode{tfm:zd/a54}{273B}
\pdfglyphtounicode{tfm:zd/a55}{273C}
\pdfglyphtounicode{tfm:zd/a56}{273D}
\pdfglyphtounicode{tfm:zd/a57}{273E}
\pdfglyphtounicode{tfm:zd/a58}{273F}
\pdfglyphtounicode{tfm:zd/a59}{2740}
\pdfglyphtounicode{tfm:zd/a6}{271D}
\pdfglyphtounicode{tfm:zd/a60}{2741}
\pdfglyphtounicode{tfm:zd/a61}{2742}
\pdfglyphtounicode{tfm:zd/a62}{2743}
\pdfglyphtounicode{tfm:zd/a63}{2744}
\pdfglyphtounicode{tfm:zd/a64}{2745}
\pdfglyphtounicode{tfm:zd/a65}{2746}
\pdfglyphtounicode{tfm:zd/a66}{2747}
\pdfglyphtounicode{tfm:zd/a67}{2748}
\pdfglyphtounicode{tfm:zd/a68}{2749}
\pdfglyphtounicode{tfm:zd/a69}{274A}
\pdfglyphtounicode{tfm:zd/a7}{271E}
\pdfglyphtounicode{tfm:zd/a70}{274B}
\pdfglyphtounicode{tfm:zd/a71}{25CF}
\pdfglyphtounicode{tfm:zd/a72}{274D}
\pdfglyphtounicode{tfm:zd/a73}{25A0}
\pdfglyphtounicode{tfm:zd/a74}{274F}
\pdfglyphtounicode{tfm:zd/a75}{2751}
\pdfglyphtounicode{tfm:zd/a76}{25B2}
\pdfglyphtounicode{tfm:zd/a77}{25BC}
\pdfglyphtounicode{tfm:zd/a78}{25C6}
\pdfglyphtounicode{tfm:zd/a79}{2756}
\pdfglyphtounicode{tfm:zd/a8}{271F}
\pdfglyphtounicode{tfm:zd/a81}{25D7}
\pdfglyphtounicode{tfm:zd/a82}{2758}
\pdfglyphtounicode{tfm:zd/a83}{2759}
\pdfglyphtounicode{tfm:zd/a84}{275A}
\pdfglyphtounicode{tfm:zd/a85}{276F}
\pdfglyphtounicode{tfm:zd/a86}{2771}
\pdfglyphtounicode{tfm:zd/a87}{2772}
\pdfglyphtounicode{tfm:zd/a88}{2773}
\pdfglyphtounicode{tfm:zd/a89}{2768}
\pdfglyphtounicode{tfm:zd/a9}{2720}
\pdfglyphtounicode{tfm:zd/a90}{2769}
\pdfglyphtounicode{tfm:zd/a91}{276C}
\pdfglyphtounicode{tfm:zd/a92}{276D}
\pdfglyphtounicode{tfm:zd/a93}{276A}
\pdfglyphtounicode{tfm:zd/a94}{276B}
\pdfglyphtounicode{tfm:zd/a95}{2774}
\pdfglyphtounicode{tfm:zd/a96}{2775}
\pdfglyphtounicode{tfm:zd/a97}{275B}
\pdfglyphtounicode{tfm:zd/a98}{275C}
\pdfglyphtounicode{tfm:zd/a99}{275D}
\pdfglyphtounicode{tfm:zpzdr-reversed/a1}{2701}
\pdfglyphtounicode{tfm:zpzdr-reversed/a10}{2721}
\pdfglyphtounicode{tfm:zpzdr-reversed/a100}{275E}
\pdfglyphtounicode{tfm:zpzdr-reversed/a101}{2761}
\pdfglyphtounicode{tfm:zpzdr-reversed/a102}{2762}
\pdfglyphtounicode{tfm:zpzdr-reversed/a103}{2763}
\pdfglyphtounicode{tfm:zpzdr-reversed/a104}{2764}
\pdfglyphtounicode{tfm:zpzdr-reversed/a105}{2710}
\pdfglyphtounicode{tfm:zpzdr-reversed/a106}{2765}
\pdfglyphtounicode{tfm:zpzdr-reversed/a107}{2766}
\pdfglyphtounicode{tfm:zpzdr-reversed/a108}{2767}
\pdfglyphtounicode{tfm:zpzdr-reversed/a109}{2660}
\pdfglyphtounicode{tfm:zpzdr-reversed/a11}{261B}
\pdfglyphtounicode{tfm:zpzdr-reversed/a110}{2665}
\pdfglyphtounicode{tfm:zpzdr-reversed/a111}{2666}
\pdfglyphtounicode{tfm:zpzdr-reversed/a112}{2663}
\pdfglyphtounicode{tfm:zpzdr-reversed/a117}{2709}
\pdfglyphtounicode{tfm:zpzdr-reversed/a118}{2708}
\pdfglyphtounicode{tfm:zpzdr-reversed/a119}{2707}
\pdfglyphtounicode{tfm:zpzdr-reversed/a12}{261E}
\pdfglyphtounicode{tfm:zpzdr-reversed/a120}{2460}
\pdfglyphtounicode{tfm:zpzdr-reversed/a121}{2461}
\pdfglyphtounicode{tfm:zpzdr-reversed/a122}{2462}
\pdfglyphtounicode{tfm:zpzdr-reversed/a123}{2463}
\pdfglyphtounicode{tfm:zpzdr-reversed/a124}{2464}
\pdfglyphtounicode{tfm:zpzdr-reversed/a125}{2465}
\pdfglyphtounicode{tfm:zpzdr-reversed/a126}{2466}
\pdfglyphtounicode{tfm:zpzdr-reversed/a127}{2467}
\pdfglyphtounicode{tfm:zpzdr-reversed/a128}{2468}
\pdfglyphtounicode{tfm:zpzdr-reversed/a129}{2469}
\pdfglyphtounicode{tfm:zpzdr-reversed/a13}{270C}
\pdfglyphtounicode{tfm:zpzdr-reversed/a130}{2776}
\pdfglyphtounicode{tfm:zpzdr-reversed/a131}{2777}
\pdfglyphtounicode{tfm:zpzdr-reversed/a132}{2778}
\pdfglyphtounicode{tfm:zpzdr-reversed/a133}{2779}
\pdfglyphtounicode{tfm:zpzdr-reversed/a134}{277A}
\pdfglyphtounicode{tfm:zpzdr-reversed/a135}{277B}
\pdfglyphtounicode{tfm:zpzdr-reversed/a136}{277C}
\pdfglyphtounicode{tfm:zpzdr-reversed/a137}{277D}
\pdfglyphtounicode{tfm:zpzdr-reversed/a138}{277E}
\pdfglyphtounicode{tfm:zpzdr-reversed/a139}{277F}
\pdfglyphtounicode{tfm:zpzdr-reversed/a14}{270D}
\pdfglyphtounicode{tfm:zpzdr-reversed/a140}{2780}
\pdfglyphtounicode{tfm:zpzdr-reversed/a141}{2781}
\pdfglyphtounicode{tfm:zpzdr-reversed/a142}{2782}
\pdfglyphtounicode{tfm:zpzdr-reversed/a143}{2783}
\pdfglyphtounicode{tfm:zpzdr-reversed/a144}{2784}
\pdfglyphtounicode{tfm:zpzdr-reversed/a145}{2785}
\pdfglyphtounicode{tfm:zpzdr-reversed/a146}{2786}
\pdfglyphtounicode{tfm:zpzdr-reversed/a147}{2787}
\pdfglyphtounicode{tfm:zpzdr-reversed/a148}{2788}
\pdfglyphtounicode{tfm:zpzdr-reversed/a149}{2789}
\pdfglyphtounicode{tfm:zpzdr-reversed/a15}{270E}
\pdfglyphtounicode{tfm:zpzdr-reversed/a150}{278A}
\pdfglyphtounicode{tfm:zpzdr-reversed/a151}{278B}
\pdfglyphtounicode{tfm:zpzdr-reversed/a152}{278C}
\pdfglyphtounicode{tfm:zpzdr-reversed/a153}{278D}
\pdfglyphtounicode{tfm:zpzdr-reversed/a154}{278E}
\pdfglyphtounicode{tfm:zpzdr-reversed/a155}{278F}
\pdfglyphtounicode{tfm:zpzdr-reversed/a156}{2790}
\pdfglyphtounicode{tfm:zpzdr-reversed/a157}{2791}
\pdfglyphtounicode{tfm:zpzdr-reversed/a158}{2792}
\pdfglyphtounicode{tfm:zpzdr-reversed/a159}{2793}
\pdfglyphtounicode{tfm:zpzdr-reversed/a16}{270F}
\pdfglyphtounicode{tfm:zpzdr-reversed/a160}{2794}
\pdfglyphtounicode{tfm:zpzdr-reversed/a161}{2192}
\pdfglyphtounicode{tfm:zpzdr-reversed/a162}{27A3}
\pdfglyphtounicode{tfm:zpzdr-reversed/a163}{2194}
\pdfglyphtounicode{tfm:zpzdr-reversed/a164}{2195}
\pdfglyphtounicode{tfm:zpzdr-reversed/a165}{2799}
\pdfglyphtounicode{tfm:zpzdr-reversed/a166}{279B}
\pdfglyphtounicode{tfm:zpzdr-reversed/a167}{279C}
\pdfglyphtounicode{tfm:zpzdr-reversed/a168}{279D}
\pdfglyphtounicode{tfm:zpzdr-reversed/a169}{279E}
\pdfglyphtounicode{tfm:zpzdr-reversed/a17}{2711}
\pdfglyphtounicode{tfm:zpzdr-reversed/a170}{279F}
\pdfglyphtounicode{tfm:zpzdr-reversed/a171}{27A0}
\pdfglyphtounicode{tfm:zpzdr-reversed/a172}{27A1}
\pdfglyphtounicode{tfm:zpzdr-reversed/a173}{27A2}
\pdfglyphtounicode{tfm:zpzdr-reversed/a174}{27A4}
\pdfglyphtounicode{tfm:zpzdr-reversed/a175}{27A5}
\pdfglyphtounicode{tfm:zpzdr-reversed/a176}{27A6}
\pdfglyphtounicode{tfm:zpzdr-reversed/a177}{27A7}
\pdfglyphtounicode{tfm:zpzdr-reversed/a178}{27A8}
\pdfglyphtounicode{tfm:zpzdr-reversed/a179}{27A9}
\pdfglyphtounicode{tfm:zpzdr-reversed/a18}{2712}
\pdfglyphtounicode{tfm:zpzdr-reversed/a180}{27AB}
\pdfglyphtounicode{tfm:zpzdr-reversed/a181}{27AD}
\pdfglyphtounicode{tfm:zpzdr-reversed/a182}{27AF}
\pdfglyphtounicode{tfm:zpzdr-reversed/a183}{27B2}
\pdfglyphtounicode{tfm:zpzdr-reversed/a184}{27B3}
\pdfglyphtounicode{tfm:zpzdr-reversed/a185}{27B5}
\pdfglyphtounicode{tfm:zpzdr-reversed/a186}{27B8}
\pdfglyphtounicode{tfm:zpzdr-reversed/a187}{27BA}
\pdfglyphtounicode{tfm:zpzdr-reversed/a188}{27BB}
\pdfglyphtounicode{tfm:zpzdr-reversed/a189}{27BC}
\pdfglyphtounicode{tfm:zpzdr-reversed/a19}{2713}
\pdfglyphtounicode{tfm:zpzdr-reversed/a190}{27BD}
\pdfglyphtounicode{tfm:zpzdr-reversed/a191}{27BE}
\pdfglyphtounicode{tfm:zpzdr-reversed/a192}{279A}
\pdfglyphtounicode{tfm:zpzdr-reversed/a193}{27AA}
\pdfglyphtounicode{tfm:zpzdr-reversed/a194}{27B6}
\pdfglyphtounicode{tfm:zpzdr-reversed/a195}{27B9}
\pdfglyphtounicode{tfm:zpzdr-reversed/a196}{2798}
\pdfglyphtounicode{tfm:zpzdr-reversed/a197}{27B4}
\pdfglyphtounicode{tfm:zpzdr-reversed/a198}{27B7}
\pdfglyphtounicode{tfm:zpzdr-reversed/a199}{27AC}
\pdfglyphtounicode{tfm:zpzdr-reversed/a2}{2702}
\pdfglyphtounicode{tfm:zpzdr-reversed/a20}{2714}
\pdfglyphtounicode{tfm:zpzdr-reversed/a200}{27AE}
\pdfglyphtounicode{tfm:zpzdr-reversed/a201}{27B1}
\pdfglyphtounicode{tfm:zpzdr-reversed/a202}{2703}
\pdfglyphtounicode{tfm:zpzdr-reversed/a203}{2750}
\pdfglyphtounicode{tfm:zpzdr-reversed/a204}{2752}
\pdfglyphtounicode{tfm:zpzdr-reversed/a205}{276E}
\pdfglyphtounicode{tfm:zpzdr-reversed/a206}{2770}
\pdfglyphtounicode{tfm:zpzdr-reversed/a21}{2715}
\pdfglyphtounicode{tfm:zpzdr-reversed/a22}{2716}
\pdfglyphtounicode{tfm:zpzdr-reversed/a23}{2717}
\pdfglyphtounicode{tfm:zpzdr-reversed/a24}{2718}
\pdfglyphtounicode{tfm:zpzdr-reversed/a25}{2719}
\pdfglyphtounicode{tfm:zpzdr-reversed/a26}{271A}
\pdfglyphtounicode{tfm:zpzdr-reversed/a27}{271B}
\pdfglyphtounicode{tfm:zpzdr-reversed/a28}{271C}
\pdfglyphtounicode{tfm:zpzdr-reversed/a29}{2722}
\pdfglyphtounicode{tfm:zpzdr-reversed/a3}{2704}
\pdfglyphtounicode{tfm:zpzdr-reversed/a30}{2723}
\pdfglyphtounicode{tfm:zpzdr-reversed/a31}{2724}
\pdfglyphtounicode{tfm:zpzdr-reversed/a32}{2725}
\pdfglyphtounicode{tfm:zpzdr-reversed/a33}{2726}
\pdfglyphtounicode{tfm:zpzdr-reversed/a34}{2727}
\pdfglyphtounicode{tfm:zpzdr-reversed/a35}{2605}
\pdfglyphtounicode{tfm:zpzdr-reversed/a36}{2729}
\pdfglyphtounicode{tfm:zpzdr-reversed/a37}{272A}
\pdfglyphtounicode{tfm:zpzdr-reversed/a38}{272B}
\pdfglyphtounicode{tfm:zpzdr-reversed/a39}{272C}
\pdfglyphtounicode{tfm:zpzdr-reversed/a4}{260E}
\pdfglyphtounicode{tfm:zpzdr-reversed/a40}{272D}
\pdfglyphtounicode{tfm:zpzdr-reversed/a41}{272E}
\pdfglyphtounicode{tfm:zpzdr-reversed/a42}{272F}
\pdfglyphtounicode{tfm:zpzdr-reversed/a43}{2730}
\pdfglyphtounicode{tfm:zpzdr-reversed/a44}{2731}
\pdfglyphtounicode{tfm:zpzdr-reversed/a45}{2732}
\pdfglyphtounicode{tfm:zpzdr-reversed/a46}{2733}
\pdfglyphtounicode{tfm:zpzdr-reversed/a47}{2734}
\pdfglyphtounicode{tfm:zpzdr-reversed/a48}{2735}
\pdfglyphtounicode{tfm:zpzdr-reversed/a49}{2736}
\pdfglyphtounicode{tfm:zpzdr-reversed/a5}{2706}
\pdfglyphtounicode{tfm:zpzdr-reversed/a50}{2737}
\pdfglyphtounicode{tfm:zpzdr-reversed/a51}{2738}
\pdfglyphtounicode{tfm:zpzdr-reversed/a52}{2739}
\pdfglyphtounicode{tfm:zpzdr-reversed/a53}{273A}
\pdfglyphtounicode{tfm:zpzdr-reversed/a54}{273B}
\pdfglyphtounicode{tfm:zpzdr-reversed/a55}{273C}
\pdfglyphtounicode{tfm:zpzdr-reversed/a56}{273D}
\pdfglyphtounicode{tfm:zpzdr-reversed/a57}{273E}
\pdfglyphtounicode{tfm:zpzdr-reversed/a58}{273F}
\pdfglyphtounicode{tfm:zpzdr-reversed/a59}{2740}
\pdfglyphtounicode{tfm:zpzdr-reversed/a6}{271D}
\pdfglyphtounicode{tfm:zpzdr-reversed/a60}{2741}
\pdfglyphtounicode{tfm:zpzdr-reversed/a61}{2742}
\pdfglyphtounicode{tfm:zpzdr-reversed/a62}{2743}
\pdfglyphtounicode{tfm:zpzdr-reversed/a63}{2744}
\pdfglyphtounicode{tfm:zpzdr-reversed/a64}{2745}
\pdfglyphtounicode{tfm:zpzdr-reversed/a65}{2746}
\pdfglyphtounicode{tfm:zpzdr-reversed/a66}{2747}
\pdfglyphtounicode{tfm:zpzdr-reversed/a67}{2748}
\pdfglyphtounicode{tfm:zpzdr-reversed/a68}{2749}
\pdfglyphtounicode{tfm:zpzdr-reversed/a69}{274A}
\pdfglyphtounicode{tfm:zpzdr-reversed/a7}{271E}
\pdfglyphtounicode{tfm:zpzdr-reversed/a70}{274B}
\pdfglyphtounicode{tfm:zpzdr-reversed/a71}{25CF}
\pdfglyphtounicode{tfm:zpzdr-reversed/a72}{274D}
\pdfglyphtounicode{tfm:zpzdr-reversed/a73}{25A0}
\pdfglyphtounicode{tfm:zpzdr-reversed/a74}{274F}
\pdfglyphtounicode{tfm:zpzdr-reversed/a75}{2751}
\pdfglyphtounicode{tfm:zpzdr-reversed/a76}{25B2}
\pdfglyphtounicode{tfm:zpzdr-reversed/a77}{25BC}
\pdfglyphtounicode{tfm:zpzdr-reversed/a78}{25C6}
\pdfglyphtounicode{tfm:zpzdr-reversed/a79}{2756}
\pdfglyphtounicode{tfm:zpzdr-reversed/a8}{271F}
\pdfglyphtounicode{tfm:zpzdr-reversed/a81}{25D7}
\pdfglyphtounicode{tfm:zpzdr-reversed/a82}{2758}
\pdfglyphtounicode{tfm:zpzdr-reversed/a83}{2759}
\pdfglyphtounicode{tfm:zpzdr-reversed/a84}{275A}
\pdfglyphtounicode{tfm:zpzdr-reversed/a85}{276F}
\pdfglyphtounicode{tfm:zpzdr-reversed/a86}{2771}
\pdfglyphtounicode{tfm:zpzdr-reversed/a87}{2772}
\pdfglyphtounicode{tfm:zpzdr-reversed/a88}{2773}
\pdfglyphtounicode{tfm:zpzdr-reversed/a89}{2768}
\pdfglyphtounicode{tfm:zpzdr-reversed/a9}{2720}
\pdfglyphtounicode{tfm:zpzdr-reversed/a90}{2769}
\pdfglyphtounicode{tfm:zpzdr-reversed/a91}{276C}
\pdfglyphtounicode{tfm:zpzdr-reversed/a92}{276D}
\pdfglyphtounicode{tfm:zpzdr-reversed/a93}{276A}
\pdfglyphtounicode{tfm:zpzdr-reversed/a94}{276B}
\pdfglyphtounicode{tfm:zpzdr-reversed/a95}{2774}
\pdfglyphtounicode{tfm:zpzdr-reversed/a96}{2775}
\pdfglyphtounicode{tfm:zpzdr-reversed/a97}{275B}
\pdfglyphtounicode{tfm:zpzdr-reversed/a98}{275C}
\pdfglyphtounicode{tfm:zpzdr-reversed/a99}{275D}
\pdfglyphtounicode{thabengali}{09A5}
\pdfglyphtounicode{thadeva}{0925}
\pdfglyphtounicode{thagujarati}{0AA5}
\pdfglyphtounicode{thagurmukhi}{0A25}
\pdfglyphtounicode{thalarabic}{0630}
\pdfglyphtounicode{thalfinalarabic}{FEAC}
\pdfglyphtounicode{thanthakhatlowleftthai}{F898}
\pdfglyphtounicode{thanthakhatlowrightthai}{F897}
\pdfglyphtounicode{thanthakhatthai}{0E4C}
\pdfglyphtounicode{thanthakhatupperleftthai}{F896}
\pdfglyphtounicode{theharabic}{062B}
\pdfglyphtounicode{thehfinalarabic}{FE9A}
\pdfglyphtounicode{thehinitialarabic}{FE9B}
\pdfglyphtounicode{thehmedialarabic}{FE9C}
\pdfglyphtounicode{thereexists}{2203}
\pdfglyphtounicode{therefore}{2234}
\pdfglyphtounicode{theta}{03B8}
\pdfglyphtounicode{theta1}{03D1}
\pdfglyphtounicode{thetasymbolgreek}{03D1}
\pdfglyphtounicode{thieuthacirclekorean}{3279}
\pdfglyphtounicode{thieuthaparenkorean}{3219}
\pdfglyphtounicode{thieuthcirclekorean}{326B}
\pdfglyphtounicode{thieuthkorean}{314C}
\pdfglyphtounicode{thieuthparenkorean}{320B}
\pdfglyphtounicode{thirteencircle}{246C}
\pdfglyphtounicode{thirteenparen}{2480}
\pdfglyphtounicode{thirteenperiod}{2494}
\pdfglyphtounicode{thonangmonthothai}{0E11}
\pdfglyphtounicode{thook}{01AD}
\pdfglyphtounicode{thophuthaothai}{0E12}
\pdfglyphtounicode{thorn}{00FE}
\pdfglyphtounicode{thothahanthai}{0E17}
\pdfglyphtounicode{thothanthai}{0E10}
\pdfglyphtounicode{thothongthai}{0E18}
\pdfglyphtounicode{thothungthai}{0E16}
\pdfglyphtounicode{thousandcyrillic}{0482}
\pdfglyphtounicode{thousandsseparatorarabic}{066C}
\pdfglyphtounicode{thousandsseparatorpersian}{066C}
\pdfglyphtounicode{three}{0033}
\pdfglyphtounicode{threearabic}{0663}
\pdfglyphtounicode{threebengali}{09E9}
\pdfglyphtounicode{threecircle}{2462}
\pdfglyphtounicode{threecircleinversesansserif}{278C}
\pdfglyphtounicode{threedeva}{0969}
\pdfglyphtounicode{threeeighths}{215C}
\pdfglyphtounicode{threegujarati}{0AE9}
\pdfglyphtounicode{threegurmukhi}{0A69}
\pdfglyphtounicode{threehackarabic}{0663}
\pdfglyphtounicode{threehangzhou}{3023}
\pdfglyphtounicode{threeideographicparen}{3222}
\pdfglyphtounicode{threeinferior}{2083}
\pdfglyphtounicode{threemonospace}{FF13}
\pdfglyphtounicode{threenumeratorbengali}{09F6}
\pdfglyphtounicode{threeoldstyle}{0033}
\pdfglyphtounicode{threeparen}{2476}
\pdfglyphtounicode{threeperiod}{248A}
\pdfglyphtounicode{threepersian}{06F3}
\pdfglyphtounicode{threequarters}{00BE}
\pdfglyphtounicode{threequartersemdash}{F6DE}
\pdfglyphtounicode{threeroman}{2172}
\pdfglyphtounicode{threesuperior}{00B3}
\pdfglyphtounicode{threethai}{0E53}
\pdfglyphtounicode{thzsquare}{3394}
\pdfglyphtounicode{tihiragana}{3061}
\pdfglyphtounicode{tikatakana}{30C1}
\pdfglyphtounicode{tikatakanahalfwidth}{FF81}
\pdfglyphtounicode{tikeutacirclekorean}{3270}
\pdfglyphtounicode{tikeutaparenkorean}{3210}
\pdfglyphtounicode{tikeutcirclekorean}{3262}
\pdfglyphtounicode{tikeutkorean}{3137}
\pdfglyphtounicode{tikeutparenkorean}{3202}
\pdfglyphtounicode{tilde}{02DC}
\pdfglyphtounicode{tildebelowcmb}{0330}
\pdfglyphtounicode{tildecmb}{0303}
\pdfglyphtounicode{tildecomb}{0303}
\pdfglyphtounicode{tildedoublecmb}{0360}
\pdfglyphtounicode{tildeoperator}{223C}
\pdfglyphtounicode{tildeoverlaycmb}{0334}
\pdfglyphtounicode{tildeverticalcmb}{033E}
\pdfglyphtounicode{timescircle}{2297}
\pdfglyphtounicode{tipehahebrew}{0596}
\pdfglyphtounicode{tipehalefthebrew}{0596}
\pdfglyphtounicode{tippigurmukhi}{0A70}
\pdfglyphtounicode{titlocyrilliccmb}{0483}
\pdfglyphtounicode{tiwnarmenian}{057F}
\pdfglyphtounicode{tlinebelow}{1E6F}
\pdfglyphtounicode{tmonospace}{FF54}
\pdfglyphtounicode{toarmenian}{0569}
\pdfglyphtounicode{tohiragana}{3068}
\pdfglyphtounicode{tokatakana}{30C8}
\pdfglyphtounicode{tokatakanahalfwidth}{FF84}
\pdfglyphtounicode{tonebarextrahighmod}{02E5}
\pdfglyphtounicode{tonebarextralowmod}{02E9}
\pdfglyphtounicode{tonebarhighmod}{02E6}
\pdfglyphtounicode{tonebarlowmod}{02E8}
\pdfglyphtounicode{tonebarmidmod}{02E7}
\pdfglyphtounicode{tonefive}{01BD}
\pdfglyphtounicode{tonesix}{0185}
\pdfglyphtounicode{tonetwo}{01A8}
\pdfglyphtounicode{tonos}{0384}
\pdfglyphtounicode{tonsquare}{3327}
\pdfglyphtounicode{topatakthai}{0E0F}
\pdfglyphtounicode{tortoiseshellbracketleft}{3014}
\pdfglyphtounicode{tortoiseshellbracketleftsmall}{FE5D}
\pdfglyphtounicode{tortoiseshellbracketleftvertical}{FE39}
\pdfglyphtounicode{tortoiseshellbracketright}{3015}
\pdfglyphtounicode{tortoiseshellbracketrightsmall}{FE5E}
\pdfglyphtounicode{tortoiseshellbracketrightvertical}{FE3A}
\pdfglyphtounicode{totaothai}{0E15}
\pdfglyphtounicode{tpalatalhook}{01AB}
\pdfglyphtounicode{tparen}{24AF}
\pdfglyphtounicode{trademark}{2122}
\pdfglyphtounicode{trademarksans}{2122}
\pdfglyphtounicode{trademarkserif}{2122}
\pdfglyphtounicode{tretroflexhook}{0288}
\pdfglyphtounicode{triagdn}{25BC}
\pdfglyphtounicode{triaglf}{25C4}
\pdfglyphtounicode{triagrt}{25BA}
\pdfglyphtounicode{triagup}{25B2}
\pdfglyphtounicode{triangle}{25B3}
\pdfglyphtounicode{triangledownsld}{25BC}
\pdfglyphtounicode{triangleinv}{25BD}
\pdfglyphtounicode{triangleleft}{25C1}
\pdfglyphtounicode{triangleleftequal}{22B4}
\pdfglyphtounicode{triangleleftsld}{25C0}
\pdfglyphtounicode{triangleright}{25B7}
\pdfglyphtounicode{trianglerightequal}{22B5}
\pdfglyphtounicode{trianglerightsld}{25B6}
\pdfglyphtounicode{trianglesolid}{25B2}
\pdfglyphtounicode{ts}{02A6}
\pdfglyphtounicode{tsadi}{05E6}
\pdfglyphtounicode{tsadidagesh}{FB46}
\pdfglyphtounicode{tsadidageshhebrew}{FB46}
\pdfglyphtounicode{tsadihebrew}{05E6}
\pdfglyphtounicode{tsecyrillic}{0446}
\pdfglyphtounicode{tsere}{05B5}
\pdfglyphtounicode{tsere12}{05B5}
\pdfglyphtounicode{tsere1e}{05B5}
\pdfglyphtounicode{tsere2b}{05B5}
\pdfglyphtounicode{tserehebrew}{05B5}
\pdfglyphtounicode{tserenarrowhebrew}{05B5}
\pdfglyphtounicode{tserequarterhebrew}{05B5}
\pdfglyphtounicode{tserewidehebrew}{05B5}
\pdfglyphtounicode{tshecyrillic}{045B}
\pdfglyphtounicode{tsuperior}{0074}
\pdfglyphtounicode{ttabengali}{099F}
\pdfglyphtounicode{ttadeva}{091F}
\pdfglyphtounicode{ttagujarati}{0A9F}
\pdfglyphtounicode{ttagurmukhi}{0A1F}
\pdfglyphtounicode{tteharabic}{0679}
\pdfglyphtounicode{ttehfinalarabic}{FB67}
\pdfglyphtounicode{ttehinitialarabic}{FB68}
\pdfglyphtounicode{ttehmedialarabic}{FB69}
\pdfglyphtounicode{tthabengali}{09A0}
\pdfglyphtounicode{tthadeva}{0920}
\pdfglyphtounicode{tthagujarati}{0AA0}
\pdfglyphtounicode{tthagurmukhi}{0A20}
\pdfglyphtounicode{tturned}{0287}
\pdfglyphtounicode{tuhiragana}{3064}
\pdfglyphtounicode{tukatakana}{30C4}
\pdfglyphtounicode{tukatakanahalfwidth}{FF82}
\pdfglyphtounicode{turnstileleft}{22A2}
\pdfglyphtounicode{turnstileright}{22A3}
\pdfglyphtounicode{tusmallhiragana}{3063}
\pdfglyphtounicode{tusmallkatakana}{30C3}
\pdfglyphtounicode{tusmallkatakanahalfwidth}{FF6F}
\pdfglyphtounicode{twelvecircle}{246B}
\pdfglyphtounicode{twelveparen}{247F}
\pdfglyphtounicode{twelveperiod}{2493}
\pdfglyphtounicode{twelveroman}{217B}
\pdfglyphtounicode{twentycircle}{2473}
\pdfglyphtounicode{twentyhangzhou}{5344}
\pdfglyphtounicode{twentyparen}{2487}
\pdfglyphtounicode{twentyperiod}{249B}
\pdfglyphtounicode{two}{0032}
\pdfglyphtounicode{twoarabic}{0662}
\pdfglyphtounicode{twobengali}{09E8}
\pdfglyphtounicode{twocircle}{2461}
\pdfglyphtounicode{twocircleinversesansserif}{278B}
\pdfglyphtounicode{twodeva}{0968}
\pdfglyphtounicode{twodotenleader}{2025}
\pdfglyphtounicode{twodotleader}{2025}
\pdfglyphtounicode{twodotleadervertical}{FE30}
\pdfglyphtounicode{twogujarati}{0AE8}
\pdfglyphtounicode{twogurmukhi}{0A68}
\pdfglyphtounicode{twohackarabic}{0662}
\pdfglyphtounicode{twohangzhou}{3022}
\pdfglyphtounicode{twoideographicparen}{3221}
\pdfglyphtounicode{twoinferior}{2082}
\pdfglyphtounicode{twomonospace}{FF12}
\pdfglyphtounicode{twonumeratorbengali}{09F5}
\pdfglyphtounicode{twooldstyle}{0032}
\pdfglyphtounicode{twoparen}{2475}
\pdfglyphtounicode{twoperiod}{2489}
\pdfglyphtounicode{twopersian}{06F2}
\pdfglyphtounicode{tworoman}{2171}
\pdfglyphtounicode{twostroke}{01BB}
\pdfglyphtounicode{twosuperior}{00B2}
\pdfglyphtounicode{twothai}{0E52}
\pdfglyphtounicode{twothirds}{2154}
\pdfglyphtounicode{u}{0075}
\pdfglyphtounicode{uacute}{00FA}
\pdfglyphtounicode{ubar}{0289}
\pdfglyphtounicode{ubengali}{0989}
\pdfglyphtounicode{ubopomofo}{3128}
\pdfglyphtounicode{ubreve}{016D}
\pdfglyphtounicode{ucaron}{01D4}
\pdfglyphtounicode{ucircle}{24E4}
\pdfglyphtounicode{ucircumflex}{00FB}
\pdfglyphtounicode{ucircumflexbelow}{1E77}
\pdfglyphtounicode{ucyrillic}{0443}
\pdfglyphtounicode{udattadeva}{0951}
\pdfglyphtounicode{udblacute}{0171}
\pdfglyphtounicode{udblgrave}{0215}
\pdfglyphtounicode{udeva}{0909}
\pdfglyphtounicode{udieresis}{00FC}
\pdfglyphtounicode{udieresisacute}{01D8}
\pdfglyphtounicode{udieresisbelow}{1E73}
\pdfglyphtounicode{udieresiscaron}{01DA}
\pdfglyphtounicode{udieresiscyrillic}{04F1}
\pdfglyphtounicode{udieresisgrave}{01DC}
\pdfglyphtounicode{udieresismacron}{01D6}
\pdfglyphtounicode{udotbelow}{1EE5}
\pdfglyphtounicode{ugrave}{00F9}
\pdfglyphtounicode{ugujarati}{0A89}
\pdfglyphtounicode{ugurmukhi}{0A09}
\pdfglyphtounicode{uhiragana}{3046}
\pdfglyphtounicode{uhookabove}{1EE7}
\pdfglyphtounicode{uhorn}{01B0}
\pdfglyphtounicode{uhornacute}{1EE9}
\pdfglyphtounicode{uhorndotbelow}{1EF1}
\pdfglyphtounicode{uhorngrave}{1EEB}
\pdfglyphtounicode{uhornhookabove}{1EED}
\pdfglyphtounicode{uhorntilde}{1EEF}
\pdfglyphtounicode{uhungarumlaut}{0171}
\pdfglyphtounicode{uhungarumlautcyrillic}{04F3}
\pdfglyphtounicode{uinvertedbreve}{0217}
\pdfglyphtounicode{ukatakana}{30A6}
\pdfglyphtounicode{ukatakanahalfwidth}{FF73}
\pdfglyphtounicode{ukcyrillic}{0479}
\pdfglyphtounicode{ukorean}{315C}
\pdfglyphtounicode{umacron}{016B}
\pdfglyphtounicode{umacroncyrillic}{04EF}
\pdfglyphtounicode{umacrondieresis}{1E7B}
\pdfglyphtounicode{umatragurmukhi}{0A41}
\pdfglyphtounicode{umonospace}{FF55}
\pdfglyphtounicode{underscore}{005F}
\pdfglyphtounicode{underscoredbl}{2017}
\pdfglyphtounicode{underscoremonospace}{FF3F}
\pdfglyphtounicode{underscorevertical}{FE33}
\pdfglyphtounicode{underscorewavy}{FE4F}
\pdfglyphtounicode{union}{222A}
\pdfglyphtounicode{uniondbl}{22D3}
\pdfglyphtounicode{unionmulti}{228E}
\pdfglyphtounicode{unionsq}{2294}
\pdfglyphtounicode{universal}{2200}
\pdfglyphtounicode{uogonek}{0173}
\pdfglyphtounicode{uparen}{24B0}
\pdfglyphtounicode{upblock}{2580}
\pdfglyphtounicode{upperdothebrew}{05C4}
\pdfglyphtounicode{uprise}{22CF}
\pdfglyphtounicode{upsilon}{03C5}
\pdfglyphtounicode{upsilondieresis}{03CB}
\pdfglyphtounicode{upsilondieresistonos}{03B0}
\pdfglyphtounicode{upsilonlatin}{028A}
\pdfglyphtounicode{upsilontonos}{03CD}
\pdfglyphtounicode{upslope}{29F8}
\pdfglyphtounicode{uptackbelowcmb}{031D}
\pdfglyphtounicode{uptackmod}{02D4}
\pdfglyphtounicode{uragurmukhi}{0A73}
\pdfglyphtounicode{uring}{016F}
\pdfglyphtounicode{ushortcyrillic}{045E}
\pdfglyphtounicode{usmallhiragana}{3045}
\pdfglyphtounicode{usmallkatakana}{30A5}
\pdfglyphtounicode{usmallkatakanahalfwidth}{FF69}
\pdfglyphtounicode{ustraightcyrillic}{04AF}
\pdfglyphtounicode{ustraightstrokecyrillic}{04B1}
\pdfglyphtounicode{utilde}{0169}
\pdfglyphtounicode{utildeacute}{1E79}
\pdfglyphtounicode{utildebelow}{1E75}
\pdfglyphtounicode{uubengali}{098A}
\pdfglyphtounicode{uudeva}{090A}
\pdfglyphtounicode{uugujarati}{0A8A}
\pdfglyphtounicode{uugurmukhi}{0A0A}
\pdfglyphtounicode{uumatragurmukhi}{0A42}
\pdfglyphtounicode{uuvowelsignbengali}{09C2}
\pdfglyphtounicode{uuvowelsigndeva}{0942}
\pdfglyphtounicode{uuvowelsigngujarati}{0AC2}
\pdfglyphtounicode{uvowelsignbengali}{09C1}
\pdfglyphtounicode{uvowelsigndeva}{0941}
\pdfglyphtounicode{uvowelsigngujarati}{0AC1}
\pdfglyphtounicode{v}{0076}
\pdfglyphtounicode{vadeva}{0935}
\pdfglyphtounicode{vagujarati}{0AB5}
\pdfglyphtounicode{vagurmukhi}{0A35}
\pdfglyphtounicode{vakatakana}{30F7}
\pdfglyphtounicode{vav}{05D5}
\pdfglyphtounicode{vavdagesh}{FB35}
\pdfglyphtounicode{vavdagesh65}{FB35}
\pdfglyphtounicode{vavdageshhebrew}{FB35}
\pdfglyphtounicode{vavhebrew}{05D5}
\pdfglyphtounicode{vavholam}{FB4B}
\pdfglyphtounicode{vavholamhebrew}{FB4B}
\pdfglyphtounicode{vavvavhebrew}{05F0}
\pdfglyphtounicode{vavyodhebrew}{05F1}
\pdfglyphtounicode{vcircle}{24E5}
\pdfglyphtounicode{vdotbelow}{1E7F}
\pdfglyphtounicode{vector}{20D7}
\pdfglyphtounicode{vecyrillic}{0432}
\pdfglyphtounicode{veharabic}{06A4}
\pdfglyphtounicode{vehfinalarabic}{FB6B}
\pdfglyphtounicode{vehinitialarabic}{FB6C}
\pdfglyphtounicode{vehmedialarabic}{FB6D}
\pdfglyphtounicode{vekatakana}{30F9}
\pdfglyphtounicode{venus}{2640}
\pdfglyphtounicode{verticalbar}{007C}
\pdfglyphtounicode{verticallineabovecmb}{030D}
\pdfglyphtounicode{verticallinebelowcmb}{0329}
\pdfglyphtounicode{verticallinelowmod}{02CC}
\pdfglyphtounicode{verticallinemod}{02C8}
\pdfglyphtounicode{vewarmenian}{057E}
\pdfglyphtounicode{vhook}{028B}
\pdfglyphtounicode{vikatakana}{30F8}
\pdfglyphtounicode{viramabengali}{09CD}
\pdfglyphtounicode{viramadeva}{094D}
\pdfglyphtounicode{viramagujarati}{0ACD}
\pdfglyphtounicode{visargabengali}{0983}
\pdfglyphtounicode{visargadeva}{0903}
\pdfglyphtounicode{visargagujarati}{0A83}
\pdfglyphtounicode{visiblespace}{2423}
\pdfglyphtounicode{visualspace}{2423}
\pdfglyphtounicode{vmonospace}{FF56}
\pdfglyphtounicode{voarmenian}{0578}
\pdfglyphtounicode{voicediterationhiragana}{309E}
\pdfglyphtounicode{voicediterationkatakana}{30FE}
\pdfglyphtounicode{voicedmarkkana}{309B}
\pdfglyphtounicode{voicedmarkkanahalfwidth}{FF9E}
\pdfglyphtounicode{vokatakana}{30FA}
\pdfglyphtounicode{vparen}{24B1}
\pdfglyphtounicode{vtilde}{1E7D}
\pdfglyphtounicode{vturned}{028C}
\pdfglyphtounicode{vuhiragana}{3094}
\pdfglyphtounicode{vukatakana}{30F4}
\pdfglyphtounicode{w}{0077}
\pdfglyphtounicode{wacute}{1E83}
\pdfglyphtounicode{waekorean}{3159}
\pdfglyphtounicode{wahiragana}{308F}
\pdfglyphtounicode{wakatakana}{30EF}
\pdfglyphtounicode{wakatakanahalfwidth}{FF9C}
\pdfglyphtounicode{wakorean}{3158}
\pdfglyphtounicode{wasmallhiragana}{308E}
\pdfglyphtounicode{wasmallkatakana}{30EE}
\pdfglyphtounicode{wattosquare}{3357}
\pdfglyphtounicode{wavedash}{301C}
\pdfglyphtounicode{wavyunderscorevertical}{FE34}
\pdfglyphtounicode{wawarabic}{0648}
\pdfglyphtounicode{wawfinalarabic}{FEEE}
\pdfglyphtounicode{wawhamzaabovearabic}{0624}
\pdfglyphtounicode{wawhamzaabovefinalarabic}{FE86}
\pdfglyphtounicode{wbsquare}{33DD}
\pdfglyphtounicode{wcircle}{24E6}
\pdfglyphtounicode{wcircumflex}{0175}
\pdfglyphtounicode{wdieresis}{1E85}
\pdfglyphtounicode{wdotaccent}{1E87}
\pdfglyphtounicode{wdotbelow}{1E89}
\pdfglyphtounicode{wehiragana}{3091}
\pdfglyphtounicode{weierstrass}{2118}
\pdfglyphtounicode{wekatakana}{30F1}
\pdfglyphtounicode{wekorean}{315E}
\pdfglyphtounicode{weokorean}{315D}
\pdfglyphtounicode{wgrave}{1E81}
\pdfglyphtounicode{whitebullet}{25E6}
\pdfglyphtounicode{whitecircle}{25CB}
\pdfglyphtounicode{whitecircleinverse}{25D9}
\pdfglyphtounicode{whitecornerbracketleft}{300E}
\pdfglyphtounicode{whitecornerbracketleftvertical}{FE43}
\pdfglyphtounicode{whitecornerbracketright}{300F}
\pdfglyphtounicode{whitecornerbracketrightvertical}{FE44}
\pdfglyphtounicode{whitediamond}{25C7}
\pdfglyphtounicode{whitediamondcontainingblacksmalldiamond}{25C8}
\pdfglyphtounicode{whitedownpointingsmalltriangle}{25BF}
\pdfglyphtounicode{whitedownpointingtriangle}{25BD}
\pdfglyphtounicode{whiteleftpointingsmalltriangle}{25C3}
\pdfglyphtounicode{whiteleftpointingtriangle}{25C1}
\pdfglyphtounicode{whitelenticularbracketleft}{3016}
\pdfglyphtounicode{whitelenticularbracketright}{3017}
\pdfglyphtounicode{whiterightpointingsmalltriangle}{25B9}
\pdfglyphtounicode{whiterightpointingtriangle}{25B7}
\pdfglyphtounicode{whitesmallsquare}{25AB}
\pdfglyphtounicode{whitesmilingface}{263A}
\pdfglyphtounicode{whitesquare}{25A1}
\pdfglyphtounicode{whitestar}{2606}
\pdfglyphtounicode{whitetelephone}{260F}
\pdfglyphtounicode{whitetortoiseshellbracketleft}{3018}
\pdfglyphtounicode{whitetortoiseshellbracketright}{3019}
\pdfglyphtounicode{whiteuppointingsmalltriangle}{25B5}
\pdfglyphtounicode{whiteuppointingtriangle}{25B3}
\pdfglyphtounicode{wihiragana}{3090}
\pdfglyphtounicode{wikatakana}{30F0}
\pdfglyphtounicode{wikorean}{315F}
\pdfglyphtounicode{wmonospace}{FF57}
\pdfglyphtounicode{wohiragana}{3092}
\pdfglyphtounicode{wokatakana}{30F2}
\pdfglyphtounicode{wokatakanahalfwidth}{FF66}
\pdfglyphtounicode{won}{20A9}
\pdfglyphtounicode{wonmonospace}{FFE6}
\pdfglyphtounicode{wowaenthai}{0E27}
\pdfglyphtounicode{wparen}{24B2}
\pdfglyphtounicode{wreathproduct}{2240}
\pdfglyphtounicode{wring}{1E98}
\pdfglyphtounicode{wsuperior}{02B7}
\pdfglyphtounicode{wturned}{028D}
\pdfglyphtounicode{wynn}{01BF}
\pdfglyphtounicode{x}{0078}
\pdfglyphtounicode{xabovecmb}{033D}
\pdfglyphtounicode{xbopomofo}{3112}
\pdfglyphtounicode{xcircle}{24E7}
\pdfglyphtounicode{xdieresis}{1E8D}
\pdfglyphtounicode{xdotaccent}{1E8B}
\pdfglyphtounicode{xeharmenian}{056D}
\pdfglyphtounicode{xi}{03BE}
\pdfglyphtounicode{xmonospace}{FF58}
\pdfglyphtounicode{xparen}{24B3}
\pdfglyphtounicode{xsuperior}{02E3}
\pdfglyphtounicode{y}{0079}
\pdfglyphtounicode{yaadosquare}{334E}
\pdfglyphtounicode{yabengali}{09AF}
\pdfglyphtounicode{yacute}{00FD}
\pdfglyphtounicode{yadeva}{092F}
\pdfglyphtounicode{yaekorean}{3152}
\pdfglyphtounicode{yagujarati}{0AAF}
\pdfglyphtounicode{yagurmukhi}{0A2F}
\pdfglyphtounicode{yahiragana}{3084}
\pdfglyphtounicode{yakatakana}{30E4}
\pdfglyphtounicode{yakatakanahalfwidth}{FF94}
\pdfglyphtounicode{yakorean}{3151}
\pdfglyphtounicode{yamakkanthai}{0E4E}
\pdfglyphtounicode{yasmallhiragana}{3083}
\pdfglyphtounicode{yasmallkatakana}{30E3}
\pdfglyphtounicode{yasmallkatakanahalfwidth}{FF6C}
\pdfglyphtounicode{yatcyrillic}{0463}
\pdfglyphtounicode{ycircle}{24E8}
\pdfglyphtounicode{ycircumflex}{0177}
\pdfglyphtounicode{ydieresis}{00FF}
\pdfglyphtounicode{ydotaccent}{1E8F}
\pdfglyphtounicode{ydotbelow}{1EF5}
\pdfglyphtounicode{yeharabic}{064A}
\pdfglyphtounicode{yehbarreearabic}{06D2}
\pdfglyphtounicode{yehbarreefinalarabic}{FBAF}
\pdfglyphtounicode{yehfinalarabic}{FEF2}
\pdfglyphtounicode{yehhamzaabovearabic}{0626}
\pdfglyphtounicode{yehhamzaabovefinalarabic}{FE8A}
\pdfglyphtounicode{yehhamzaaboveinitialarabic}{FE8B}
\pdfglyphtounicode{yehhamzaabovemedialarabic}{FE8C}
\pdfglyphtounicode{yehinitialarabic}{FEF3}
\pdfglyphtounicode{yehmedialarabic}{FEF4}
\pdfglyphtounicode{yehmeeminitialarabic}{FCDD}
\pdfglyphtounicode{yehmeemisolatedarabic}{FC58}
\pdfglyphtounicode{yehnoonfinalarabic}{FC94}
\pdfglyphtounicode{yehthreedotsbelowarabic}{06D1}
\pdfglyphtounicode{yekorean}{3156}
\pdfglyphtounicode{yen}{00A5}
\pdfglyphtounicode{yenmonospace}{FFE5}
\pdfglyphtounicode{yeokorean}{3155}
\pdfglyphtounicode{yeorinhieuhkorean}{3186}
\pdfglyphtounicode{yerahbenyomohebrew}{05AA}
\pdfglyphtounicode{yerahbenyomolefthebrew}{05AA}
\pdfglyphtounicode{yericyrillic}{044B}
\pdfglyphtounicode{yerudieresiscyrillic}{04F9}
\pdfglyphtounicode{yesieungkorean}{3181}
\pdfglyphtounicode{yesieungpansioskorean}{3183}
\pdfglyphtounicode{yesieungsioskorean}{3182}
\pdfglyphtounicode{yetivhebrew}{059A}
\pdfglyphtounicode{ygrave}{1EF3}
\pdfglyphtounicode{yhook}{01B4}
\pdfglyphtounicode{yhookabove}{1EF7}
\pdfglyphtounicode{yiarmenian}{0575}
\pdfglyphtounicode{yicyrillic}{0457}
\pdfglyphtounicode{yikorean}{3162}
\pdfglyphtounicode{yinyang}{262F}
\pdfglyphtounicode{yiwnarmenian}{0582}
\pdfglyphtounicode{ymonospace}{FF59}
\pdfglyphtounicode{yod}{05D9}
\pdfglyphtounicode{yoddagesh}{FB39}
\pdfglyphtounicode{yoddageshhebrew}{FB39}
\pdfglyphtounicode{yodhebrew}{05D9}
\pdfglyphtounicode{yodyodhebrew}{05F2}
\pdfglyphtounicode{yodyodpatahhebrew}{FB1F}
\pdfglyphtounicode{yohiragana}{3088}
\pdfglyphtounicode{yoikorean}{3189}
\pdfglyphtounicode{yokatakana}{30E8}
\pdfglyphtounicode{yokatakanahalfwidth}{FF96}
\pdfglyphtounicode{yokorean}{315B}
\pdfglyphtounicode{yosmallhiragana}{3087}
\pdfglyphtounicode{yosmallkatakana}{30E7}
\pdfglyphtounicode{yosmallkatakanahalfwidth}{FF6E}
\pdfglyphtounicode{yotgreek}{03F3}
\pdfglyphtounicode{yoyaekorean}{3188}
\pdfglyphtounicode{yoyakorean}{3187}
\pdfglyphtounicode{yoyakthai}{0E22}
\pdfglyphtounicode{yoyingthai}{0E0D}
\pdfglyphtounicode{yparen}{24B4}
\pdfglyphtounicode{ypogegrammeni}{037A}
\pdfglyphtounicode{ypogegrammenigreekcmb}{0345}
\pdfglyphtounicode{yr}{01A6}
\pdfglyphtounicode{yring}{1E99}
\pdfglyphtounicode{ysuperior}{02B8}
\pdfglyphtounicode{ytilde}{1EF9}
\pdfglyphtounicode{yturned}{028E}
\pdfglyphtounicode{yuhiragana}{3086}
\pdfglyphtounicode{yuikorean}{318C}
\pdfglyphtounicode{yukatakana}{30E6}
\pdfglyphtounicode{yukatakanahalfwidth}{FF95}
\pdfglyphtounicode{yukorean}{3160}
\pdfglyphtounicode{yusbigcyrillic}{046B}
\pdfglyphtounicode{yusbigiotifiedcyrillic}{046D}
\pdfglyphtounicode{yuslittlecyrillic}{0467}
\pdfglyphtounicode{yuslittleiotifiedcyrillic}{0469}
\pdfglyphtounicode{yusmallhiragana}{3085}
\pdfglyphtounicode{yusmallkatakana}{30E5}
\pdfglyphtounicode{yusmallkatakanahalfwidth}{FF6D}
\pdfglyphtounicode{yuyekorean}{318B}
\pdfglyphtounicode{yuyeokorean}{318A}
\pdfglyphtounicode{yyabengali}{09DF}
\pdfglyphtounicode{yyadeva}{095F}
\pdfglyphtounicode{z}{007A}
\pdfglyphtounicode{zaarmenian}{0566}
\pdfglyphtounicode{zacute}{017A}
\pdfglyphtounicode{zadeva}{095B}
\pdfglyphtounicode{zagurmukhi}{0A5B}
\pdfglyphtounicode{zaharabic}{0638}
\pdfglyphtounicode{zahfinalarabic}{FEC6}
\pdfglyphtounicode{zahinitialarabic}{FEC7}
\pdfglyphtounicode{zahiragana}{3056}
\pdfglyphtounicode{zahmedialarabic}{FEC8}
\pdfglyphtounicode{zainarabic}{0632}
\pdfglyphtounicode{zainfinalarabic}{FEB0}
\pdfglyphtounicode{zakatakana}{30B6}
\pdfglyphtounicode{zaqefgadolhebrew}{0595}
\pdfglyphtounicode{zaqefqatanhebrew}{0594}
\pdfglyphtounicode{zarqahebrew}{0598}
\pdfglyphtounicode{zayin}{05D6}
\pdfglyphtounicode{zayindagesh}{FB36}
\pdfglyphtounicode{zayindageshhebrew}{FB36}
\pdfglyphtounicode{zayinhebrew}{05D6}
\pdfglyphtounicode{zbopomofo}{3117}
\pdfglyphtounicode{zcaron}{017E}
\pdfglyphtounicode{zcircle}{24E9}
\pdfglyphtounicode{zcircumflex}{1E91}
\pdfglyphtounicode{zcurl}{0291}
\pdfglyphtounicode{zdot}{017C}
\pdfglyphtounicode{zdotaccent}{017C}
\pdfglyphtounicode{zdotbelow}{1E93}
\pdfglyphtounicode{zecyrillic}{0437}
\pdfglyphtounicode{zedescendercyrillic}{0499}
\pdfglyphtounicode{zedieresiscyrillic}{04DF}
\pdfglyphtounicode{zehiragana}{305C}
\pdfglyphtounicode{zekatakana}{30BC}
\pdfglyphtounicode{zero}{0030}
\pdfglyphtounicode{zeroarabic}{0660}
\pdfglyphtounicode{zerobengali}{09E6}
\pdfglyphtounicode{zerodeva}{0966}
\pdfglyphtounicode{zerogujarati}{0AE6}
\pdfglyphtounicode{zerogurmukhi}{0A66}
\pdfglyphtounicode{zerohackarabic}{0660}
\pdfglyphtounicode{zeroinferior}{2080}
\pdfglyphtounicode{zeromonospace}{FF10}
\pdfglyphtounicode{zerooldstyle}{0030}
\pdfglyphtounicode{zeropersian}{06F0}
\pdfglyphtounicode{zerosuperior}{2070}
\pdfglyphtounicode{zerothai}{0E50}
\pdfglyphtounicode{zerowidthjoiner}{FEFF}
\pdfglyphtounicode{zerowidthnonjoiner}{200C}
\pdfglyphtounicode{zerowidthspace}{200B}
\pdfglyphtounicode{zeta}{03B6}
\pdfglyphtounicode{zhbopomofo}{3113}
\pdfglyphtounicode{zhearmenian}{056A}
\pdfglyphtounicode{zhebrevecyrillic}{04C2}
\pdfglyphtounicode{zhecyrillic}{0436}
\pdfglyphtounicode{zhedescendercyrillic}{0497}
\pdfglyphtounicode{zhedieresiscyrillic}{04DD}
\pdfglyphtounicode{zihiragana}{3058}
\pdfglyphtounicode{zikatakana}{30B8}
\pdfglyphtounicode{zinorhebrew}{05AE}
\pdfglyphtounicode{zlinebelow}{1E95}
\pdfglyphtounicode{zmonospace}{FF5A}
\pdfglyphtounicode{zohiragana}{305E}
\pdfglyphtounicode{zokatakana}{30BE}
\pdfglyphtounicode{zparen}{24B5}
\pdfglyphtounicode{zretroflexhook}{0290}
\pdfglyphtounicode{zstroke}{01B6}
\pdfglyphtounicode{zuhiragana}{305A}
\pdfglyphtounicode{zukatakana}{30BA}

\renewcommand{\L}{\mathcal{L}}
\newcommand{\M}{\mathcal{M}}
\renewcommand{\O}{\mathcal{O}}
\renewcommand{\P}{\mathcal{P}}

\renewcommand{\S}{\mathcal{S}}

\def\<{\langle}
\def\>{\rangle}

\newcommand{\Until}{\mathcal{U}}

\newcommand{\BB}{\mathbb{B}}

\newcommand{\ExpCostOp}{\mathsf{E}}

\newcommand{\ProbOp}{\mathsf{P}}

\newcommand{\Distr}{\mathit{Distr}}

\newcommand{\Dirac}{\mathit{Dirac}}

\newcommand{\Support}{\mathit{supp}}

\newcommand{\ExpCost}{\mathbb{E}}

\renewcommand{\Pr}{\mathbb{P}}

\newcommand{\Nat}{\mathbb{N}}

\newcommand{\prism}{\textsc{Prism}}

\newcommand{\FeatureSet}{F}

\newcommand{\ValidFeat}{\mathcal{V}}

\newcommand{\ValidFeatInit}{\ValidFeat^{\mathit{init}}}

\newcommand{\Moves}{\mathsf{Moves}}

\newcommand{\SInit}{S^\mathit{init}}

\newcommand{\Paths}{\mathit{Paths}}
\newcommand{\FPaths}{\mathit{FPaths}}

\newcommand{\cost}{\mathrm{cost}}
\newcommand{\probability}{\mathrm{Pr}}

\newcommand{\sched}{\mathfrak{S}}

\newcommand{\move}[1]{\stackrel{#1}{\hookrightarrow}}

\newcommand{\overto}[1]{\stackrel{#1}{\longrightarrow}}

\newcommand{\Controller}{\mathsf{Con}}

\newcommand{\SwitchRel}{\mathsf{SwRel}}

\newcommand{\switchevent}{\mathsf{e}}

\newcommand{\Module}{\mathsf{Mod}}

\newcommand{\Act}{\mathsf{Act}}
\newcommand{\Loc}{\mathsf{Loc}}
\newcommand{\LocInit}{\Loc^{\mathit{init}}}

\newcommand{\Feat}{\mathsf{F}}
\newcommand{\OwnFeat}{\mathsf{OwnF}}
\newcommand{\ExtFeat}{\mathsf{ExtF}}

\newcommand{\Trans}{\mathsf{Trans}}
\newcommand{\TransAct}{\mathsf{TrAct}}
\newcommand{\TransSwitch}{\mathsf{TrSw}}

\newcommand{\Var}{\mathsf{Var}}
\newcommand{\LocVar}{\mathsf{LocV}}
\newcommand{\ExtVar}{\mathsf{ExtV}}
\newcommand{\VarModule}{\mathsf{V}}

\newcommand{\Val}{\mathsf{Val}}
\newcommand{\val}{\mathsf{v}}

\newcommand{\probupd}{\mathit{prob\_upd}}
\newcommand{\upd}{\mathsf{upd}}

\newcommand{\expr}{\mathit{expr}}

\newcommand{\guard}{\mathit{guard}}

\newcommand{\valcond}{\mathit{val\_cond}}

\newcommand{\init}{S^\mathit{init}}

\newcommand{\next}{\textbf{\textup{\textsf{X}}}}

\theoremstyle{plain}
\newtheorem{theorem}{Theorem}[section]

\newtheorem{definition}[theorem]{Definition}

\theoremstyle{definition}

\newtheorem{example}[theorem]{Example}

\begin{document}
\conferenceinfo{WOODSTOCK}{'97 El Paso, Texas USA}

\title{Probabilistic Model Checking for \\
       Energy Analysis in Software Product Lines%
       \titlenote{This work is partly supported by the 
                  German Research Foundation (SFB 912 HAEC, the
                  DFG-project QuaOS and the DFG/NWO-project ROCKS)
                  and the 
                  EU 7th Framework Programme under grant no. 295261 (MEALS).}\\[-.2em]
       {\small Technical Report}\\[-.6em]
 }

\authorinfo{Clemens Dubslaff\titlenote{supported by Deutsche Telekom Stiftung}\and Sascha Kl\"uppelholz \and Christel Baier}
  {Technische Universit\"at Dresden, Faculty of Computer Science, Germany}
  {$\{$dubslaff,klueppel,baier$\}$@tcs.inf.tu-dresden.de}

\maketitle

\begin{abstract}
In a \emph{software product line (SPL)}, a collection of 
software products is defined by their commonalities 
in terms of features rather than explicitly 
specifying all products one-by-one.
Several verification techniques were adapted to establish
temporal properties of SPLs. Symbolic and family-based
model checking have been proven to be successful for tackling the 
combinatorial blow-up arising when reasoning 
about several feature combinations.
However, most formal verification approaches for SPLs presented in the literature
focus on the \emph{static} SPLs, where the features of a product are
fixed and cannot be changed during runtime. This is in
contrast to \emph{dynamic} SPLs, allowing to adapt feature combinations 
of a product dynamically after deployment.

The main contribution of the paper is a compositional 
modeling framework for dynamic SPLs, which supports
probabilistic and nondeterministic choices and allows for
\emph{quantitative analysis}.
We specify the feature changes during runtime
within an automata-based coordination component, 
enabling to reason over strategies how to 
trigger dynamic feature changes for optimizing
various quantitative objectives, e.g., energy or monetary costs and reliability.
For our framework there is a natural and 
conceptually simple translation into the 
input language of the prominent probabilistic model checker $\prism$.
This facilitates the application of $\prism$'s powerful symbolic engine 
to the operational behavior of dynamic SPLs and their 
family-based analysis against various quantitative queries.
We demonstrate feasibility of our approach by a case study
issuing an energy-aware bonding network device.
\end{abstract}

\section{Introduction}
In order to meet economic requirements and to provide customers 
individualized solutions, the development and marketing of 
modern hardware and software products often follows the concept of 
\emph{product lines}. Within this concept, customers purchase a base system extendible and customizable with additional functionalities, called \emph{features}.
Although product lines are commonly established in both, 
hardware and software development, they have been first and 
foremost considered in the area of software engineering. 
A \emph{software product line (SPL)} (see, e.g., \cite{CleNor2001}) 
specifies a collection of software systems built from 
features according to rules describing 
realizable feature combinations.
Such rules for the composition of features are typically provided using
\emph{feature diagrams} \cite{Kang1990,Benavides2010}.
Feature combinations are often assumed to be static, i.e.,
some realizable feature combination is fixed when the product is 
purchased by a customer and is never changed afterwards. 
However, this do not faithfully reflect adaptations
of modern software during its lifetime. For instance, when a software is 
updated or when a free trial version expires, features are activated 
or deactivated during runtime of the system.
SPLs which model such adaptations are called \emph{dynamic SPLs} 
\cite{GomHus2003}, for which the design of specification formalisms 
is an active and emerging field in SPL engineering \cite{HalHinParSch2008,DinMitFetMez2010,Ros2011,DamSch2011}.

The goal of this paper is provide a \emph{compositional framework} for modeling \emph{dynamic SPLs} which allows for a \emph{quantitative analysis} in order to reason, e.g., about system's resource requirements.

\paragraph{Verification of SPLs.} In order to meet requirements in safety-critical parts of SPLs or to guarantee overall quality, verification is highly desirable. This is especially the case for dynamic SPLs, where side-effects arising from dynamic feature changes are difficult to predict in development phases.
Model checking \cite{ClaEmeSis1986,BaiKat2008} is a fully automatic verification technique for establishing temporal properties of systems (e.g., safety or liveness properties).
Indeed, it has been successively applied to integrate features in 
components and to detect feature interactions
\cite{PlaRya2001}.
However, as observed by Classen et al. \cite{Cla2010,Cla2011}, 
the typical task for reasoning about static SPLs is to 
solve the so-called \emph{featured model-checking problem}:
\begin{enumerate}
\item []
  Compute the set of all feature combinations such that the considered 
  temporal property $\varphi$ holds for the corresponding software products.
\end{enumerate}
This is in contrast to the classical model-checking problem that amounts
to prove that $\varphi$ holds for some fixed system, such as 
one software product obtained from a feature combination.
The standard approach solving the featured model-checking problem is to verify the products in the SPL one-by-one (see, e.g., the product-line analysis taxonomy in \cite{Pla2013}). However, already within static SPLs this approach certainly suffers from an exponential blow-up, since the number of different software products may 
rise exponentially in the number of features. To tackle this potential combinatorial blow-up, family-based \cite{Pla2013} and symbolic approaches \cite{McM1993} are very successful. Within \emph{family-based analysis}, all products in an SPL are checked at once rather than one-by-one. This requires a model which represents all behaviors of all the products of the SPL.
In \cite{Cla2010,Cla2011}, the concept of \emph{featured transition systems (FTSs)} has been introduced to encode the operational behaviors of all products in an SPL. The transitions in an FTS are annotated by feature combinations within which the transition can be taken. Based on symbolic techniques \cite{McM1993}, the featured model-checking problem for SPLs represented by FTSs could be solved efficiently for both linear-time \cite{Cla2010} and branching-time properties \cite{Cla2011}. An extension of FTSs allowing for dynamic adaptions of feature combinations was presented by Cordy et al.~\cite{CorClaHey2013}, annotating further transition guards with possible feature combination switches.

Besides classical temporal properties, the quality of software products
crucially depends on quantitative (non-functional) properties.
While measurement-based approaches for reasoning about feature-oriented
software have been studied intensively (see e.g. \cite{SRKKS08,SRKGAK13,NooBaDu12}),
probabilistic model-checking techniques have been studied only recently.
These use purely probabilistic operational models based on discrete-time
Markov chains and probabilistic computation tree logic.
The approach by Ghezzi and Sharifloo \cite{GheMol13} relies on
parametric sequence diagrams 
analyzed using the probabilistic model-checking 
tool \textsc{Param}.
Recently, a family-based approach for Markov chains 
has been presented by \cite{VarKho13}.
\paragraph{Our Compositional Framework.} 
For the compositional design of software products with parallel components, Markov chains are known to be less adequate than operational models supporting both, nondeterministic and probabilistic choices (see, e.g., \cite{Segala95}). A \emph{Markov decision process (MDP)} is such a formalism, extending labelled transition systems by internal probabilistic choices taken after resolving nondeterminism between actions of the system.
In this paper, we present a compositional framework for dynamic SPLs relying on MDPs with annotated costs, used, e.g., to reason about resource requirements, energy consumption or monetary costs. In particular, our contribution consists of
\begin{enumerate}
\item [(1)]
  feature modules:
  MDP-like models for the operational 
  feature-dependent behavior of the
  components and their interactions,
  
\item [(2)]
  a parallel operator for feature modules that represents
  the parallel execution of independent actions by interleaving,
  supporting communication according to the handshaking principle
  and over shared variables, and
  
\item [(3)]
  a feature controller:
  an MDP-like model for the potential dynamic switches of feature combinations.
\end{enumerate}
An SPL naturally induces a compositional structure over features, where features or collections thereof correspond to components. In our framework, these components are called feature modules (1), which can
contain both, nondeterministic and probabilistic choices. 
The former might be useful in early design stages, whereas probabilistic
choices can be used to model the likelihood of exceptional behaviors
(e.g., if some failure appears) or to represent randomized activities
(e.g., coin tossing actions to break symmetry). Both kinds of choices may depend on other features -- for instance, whether another feature is activated during runtime or not.

Feature Modules are composed using a parallel operator (2), which combines the operational behaviors of all features represented by the feature modules into another feature module. This composition is defined upon compatible feature interfaces of the feature modules, which keep track of the features owned by the feature modules and those which the behavior of the feature modules depends on.
Closest to our compositional approach with MDP-like models is
the approach by \cite{MRKN13} that works with nonprobabilistic
finite-state machines and addresses conformance checking.

Feature activation and deactivation is described through feature controllers (3), which is a state-based model controlling valid changes in the feature combinations. As within feature modules, choices between feature combinations can be probabilistically (e.g., on the basis of
statistical information on feature combinations and their adaptations
over time) or nondeterministically (e.g., if feature changes
rely on internal choices of the controller or are triggered from outside 
by an unknown or unpredictable environment) and combinations thereof.
To the best of our knowledge, this concept is novel in the probabilistic setting and has also been only merely considered in the nonprobabilistic case \cite{DamSch2011}. 

The semantics of a feature module under a given feature controller is defined as a parallel composition of both formalisms, providing an elegant formalization of the feature module's behavior within the dynamic SPL represented by the feature controller. This parallel composition roughly arises by augmenting probabilistic automata \cite{Segala95} with feature interfaces. Note that our approach separates between computation and coordination \cite{GelCar92,PapArb98,SLN01}, which allows for specifying features in the context of various different dynamic SPLs. Feature-oriented extensions of programming languages and specialized composition operators such as
\emph{superimposition} are an orthogonal approach \cite{Kat1993,AJTK09,ApelHut10,Apel2009}. The effect of superimposition can be encoded into our framework, e.g., using techniques proposed by Plath and Ryan \cite{PlaRya2001}, but there is no direct support for composing feature modules using superimposition.

\paragraph{Quantitative Analysis.} Fortunately, the semantics of feature modules under feature controllers rise a standard MDP, such that our approach permits the application of standard but sophisticated 
probabilistic model-checking techniques to reason about quantitative
properties. This is in contrast to existing (nonprobabilistic) approaches, which require model-checking algorithms specialized for SPLs. Within our approach, temporal or quantitative queries
such as ``minimize the energy consumption until reaching a target state''
or ``maximize the utility value to reach a target state for a given
initial energy budget'' can be answered. Corresponding to the nonprobabilistic case, the solution of the featured model-checking problem would then provide answers of these queries for all initial feature combinations. In the setting of dynamic SPLs, we go a step further and define the \emph{strategy synthesis problem} aiming to find an optimal strategy of resolving the nondeterminism between feature combination switches in the feature controller. This strategy includes the initial step of the dynamic SPL by selecting an initial feature combination, which suffices to solve the featured model-checking problem. However, our approach additionally provides the possibility to reason over worst-case scenarios concerning feature changes during runtime. Note that solving the strategy synthesis problem imposes a family-based analysis approach of the dynamic SPL, which is also novel in the nonprobabilistic setting.

As in the nonprobabilistic case, symbolic techniques can help to avoid the exponential blow-up when analyzing probabilistic SPLs. This is even more crucial for dynamic SPLs, since the number of feature changes during runtime also yield an exponential blow-up. Our compositional framework %
nicely fits with guarded-command languages such as the input language of the symbolic probabilistic model checker $\prism$ \cite{HinKwiNorPar2006}. $\prism$ uses multi-terminal binary decision diagrams for the symbolic encoding of the probabilistic model and thus ensures a compact representation. We expressed a case study based on a real-case scenario from the hardware domain according to our framework to demonstrate applicability of $\prism$. This case study details the energy-aware network device \textsc{eBond+}, an extension of the recently presented \textsc{eBond} device \cite{HahnelDVH13}. We explain how $\prism$ can be used to solve the aforementioned strategy synthesis problem w.r.t. to several quantitative queries formalizing requirements, e.g., on the energy consumption of the \textsc{eBond+} device. Our case study also illustrates that our approach is not restricted to SPLs, but can also be applied to product lines in general.
\paragraph{Outline.} In Section \ref{sec:prelim} we briefly summarize basics on SPLs,
feature models and relevant principles of MDPs and their quantitative analysis.
The compositional framework for specifying feature combinations by means of feature modules and feature controllers as a formal operational model for dynamic features changes is presented in Section \ref{sec:comp}.
We illustrate applicability of our approach within our energy-aware case study in Section \ref{sec:appl}.
The paper ends with some concluding remarks in Section \ref{sec:concl}. %

\section{Preliminaries}

\label{sec:prelim}

\paragraph{Notations for Sets and Boolean Expressions.}
The powerset of a set $X$ is denoted by $2^X$. 
For convenience, we sometimes use symbolic notations based on Boolean expressions
(propositional formulas)
for the elements of $2^X$, i.e., the subsets of $X$.
Let $\BB(X)$ denote the set of all Boolean expressions
$\rho$ built by elements $x\in X$ as atoms (Boolean variables)
and the usual connectives of propositional logic (negation $\neg$,
conjunction $\wedge$,
 etc.).
The satisfaction relation $\models \, \subseteq 2^X{\times}\ \BB(X)$ 
is defined in the obvious way.
E.g., if $X = \{x_1,x_2,x_3\}$ and 
$\rho = x_1 \wedge \neg x_2$,
then $Y \models \rho$ iff $Y = \{x_1\}$ or $Y = \{x_1,x_3\}$.
To specify binary relations on $2^X$
symbolically, we use Boolean expressions $\rho \in \BB(X\cup X')$, where
$X'$ is the set consisting of pairwise distinct, fresh
copies of the elements of $X$. 
Then, the relation $R_{\rho} \subseteq 2^X{\times}\ 2^X$ is given by:
\begin{center}
   $(Y,Z) \in R_{\rho}$ \ \ iff \ \ $Y \cup \{ z' : z \in Z \} \models \rho$
\end{center}
E.g., the Boolean expression $\rho = (x_1 \vee x_3') \wedge \neg x_2$
represents the relation $R_{\rho}$ 
consisting of all pairs $(Y,Z)\in 2^X{\times}\ 2^X$,
where %
(1) $x_1 \in Y$ or $x_3 \in Z$ and (2) $x_2 \notin Y$.
For $Y \subseteq X$, we use $Y=Y'$ as a shortform notation for the
Boolean expression $\bigwedge\nolimits_{y\in Y} y \leftrightarrow y'$.

\paragraph{Distributions.}
Let $S$ be a countable nonempty set. 
A \emph{distribution over $S$}
is a function $\sigma: S\rightarrow [0,1]$ with
$\sum_{s\in S} \sigma(s) = 1$. 
The set $\{s\in S : \sigma(s) > 0\}$ is called the \emph{support of $\sigma$}
and is denoted by $\Support(\sigma)$.
$\Distr(S)$ denotes the set of distributions over $S$. 
Given $t\in S$, the distribution $\Dirac[t]\in\Distr(S)$ defined by 
\begin{center}
    $\Dirac[t](t) = 1$ and $\Dirac[t](s)=0$ for all $s\in S\setminus\{t\}$ 
\end{center}
is called the \emph{Dirac distribution of $t$ over $S$}. 
The \emph{product} of two distributions 
$\sigma_1\in\Distr(S_1)$ and $\sigma_2\in\Distr(S_2)$ 
is defined as the distribution $\sigma_1*\sigma_2\in\Distr(S_1{\times}S_2)$, 
where 
    $(\sigma_1*\sigma_2)(s_1,s_2) \ =\ \sigma_1(s_1)\cdot\sigma_2(s_2)$ 
for all $s_1\in S_1$ and $s_2\in S_2$.

\subsection{Feature Models}

According to \cite{CleNor2001}, 
a \emph{software product line (SPL)} is a collection of software products,
which have commonalities w.r.t. assets called \emph{features}. 
When $\FeatureSet$ denotes the set of all such features in an SPL, 
a \emph{feature combination} is a subset $C$ of $\FeatureSet$, 
which is said to be \emph{valid} if there is a corresponding product 
in the SPL consisting exactly of the features in $C$. 
An SPL can hence be formalized in terms of a \emph{feature signature} 
$(\FeatureSet,\ValidFeat)$, where $\ValidFeat$ is the set of 
valid feature combinations.
\emph{Feature diagrams} \cite{Kang1990} provide a compact representation of feature signatures via a tree-like hierarchical diagram (see, e.g., Figure \ref{fig:sfd}). Nodes in feature diagrams correspond to features of $\FeatureSet$, where nodes with a circle on top denote optional features. 
If the node for feature $f'$ is a son of the node for feature $f$, then 
feature $f'$ requires $f$. Several types of branchings from a node for feature $f$ towards its sons $f'_1,\ldots,f'_n$ are possible. 
Standard branchings denote that all nonoptional sons are required by $f$ 
(AND connective), connected branchings indicate that exactly one son 
is required by $f$ (XOR connective) and solid ones require at least one son 
(OR connective). 
An additional arrow from a node for feature $f$ towards a node for feature $f'$ can be used to indicate that $f'$ is required by $f$.
Boolean expressions over $\FeatureSet$ may be further used as constraints on possible feature combinations.
For analyzing SPLs, various approaches annotating additional data to feature models were considered. 
E.g., \cite{CzaSheWas2008} amends feature diagrams with statistical data, 
which yields probability distribution over valid feature combinations.

\paragraph{Static vs.~Dynamic SPL.}
Usually, SPLs are static in the sense that a valid feature combination 
is fixed prior the execution of the system. 
SPLs allowing for activation and deactivation of features during runtime 
of a system are called \emph{dynamic SPLs}. 
The common approach towards dynamic SPLs is to indicate disjoint sets 
of \emph{dynamic features} $D$ and \emph{environment features} $E$, 
which respectively include features that
can be activated or deactivated at runtime either by the 
system itself (features of $D$) or by the environment 
(features of $E$). 
Intuitively, an activation and deactivation of an environment feature may 
impose (de-)activations of dynamic features \cite{CorClaHey2013}. 
In \cite{DinMitFetMez2010} dynamic SPLs are formalized using 
a generalization of feature diagrams where dashed nodes represent 
elements of $D\cup E$. Costs for feature activations in dynamic SPLs have been considered in \cite{WhiteDSB2009}.
The following example details a dynamic SPL for a productivity system, provided by a feature diagram with annotated costs.
\begin{example}
\label{ex:productivity}
Features of the dynamic SPL represented by the feature diagram shown in Figure \ref{fig:sfd}
\begin{figure}[h]%
    \includegraphics[width=\linewidth]{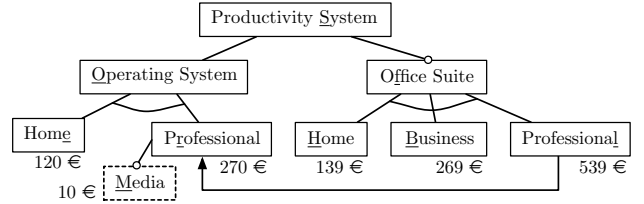}%
    \nocaptionrule\caption{A feature diagram representing an SPL}%
    \label{fig:sfd}%
\end{figure}%
have underlined symbols used as abbreviations, 
i.e., the set of features in the SPL would be 
$\FeatureSet =\{\mathrm{s,o,e,r,m,f,h,b,l}\}$. 
According to the semantics of feature diagrams, $\{\mathrm{s,o,e}\}$ 
(briefly written $\mathrm{soe}$) is the smallest valid feature combination 
and, e.g., $\mathrm{soefb}$ describes a valid feature combination 
with a business feature in the office suite.
The media center feature 
$\mathrm{m}$ is an optional environment feature, i.e., 
if the customer is unsatisfied with the media functionalities, 
she can downgrade to the plain professional version of 
the operating system and allowed to upgrade again if she changed 
her mind.
Note that the professional office suite requires the 
professional operating system. Thus, the feature 
combination $\mathrm{soefl}$ is invalid but $\mathrm{sorfl}$ is valid.
\end{example}

\subsection{Markov Decision Processes}

\label{sec:MDP}

The operational model used in this paper for modeling and analyzing
the behavior of the instances represented by a dynamic SPL
is given in terms of \emph{Markov decision processes (MDPs)} \cite{Puterman}.
We deal here with MDPs where transitions are labeled with
decision identifiers and a cost value. MDPs with multiple cost functions of different types (e.g. for
reasoning energy and memory requirements and utility values) can
be defined accordingly.
Formally, the notion of an MDP
is a tuple
\begin{center}
      $\M\ =\ (S,\SInit,\Moves)$,
\end{center}
where $S$ is a finite set of states,
$\SInit\subseteq S$ is the set of initial states
and
$
  \Moves \subseteq S \times \Nat \times \Distr(S)
$
specifies the possible moves of $\M$ and their costs. We require $\Moves$ to be finite and often write $s \overto{c} \sigma$ iff $(s,c,\sigma)\in \Moves$.
Intuitively, the operational behavior of $\M$ is as follows.
The computations of $\M$ start in some nondeterministically chosen
initial state of $\SInit$.
If during $\M$'s computation the current state is $s$,
one of the moves $s \overto{c} \sigma$ is selected nondeterministically first, before there is an internal probabilistic choice, selecting a successor state $s'$ with probability $\sigma(s')$. 
Value $c$ specifies the cost for taking the move $s \overto{c} \sigma$.

\emph{Steps of $\M$}, written in the form $s \move{c}_p s'$,
arise from moves when resolving the probabilistic choice by
plugging in some state $s'$ with positive probability, i.e.,
$p = \sigma(s') > 0$.
\emph{Paths} in $\M$ are sequences of consecutive steps.
In the following, we assume a finite path $\pi$ having the form
\begin{center}
 $\pi\ =\ 
   s_0 \move{c_1}_{p_1} 
   s_1 \move{c_2}_{p_2} 
   s_2 \move{c_3}_{p_3} \ldots \move{c_n}_{p_n} s_n$. \ \ \ \ \ \ \ \ ($*$)
\end{center}
We refer to the number $n$ of steps as the length of $\pi$.
If $0 \leq k \leq n$, 
we write $\pi[k]$ for the prefix of $\pi$
consisting of the first $k$ steps (then, $\pi[k]$ ends in state $s_k$).
Given a finite path $\pi$, the probability $\probability(\pi)$
is defined as the product of the probabilities in the steps of $\pi$
and the accumulated costs $\cost(\pi)$ are defined as the sum of the costs of $\pi$'s steps.
Formally,
\begin{center}
     $\probability(\pi)= p_1 \cdot p_2 \cdot \ldots \cdot p_n$
     and 
     $\cost(\pi) = c_1 + c_2 + \ldots + c_n$.
\end{center}
State $s\in S$ is called \emph{terminal} if there is no
move $s \overto{c} \sigma$.
A path is \emph{maximal}, if it is either infinite or ends in a 
terminal state. 
The set of finite paths starting in state $s$ is denoted by 
$\FPaths(s)$. Likewise, we write $\Paths(s)$ for the set of all maximal 
paths starting in $s$.

\paragraph{Schedulers and Probability Measure.}
Reasoning about probabilities in MDPs
requires the selection of an initial state and
resolution of the nondeterministic choices between
possible moves.
The latter is formalized via \emph{schedulers}, 
also called policies or adversaries, 
which take as input a finite path and decide which move to take next.
For the purposes of this paper, it suffices to consider 
deterministic, possibly history-dependent schedulers, i.e., partial
functions 
\begin{center}
   $\sched: \, \FPaths  \, \to \, \Nat \times \Distr(S)$,
\end{center}
which are undefined for finite maximal paths and for which if $\sched(\pi) = (c,\sigma)$, then $s \overto{c} \sigma$ for all finite paths $\pi$ that end in a nonterminal state $s$.
A \emph{$\sched$-path} is any path that arises when the nondeterministic
choices in $\M$ are resolved by $\sched$. Thus, a finite path $\pi$ as in ($*$) is a $\sched$-path iff there are distributions $\sigma_1,\ldots,\sigma_k\in\Distr(S)$
such that
$\sched\bigl(\, \pi[k{-}1] \, \bigr)$ $=$ $(c_k,\sigma_k)$
and $p_k = \sigma_k(s_k)$ for all $1 \leq k \leq n$. 
Infinite $\sched$-paths are defined accordingly.

Given a scheduler $\sched$ and some initial state $s \in \SInit$,
the behavior of $\M$ under $\sched$ is 
purely probabilistic and can be formalized 
by a tree-like infinite-state Markov chain $\M^\sched_{s}$.%
\footnote{Markov chains are MDPs that do not have any nondeterministic choices, i.e,
          where $\init$ is a singleton and 
          $|\Moves(s)|\leq 1$ for all states $s\in S$. 
}
Using standard concepts, 
a probability measure $\Pr_{s}^\sched$ for measurable sets 
of maximal branches in the Markov chain $\M_{s}^\sched$ is defined and 
can be transferred to maximal $\sched$-paths in $\M$ starting in $s$. 
For further details we refer to standard text books such as 
\cite{Haverkort,Kulkarni,Puterman}.

\paragraph{Quantitative Properties and Queries.}
The concept of schedulers permits to talk about the probability 
of a measurable path property 
$\varphi$  for fixed starting state $s$ under a given scheduler $\sched$. 
Typical examples for such a property $\varphi$ are reachability conditions of the following
type, where $T$ and  $V$ are sets of states:
\begin{itemize}
\item
    ordinary reachability:
    $\varphi=\Diamond T$ states that eventually some state in $T$
    will be visited
\item
    constrained reachability:
    $\varphi=V\, \Until\, T$ 
    imposes the same constraint as $\Diamond T$
    with the side-condition  that all states visited 
    before reaching $T$ belong to $V$
\end{itemize}
	
\noindent For a worst-case analysis of a system modeled by an MDP $\M$, one ranges 
over all initial states and all schedulers
(i.e., all possible resolutions of the nondeterminism) 
and considers the maximal or minimal probabilities for $\varphi$.
If $\varphi$ represents a desired path property, then 
$\Pr^{\min}_s(\varphi) = \inf_{\sched} \Pr^{\sched}_s(\varphi)$ 
is the probability for $\M$ satisfying $\varphi$ 
that can be guaranteed even for the worst-case scenarios.
Similarly, %
$\Pr^{\max}_s(\varphi) =  \sup_{\sched} \Pr^{\sched}_s(\varphi)$ 
is the least upper bound that can be guaranteed for the 
likelihood of $\M$ to satisfy $\varphi$.

One can also reason about bounds for expected costs of paths in $\M$. 
We consider here accumulated costs to reach a set 
$T\subseteq S$ of target states from a state $s\in S$.
Formally, if $\sched$ is a scheduler
such that $\Pr^{\sched}_s(\Diamond T)=1$, then
the \emph{expected accumulated costs} for 
reaching $T$ from $s$ under $\sched$ are defined by:
\begin{center}
   $\ExpCost^{\sched}_s(\Diamond T) 
    \ \ = \ \ 
    \sum\nolimits_{\pi} \cost(\pi) \cdot \probability(\pi)$,
\end{center}
where $\pi$ ranges over all finite $\sched$-paths with $s_n\in T$, $s_0 = s$ and $\{s_0,\ldots,s_{n-1}\}\cap T = \emptyset$. 
If $\Pr^{\sched}_s(\Diamond T)<1$, 
i.e., with positive probability $T$ will never be visited,
then $\ExpCost^{\sched}_s(\Diamond T) = \infty$.
Furthermore, 
\begin{center}
   $\ExpCost^{\min}_s(\Diamond T)$  =  
   $\inf\nolimits_{\sched} \ExpCost^{\sched}_s(\Diamond T)$ \ \ and\ \ 
   $\ExpCost^{\max}_s(\Diamond T)$ =  
   $\sup\nolimits_{\sched} \ExpCost^{\sched}_s(\Diamond T)$ 
\end{center}
specify the greatest lower bound (least upper bound, respectively)
for the expected accumulated costs reaching $T$ from $s$ in $\M$.

There are several powerful probabilistic model-checking tools
that support the algorithmic quantitative analysis of MDPs against
temporal specifications, such as formulas of 
linear temporal logic (LTL) or probabilistic computation-tree logic (PCTL)
\cite{BdAlf95,BaierKwi98}. 
In our case study, we will use the prominent probabilistic model checker
$\prism$ \cite{HinKwiNorPar2006}
that offers a symbolic MDP-engine for PCTL, dealing with a compact
internal representation of the MDP using multi-terminal binary decision
diagrams.
PCTL provides an elegant formalism to specify various temporal properties,
reliability and resource conditions.
For the purpose of the paper, the precise syntax and semantics 
of PCTL over MDPs is not relevant. We only give brief explanations for
PCTL formula patterns and queries that will be used in our case study.

Let $q\in [0,1]$ be a rational number that serves as a probability bound
and let $\varphi$ be a path property, e.g., one of the 
reachability conditions stated above.
Then, the formula $\Phi = \exists \ProbOp_{>q}(\varphi)$ 
holds for a state $s$, denoted 
$s \models \Phi$, if $\Pr^{\sched}_s(\varphi) > q$ for some scheduler
$\sched$. This is equivalent to $\Pr^{\max}_s(\varphi)>q$ for
reachability as above (and all path conditions expressible in PCTL). 
Likewise, the $\ProbOp$-operator can be used with nonstrict lower 
or upper probability bounds and universal rather than existential
quantification over schedulers.
We write $\M \models \Phi$ to indicate that all initial states
of $\M$ satisfying $\Phi$.
$\ProbOp^{\max = ?}[\varphi]$, respectively 
$\ProbOp^{\min = ?}[\varphi]$
denote the PCTL-queries to compute for all states $s$ the maximal, respectively minimal probability
for $\varphi$.
In our case study, we will also use queries of the form
$\ExpCostOp^{\min = ?}[\Diamond T]$, which amount computing the
values $\ExpCost^{\min}_s(\Diamond T)$ for all states $s$ defined above.

\section{Compositional Framework}

\label{sec:comp}
\label{sec:framework}
An SPL naturally induces a compositional structure, 
where features correspond to modules composed, e.g., 
along the hierarchy of features provided by feature diagrams. 
Thus, it is rather natural that our modeling framework for dynamic SPLs relies on a compositional
approach. We formalize feature implementations by so-called
\emph{feature modules} that might interact with each other and can depend 
on the presence of other features and their current configurations.
Dependencies between feature modules are represented in form 
of guarded transitions in the feature modules, which can impose constraints 
on the current feature combination and ask for synchronizing actions.
The interplay of the feature modules can be also described by a
single feature module, which arises from the feature implementations
via parallel composition and hence only depends on the dynamic
feature changes.
Unlike other models for dynamic SPLs, 
there is no explicit representation of the
dynamic feature combination changes inside the feature modules. 
Instead, we adopt the clear separation between computation and coordination
as it is central for coordination languages \cite{GelCar92,PapArb98,SLN01}.
In our approach, the dynamic activation and deactivation of features is
represented in a separate module, called \emph{feature controller}.
This separation yields the usual advantages:
feature modules can be replaced and reused for many scenarios
that vary in constraints for switching feature combinations
and that might even rely on different feature signatures.

We model both, feature modules and feature controllers, as MDP-like 
automata models with annotations for (possibly feature-dependent)
interactions between modules and the controller.
To reason about resource constraints, cost functions are attached
to the transitions of both, the feature modules and the feature controller.
Through parallel composition operators, the complete dynamic
SPL has a standard MDP semantics, which facilitates the use of standard model-checking 
techniques for the functional and quantitative analysis.
This is in contrast do other but similar but nonprobabilistic
and noncompositional approaches, which require
specialized feature-dependent analysis algorithms.
We show that our approach towards dynamic SPLs is more expressive than 
existing approaches by providing embeddings into our framework.
The compositional framework we present here aims also to provide a link between abstract 
models for feature implementations and the guarded command languages supported by state-of-the art probabilistic model checkers. As stated in the introduction, this approach is orthogonal to the compositional approaches 
for SPLs that have been proposed in the literature (see, e.g.,
\cite{HayAtl2000,PlaRya2001,Apel2009,MRKN13}) presenting an algebra for the nonprobabilistic feature-oriented composition of modules that covers several subtle implementation details.

\subsection{Feature Modules}

\label{sec:feature-modules}

For our definitions, let us fix some feature signature $(\FeatureSet,\ValidFeat)$.
To keep the mathematical model simple, we put the emphasis on the compositional treatment of features 
and therefore present first a data-abstract lightweight formalism
for the feature modules. In this setting, feature modules can be seen as 
labeled transition systems, where the transitions have guards
that formalize feature-dependent behaviors and are annotated with
probabilities and costs to model stochastic phenomena and resource
constraints.

We start with the definition of a feature interface that declares
which features are ``implemented'' by the given feature module 
(called \emph{own features}) and on which \emph{external} 
features the behavior of the module depends on.

\begin{definition}[Feature interface]
\label{def:featinterface}
A feature interface is a pair 
$\Feat = \<\OwnFeat,\ExtFeat\>$
consisting of two subsets $\OwnFeat$ and $\ExtFeat$ of $\FeatureSet$
such that $\OwnFeat \cap \ExtFeat=\varnothing$.
\end{definition}

With abuse of notations, we often write $\Feat$ to also denote the set $\OwnFeat \cup \ExtFeat$ of features affected by the feature interface $\Feat$. We now define feature modules as an MDP-like formalism according to a feature interface, where moves may depend on features of the feature interface and the change of own features can be triggered, e.g., from the environment.

\begin{definition}[Feature module]
\label{def:featmodul}
A \emph{feature module} is a tuple
     $\Module = (\Loc,\LocInit,\Feat,\Act,\Trans)$, where 
\begin{itemize}
\item 
    $\Loc$ is a set of locations,
\item 
    $\LocInit\subseteq\Loc$ is the set of initial locations,	
\item
    $\Feat = \<\OwnFeat,\ExtFeat\>$ is a feature interface,
\item
    $\Act$ is a finite set of actions, and
\item
	$\Trans = \TransAct \cup \TransSwitch$ is a finite transition relation.
\end{itemize}
The operational behavior of $\Module$ specified through $\Trans$ is given by feature-guarded transitions that are either labeled by an action $(\TransAct)$ or by a switch event describing own features changes $(\TransSwitch)$. Formally:\\[.5em]
{\centering
  \begin{tabular}{@{\hspace*{0cm}}l@{\hspace*{0.15cm}}c@{\hspace*{0.15cm}}l}
       $\TransAct$ & $\subseteq$ &
       $\Loc\times \BB(\Feat) \times\Act \times \Nat \times \Distr(\Loc)$
       \\[1ex]

       $\TransSwitch$ & $\subseteq$ &
       $\Loc\times \BB(\Feat) \times\BB(\OwnFeat \cup \OwnFeat') 
            \times \Nat \times \Distr(\Loc)$
   \end{tabular}}\\[.5em]
Recall that $\BB(\cdot)$ stands for the set of Boolean expressions over the augmented set of features.
\end{definition}
Let us go more into detail concerning the operational behavior of feature modules. Both types of transitions in $\Module$, action-labeled transitions and switch transitions, have the form
$\theta = (\ell,\phi,*,c,\lambda)$, where 
\begin{itemize}
\item
   $\ell$ is a location, called \emph{source location} of $\theta$,
\item
   $\phi \in \BB(\Feat)$ is a Boolean expression, called 
   \emph{feature guard},
\item 
   $c \in \Nat$ specifies the cost caused by executing $\theta$,%
   \footnote{For simplicity, we deal here a single cost value 
             for each guarded transition.
             Feature modules with multiple cost values will be 
             considered in the
             case study and can be defined accordingly.} and

\item
   $\lambda$ is a distribution over $\Loc$ specifying an internal %
   choice that determines the probabilities for the successor locations.
\end{itemize}
For action-labeled transitions,
the third component $*$ is an action $\alpha \in \Act$ 
representing some computation of $\Module$. Hence, wether an action-labeled is enabled or not depends on the current feature combination (fulfilling the feature guard or not) and on the interaction with other feature modules
(see Section \ref{sec:parallel-comp}).
For switch transitions, $*$ is a Boolean expression $\rho \in \BB(\OwnFeat \cup \OwnFeat')$, enabling $\Module$ to react or impose constraints on dynamic changes of features owned by $\Module$. In Section \ref{sec:fc}, we introduce feature controllers to describe the operational behavior of feature changes during runtime. A switch transition is then only enabled if the feature guard is fulfilled and the controller permits a change of own features of $\Module$ as described by $\rho$. The precise meaning of switch transitions will become more clear from the operational behavior of $\Module$ in the context of such controllers presented in Section \ref{sec:MDP-semantics}.

Note that we defined feature modules in a generic way, such that feature modules need not to be aware of the feature signature and realizable feature switches, which makes them reusable for different dynamic SPLs.

\begin{figure*}[t]
$$
 \begin{array}{c}
  \begin{array}{c}
     \alpha \in \Act_1 \setminus \Act_2, \ \ 
     (\ell_1,\phi,\alpha,c,\lambda_1) \in \TransAct_1
     \\[0.5ex]
     \hline
     \\[-1.5ex]
     (\<\ell_1,\ell_2\>,\phi,\alpha,c,\lambda_1 * \Dirac[\ell_2])
     \in \TransAct
  \end{array}
  \hspace*{2cm}

  \begin{array}{c}
     \alpha \in  \Act_2 \setminus \Act_1, \ \ 
     (\ell_2,\phi,\alpha,c,\lambda_2) \in \TransAct_2
     \\[0.5ex]
     \hline
     \\[-1.5ex]
     (\<\ell_1,\ell_2\>,\phi,\alpha,c, \Dirac[\ell_1] * \lambda_2)
      \in \TransAct
  \end{array}
  \\
  \\[0ex]

  \begin{array}{c}
     \alpha \in \Act_1 \cap \Act_2, \ \ 
     (\ell_2,\phi_1,\alpha,c_1,\lambda_1) \in \TransAct_1, \ \
     (\ell_2,\phi_2,\alpha,c_2,\lambda_2) \in \TransAct_2
     \\[0.5ex]
     \hline
     \\[-1.5ex]
     (\<\ell_1,\ell_2\>,\phi_1 \wedge \phi_2,\alpha,
      c_1 + c_2, \lambda_1 * \lambda_2) \in \TransAct
  \end{array}
  \\
  \\[0ex]

  \begin{array}{c}
     (\ell_1,\phi,\rho,c,\lambda_1) \in \TransSwitch_1
     \\[0.5ex]
     \hline
     \\[-1.5ex]
     (\<\ell_1,\ell_2\>,\phi,\rho \wedge \OwnFeat_2 = \OwnFeat_2',c ,
       \lambda_1 * \Dirac[\ell_2]) \in \TransSwitch
  \end{array}
  \hspace*{1cm}

  \begin{array}{c}
     (\ell_2,\phi,\rho,c,\lambda_2) \in \TransSwitch_2
     \\[0.5ex]
     \hline
     \\[-1.5ex]
     (\<\ell_1,\ell_2\>,\phi,\rho \wedge \OwnFeat_1 = \OwnFeat_1',c ,
       \Dirac[\ell_1] * \lambda_2) \in \TransSwitch
  \end{array}
  \\
  \\[0ex]

  \begin{array}{c}
     (\ell_1,\phi_1,\rho_1,c_1,\lambda_1) \in \TransSwitch_1,  \ \
     (\ell_2,\phi_2,\rho_2,c_2,\lambda_2) \in \TransSwitch_2
     \\[0.5ex]
     \hline
     \\[-1.5ex]
     (\<\ell_1,\ell_2\>,\phi_1 \wedge \phi_2,
      \rho_1 \wedge \rho_2, c_1 + c_2, \lambda_1 * \lambda_2)
      \in \TransSwitch
  \end{array}
 
 \end{array}
$$
 \nocaptionrule\caption{Rules for the parallel composition of feature modules}%
          \label{fig:parallel}
         \vspace{-1em}
\end{figure*}

\subsection{Parallel Composition}

\label{sec:parallel-comp}

We formalize the interactions of feature modules 
by introducing a parallel operator on feature modules. 
Thus, starting with separate feature modules for all features
$f\in \FeatureSet$ one might generate feature modules that 
``implement'' several features, and eventually obtain a feature model
that describes the behavior of all ``controllable'' features of
the SPL over the feature signature $(\FeatureSet,\ValidFeat)$. 
Additionally, there might be some features in the set of features 
$\FeatureSet$ provided by an unknown environment, where no feature
modules are given.

We now consider a  parallel operator for two composable
feature modules in the style of parallel composition 
of probabilistic automata \cite{SegLynch94,Segala95} 
using \emph{synchronization} over shared actions (handshaking)
and interleaving for all other actions. 
Let
\begin{center}
\begin{tabular}{lcl}
   $\Module_1$ & = &
   $(\Loc_1,\LocInit_1,\Feat_1,\Act_1,\Trans_1)$
   \\
   $\Module_2$ & = &
   $(\Loc_2,\LocInit_2,\Feat_2,\Act_2,\Trans_2)$,
\end{tabular}
\end{center}
where $\Feat_i = \<\OwnFeat_i,\ExtFeat_i\>$ and
$\Trans_i = \TransAct_i \cup \TransSwitch_i$.
Composability of $\Module_1$ and $\Module_2$ means that
$\OwnFeat_1\cap\OwnFeat_2=\varnothing$.
Own features of $\Module_1$ might be  external for
$\Module_2$ and vice versa, influencing each others behavior. 

\begin{definition}[Parallel composition]
\label{def:parallel}
Let $\Module_1$, $\Module_2$
be two composable feature modules as above.
The \emph{parallel composition} of 
$\Module_1$ and $\Module_2$ is defined as the feature module
\begin{center}
    $\Module_1 \| \Module_2 \ = \ 
     (\Loc,\LocInit,\Feat,\Act,\Trans)$,
\end{center}
where the feature interface 
$\Feat = \<\OwnFeat,\ExtFeat\>$ and 
the other components are
defined as follows:
\begin{center}
  \begin{tabular}{rcl}
     $\Loc$ & = & $\Loc_1\times \Loc_2$
     \\[.5ex]
     
     $\LocInit$ & = & $\LocInit_1{\times}\ \LocInit_2$
     \\[.5ex]

     $\OwnFeat$ & = & $\OwnFeat_1 \cup \OwnFeat_2$
     \\[.5ex]

     $\ExtFeat$ & = & $ (\ExtFeat_1\cup\ExtFeat_2)\setminus\OwnFeat$
     \\[.5ex]

     $\Act$ & = & $\Act_1\cup \Act_2$
 \end{tabular}
\end{center}
The treansition relation $\Trans = \TransAct \cup \TransSwitch$ 
is defined by the rules shown in Figure \ref{fig:parallel}.
\end{definition}

Obviously, $\Module_1 \| \Module_2$ is again a feature module.  
In contrast to the (nonprobabilistic) superimposition approach 
for composing modules representing feature implementations 
\cite{Kat1993,PlaRya2001}, 
the parallel operator $\|$  is commutative and associative.
More precisely, if  $\Module_i$ for $i\in\{1,2,3\}$ %
are pairwise composable feature modules, then:
\begin{center}
\begin{tabular}{rcl}
      $\Module_1\|\Module_2$ & = & $\Module_2\|\Module_1$
      \\[.5ex]

      $(\Module_1\|\Module_2)\|\Module_3$ & = & 
      $\Module_1\|(\Module_2\|\Module_3)$
\end{tabular}
\end{center}

\noindent For the parallel composition of feature modules with multiple
cost functions, one has to declare which cost functions are combined.
This can be achieved by dealing with types 
(e.g., energy, money, memory requirements) of cost functions
and accumulate costs of the same type.

\begin{figure*}[t]
$$
 \begin{array}{c}
  \begin{array}{c}
     (\ell,\phi,\alpha,c,\lambda)\in \TransAct, \ \
     C \models \phi
     \\[0.5ex]
     \hline
     \\[-1.5ex]
     (\<\ell,C\>,c,\lambda * \Dirac[C]) \in \Moves
  \end{array}
  \hspace*{2cm}

  \begin{array}{c}
     C \overto{d} \gamma, \ \ 
     C \models \phi, \ \
     \forall C' \in \Support(\gamma).
     C \cap \OwnFeat = C' \cap \OwnFeat 
     \\[0.5ex]
     \hline
     \\[-2ex]
     (\<\ell,C\>,d,\Dirac[\ell]*\gamma) \in \Moves
  \end{array}
  \\
  \\[-1ex]

  \begin{array}{c}
     (\ell,\phi,\rho,c,\lambda)\in \TransSwitch, \ \ 
     C \models \phi, \ \ 
     C \overto{d} \gamma, \ \
     \exists C' \in \Support(\gamma).
     (C \cap \OwnFeat \not= C' \cap \OwnFeat), \ \
     \forall C' \in \Support(\gamma).
     (C,C') \in R_{\rho}
     \\[0.5ex]
     \hline
     \\[-2ex]
     (\<\ell,C\>,c+d,\lambda*\gamma) \in \Moves
  \end{array}
 \end{array}
$$
 \nocaptionrule\caption{Rules for the moves in the MDP $\Module \Join \Controller$
          \label{fig:semantics}}
          \vspace{-1em}
\end{figure*}

\subsection{Feature Controller}

\label{sec:fc}
\label{sec:controller}

After we defined feature modules and described how their operational behavior is influenced via interacting with other feature modules, we now turn to feature controllers, which specify the rules for the possible changes of feature combinations during runtime of the system. We start with purely nondeterministic controllers
switching feature combinations similar to \cite{DamSch2011} (Definition~\ref{def:npcontroller}). Then, we extend such simple controllers by assigning probabilities to the feature switch events (Definition~\ref{def:controller}).

\begin{definition}
\label{def:npcontroller}
A \emph{simple feature controller} 
over the feature signature $(\FeatureSet,\ValidFeat)$ is a tuple 
\begin{center}
   $\Controller\ =\ (\ValidFeat,\ValidFeatInit,\SwitchRel)$, 
\end{center}
where 
$\ValidFeatInit \subseteq\ValidFeat$ is the set of 
\emph{initial feature combinations} and
$\SwitchRel \subseteq \ValidFeat{\times}\Nat{\times}\ValidFeat$ is a 
relation, called \emph{(feature) switch relation}, 
that formalizes the possible dynamic changes
of the feature combinations and their cost.
We refer to elements in $\SwitchRel$ as \emph{(feature) switch events}
and require that $(C,d_1,C'), (C,d_2,C')\in \SwitchRel$ implies $d_1 = d_2$.
\end{definition}

If there are several switch events $(C,d_1,C_1), (C,d_2,C_2), \ldots$
that are enabled for the feature
combination $C$, then the choice which switch event fires is chosen
nondeterministically.
This is adequate, e.g., to represent potential upgrades or downgrades of a software product or express environmental influences.

Although our focus is on reasoning about dynamic SPLs,
we like to mention that our framework is also applicable
for static SPLs, where one valid feature combination is selected
initially and is never be changed at runtime.
Static SPLs can easily be modeled using 
the simple feature controller 
   $\Controller_{\textsf{static}} = (\ValidFeat,\ValidFeat,\varnothing)$, 
where the switch relation  is  empty.

The concept of simple feature controllers also covers the approach of
\cite{CorClaHey2013,DinMitFetMez2010},
where dynamic SPLs are represented by feature signatures $(\FeatureSet,\ValidFeat)$ 
extended with disjoint sets of dynamic features $D\subseteq \FeatureSet$ and environment features $E \subseteq \FeatureSet$.
The features in $D \cup E$ can be activated or deactivated at any time,
while the modes of all other features remain unchanged.
This dynamic behavior of the feature combinations 
is formalized using the controller 
(we omit the cost values of switch events):
\begin{center}
   $\Controller_{D,E} \ = \ (\ValidFeat,\ValidFeat,\SwitchRel_{D,E})$,
\end{center}
where
     $(C,C') \in \SwitchRel_{D,E}$
     iff %
     $\varnothing \, \not= \, C \ominus C' \, \subseteq \, D\cup E$
for all $C, C'\in \ValidFeat$. Here, $C \ominus C'$ denotes the symmetric
difference of $C$ and $C'$, i.e., $C \ominus C' \ = \ C \setminus C' \cup C' \setminus C$.

As already mentioned when detailing feature modules, switch events can require interactions between the feature controller and the feature modules. Thus, feature modules can trigger or prevent switch events
by offering or refusing the required interactions with the feature controller.
For example, suppose some software product is only distributed in a 
basic version. Potential upgrades after purchasing the software product will be triggered
by the user, represented in our framework by some feature module.

There might be other switch events that are uncontrollable 
by the feature modules, e.g., the deactivation of features that are damaged
due to environmental influences 
(electrical power outage, extreme hotness, etc.).
Such switch events in the controller do not rely on interactions
with the feature modules.
Instead, statistical data might be available that permits to 
model the frequency of
such uncontrollable switch events by probabilities.
This leads to the more general concept 
of \emph{probabilistic feature controllers},
where switch events are pairs $(C,d,\gamma)$ consisting of a feature
combination $C$, a cost value $d \in \Nat$ and
a distribution $\gamma$ over $\ValidFeat$.
Thus, probabilistic feature controllers can be seen MDPs with switch events as moves.

\begin{definition}[Controller]
   \label{def:controller}
A probabilistic feature controller over the signature $(\FeatureSet,\ValidFeat)$, briefly called \emph{controller}, 
is a tuple $\Controller$ $=$ $(\ValidFeat,\ValidFeatInit,\SwitchRel)$
as in Definition \ref{def:npcontroller}, but
\begin{center}
   $\SwitchRel \ \subseteq \ 
    \ValidFeat \times \Nat \times \Distr(\ValidFeat)$.
\end{center}
Again, we require that the switch relation is finite
and that $(C,d_1,\gamma)$, $C,d_2,\gamma)\in \SwitchRel$
implies $d_1=d_2$. 
\end{definition}

Clearly, each simple feature controller $\Controller$ can be seen as a 
(probabilistic feature) controller. For this, we just have to
identify each switch event $(C,d,C')$ with $(C,d,\Dirac[C'])$.
The following example shows a controller of our productivity system 
detailed already in Example \ref{ex:productivity}.

\begin{example}
\label{ex:pdfm}
Let us consider the feature signature $(\FeatureSet,\ValidFeat)$,
where $\FeatureSet = \{\mathrm{s,o,e,r,m,f,h,b,l}\}$ 
given by the  feature diagram in Example \ref{ex:productivity}
and the controller $\Controller_\mathrm{ps}$ depicted in Figure \ref{fig:pdfm}
with the initial feature combinations $\mathrm{soe,sor,sorm}$.
States are valid feature combinations in $\ValidFeat$ and arrows describe 
feature combination switches. 
These switches are amended with a probability, 
which is supposed to be estimated from statistical user data and 
costs for taking the switch (upgrade/downgrade). 
For instance, the step
    $\mathrm{sor} \, \move{\switchevent,269}_{0.15} \, \mathrm{sorfb}$
indicates that with probability 15\%, a user is 
buying a business office feature $(\mathrm{b})$ for 269 \euro, 
given she has a professional operating system $(\mathrm{r})$.
Although $\Controller_\mathrm{ps}$ is purely probabilistic 
(i.e., $\Controller_{\mathrm{ps}}$ 
can be seen as a Markov chain) in the sense that
in all states precisely one move is enabled, it also formalizes the rules for upgrade or downgrade features.
For better readability, self-loops with the remaining probability value
are not depicted in this figure.
\begin{figure}[h]%
   \includegraphics[width=\linewidth]{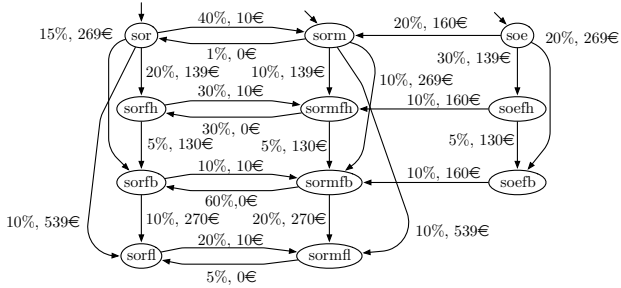}%
   \nocaptionrule\caption{A probabilistic controller}%
   \label{fig:pdfm}%
\end{figure}%
Note that the media center feature $(\mathrm{m})$ can be activated and 
deactivated at any time if the professional operating system 
feature $(\mathrm{r})$ is activated. 
Upgrades from the home edition $(\mathrm{e})$ of the operating system 
are only possible to the professional edition 
including the media center feature $(\mathrm{m})$. 
However, downgrading from the professional operating 
system $(\mathrm{r})$ to the home edition $(\mathrm{e})$ is prohibited. 
The home edition $(\mathrm{h})$ of the office suite can only be 
upgraded to the professional one $(\mathrm{l})$ if the operating system is professional $(\mathrm{r})$.
\end{example}

\subsection{MDP-semantics of Feature Modules}

\label{sec:MDP-semantics}

The semantics of a feature module $\Module$ under some
controller $\Controller$ is given in terms of an MDP.
If $\Module$ stands for the parallel composition of all
modules that implement one of the ``internal'' features
of a given SPL and the controller $\Controller$ specifies the dynamic adaptions
of the feature combinations, then this MDP formalizes the operational behavior of the composite system.
In what follows, we fix a feature module and a controller
\begin{center}
  \begin{tabular}{rcl}
      $\Module$ & = & $(\Loc$, $\LocInit$, $\Feat$, $\Act$, $\Trans)$
      \\[.5ex]

      $\Controller$ & = & $(\ValidFeat,\ValidFeatInit,\SwitchRel)$
  \end{tabular}
\end{center}
as in Definition \ref{def:featmodul} and Definition \ref{def:controller}
where $\Feat \subseteq \FeatureSet$.
Intuitively, taking an action-labeled 
transition $(\ell,\phi,\alpha,c,\lambda)$ of $\Module$
is a possible behavior of $\Module$ in location $\ell$,
provided that the current state $C$ of the controller $\Controller$ 
(which is simply the current feature combination) 
meets the guard $\phi$.
Switch events of the controller can be performed
independently from $\Module$ if they do not affect the own features
of $\Module$, whereas if they affect at least one feature in $\OwnFeat$, the changes of the mode have to be executed synchronously.

\begin{definition}[Semantics of feature modules]
\label{def:semantics}
Let $\Module$ and $\Controller$ be as before.
The behavior of $\Module$ under the controller $\Controller$ 
is formalized by the MDP 

\begin{center}
     $\Module \Join \Controller\ = \ (S,\SInit,\Moves)$,
\end{center}

\noindent where 
$S=\Loc\times\ValidFeat$,
$\SInit=\LocInit\times\ValidFeatInit$
and where the move relation $\Moves$ is defined
by the rules in Figure \ref{fig:semantics}.
In the last rule, $\rho \in \BB(\OwnFeat \cup \OwnFeat')$ is viewed as 
a Boolean expression over $\FeatureSet \cup \FeatureSet'$. 
Thus, $\rho$ specifies a binary relation $R_{\rho}$ over 
$2^{\FeatureSet} \times 2^{\FeatureSet}$.
\end{definition}

Observe that due to the MDP semantics of feature modules under a controller, standard 
probabilistic model-checking techniques for the quantitative analysis can be directly applied. This includes properties about current feature combinations, since they are encoded into the states of the arising MDP.

\subsection{Remarks on our Framework}

\label{sec:feature-control-graphs}
\label{sec:variants}

\paragraph{Feature Modules with Variables.}
So far, we presented a light-weight data-abstract formalism for feature modules
with abstract action and location names.
This simplified the presentation of the mathematical model.
From the theoretical point of view, feature modules in the sense
of Definition~\ref{def:featmodul} 
are powerful enough to encode systems where the modules
operate on variables with finite domains. Even communication over
shared variables can be mimicked by dealing with handshaking
and local copies of shared variables.
However, in case studies  the explicit use of assignments for variables
and guards for the transitions that impose constraints for local and shared
variables is desirable; not only to avoid unreadable encodings, 
but also for performance reasons of the algorithmic analysis. 
Although message passing via channels would be more in the spirit of
coordination paradigms, the concept of shared variables can help to  
generate more compact representations of the MDP for the composite system,
which makes it useful for the application of model-checking tools.
The formal definition of an extension of 
\emph{feature modules by variables}
is rather technical, but fairly standard.
We present their syntax and MDP-semantics under a given controller in the Appendix.
These extended feature modules directly yield a translation
in $\prism$'s input language that we used in our case study
described in the next section.

\paragraph{Other Variants.}
Besides amending feature modules by variables, the basic formalisms of our framework
can be refined in various directions. We briefly mention here a few of them. 

With the presented formalism the switch events appear 
as nondeterministic choices and require interactions between the controller 
and all modules that provide implementations for the affected features.
Employing the standard semantics of MDPs, where
one of the enabled moves is selected nondeterministically,
this rules out the possibility to express that certain switch events
might be unpreventable.
Unpreventable switch events can be included into our framework, refining the concept of feature controllers by explicitly specifying which switch events must be taken whenever they are enabled in the controller.
This could modeled by adding an extra transition relation for 
\emph{urgent} switch events or prioritizing switches.

Instead of urgency or priorities, one might also keep the presented
syntax of feature modules and controllers, but refine the MDP-semantics
by adding \emph{fairness conditions} that rule out computations where 
enabled switch events are postponed ad infinitum.

Another option for refining the nondeterministic choices in the
controller is the distinction between
switch events that are indeed controllable by the controller
and those that are triggered by the environment.
This naturally leads to a game-based view of the MDP for
the composite system (see also Section \ref{sec:conclusions}).

\paragraph{Controllers as Feature Modules.}
To emphasize the feature-oriented aspects of our framework,
we used a different syntax for controllers and feature modules.
Nevertheless, controllers can be viewed as special feature modules when we discard the concept of switch events and
switch transitions and rephrase them as action-labeled transitions. 
To transform controllers syntactically to feature modules,
we have to add the trivial guard 
and introduce names for all switch events.
When turning the switch transitions of the feature modules into 
action-labeled transitions, matching names must be introduced to 
align the parallel operators $\|$ and $\bowtie$. Note that in the constructed feature modules, all features are external and the locations coincide with feature combinations. However, an extended version of controllers can also be considered, where in addition to feature combinations, arbitrary other internal locations of the controller can be specified.

\section{Quantitative Feature Analysis}

\label{sec:fmc}

\label{sec:appl}

Within the compositional framework presented in the last section, let us assume that we are given feature modules $\Module_1,\ldots\Module_n$ which stand for abstract models of certain features $f \in \FeatureSet$ and a feature controller $\Controller$ specifying the rules for feature combination changes.
The feature set $\FeatureSet$ might still contain other features 
where no implementations are given, which are external features
controlled by the environment. Alternatively, one of the feature modules can formalize the
interference of the feature implementations with 
a partially known environment, e.g., in form of stochastic
assumptions on the workload or the frequency of user interactions.
Applying the compositional construction by putting feature modules in parallel and joining them with the feature controller, we obtain an MDP of the form
\begin{center}
     $\M \ = \ (\Module_1 \| \ldots \| \Module_n) \Join \Controller$.
\end{center}
This MDP $\M$ formalizes the operational behavior of a dynamic SPL and can now be used for quantitative analysis.
Hence, the task of a quantitative analysis of dynamic SPLs is reduced to standard algorithmic problems for MDP and permits the use of generic probabilistic model-checking techniques. This is in contrast to other family-based model-checking approaches for SPLs, where feature-adapted algorithms were constructed \cite{Cla2010,Cla2011}.

\subsection{Quantitative Analysis and Strategy Synthesis Problem}

A \emph{quantitative worst-case analysis} in the MDP $\M$ 
that establishes least upper or greatest
lower bounds for the probabilities of certain properties
or for the expected accumulated costs
by means of the queries 
$\ProbOp^{\max = ?}[\varphi]$, $\ProbOp^{\min = ?}[\varphi]$
or $\ExpCostOp^{\min = ?}[\Diamond T]$
(see Section \ref{sec:MDP}) can be carried out with standard
probabilistic model-checking tools. 
These values provide guarantees on the probabilities under all
potential resolutions of the nondeterministic choices in $\M$,
possibly imposing some fairness constraints to ensure that
continuously enabled dynamic adaptions of the feature
combinations (switch events) cannot be superseded forever
by action-labeled transitions of the feature modules.

Although the quantitative worst-case analysis can give important
insights in the correctness and quality of an SPL,
in our framework with separate specifications of the potential
dynamic adaptions of feature combinations (the controller)
and the implementations of the features (the feature modules), 
it appears naturally to go one step further by asking for 
\emph{optimal strategies} for triggering switch events.
Optimality can be understood with respect to queries like
minimizing the probability for undesired behaviors or
minimizing the expected energy consumption while meeting given
deadlines, or maximizing the utility value when an initial
energy budget is given.

Several variants of this problem can be considered.
The basic variant that we address in our case study
relies on the assumption that the nondeterminism in the MDP $\M$
for the composite system stands for decisions to be made by the
controller, i.e., only the switch events appear nondeterministically,
whereas the feature modules behave purely probabilistically 
(or deterministically)
when putting them in parallel with the controller.
More formally, we suppose that in each state $s$ of $\M$,
either there is a single enabled move representing some
action-labeled transition of one or more feature modules
or all enabled moves stand for switch events.
In this case, an optimal strategy for the controller is just a scheduler
for $\M$ that optimizes the quantitative measure of interest.
Thus, the natural task that we address is 
the \emph{strategy synthesis problem},
where $\M$ and some PCTL-query $\Phi$
as in Section \ref{sec:prelim} are given 
and the task is to construct a scheduler $\sched$ for $\M$
that optimizes the solution of the query $\Phi$.
Indeed, the standard probabilistic model-checking
algorithms for PCTL are applicable to solve the
strategy synthesis problem.

\subsection{Case Study}

In this section, we describe a case study to show the 
applicability of our framework to a real-case scenario. 
Our case study is based on \textsc{eBond}, 
which is an energy-aware network device allowing for energy savings 
on the server-side \cite{HahnelDVH13}. 
The \textsc{eBond} device supports bonding of (heterogenous)
network interface cards (NICs) with different performance 
and energy characteristics 
into a single device. Individual NICs can be switched on at any time 
whenever more bandwidth is needed and switched off otherwise. 
In \cite{HahnelDVH13}, simulation-based techniques were used to
show that within \textsc{eBond}, energy savings up to 75\% can be 
achieved when demands for bandwidth varies, e.g., between day and night time. 

\begin{figure}[h]%
   \centering\includegraphics[width=\linewidth]{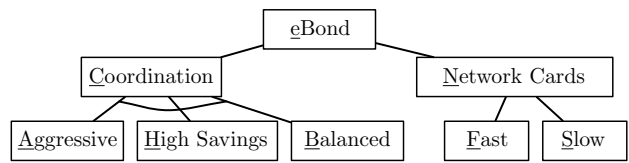}%
   \nocaptionrule\caption{Feature diagram for an eBond product line}%
   \label{fig:eBond_featurediagram}%
\end{figure}

\paragraph*{Original \textsc{eBond}.} The simulation in \cite{HahnelDVH13} was carried out for a fixed \textsc{eBond} device with exactly two NICs. The first NIC requires much energy but supports
up to 10 GBit bandwidth, whereas the second NIC is a slow 1 GBit NIC with 
low energy consumption. The NICs were only allowed to be used exclusively, 
i.e., the bonding of the two devices was not considered.
Furthermore, three energy saving algorithms have been detailed:
\begin{enumerate}
\item[(1)] an \emph{aggressive} algorithm, in which the 10GBit NIC is 
switched off whenever possible (i.e., the last observed bandwidth request is at most 1GBit), 
\item[(2)] a \emph{high saving} algorithm, which assumes a higher requested bandwidth, thus switching later to the slow NIC and earlier to the fast NIC, 
and 
\item[(3)] a \emph{balanced} algorithm, which behaves as the high saving algorithm, but introduces an additional cool-down phase delaying card switches even further.
\end{enumerate}
The setting from \cite{HahnelDVH13} can be interpreted in terms of features, 
where we assume the energy saving algorithms to be enclosed in a 
coordination feature. The arising feature signature of this static SPL could 
be specified as a feature diagram shown in Figure \ref{fig:eBond_featurediagram}. Note that the energy saving algorithm is chosen initially when the \textsc{eBond} device is deployed.

\noindent The \textsc{eBond} model operates in two phases, where a 5 minutes operating
phase alternates with a reconfiguration phase, in which the active NIC is chosen by the energy saving algorithm.
The analysis carried out in \cite{HahnelDVH13} issued 
the measurement of the energy consumption for the different NICs in their 
sleeping mode and under load, as well as counting the number of service-level 
agreement (SLA) violations. An SLA violation was assumed to happen 
whenever the demanded bandwidth could not be delivered by the server, i.e., when the 1Gbit NIC has been activated by the energy saving algorithm but the requested bandwidth exceeds 1Gbit.
The different energy saving algorithms have been simulated using bandwidth requirements from two real-case scenarios. In particular, the total energy consumption and number of SLA violations over 43 days have been detailed.

\begin{figure}[h]%
   \centering\includegraphics[width=\linewidth]{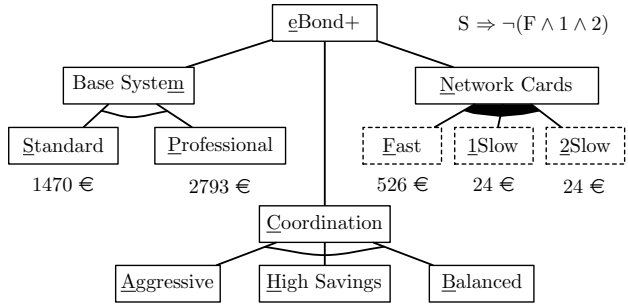}%
   \nocaptionrule\caption{Feature diagram of the \textsc{eBond+} product line}%
   \label{fig:eBondplus_featurediagram}%
\end{figure}

\paragraph*{Dynamic \textsc{eBond+}.} We extend the static SPL setting of \textsc{eBond} 
towards a dynamic SPL, gaining more flexibility in bonding NICs. Our extended version, called \textsc{eBond+}, allows for more than one NIC being active at the same time and involves dynamics
by supporting to change the NIC combinations at runtime. We furthermore distinguish 
between a standard and a professional bundle which are for sale. In the standard
bundle a costumer can plug up to two NICs, whereas the professional bundle supports up to three NICs. 
When buying an \textsc{eBond+} device, the costumer decides for either the standard or the 
professional bundle. We assume that this decision if fixed and that there is no upgrade option 
later on. Also the energy saving algorithm is fixed on purchase. For the NICs we support 
the same two types of cards as in the original \textsc{eBond}. The customer selects on the number 
and type of NICs the \textsc{eBond+} device will be shipped with. The NICs can be bought 
or dropped also after the purchase. Interpreting each of the described functionality as features, 
the feature signature of \textsc{eBond+} can be specified by a feature diagram 
(see Figure \ref{fig:eBondplus_featurediagram}). Note the additional constraint on the upper right 
of Figure \ref{fig:eBondplus_featurediagram}, indicating that in the standard bundle only two NICs 
can be plugged into the system, i.e., if the standard bundle $\mathrm{s}$ is selected, it is not possible to
purchase all three NICs.

\noindent We formalized the dynamic feature combination switches, i.e., plugging or unplugging NICs of the 
system, via a feature controller according to the framework developed in the previous section 
(see Section \ref{sec:controller}). The controller implements the constraints on plugging 
and unplugging NICs. We made the following assumptions on the dynamics of the feature switches: 
a NIC can only be bought (dropped) and plugged into (unplugged from) the device when there is a need, 
meaning that whenever the required bandwidth becomes either too high or too low w.r.t. the current 
configuration of the \textsc{eBond+} device. Furthermore, we assume that any change of the NICs 
requires a minimal amount of time.
In contrast to the \textsc{eBond} model, the \textsc{eBond+} model operates in three phases
rather than two, as we introduce an additional feature controller step allowing 
dynamic feature switches. The initial phase is controlled by the costumer, who decides 
on the initial configuration of the \textsc{eBond+} device.

For the NICs, the energy saving algorithms and the system environment in terms of the requested bandwidth
are formalized as feature modules in the spirit of our compositional framework. The standard or professional 
system features are only influencing the number of NIC features activated and are hence specified within the feature 
controller. For the operational behavior of the NICs we introduced a probabilistic choice with low probability modeling 
the possibility of failing network cards. In this case, the respective NIC feature is active but does not provide any 
functionality. The coordination features are implemented as for \textsc{eBond} \cite{HahnelDVH13}, where
waking up and putting NICs into sleep follows a purely deterministic strategy (without any probabilistic or 
nondeterministic behavior). The environment feature, which models the requested bandwidth, is present in all
valid feature combinations and behaves probabilistically. The exact distribution is derived from statistical user data.

\paragraph*{Reasoning over \textsc{eBond+}.}
With the above model of the \textsc{eBond+} device, we obtain a standard MDP
\begin{center}
     $\M_+ \ = \ (\ \underbrace{\textsf{Fast} \| \textsf{1Slow} \| \textsf{2Slow}}_\text{network cards}\ \| \underbrace{\textsf{A} \| \textsf{H} \| \textsf{B}}_\text{coordination} \| \underbrace{\textsf{Env}\phantom{\|\!\!}}_\text{environment}) \Join \Controller$ 
\end{center}
having exactly one starting state, which yields the basis for any kind of quantitative analysis.
We equip $\M_+$ with three different cost functions.
Beyond the cost measures for the energy consumption of the active NICs
and the number of SLA violations as considered in \cite{HahnelDVH13}, we 
introduce here a third cost measure for money. Costs in terms of money include 
purchasing costs of the initial system, money spent for buying new 
NICs, paying the card switches as well as the cost of SLA violations. 
As SLA violations are rather expensive, it is clear that a customer 
tries to avoid SLA violations by purchasing a device whose reliability guarantees 
the desired throughput functionality. On the other hand, a customer also tries to save 
initial costs when buying the device. The strategy synthesis problem for $\M_+$ thus 
aims to find an optimal strategy (w.r.t. the introduced costs) for the customer resolving 
nondeterminism in the feature controller $\Controller$ (plugging/unplugging NICs)
to fulfill her needs assuming that the workload behaves as modeled by the environment 
feature $\textsf{Env}$. 

In the analysis part of our \textsc{eBond+} case study,
we consider four different strategy synthesis problems for $\M_+$ 
w.r.t. queries
$$
	\begin{array}{rcllrcl}
		\Phi_p&=&\ProbOp^{\max = ?}[(\neg \textit{Sla})\ \Until\ T]&&\Phi_e&=&\ExpCostOp^{\min = ?}_\text{energy}[\Diamond T]\\
		\Phi_m&=&\ExpCostOp^{\min = ?}_\text{money}[\Diamond T]&&\Phi_s&=&\ExpCostOp^{\min = ?}_\text{slavio}[\Diamond T]
	\end{array}
$$
Here, the type of the expected minimal costs is annotated to the query (i.e., energy, money and slavio). Furthermore, $\textit{Sla}$ stands for the set of states in $\M_+$ where an SLA violation occurred and $T$ 
for the set of states in $\M_+$ where some fixed time horizon is reached. Hence, the strategy 
synthesis problem for $\M_+$ corresponds to the optimization problem of maximizing the probability 
of not raising an SLA violation (i.e., reliability of the device), minimizing the expected energy consumption, money spent or 
percentage of SLA violations, respectively, all within the fixed time horizont.

\subsection{Quantitative analysis of \textsc{eBond+} in \textsc{Prism}}
Using the compositional framework presented in Section \ref{sec:comp}, 
we modeled a parameterized version of \textsc{eBond+} within $\prism$ 
as MDP $\M_+$.%
\footnote{The full $\prism$ model and the selected queries 
          are available online at 
          \url{http://wwwtcs.inf.tu-dresden.de/ALGI/features/ebond.zip}.}
All features were translated into individual $\prism$ modules, which results in $\M_+$ when parallel composed by $\prism$. The types of NICs and their energy consumption profile are according to \cite{HahnelDVH13}, i.e., the fast 10 GBit NIC corresponds to an Intel Ethernet Server Adapter X520-T featuring an E76983 CPU, whereas the remaining (at most two) NICs are supposed to be 1 GBit Intel EXPI9301CTBLK NICs with an E25869 CPU. The purchase costs for the system and the network cards (in \euro) are taken from a leading 
vendor's online store and an SLA violation is assumed to cost 200 \euro\ each.
Whereas the coordination features were implemented 
according to the energy saving algorithms of \textsc{eBond}, the environment feature modeling 
the bandwidth requirements differs. Instead of employing the statistical user data from one of the 
two setting addressed in \cite{HahnelDVH13}, we assume a maximal bandwidth bound 
$b$ [GBit/s]. Dependent on the current bandwidth requirements the bandwidth requirement rises 
and falls -- the lower (higher) the current bandwidth is below (above) $b/2$, the higher is the 
probability that the environment requires more (less) bandwidth in the next phase.

\paragraph*{Model Parameters.}
For the case study we fixed certain model parameters. First, we chose a time horizon of $T$=12 hours and a delay of 
20 minutes for reconfiguring the system. Other timing constraints 
are taken from the \textsc{eBond} case study, involving a reconfiguration timer of 5 minutes and a 
cool-down timer of 30 minutes for the balanced coordination. For the high savings and 
balanced coordination feature, we assumed a predictor 10\% hysteresis (also taken from \textsc{eBond}). 
The probability that a NIC fails is set to 0.1\%. Bandwidth values are evaluated with an accuracy 
of 100 MBit/s.

\subsection{Empirical Evaluation.}
In our experiments, we parameterized over the maximal bandwidth bound from values 
between 200 MBit/s and 7200 MBit/s, solving the strategy synthesis problem for 
$\M_+$ w.r.t. each query $\Phi_p$, $\Phi_e$, $\Phi_s$ and $\Phi_m$ as detailed above. 
The figures illustrate the influence the initial \textsc{eBond+} configuration when purchasing the system. The encoding is of the initial configurations is of the form ``XY\_B\_A'', where X stands for the number of 10 GBit NICs, Y stands for the number of 1 Gbit NICs, $\text{B}\in\{\text{S,P}\}$ stands for either
the standard or the professional bundle, and $\text{A}\in\{A,H,B\}$ stands for either
the aggressive, high saving, or balanced energy saving algorithm.

Our results for all queries show that the chosen energy saving algorithm has a very similar influence on the results as determined in \cite{HahnelDVH13}. Although, our results indicate that there is no significant difference between the high savings and aggressive energy saving algorithm. The reason is that the aggressive energy saving 
algorithm relies mainly on switching cards as in \textsc{eBond}, whereas in \textsc{eBond+} 
also bonding of two and more cards is supported.

\begin{figure}[h]%
\centering\includegraphics[width=\linewidth]{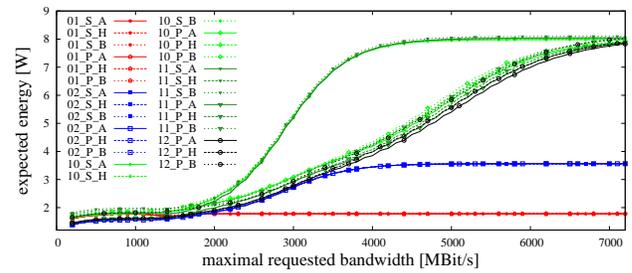}%
\nocaptionrule\caption{Evaluation of $\Phi_e$ for the different \textsc{eBond+} variants}%
\label{fig:emin_time}%
\end{figure}

\paragraph{Utility Analysis.}
We first look at $\Phi_p$, i.e., the maximum probability of avoiding SLA violation 
within the given fixed time period, corresponding to a measure of reliability for an \textsc{eBond+} device.
\begin{figure}[h]%
\centering\includegraphics[width=\linewidth]{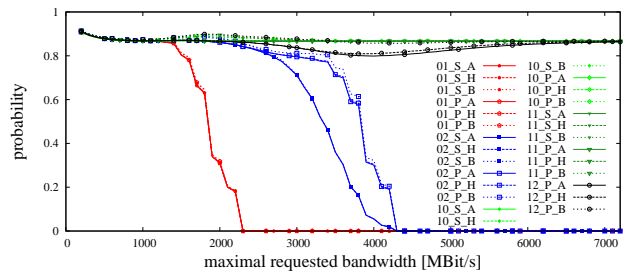}%
\nocaptionrule\caption{Evaluation of $\Phi_p$ for the different \textsc{eBond+} variants}%
\label{fig:pmax_time}%
\end{figure}%
In Figure \ref{fig:pmax_time} it can be seen that when the maximal required bandwidth is low, the probability 
of avoiding an SLA violation within the considered time bound is nearly 90\%,
independent from the initial feature configuration. For initial feature combinations 
that have only one 1 GBit NIC activated, an SLA violation can 
hardly be avoided for maximal requested bandwidths greater than $b$=2 GBit/s. 
This is due to the fact that the expected average bandwidth is $b/2$=1 GBit/s,
which agrees with the maximal available bandwidth of the NIC. Within the
20 minutes required to change the initial feature combination 
and upgrade to more NICs, an SLA violation becomes very likely. The same 
phenomenon appears with only two 1 GBit NICs activated, but there the probability 
value drops below 50\% at a maximal bandwidth of 4 GBit/s. %
Note that in this setting with the two 1 GBit NICs activated and being under load,
only with the professional bundle the additional 10 GBit NIC can be bought 
and plugged, such that the impact of the slower NICs is superseded. %
With the fast NIC initially activated, the probability can be maximized always at 
around 88\%, since the required bandwidth can always be complied up to the case the 
10 GBit NIC fails. Note that whenever the balanced coordination feature is activated, 
the maximized probability avoiding an SLA violation is higher than within the other 
coordination features.
\paragraph{Energy Analysis.}
When turning to the minimization of the expected energy consumption, i.e., solving 
query $\Phi_e$ for $\M_+$, things are different as shown in Figure \ref{fig:emin_time}.
Since the other cost measures are independent from the energy costs,
the smallest configuration with only one slow card initially activated performs 
best with only 1.78 W energy consumption\footnote{according to \cite{HahnelDVH13}
the NIC requires 1.92 W on full load}. However, the standard bundles where 
the fast card is activated at the beginning have significant higher energy 
consumption for increasing maximal required bandwidth. This is due to the fact
that the feature controller cannot unplug the fast NIC under load. For the same reason
when activating a slow card in situations with maximal requested bandwidth above 2 GBit/s 
the fast card is very likely to be under load. Within a professional bundle, the feature controller 
is more flexible, allowing to plug the fast NIC on demand, such that the corresponding feature 
combinations have similar expected energy consumptions. 
\begin{figure}[h]%
\centering\includegraphics[width=\linewidth]{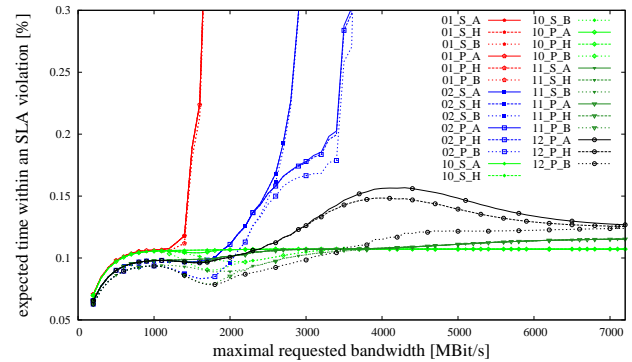}%
\nocaptionrule\caption{Evaluation of $\Phi_s$ for the different \textsc{eBond+} variants}%
\label{fig:smin_time}%
\end{figure}%
\paragraph{SLA violation analysis.}
When minimizing the expected number of SLA violations, i.e., solving query $\Phi_s$ for $\M_+$, 
similar phenomena as within our utility analysis can be observed. In Figure \ref{fig:smin_time} 
it can be seen that when choosing initial configurations with slow NICs, the expected percentage 
of time within an SLA violation rises significantly when the maximal required bandwidth exceeds 
the supported bandwidth of the activated NICs.
When choosing an appropriate initial feature combination, the minimal expected time run with 
SLA violations is between 0.06\% and 0.11\%, which is in the range of the values from the \textsc{eBond} 
case study \cite{HahnelDVH13}. Note that as in \textsc{eBond} case study, the balanced energy saving algorithm minimizes 
SLA violations always best, followed by the high savings and aggressive energy saving algorithms.

\begin{figure}[h]%
\centering\includegraphics[width=\linewidth]{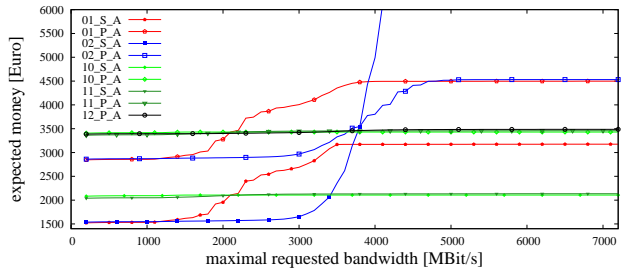}%
\nocaptionrule\caption{Evaluation of $\Phi_m$ for the different \textsc{eBond+} variants}%
\label{fig:mmin_time}%
\end{figure}

\paragraph{Monetary Analysis.}
A novel aspect not considered in the case study by \cite{HahnelDVH13} is the expected run-time costs 
in terms of money. Figure \ref{fig:mmin_time} shows the results of evaluating query $\Phi_m$ for $\M_+$ 
minimizing the expected monetary costs for all initial feature combinations.
As one expects, choosing a system with a fast 10 GBit NIC does not yield to additional costs after the purchase, 
since SLA violations are unlikely (see utility analysis with evaluating $\Phi_p$). However, when purchasing only 
slow cards, increasing the maximal required bandwidth leads to additional costs for SLA violation which may even 
supersede system configurations with higher initial costs. Thus, the customer may purchase a better performing 
but more expensive system if the maximal required bandwidth is high. However, if the maximal required bandwidth 
is below 2 GBit/s, it is always advisable to purchase the standard bundle with only one 1 GBit NIC, eventually 
plugging an additional 1 GBit NIC.

\paragraph{Statistical Evaluation.} We analyzed the above queries on an Intel Xeon X5650 @ 2.67 GHz using 
$\prism$ 4.1 and employing the sparse engine with a precision of $10^{-5}$. 
It is well-known that an explicit engine is usually faster than a symbolic
one when many different probability values appear in the model. Due to the
dynamic changes of bandwidth probabilities in the environment feature, 
this is also the case for our model.
Hence, symbolic approaches are 
only used for the construction of the model and reachability 
analysis, which however have great impact on the instance of the 
strategy synthesis problem we considered in our 
case study. Due to the family-based symbolic representation, the 
complete model is small compared to the accumulated 
model size when constructing models for all initial feature
combinations one-by-one. The model size also influences the time spent
for the evaluation of the queries, as well as its maximal memory consumption.
\begin{figure}[h]%
\centering\includegraphics[width=\linewidth]{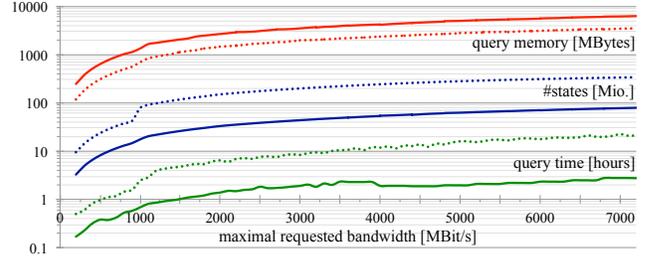}%
\nocaptionrule\caption{Statistical evaluation of the experiments}%
\label{fig:stats}%
\end{figure}%
The logarithmically scaled Figure \ref{fig:stats}
shows a comparison of these characteristics, where solid curves stand
for our family-based approach and the dashed ones for the one-by-one approach.
In Table \ref{tab:stats}, these characteristics are exemplified with 
a fixed maximal bandwidth of 2.4 GBit/s. The entire computation for bandwidth constraints of 0.2 till 7.2 GBit/s in steps of 0.1 GBit/s took 123 hours of CPU time and consumed at most 6244 MBytes of memory using our approach, whereas the one-by-one approach took more than 782 hours with a maximal memory consumption of 3482 MBytes. All these statistics illustrate that the family-based approach for our framework outperforms the one-by-one approach and is around six times faster. Note that in our case study, we only considered 27 different feature combinations. Due to the exponential blow-up in the number of feature combinations, an even greater speed-up can be expected for bigger SPLs.
\begin{table}[h]%
	\centering\begin{tabular}{|l|r|r|}%
		\hline & family-based ($\M_+$) & one-by-one\\\hline\hline%
		\#states&37 Mio.&427 Mio.\\\hline%
		query time& 1.3 hours & 6.9 hours\\\hline%
		query memory& 2970 MBytes&1680 MBytes\\\hline%
	\end{tabular}%
	\nocaptionrule\caption{Evaluation statistics (maximal bandwidth = 2.4 GBit/s)}%
	\label{tab:stats}%
\end{table}%
\section{Conclusions}

\label{sec:concl}
\label{sec:conclusions}

We presented a compositional modeling framework for dynamic SPLs 
that relies on dynamic adaptions
of the feature combinations expressed by means of an MDP-like model.
The feature implementations and the behavior
of possibly unknown or only partially known implementations of external
features are represented by separate automata with feature guards 
for the action-labeled transitions and special switch transitions
for the dynamic activation or deactivation of own features.
With the MDP-semantics of a dynamic SPLs, many feature-oriented
problems are reducible to well-known algorithmic problems for MDPs
and solvable with standard techniques.
We illustrated this by means of an energy-aware network protocol. 
In this case study, we used probabilistic model checking
for establishing several quantitative properties
and addressed the strategy synthesis problem
to generate an energy-efficient strategy for triggering feature
combination changes.

There are many other interesting variants of the task to synthesize
optimal strategies that are also solvable by known algorithms.
One might distinguish between switch events that are indeed
controllable and those that cannot be enforced or prevented, but are
triggered by the environment.
In this case, the MDP $\M$ can be seen as stochastic game-structure,
where the controller and the environment are opponents and the task to
generate an optimal strategy for the controller reduces to
well-known game-based problems 
\cite{Condon92,FilarVrieze97,deAlfMaj01,CJH04}.
Similarly, one might take into account that also the feature modules
can behave nondeterministically. Depending on the meaning of the
nondeterminism in the feature modules (e.g.,
implementation freedom or interactions with the environment),
the nondeterministic choices in the feature modules can be classified
into controllable and uncontrollable ones.
Assuming that the controller and all feature modules build one
coalition that aims to achieve some optimal value for a quantitative objective,
no matter how the environment behaves,
then again well-known algorithms for stochastic two-player games 
are applicable. This, and investigations on the scalability of
our approach towards real-case dynamic SPLs with more features are
left for further work.

~\\\textbf{Acknowledgements.} 
We thank Marcus Daum and Steffen M{\"a}rcker for their support 
concerning the case study.

\bibliographystyle{abbrv}

\bibliography{paper-short}

\begin{thebibliography}{10}

\bibitem{ApelHut10}
S.~Apel and D.~Hutchins.
\newblock A calculus for uniform feature composition.
\newblock {\em ACM Transactions on Programming Languages and Systems}, 32(5),
  2010.

\bibitem{AJTK09}
S.~Apel, F.~Janda, S.~Trujillo, and C.~K{\"a}stner.
\newblock Model superimposition in software product lines.
\newblock In {\em ICMT'09}, volume 5563 of {\em LNCS}, pages 4--19, 2009.

\bibitem{Apel2009}
S.~Apel, F.~Janda, S.~Trujillo, and C.~K\"{a}stner.
\newblock Model superimposition in software product lines.
\newblock In {\em ICMT'09}, pages 4--19, Berlin, Heidelberg, 2009.
  Springer-Verlag.

\bibitem{BaiKat2008}
C.~Baier and J.-P. Katoen.
\newblock {\em Principles of model checking}.
\newblock The MIT Press, 2008.

\bibitem{BaierKwi98}
C.~Baier and M.~Kwiatkoswka.
\newblock Model checking for a probabilistic branching time logic with
  fairness.
\newblock {\em Distributed Computing}, 11(3):125--155, 1998.

\bibitem{Benavides2010}
D.~Benavides, S.~Segura, and A.~Ruiz-Cort{\'e}s.
\newblock Automated analysis of feature models 20 years later: A literature
  review.
\newblock {\em Information Systems}, 35(6):615 -- 636, 2010.

\bibitem{BdAlf95}
A.~Bianco and L.~de~Alfaro.
\newblock Model checking of probabilistic and non-deterministic systems.
\newblock In {\em FSTTCS'95}, volume 1026 of {\em LNCS}, pages 499--513, 1995.

\bibitem{CJH04}
K.~Chatterjee, M.~Jurdzinski, and T.~Henzinger.
\newblock Quantitative simple stochastic parity games.
\newblock In {\em SODA'04}, pages 121--130. SIAM, 2004.

\bibitem{ClaEmeSis1986}
E.~M. Clarke, E.~A. Emerson, and A.~P. Sistla.
\newblock Automatic verification of finite-state concurrent systems using
  temporal logic specifications.
\newblock {\em ACM Trans. Program. Lang. Syst.}, 8:244--263, 1986.

\bibitem{Cla2011}
A.~Classen, P.~Heymans, P.-Y. Schobbens, and A.~Legay.
\newblock Symbolic model checking of software product lines.
\newblock In {\em ICSE'2011}, pages 321--330. ACM, 2011.

\bibitem{Cla2010}
A.~Classen, P.~Heymans, P.-Y. Schobbens, A.~Legay, and J.-F. Raskin.
\newblock Model checking lots of systems: Efficient verification of temporal
  properties in software product lines.
\newblock In {\em ICSE'2010}, pages 335--344. ACM, 2010.

\bibitem{CleNor2001}
P.~Clements and L.~Northrop.
\newblock {\em Software Product Lines : Practices and Patterns}.
\newblock Addison-Wesley Professional, 2001.

\bibitem{Condon92}
A.~Condon.
\newblock The complexity of stochastic games.
\newblock {\em Information and Computation}, 96(2):203--224, 1992.

\bibitem{CorClaHey2013}
M.~Cordy, A.~Classen, P.~Heymans, A.~Legay, and P.-Y. Schobbens.
\newblock {\em Model Checking Adaptive Software with Featured Transition
  Systems}, volume 7740, pages 1--29.
\newblock Springer Berlin Heidelberg, 2013.

\bibitem{CzaSheWas2008}
K.~Czarnecki, S.~She, and A.~Wasowski.
\newblock Sample spaces and feature models: There and back again.
\newblock In {\em SPLC'08}, pages 22--31, 2008.

\bibitem{DamSch2011}
F.~Damiani and I.~Schaefer.
\newblock Dynamic delta-oriented programming.
\newblock In {\em Proceedings of the 15th International Software Product Line
  Conference, Volume 2}, SPLC '11, pages 34:1--34:8, New York, NY, USA, 2011.
  ACM.

\bibitem{deAlfMaj01}
L.~de~Alfaro and R.~Majumdar.
\newblock Quantitative solution of omega-regular games.
\newblock In {\em STOC'01}, pages 675--683. ACM, 2001.

\bibitem{DinMitFetMez2010}
T.~Dinkelaker, R.~Mitschke, K.~Fetzer, and M.~Mezini.
\newblock A dynamic software product line approach using aspect models at
  runtime.
\newblock In {\em Proceedings of the 1st Workshop on Composition and
  Variability}, M{\"a}rz 2010.

\bibitem{FilarVrieze97}
J.~Filar and K.~Vrieze.
\newblock {\em Competitive Markov Decision Processes}.
\newblock Springer, 1997.

\bibitem{GelCar92}
D.~Gelernter and N.~Carriero.
\newblock Coordination languages and their significance.
\newblock {\em Communications of the ACM}, 35(2):96--107, 1992.

\bibitem{GheMol13}
C.~Ghezzi and A.~M. Sharifloo.
\newblock Model-based verification of quantitative non-functional properties
  for software product lines.
\newblock {\em Information {\&} Software Technology}, 55(3):508--524, 2013.

\bibitem{GomHus2003}
H.~Gomaa and M.~Hussein.
\newblock Dynamic software reconfiguration in software product families.
\newblock In {\em PFE}, pages 435--444, 2003.

\bibitem{HahnelDVH13}
M.~H\"{a}hnel, B.~D\"{o}bel, M.~V\"{o}lp, and H.~H\"{a}rtig.
\newblock ebond: Energy saving in heterogeneous r.a.i.n.
\newblock In {\em Proceedings of the Fourth International Conference on Future
  Energy Systems}, e-Energy '13, pages 193--202, New York, NY, USA, 2013. ACM.

\bibitem{HalHinParSch2008}
S.~Hallsteinsen, M.~Hinchey, S.~Park, and K.~Schmid.
\newblock Dynamic software product lines.
\newblock {\em Computer}, 41(4):93--95, Apr. 2008.

\bibitem{Haverkort}
B.~Haverkort.
\newblock {\em Performance of Computer Communication Systems: A Model-Based
  Approach}.
\newblock Wiley, 1998.

\bibitem{HayAtl2000}
J.~D. Hay and J.~M. Atlee.
\newblock Composing features and resolving interactions.
\newblock In {\em SIGSOFT'00}, pages 110--119, New York, NY, USA, 2000. ACM.

\bibitem{HinKwiNorPar2006}
A.~Hinton, M.~Kwiatkowska, G.~Norman, and D.~Parker.
\newblock {PRISM}: A tool for automatic verification of probabilistic systems.
\newblock In H.~Hermanns and J.~Palsberg, editors, {\em TACAS'06}, volume 3920
  of {\em LNCS}, pages 441--444. Springer, 2006.

\bibitem{Kang1990}
K.~C. Kang, S.~G. Cohen, J.~A. Hess, W.~E. Novak, and A.~S. Peterson.
\newblock Feature-oriented domain analysis (foda) feasibility study.
\newblock Technical report, Carnegie-Mellon University Software Engineering
  Institute, November 1990.

\bibitem{Kat1993}
S.~Katz.
\newblock A superimposition control construct for distributed systems.
\newblock {\em ACM Trans. Program. Lang. Syst.}, 15(2):337--356, Apr. 1993.

\bibitem{Kulkarni}
V.~Kulkarni.
\newblock {\em Modeling and Analysis of Stochastic Systems}.
\newblock Chapman \& Hall, 1995.

\bibitem{McM1993}
K.~L. McMillan.
\newblock {\em Symbolic Model Checking}.
\newblock Kluwer Academic Publishers, 1993.

\bibitem{MRKN13}
J.-V. Millo, S.~Ramesh, S.~N. Krishna, and G.~K. Narwane.
\newblock Compositional verification of software product lines.
\newblock In {\em IFM'13}, volume 7940 of {\em LNCS}, pages 109--123. Springer,
  2013.

\bibitem{NooBaDu12}
M.~Noorian, E.~Bagheri, and W.~Du.
\newblock Non-functional properties in software product lines: A taxonomy for
  classification.
\newblock In {\em SEKE'12}, pages 663--667. Knowledge Systems Institute
  Graduate School, 2012.

\bibitem{PapArb98}
G.~A. Papadopoulos and F.~Arbab.
\newblock Coordination models and languages.
\newblock {\em Advances in Computers}, 46:329--400, 1998.

\bibitem{PlaRya2001}
M.~Plath and M.~Ryan.
\newblock Feature integration using a feature construct.
\newblock {\em Science of Computer Programming}, 41(1):53 -- 84, 2001.

\bibitem{Puterman}
M.~Puterman.
\newblock {\em Markov Decision Processes: Discrete Stochastic Dynamic
  Programming}.
\newblock John Wiley \& Sons, Inc., New York, NY, 1994.

\bibitem{Ros2011}
M.~Rosenm\"{u}ller, N.~Siegmund, S.~Apel, and G.~Saake.
\newblock Flexible feature binding in software product lines.
\newblock {\em Automated Software Engg.}, 18(2):163--197, June 2011.

\bibitem{SLN01}
J.-G. Schneider, M.~Lumpe, and O.~Nierstrasz.
\newblock Agent coordination via scripting languages.
\newblock In {\em Coordination of Internet Agents: Models, Technologies, and
  Applications}, pages 153--175, 2001.

\bibitem{Segala95}
R.~Segala.
\newblock {\em Modeling and Verification of Randomized Distributed Real-Time
  Systems}.
\newblock PhD thesis, Mass\-a\-chu\-setts Insti\-tute of Tech\-no\-logy, 1995.

\bibitem{SegLynch94}
R.~Segala and N.~A. Lynch.
\newblock Probabilistic simulations for probabilistic processes.
\newblock {\em Nordic Journal of Computing}, 2(2):250--273, 1995.

\bibitem{SRKGAK13}
N.~Siegmund, M.~Rosenm{\"u}ller, C.~K{\"a}stner, P.~G. Giarrusso, S.~Apel, and
  S.~S. Kolesnikov.
\newblock Scalable prediction of non-functional properties in software product
  lines: Footprint and memory consumption.
\newblock {\em Information {\&} Software Technology}, 55(3):491--507, 2013.

\bibitem{SRKKS08}
N.~Siegmund, M.~Rosenm{\"u}ller, M.~Kuhlemann, C.~K{\"a}stner, and G.~Saake.
\newblock Measuring non-functional properties in software product line for
  product derivation.
\newblock In {\em APSEC'08}, pages 187--194. IEEE, 2008.

\bibitem{VarKho13}
M.~Varshosaz and R.~Khosravi.
\newblock Discrete time {M}arkov chain families: modeling and verification of
  probabilistic software product lines.
\newblock In {\em SPLC'13}, pages 34--41. ACM, 2013.

\bibitem{Pla2013}
A.~von Rhein, S.~Apel, C.~K\"{a}stner, T.~Th\"{u}m, and I.~Schaefer.
\newblock The {PLA} model: On the combination of product-line analyses.
\newblock In {\em Proceedings of the Seventh International Workshop on
  Variability Modelling of Software-intensive Systems}, VaMoS '13, pages
  14:1--14:8, New York, NY, USA, 2013. ACM.

\bibitem{WhiteDSB2009}
J.~White, B.~Dougherty, D.~C. Schmidt, and D.~Benavides.
\newblock Automated reasoning for multi-step feature model configuration
  problems.
\newblock In {\em SPLC'09}, pages 11--20, 2009.

\end{thebibliography}

\clearpage
\section{Appendix}

In Section \ref{sec:framework}, we presented our compositional framework towards feature modules and its MDP semantics under feature controllers in a lightweight fashion, detailing parallel operators which synchronize over common actions. Guarded-command languages as the input language of $\prism$ support further variables over which modules can communicate. It is well-known that variables do not add further expressivity to the model. However, to illustrate the connection to guarded-command languages, we discuss our framework refined supporting variables more in detail.

~\\\textbf{Variables and Valuations.}
Let use suppose that $\Var$ is a finite set of typed variables,
where the types are assumed to be finite as well (e.g., Boolean variables
or integers with some fixed number of digits).
We denote furthermore by $\Val$ the set of valuation functions 
for the variables, i.e.,  type-consistent 
mappings that assign to each variable $x \in \Var$ a value.
In analogy to the symbolic representation sets by Boolean expressions,
we can represent subsets of $\Val$ by Boolean expressions, where
the atoms are assertions on the values of the variables. 
Let $\BB(\Var)$ denote the set of these Boolean expressions.
For example, if $x$ and $y$ are variables with domain $\{0,1,2,3\}$ and
$z$ a variable with domain 
$\{\mathrm{red},\mathrm{green},\mathrm{blue}\}$,
then the Boolean expression 
$\phi = (x < y) \wedge (y > 2) \wedge (z \not= \mathrm{green})$
represents all valuations $\val\in \Val$ with $\val(x) < \val(y)=3$
and either $\val(z)\in \{\mathrm{red},\mathrm{blue}\}$.

~\\\textbf{Interface.}
The interface of a feature module $\Module$ 
now consists of a feature interface
$\Feat = \<\OwnFeat,\ExtFeat\>$ as in Def.~\ref{def:featinterface}
and a declaration which variables from $\Var$ are local 
and which one are external. 
The local variables can appear in guards and can be modified by $\Module$,
while the external variables can only appear in guards, 
but cannot be written by $\Module$. Instead, the external variables
of $\Module$ are supposed to be local for some other module.
We denote these sets by $\LocVar$ and
$\ExtVar$, write $\VarModule$ for $\LocVar \cup \ExtVar$
and extend the notion of composability of two feature
modules by the (natural) requirement that there are 
no shared local variables.

~\\\textbf{Locations and Initial Condition.}
One can think of the variable valuations for the local variables
to serve as locations in the module $\Module$.
However, there is no need for an explicit reference to locations
since all transitions will be described symbolically (see below).
Instead of initial locations, we deal with an initial condition
for the local variables.

~\\\textbf{Updates.}
Transitions in $\Module$ might update the values of the local variables.
The updates are given by sequences of assignments
$x_1 :=\expr_1; \ldots ; x_n := \expr_n$,
where $x_1,\ldots,x_n$ are pairwise distinct variables in
$\LocVar$ and $\expr_i$ are type-consistent expressions that might refer
to all variables in $\VarModule = \LocVar \cup \ExtVar$.
The precise syntax of expressions is irrelevant here. 
Instead, we formalize the effect of the updates that might appear in
$\Module$ 
by functions $\upd : \Val \to \Val$ with
$\upd(\val)(y) = \val(y)$ for all non-local variables, i.e., all
variables $y \in \Var \setminus \LocVar$.

~\\\textbf{Symbolic Transitions.}
Instead of explicit references to the
variable valuations in the transitions, we use a symbolic approach
based on symbolic transitions. They represent sets of guarded 
transitions, possibly emanating from different locations,
of the following form:
$$
  \theta \ \ = \ \ 
  (\guard,\phi,*,c,\probupd),
$$
where $\guard \in \BB(\VarModule)$ is variable guard
imposing conditions on the local and external variables, 
and $\phi \in \BB(\Feat)$ is a feature guard as before.
The third and fourth component $*$ and $c$ are as in the data-abstract setting.
That is, $*$ stands for an action label $\alpha \in \Act$ 
or a Boolean expression $\rho \in \BB(\OwnFeat \cup \OwnFeat')$ for
the switch events, while $c \in \Nat$ stands for the cost caused by taking
transition $\theta$.
The last component $\probupd$ is a \emph{probabilistic update},
i.e., a distribution over updates for the variables in $\LocVar$
with finite support. These are written in the form
\begin{center}
   $p_1 : \upd_1$ + 
   $p_2 : \upd_2$ + 
   $\ldots$ +
   $p_k : \upd_k$,
\end{center}
where $p_i$ are positive real numbers with $p_1 + \ldots + p_k=1$
and the $\upd_i$'s are updates for the local variables.
That is, $p_i$ is the probability for update $\upd_i$.

~\\\textbf{MDP-semantics.}
In the data-abstract setting, a reasonable MDP-semantics of a feature
module $\Module$ under controller 
$\Controller = (\ValidFeat,\ValidFeatInit,\SwitchRel)$ has been defined,
no matter whether $\Module$ is just a fragment of the SPL and
interacts with other modules.
An analogous definition for the data-aware setting
can be provided either for modules without external variables
or by modelling the changes of the values of the external variables
by nondeterminism. 

Let us here consider the first case where we are given the ``final'' module
$\Module = \Module_1 \| \ldots \| \Module_n$ that arises through the 
parallel composition of several modules such that all variables
$x\in \Var$ are local for some module $\Module_i$.
Then, $\Module$ has no external variables and $\Var = \LocVar = \VarModule$.
Furthermore, $\OwnFeat$ is the set of all features of the given SPL 
for which implementations are given, 
while $\ExtFeat$ stands for the set of features
controlled by the environment.
The MDP $\Module \Join \Controller$ has the state space
$S = \Val \times \ValidFeat$.
The initial states are the pairs $\<\val,C\>$ where $\val$ satisfies
the initial variable condition of $\Module$ and $C\in \ValidFeatInit$.
The moves in $\Module \Join \Controller$ 
arise through rules
that are analogous to the rules shown in Figure \ref{fig:semantics}.
More precisely, $\Moves$ is the smallest set of moves that arise
through the following three cases, where $s = \<\val,C\>$ is an arbitrary
state in $\Module \Join \Controller$:
\begin{itemize}
\item
    An action-labeled transition 
    $(\guard, \phi, \alpha, c, \probupd)$
    in $\Module$ is enabled in state $s$ if $\valcond$ and $\guard$ are
    satisfied in $s$. 
    This means $C \models \phi$ and 
    $\val \models \guard$.
    Assuming $\upd_i(\val) \not= \upd_j(\val)$ for $i \not= j$, we have:
    \begin{center}
       $(\<\val,C\>,c, \lambda * \Dirac[C]) \in \Moves$,
    \end{center}
    where $\lambda\bigl(\, \upd_i(\val)\, \bigr)=p_i$ for $i=1,\ldots,k$ and
    $\lambda(\val')=0$ for all other valuation functions $\val'$.
   
\item
    If $C \overto{d} \gamma$ is a switch transition in $\Controller$
    that does affect at most the features of the environment,
    i.e., $(C \ominus C') \cap \OwnFeat=\varnothing$ 
    for all $C'\in \Support(\gamma)$, then:
    \begin{center}
      $(\<\val,C\>,d,\Dirac[\ell]* \gamma) \in \Moves$
    \end{center}
\item
    Suppose now that 
    $(\guard,\phi,\rho,c,\probupd)$
    is a switch transition in $\Module$ and
    $\guard$  holds in state $s$.
    Again, $\rho \in \BB(\OwnFeat \cup \OwnFeat')$ is viewed as 
    a Boolean expression over $\FeatureSet \cup \FeatureSet'$ 
    and
    specifies a binary relation $R_{\rho}$ over 
    $2^{\FeatureSet} \times 2^{\FeatureSet}$.
    If $(C,C') \in R_{\rho}$ for all $C'\in \Support(\gamma)$ then:
    \begin{center}
         $(\<\val,C\>,c+d,\lambda * \gamma)\in \Moves$,
    \end{center}
    where $\lambda$ is defined as in the first (action-labeled) case.
\end{itemize}

~\\\textbf{Parallel Composition.}
The extension of the parallel operator $\|$ for composable
feature modules with variables is rather tedious, but straightforward.
As stated above, composability requires that there no common
own features and no common local variables.
The local variables of the composite module $\Module_1 \| \Module_2$
are the variables that are local for one module $\Module_i$.
The feature interface of $\Module_1 \| \Module_2$ is defined as in the
data-abstract setting.
The initial variable condition of $\Module_1 \| \Module_2$ 
arises by the conjunction of
the initial conditions for $\Module_1$ and $\Module_2$.
Let us turn now to the transitions in $\Module_1 \| \Module_2$.
\begin{enumerate}
\item [1.]
  All action-labeled symbolic 
  transitions in $\Module_1$ or $\Module_2$ with some 
  non-shared action $\alpha \in \Act_1 \ominus \Act_2$
  are also transitions in $\Module_1 \| \Module_2$.
\item [2.]
  Given action-labeled symbolic 
  transitions in $\Module_1$ and $\Module_2$
  with the same action 
  $\alpha \in \Act_1 \cap \Act_2$
  \begin{center}
      \begin{tabular}{l@{\hspace*{0.2cm}}c@{\hspace*{0.2cm}}l}
         $\theta_1$ & = &
         $(\guard_1,\phi_1,\alpha,c_1,\probupd_1)$
         \\[1ex]

         $\theta_2$ & = &
         $(\guard_2,\phi_2,\alpha,c_2,\probupd_2)$
      \end{tabular}
  \end{center}
  are combined into a symbolic transition of $\Module_1 \| \Module_2$:
  \begin{center}
      \begin{tabular}{l@{\hspace*{0.2cm}}c@{\hspace*{0.2cm}}l}
         $\theta_1 \| \theta_2$ & = &
         $(\guard,\phi,\alpha,c_1+c_2,\probupd)$,
      \end{tabular}
  \end{center}
  where $\guard = \guard_1 \wedge \guard_2$, $\phi = \phi_1 \wedge \phi_2$ 
  and $\probupd$ combines the probabilistic update functions
  $\probupd_1$ and $\probupd_2$.
  That is, if $\upd_i$ has probability $p_i$ under distribution $\probupd_i$ 
  for $i=1,2$, then the combined update 
  that performs the assignments in $\upd_1$ and $\upd_2$ simultaneously
  has probability $p_1 \cdot p_2$ under $\probupd$.
\item [4.]
  The adaption of the rules for switch transitions in
  $\Module_1 \| \Module_2$ is analogous and omitted here.
\end{enumerate}

\end{document}